\documentclass[aps,prf,preprint,groupedaddress]{revtex4-2}
\usepackage[margin=1in]{geometry}
\usepackage{lipsum}
\usepackage{amsmath,accents}
\usepackage[labelfont={small}, font={normal}]{subfig}
\usepackage{graphicx}
\usepackage{epstopdf}
\usepackage{amssymb}
\usepackage{lineno}
\usepackage{enumitem}
\usepackage[utf8]{inputenc}
\usepackage[dvipsnames]{xcolor}
\graphicspath{{./Figures/}}

\begin{document}
\title{Steady state extensional rheology of a dilute suspension of spheres in a dilute polymer solution}
\author{Arjun Sharma}
\affiliation{Sibley School of Mechanical and Aerospace Engineering, Cornell University, Ithaca, NY, 14853, USA}
\author{Donald L. Koch}
\affiliation{Robert Frederick Smith School of Chemical and Biomolecular Engineering, Cornell University, Ithaca, NY, 14853, USA}	
\date{\today}
\begin{abstract}
We investigate the steady-state extensional rheology of a dilute suspension of spherical particles in a dilute polymer solution modeled by the FENE-P constitutive relation. The ensemble-averaged suspension stress uses a recent construct of Koch et al. [Physical Review Fluids, \textbf{1}, 013301 (2016)], based on perturbation for small polymer concentration and the generalized reciprocal theorem, to determine the polymers' influence on the particle stresslet and the particles' influence on the polymer stress. The extensional viscosity is defined as half of the constant of proportionality between the deviatoric stress and the imposed rate of strain tensor in uniaxial extensional flow. For a particle-free polymeric fluid, the extensional viscosity (non-dimensionalized by the solvent viscosity) is 1+$\mu^\text{poly}$, where $\mu^\text{poly}$ is the polymer contribution to the extensional viscosity. When a small volume fraction, $\phi$, of spheres is added to a polymeric fluid, we find that the stress is altered by the Einstein viscosity of 2.5$\phi$ and two additional stress contributions: the polymer influence on the stresslet and the particle-induced polymer stress (PIPS). At lower Deborah numbers (defined as the product of extension rate and polymer relaxation time), $De\lesssim0.5$, the net interaction stress is positive, while it becomes negative at large $De$. Relative to undisturbed flow, the presence of spheres in uniaxial extensional flow creates regions of both larger and smaller local stretching. Below the coil-stretch transition, the polymers far from the particles are in a coiled state while they are stretched more than their undisturbed state by large stretching regions around the particle. Due to their finite relaxation time, they also form a wake of stretched polymers downstream of the particle. This leads to a positive contribution to the suspension stress from both the stresslet (surface) and the PIPS (stretched wake). Beyond the coil-stretch transition, polymers far from the particle are highly stretched but they collapse closer to the coiled state as they arrive at the low stretching regions near the particle surface. Therefore, a negative PIPS results from the regions of collapsed polymers. At sufficiently high Deborah numbers, $De\gtrsim 1.5$, this region is very thin and it becomes thinner and more intense upon further increasing $De$. For large maximum polymer extensibility, $L$, the particle-polymer contribution to the suspension rheology is independent of $L$ below the coil-stretch transition, whereas it scales as $L^2$ above the coil-stretch transition. When $De\gtrsim0.6$, the changes in extensional viscosity from the stresslet and the PIPS are $\phi\mu^\text{poly}$ and approximately -1.85$\phi\mu^\text{poly}$, respectively. At large $De$, the polymer extensional viscosity, $\mu^\text{poly}$, is orders of magnitude larger than that of Newtonian solvent. Hence, adding particles reduces the extensional viscosity of the suspension as $(2.5-0.85\mu^\text{poly})\phi<0$.
\end{abstract}
\maketitle
\section{Introduction}\label{sec:Introduction}

Particle-filled viscoelastic polymeric fluids are used in a variety of processes and products such as drug delivery \cite{deng2008understanding}, paints, extrusion molding, fiber spinning \cite{breitenbach2002melt,huang2003review,nakajima1994advanced} and hydraulic fracturing \cite{barbati2016complex}. As the fluid enters a narrow pore during hydraulic fracturing or is pulled by the drawing mechanism as it leaves the dye in fiber spinning, it undergoes a strong uniaxial extensional flow. Solid particles are added to the polymeric fluid for various reasons. In hydraulic fracturing they act as proppants during the fracturing of the rock, and in fiber spinning they impart further strength to the fiber. Therefore, understanding the extensional rheology of suspensions of particles in polymeric liquids is industrially relevant. It also leads to new and interesting physical mechanisms. An understanding of these mechanisms can allow one to not only tune the operating conditions of an industrial process, but also to design new fluids for specific applications \cite{ewoldt2022designing}.

Due to recent mathematical and computational advances, the study of suspension rheology of polymeric fluids is an active area of research \cite{shaqfeh2019rheology}. Shear and extensional flow are some of the most basic and predominant local flows in applications and lab experiments \cite{bird2016polymer}. While shear rheology of particle suspensions in viscoelastic fluids has been studied extensively, less attention has been paid to their extensional rheology \cite{shaqfeh2019rheology} despite the industrial importance mentioned above. Computational and theoretical predictions \cite{de1974coil,perkins1997single} as well as experiments \cite{anna2008effect} reveal that dilute polymeric solutions undergo strain hardening as the addition of polymers to the solvent increases the extensional viscosity by orders of magnitude. In this paper we study the effect of particles on extensional rheology of dilute polymeric liquids (low polymer concentration) as the addition of particles is expected to have a large influence on the suspension even at modest volume fractions \cite{scirocco2005shear,dai2014viscometric,koch2006stress,KOCH2008202,koch2016stress,yang2016numerical,yang2018mechanism,einarsson2018einstein,jain2019extensional}. At very small extension rates an increase in extensional stress upon addition of a small volume fraction of particles to polymeric liquids is found \cite{koch2006stress,KOCH2008202,einarsson2018einstein} (see \cite{shaqfeh2019rheology,koch2016stress} for a review on these asymptotic expansions in extension rate). In a recent numerical study, Jain et al. \cite{jain2019extensional} investigate the transient rheology of a dilute suspension of spheres in a concentrated polymeric solution at moderate extension rates. In their study, a non-monotonic effect of particle-polymer interaction is observed at the highest extension rates explored as the imposed Hencky strain grows, but a steady-state is not achieved in all but the smallest extension rates. Our semi-analytical study at low polymer concentration aims to reveal the steady state extensional rheology for a much wider range of extension rates,
 while thoroughly investigating the underlying physical mechanisms.

The particle surface traction and the change in the polymer stress (compared to the particle-free case) in the surrounding fluid leads to the stresslet and particle-induced polymer stress (PIPS) as two additional stresses in particle suspensions of polymeric/ viscoelastic fluids, together termed as the particle-polymer interaction stress. The stress obtained from bulk rheological measurements of a suspension of homogeneously distributed particles is the stress ensemble-averaged over particle configurations \cite{koch2016stress, hinch1977averaged}. In dilute particle suspensions, this ensemble average can be expressed in terms of the fields near an isolated particle \cite{koch2016stress}. Therefore, an investigation of the flow around an isolated particle in a polymeric fluid facilitates characterization of the impact of particle-polymer interaction on the rheology of dilute particle suspensions. In our study, we consider particles to be large enough that the Brownian effects are negligible and the polymers see the polymer solution as a continuum. See \cite{koch2016stress} for a brief discussion about how Brownian motion of the particles and polymers, and the finite size of polymers may affect the particle-polymer interaction.

We will use the  semi-analytical methodology that Koch et al. \cite{koch2016stress} formulated and applied to investigate the shear rheology of a dilute suspension of spheres in a dilute polymeric liquid.  This approach  employs  the method of ensemble averaged equations \cite{hinch1977averaged}, a regular perturbation for small polymer concentration, $c$, and a generalized reciprocal theorem \cite{kim2013microhydrodynamics}. Ensemble averaging has been used to correctly model the non-Newtonian effects in second order fluids \cite{koch2006stress,rallison2012stress}, third order fluids \cite{einarsson2018einstein} and a fully computational study of dilute suspensions of spheres in shear flow of viscoelastic fluids \cite{yang2018mechanism}. In contrast, volume averaging methods \cite{greco2007rheology,housiadas2009rheology} lead to non-convergent integrals.

In a low $c$ polymeric fluid, the leading order velocity around the particle (i.e. the Newtonian velocity field) affects the polymer configuration at leading order. The divergence of this  polymer configuration forces the fluid velocity and pressure disturbance in the $\mathcal{O}(c)$ momentum equation. The $\mathcal{O}(c)$ stresslet depends on the fluid velocity and pressure disturbance at $\mathcal{O}(c)$, that are governed by the $\mathcal{O}(c)$ momentum and mass conservation equation. These $\mathcal{O}(c)$ partial differential equations should be discretized using the techniques of traditional computational fluid dynamics such as finite difference, finite volume or finite element methods if the $\mathcal{O}(c)$ fluid velocity and pressure disturbance are required. This will be computationally expensive and would restrict the parameter regime that one can explore. As demonstrated by Koch et al. \cite{koch2016stress}, a generalized reciprocal theorem allows the $\mathcal{O}(c)$ stresslet to be obtained directly from the leading order polymer configuration, thus circumventing the numerical evaluation of the $\mathcal{O}(c)$ momentum and mass conservation equations. Furthermore, in the semi-analytical technique, using the method of characteristics, the coupled partial differential equations representing polymer constitutive equations are converted into coupled ordinary differential equations  with the streamlines of the leading order Newtonian velocity field acting as the characteristics.

We find the low $c$ assumption allows us to extract novel physics of particle-polymer interaction and our calculations of the polymer stress field capture all the major qualitative aspects seen in the previous numerical study of Jain et al. \cite{jain2019extensional} conducted at moderate $c$. They used a finite volume method to numerically integrate the governing partial difference equations and volume averaging to obtain the rheology of a dilute suspension. The limited size of the computational domain and small Hencky strain allowed them to obtain a finite value of the particle-polymer interaction stress from inappropriate volume averaging. In our semi-analytical method we calculate the characteristics/ streamlines of Newtonian velocity field around a sphere once. A straightforward numerical integration of the coupled ordinary differential equations representing the polymer constitutive equation on these predetermined characteristics around a sphere allows us to span a wide range of imposed extension rates, polymer relaxation times and maximum polymer extensions. Numerical instability issues such as the so-called high Weissenberg number problem \cite{fattal2004constitutive} do not arise. The spatial resolution depends on the spacing of these characteristics that are inexpensive to obtain. Hence we are able to obtain the polymer stress field at a much greater resolution and a fraction of the computational cost as compared to traditional computational fluid dynamics techniques such as that used in \cite{jain2019extensional}. Furthermore, we obtain several analytical expressions related to the contribution of particle-polymer interactiond to the suspension rheology.

 The study of polymeric fluids is an active area of research even without the particles. Various constitutive equations have been developed over the years that aim to model their dynamics. Many of these models faithfully mimic the polymer behavior in weak flows i.e. when the largest eigenvalue of the velocity gradient tensor is smaller than the relaxation rate of the polymer \cite{hinch1977mechanical}. In the special case of simple shear flow, this happens at all shear rates as the vorticity rotates the polymer away from the principle strain axis before it can fully stretch  \cite{hinch1977mechanical}. In that case, modeling the polymer as a Hookean spring with Brownian beads attached to its ends leads to the widely used, simple, Oldroyd B model, which matches well with the experimental observations of Boger fluids \cite{chai1988modelling}. However, in a strong flow such as uniaxial extension, beyond a critical extension rate, the Hookean dumbbell of the Oldroyd-B model stretches indefinitely. A simple means of removing this unphysical feature is to replace the Hookean spring with a non-linear spring with a finite maximum extensibility \cite{rallison1988we}, leading to the finitely extensible nonlinear elastic (FENE) model.

Similar to the Oldroyd-B constitutive equation, the FENE model involves representing the stress in terms of the configuration tensor, $\mathbf{q}\mathbf{q}$, where $\mathbf{q}$ is the end-to-end  vector of the dumbbell. To obtain a deterministic equation, $\mathbf{q}\mathbf{q}$ is averaged over all possible polymer orientations. This leads to a closure problem due to the terms arising from the non-linear spring force. One of the most successful methods to obtain closure is by using the averaged configuration to model the spring force \cite{bird2016polymer}, i.e., $\langle L^2/(L^2-\text{tr}(\mathbf{q}\mathbf{q}))\rangle_\text{polymer orientation}$ is approximated as $ L^2/(L^2-\langle\text{tr}(\mathbf{q}\mathbf{q})\rangle_\text{polymer orientation})=L^2/(L^2-\langle ||\mathbf{q}||_2^2\rangle_\text{polymer orientation})$, where $L$ is the maximum polymer extensibility and the angle brackets represent an average over polymer orientations. This is known as the Peterlin closure \cite{peterlin1968non}, whose implementation yields the FENE-P model. Steady-state uniaxial extension is modeled well by the FENE-P model \cite{keunings1997peterlin}. During the transient phase of the uniaxial extension when the polymer stretch increases from its equilibrium state, the FENE-P constitutive equation is known to over-predict the extension as compared to Brownian simulations of the FENE model \cite{keunings1997peterlin,herrchen1997detailed,van1998selection}. Thus the steady-state for the FENE-P model is achieved faster than that for the FENE model, as the mean-squared extension obtained during the transient phase in the former is larger \cite{keunings1997peterlin}. Additionally, for an extensional flow followed by relaxation, the FENE-P model does not predict the hysteresis observed in the FENE model \cite{lielens1999fene}. Several closure models have been proposed over the years which better predict certain time-varying properties. These either involve higher-order moments and hence extra equations for modeling the polymer stress \cite{herrchen1997detailed,lielens1999fene}, or, do not predict the correct steady-state stress for high extension rates \cite{van1998selection}. Closure modeling of the FENE dumbbell model is a separate research avenue. For simplicity, we will consider the FENE-P model in the rest of this study. It is a continuum constitutive equation derived from a simple molecular-level model \cite{bird2016polymer}. It matches the qualitative trends observed in extensional rheology experiments without the particles, and since it is a dumbbell model it allows an interpretation of polymer stress in terms of polymer stretch \cite{anna2008effect,anna2000interlaboratory}.

In suspensions with low particle volume fraction, $\phi$, where the effect of particle-particle interactions is negligible, capturing the interaction of polymers with a single particle reveals the suspension rheology at $\mathcal{O}(\phi)$ through the ensemble averaging method mentioned earlier. Therefore, we begin our investigation at the particle level, before studying the suspension rheology. Section \ref{sec:Formulation} introduces the governing equations in the fluid and the regular perturbation expansion in the polymer concentration, $c$. We find the particle-polymer interaction is qualitatively dependent upon the state of the polymers far from the particle. Hence, we review the polymer stress in a uniaxial extensional flow without the particles, as predicted by the FENE-P model in section \ref{sec:UndisturbedConfiguration}. The leading order velocity field, i.e., the Newtonian velocity field around a sphere, drives the polymer configuration at the leading order in a low $c$ polymer solution. Therefore, in order to build a basis for understanding the changes in polymer configuration, presented in section \ref{sec:PolymerConfiguration}, due to the presence of a spherical particle, we describe the kinematics of the velocity field in section \ref{sec:Kinematics}. In section \ref{sec:Rheology}, we describe the formulation for ensemble averaging and present the suspension rheology results. Finally, we summarize the conclusions in section \ref{sec:Conclusion}, where we also discuss the benefits and drawbacks of the FENE-P model pertaining to our findings.

\section{Governing equations}\label{sec:Formulation}
The equations governing mass and momentum conservation throughout the viscoelastic suspension of spheres in the inertia-less (zero Reynolds number) limit are,
\begin{eqnarray}
\nabla\cdot \mathbf{u}=0, \hspace{0.2in}
\nabla\cdot \boldsymbol{\sigma}=0,
\end{eqnarray}
where $\mathbf{u}$ and $\boldsymbol{\sigma}$ are the velocity vector and the stress tensor fields. In the fluid region the stress at any location is the sum of Newtonian solvent, $\boldsymbol{\tau}$, and polymer, $\mathbf{\Pi}$, stress,
\begin{equation}\label{eq:constitutive1}
\boldsymbol{\sigma}=\boldsymbol{\tau}+\mathbf{\Pi}=-p\boldsymbol{\delta}+2\boldsymbol{e}+\boldsymbol{\Pi},
\end{equation}
where $p$ is the hydrodynamic pressure and $\boldsymbol{e}= (\nabla\mathbf{u}+(\nabla\mathbf{u})^\text{T} )/2$ is the strain rate tensor at that location. For a polymer with concentration, $c$, and maximum extensibility, $L$, the polymer stress, $\boldsymbol{\Pi}$, and configuration, $\boldsymbol{\Lambda}$ ($=\langle\mathbf{q}\mathbf{q}\rangle_\text{polymer orientation}$ from section \ref{sec:Introduction}), tensors are modeled with the FENE-P relations,
\begin{eqnarray}
\boldsymbol{\Pi}=\frac{c}{De}(f\boldsymbol{\Lambda}-b\boldsymbol{\delta}), \hspace{0.2in} f=\frac{L^2}{L^2-\text{tr}(\boldsymbol{\Lambda})}, \hspace{0.2in} b=\frac{L^2}{L^2-\text{tr}(\boldsymbol{\delta})},\label{eq:constitutive2}\\
\frac{\partial \boldsymbol{\Lambda}}{\partial t}+\mathbf{u}\cdot \nabla \boldsymbol{\Lambda}=\nabla \mathbf{u}^\text{T}\cdot\boldsymbol{\Lambda}+\boldsymbol{\Lambda}\cdot\nabla\mathbf{u}+\frac{1}{De}(b\boldsymbol{\delta}-f\boldsymbol{\Lambda}),\label{eq:Configuration}
\end{eqnarray}
where $De$ is the Deborah number of the imposed flow with extension rate, $\dot{\epsilon}$, and polymer relaxation time, $\lambda$,
\begin{equation}
De={\lambda}\dot{\epsilon}.\label{eq:DeEquation}
\end{equation}
In the FENE-P constitutive relation and other dumbell models such as Oldroyd-B, FENE-CR and Giesekus \cite{bird2016polymer} $\sqrt{\text{tr}(\boldsymbol{\Lambda})}$ represents the mean-squared polymer stretch.
In the rest of the paper, steady state is assumed,
\begin{equation}\label{eq:steadystate}
\frac{\partial \boldsymbol{\Lambda}}{\partial t}=0.
\end{equation}
We expand the stress, pressure, velocity and polymer configuration using a regular perturbation in the polymer concentration: $\boldsymbol{\sigma}=\boldsymbol{\sigma}^{(0)}+c\boldsymbol{\sigma}^{(1)}+\mathcal{O}(c^2)$, $\boldsymbol{\tau}=\boldsymbol{\tau}^{(0)}+c\boldsymbol{\tau}^{(1)}+\mathcal{O}(c^2)$, ${p}={p}^{(0)}+c{p}^{(1)}+\mathcal{O}(c^2)$, $\boldsymbol{u}=\boldsymbol{u}^{(0)}+c\boldsymbol{u}^{(1)}+\mathcal{O}(c^2)$, and $\boldsymbol{\Lambda}=\boldsymbol{\Lambda}^{(0)}+c\boldsymbol{\Lambda}^{(1)}+\mathcal{O}(c^2)$. Since the polymer stress , $\boldsymbol{\Pi}$, is pre-multiplied with $c$ in equation \eqref{eq:constitutive2}, the leading order fluid velocity and pressure fields satisfy the Newtonian equations of motion. In a dilute suspension of spheres, to get the stress up to $\mathcal{O}(c)$ we only need to compute the flow around an isolated sphere (ref. \cite{koch2016stress} and section \ref{sec:Rheology}). The leading order velocity and pressure around a force- and torque-free unit sphere in an imposed extensional flow (fluid velocity approaching extensional flow at large distances from the particle) is
\begin{eqnarray}
&u_i^{(0)}=\begin{cases}
\text{E}_{ij}r_j+\frac{5}{2}\Big(\frac{1}{r^7}-\frac{1}{r^5}\Big)\text{E}_{jk}r_jr_kr_i-\frac{1}{r^5}\text{E}_{ji}r_j,& r\ge 1,\\
0,& r<1,
\end{cases}\label{eq:u_field}\\
&p^{(0)}=-\frac{5}{r^5}\text{E}_{jk}r_jr_k , \hspace{2.2in}r\ge 1,
\end{eqnarray}
where
\begin{equation}\label{eq:UndisturbedStrainRate}
\text{E}_{ij}=\delta_{i1}\delta_{j1}-\frac{1}{2}(\delta_{i2}\delta_{j2}+\delta_{i3}\delta_{j3}).
\end{equation}
Using this velocity field we solve equation \eqref{eq:Configuration} for the leading order configuration,  $\boldsymbol{\Lambda}^{(0)}$, and use equation \eqref{eq:constitutive2} to obtain the polymer stress, $\boldsymbol{\Pi}$, up to $\mathcal{O}(c)$, \begin{equation}
\boldsymbol{\Pi}=c\boldsymbol{\Pi}^{(0)}+\mathcal{O}(c^2),
\end{equation}
where,
\begin{equation}
\boldsymbol{\Pi}^{(0)}=\frac{1}{De}(f^{(0)}\boldsymbol{\Lambda}^{(0)}-b\boldsymbol{\delta}),\hspace{0.2in}  f^{(0)}=\frac{L^2}{L^2-\text{tr}(\boldsymbol{\Lambda}^{(0)})}.
\end{equation}
Similar to Koch et al. \cite{koch2016stress}, we do so numerically using the method of characteristics, where the characteristic curves are the streamlines of the steady-state velocity given by equation \eqref{eq:u_field}. For the FENE-P equations, the calculation of polymer configuration in an extensional flow with constant strain rate ($u_i=\text{E}_{ij}r_j$) is not trivial and we consider this next. Besides fully characterizing the stress in the particle-less viscoelastic fluid up to $\mathcal{O}(c)$, this configuration is the initial condition for the aforementioned method of characteristics.
\section{ Polymer configuration and stress without the particles}\label{sec:UndisturbedConfiguration}
In this paper we consider a homogeneous suspension of dilute particle concentration. Therefore, the particle-particle interactions are negligible and each particle effectively experiences a region of infinite expanse of polymeric fluid around itself before it observes the presence of another particle. Therefore, relative to each particle, a far-field or undisturbed flow region exists at large distances from the particle, where the flow approaches the one without the particles. For uniaxial extension it is $u_i=\text{E}_{ij}r_j$, with $\text{E}_{ij}$ given in equation \eqref{eq:UndisturbedStrainRate}. A homogeneous, steady-state polymer stress, $\boldsymbol{\Pi}^{(0U)}$, due to this flow is governed by,
\begin{equation}\label{eq:ConfigurationUndisturbed_simple}
\begin{split}
&\mathbf{E}\cdot\boldsymbol{\Lambda}^{(0U)}+\boldsymbol{\Lambda}^{(0U)}\cdot\mathbf{E}+\frac{1}{De}(b\boldsymbol{\delta}-f^{(0U)}\boldsymbol{\Lambda}^{(0U)})=0,\\
&f^{(0U)}=\frac{L^2}{L^2-\text{tr}(\boldsymbol{\Lambda}^{(0U)})},\hspace{0.01in} \boldsymbol{\Pi}^{(0U)}=\frac{1}{De}(f^{(0U)}\boldsymbol{\Lambda}^{(0U)}-b\boldsymbol{\delta}).
\end{split}\end{equation}
The undisturbed polymer configuration, $\boldsymbol{\Lambda}^{(0U)}$, is the result of the balance between the stretching applied by constant strain rate, $\mathbf{E}$, and the relaxation of the polymer (the term with coefficient $1/De$), in infinite time.

For steady-state planar extensional flow Becherer et al. \cite{becherer2008scaling} showed that to match boundary conditions of a realistic experiment, a spatially varying configuration must be allowed. For an Oldroyd-B fluid it is well known that this flow admits a singularity at $De=0.5$, but it loses smoothness at even smaller $De$ \cite{renardy2006comment}. The singularity is removed by using, for example, the Giesekus model, but smoothness is still lost for $De<0.5$ under certain conditions, thus leading to infinite stress gradients \cite{renardy2006comment}. Similar behavior is likely to occur for the undisturbed solution with spatial inhomogeneity, for the uniaxial extensional flow modeled with the FENE-P equations.

As pointed out by Becherer et al. \cite{becherer2008scaling}, there is a strongly stretched central region around the extensional axis of the planar extensional flow. The configuration tensor is spatially homogeneous in that region, which is the experimentally detected birefringence region \cite{becherer2008scaling}. For uniaxial extensional flow of a dilute polymer solution, obtained via a filament stretching rheometer, this is expected to occur away from the circular end-plates \cite{mckinley2002filament}. At the end-plates, the no-slip/ no-penetration condition leads to inhomogeneity \cite{mckinley2002filament}. In the planar extensional flow described by Becherer et al. \cite{becherer2008scaling}, the spatially uniform central region is numerically observed only for large $De$; its extent increases with $L$. However, in the filament stretching rheometer experiments, a long central region of uniform diameter and spatially homogeneous flow is obtained by separating the circular plates, connected by a liquid bridge of the fluid being tested, at a prescribed exponential rate \cite{mckinley2002filament}.

\begin{figure}[h!]
	\centering	\subfloat{\includegraphics[width=0.4\textwidth]{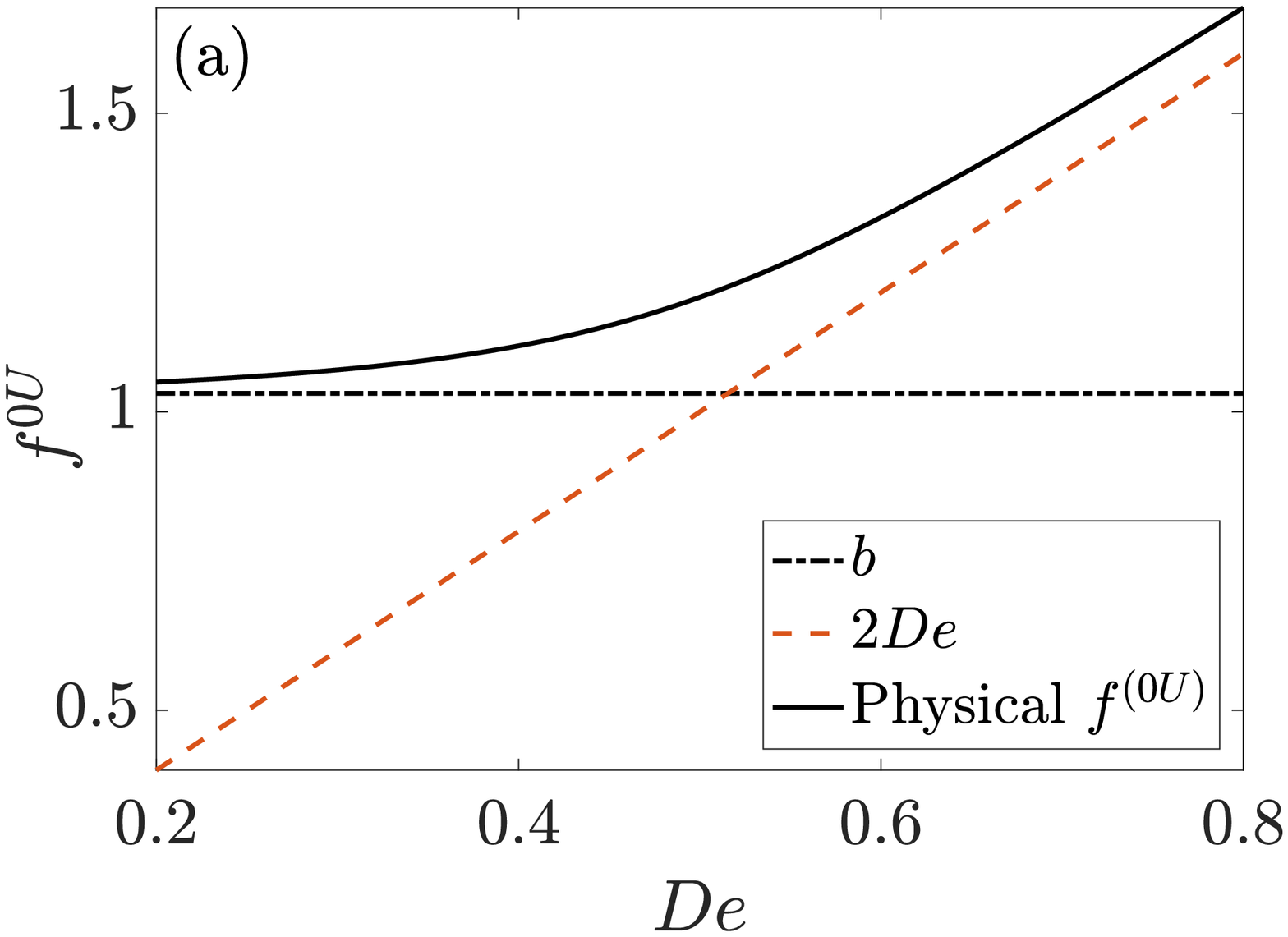}\label{fig:Roots_undisturbed_L_10.eps}}\hspace{0.2in}	\subfloat{\includegraphics[width=0.4\textwidth]{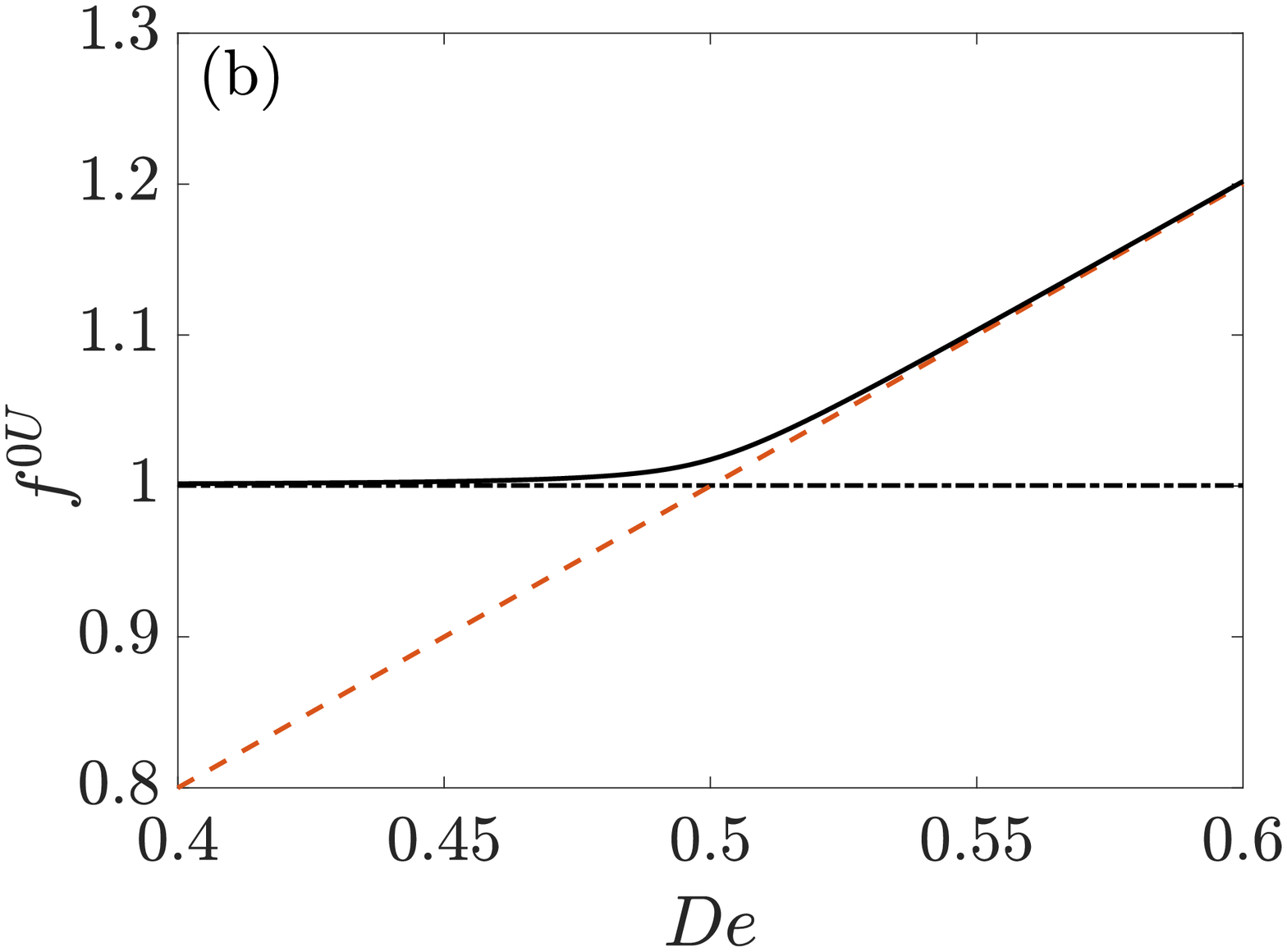}\label{fig:Roots_undisturbed_L_100.eps}}
	\caption {The physical root of the cubic equation \eqref{eq:cubicalfa} for (a) $L=10$ and (b) $L=100$. Both figures share the same legend. \label{fig:UndisturbedRoots}}
\end{figure}

For the strain tensor, $\mathbf{E}$, given by equation \eqref{eq:UndisturbedStrainRate}, the components of the spatially homogeneous, undisturbed configuration tensor, $\boldsymbol{\Lambda}^{(0U)}$, in the cylindrical coordinates ($r,z,\theta$ with $z$ measured along the extensional axis) follow the relations,
\begin{equation}
{\Lambda}^{(0U)}_{ij}=0, \text{ for } i\ne j \text{ and } {\Lambda}^{(0U)}_{rr}={\Lambda}^{(0U)}_{\theta \theta}.
\end{equation}
Therefore,
\begin{equation}
f^{(0U)}=\frac{1}{1-(2{\Lambda}^{(0U)}_{rr}+{\Lambda}^{(0U)}_{zz})/L^2},
\end{equation}
and the equations for the components ${\Lambda}^{(0U)}_{rr}$ and ${\Lambda}^{(0U)}_{zz}$ are combined to yield,
\begin{equation}\label{eq:cubicalfa}
(f^{(0U)})^3-\Big({1+De}+\frac{3}{L^2-3}\Big)(f^{(0U)})^2+\Big(De-2De^2+\frac{3De}{L^2-3}\Big)f^{(0U)}+2{De}^{2}=0.
\end{equation}
The three roots of this cubic polynomial are functions of the parameters $De$ and $L$. We find all three to be real for a range of parameters, but as we discuss next only one of the roots represents a physically valid solution. For $De\ll 1$, the FENE-P equations are equivalent to the Oldroyd-B constitutive model. Performing an expansion of FENE-P in the small parameter $De$ leads to
\begin{equation}\label{eq:smallDeEstimate}
f^{(0U)}=b+\mathcal{O}(De^2).
\end{equation}
For $De\ge0.5$, polymers with large maximum extensibility, $L$, suddenly transition from the coiled state ($\text{tr}(\boldsymbol{\Lambda}^{(0U)})\ll L^2$) to being almost fully stretched ($\text{tr}(\boldsymbol{\Lambda}^{(0U)})\approx L^2$). This is called the coil-stretch transition and is well documented, both theoretically \cite{de1974coil,hinch1977mechanical} and experimentally \cite{perkins1997single,smith1998response}. After the coil-stretch transition ($De\ge0.5$), the polymer is highly stretched in the extensional direction \cite{becherer2008scaling},
\begin{equation}\label{eq:largeDeApprox}
\Lambda^{(0U)}_{zz}\gg \Lambda^{(0U)}_{rr},
\end{equation} which leads to,
\begin{equation}\label{eq:largeDeEstimate}
f^{(0U)}=2De, \hspace{0.2in}\Lambda^{(0U)}_{zz}=L^2\Big(1-\frac{1}{2De}\Big),  \hspace{0.2in}\Lambda^{(0U)}_{rr}=\Lambda^{(0U)}_{\theta\theta}=\frac{b}{De}.
\end{equation}
For the range of $L$ considered in the rest of this paper, $L\ge10$, one of the roots is approximately equal to $-De$ for all $De$. This can be checked from equation \eqref{eq:cubicalfa} by assuming large $L$. $f^{(0U)}<0$ is unphysical as it implies $\sqrt{\text{tr}(\boldsymbol{\Lambda}^{(0U)})}>L$ i.e. polymer extension larger than $L$. The second root is always less than one and $f^{(0U)}<1$ implies $(2{\Lambda}^{(0U)}_{rr}+{\Lambda}^{(0U)}_{zz})/L^2<0$; this is unphysical as well since $2{\Lambda}^{(0U)}_{rr}+{\Lambda}^{(0U)}_{zz}$ represents the mean-square polymer stretch. The only physical root, which satisfies $f^{(0U)}>1$, is shown in figure \ref{fig:UndisturbedRoots} for $L=10$ and 100 along with the asymptotic limits mentioned above in equations \eqref{eq:smallDeEstimate} and \eqref{eq:largeDeEstimate}. 
This root of the cubic equation \eqref{eq:cubicalfa} (bold black solid line) closely follows the corresponding limits before and after the coil-stretch transition at $De=0.5$. The coil-stretch transition is evident at $De=0.5$ for large $L$ (figure \ref{fig:Roots_undisturbed_L_100.eps}), where $f^{(0U)}$ rapidly approaches the asymptote corresponding to highly stretched polymers beyond $De=0.5$. The non-linear spring force makes the transition more gradual for smaller $L$ (figure \ref{fig:Roots_undisturbed_L_10.eps}).  

\begin{figure}[h!]
	\centering
	\subfloat{\includegraphics[width=0.33\textwidth]{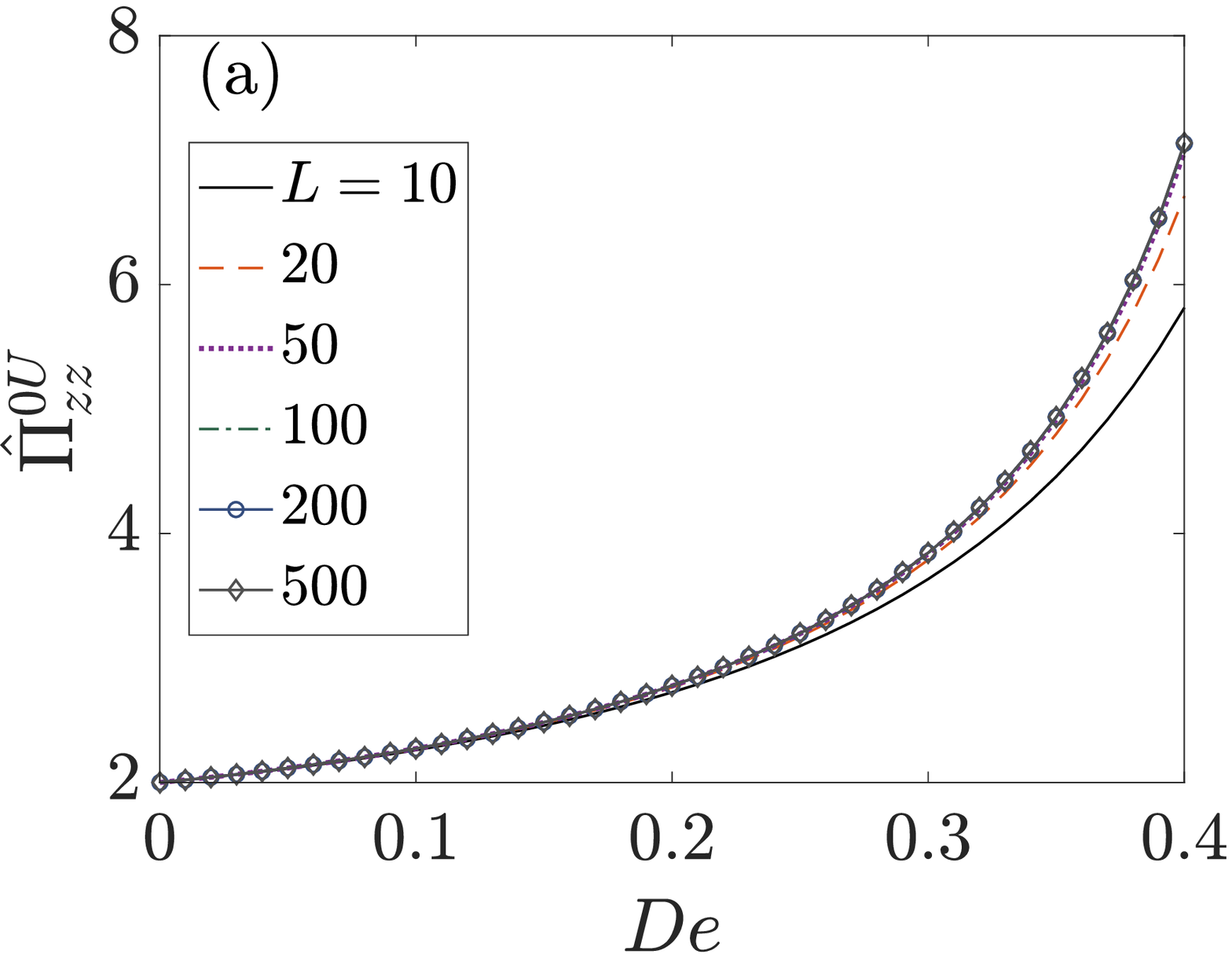}\label{fig:Stresses_SmallDe_undisturbed}}\hfill
	\subfloat{\includegraphics[width=0.33\textwidth]{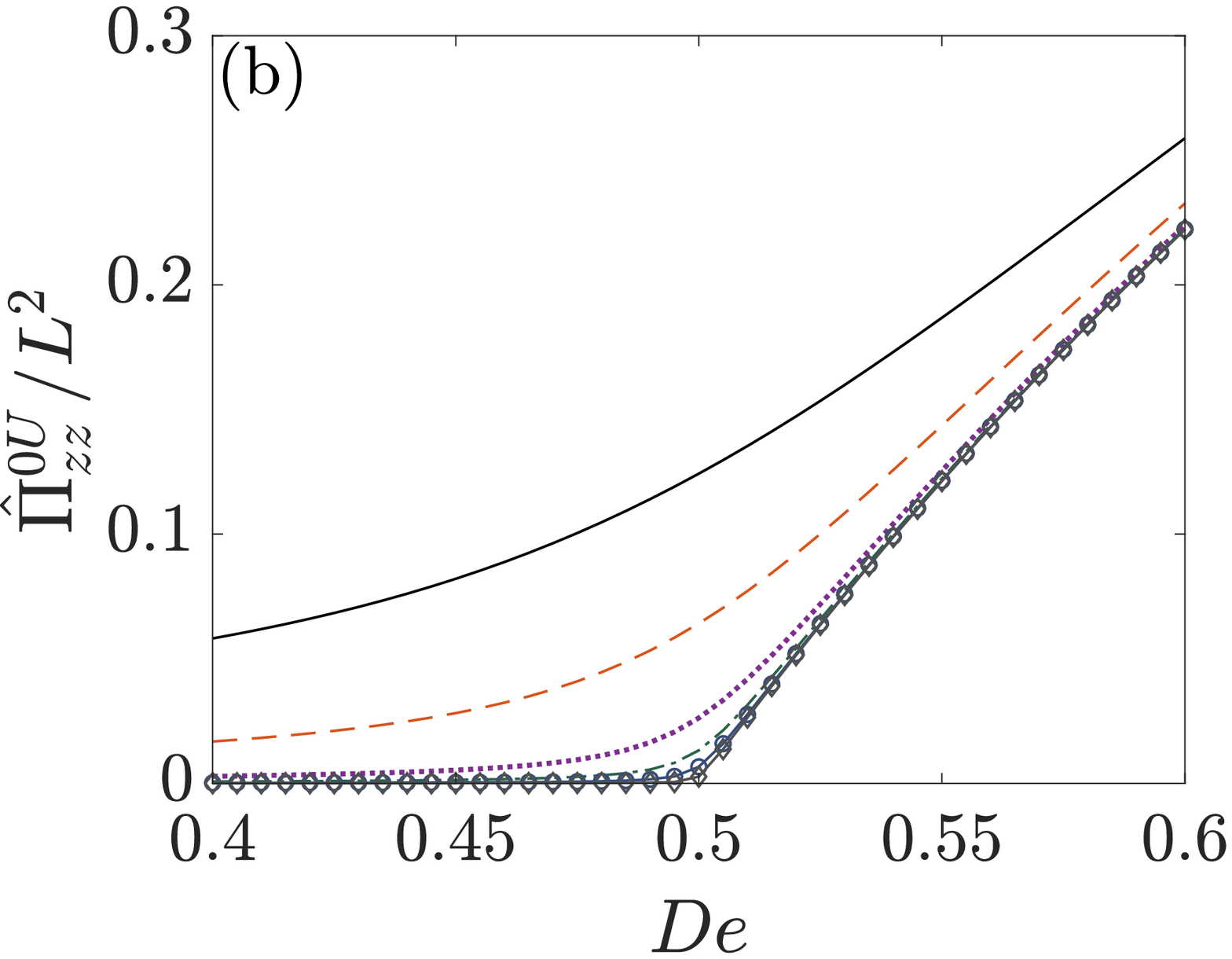}\label{fig:Stresses_MediumDe_undisturbed}}\hfill
	\subfloat{\includegraphics[width=0.33\textwidth]{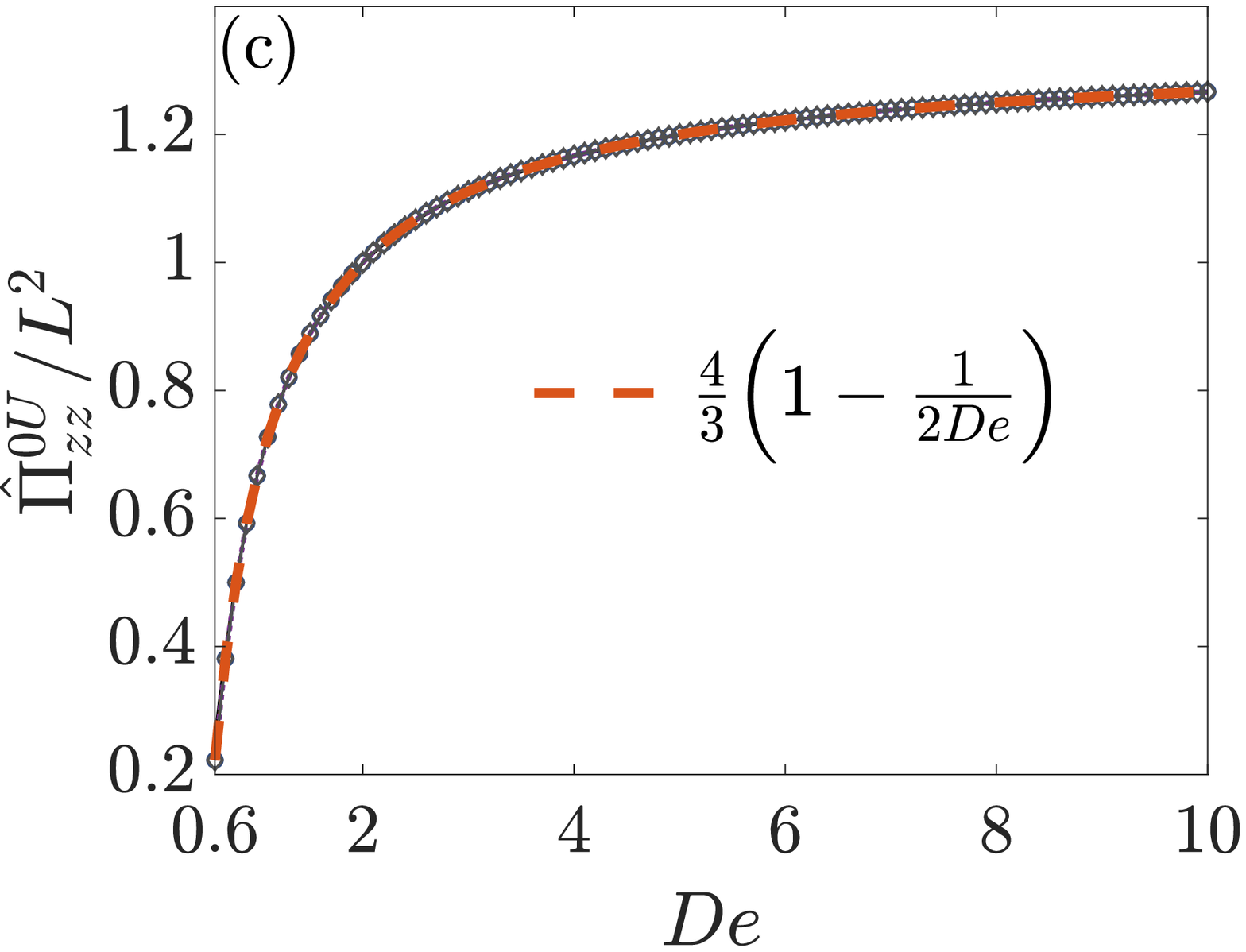}\label{fig:Stresses_LargeDe_undisturbed}}
	\caption {$\hat{\Pi}^{(0U)}_{zz}$ for various $L$ at: (a) $De<0.4$, (b) $0.4<De<0.6$ and (c) $De>0.6$. $\hat{\Pi}^{(0U)}_{zz}$ in (b) and (c) is normalized with $L^2$ and all three figures share the same legend. An additional curve corresponding to the approximate analytical solution, equation \eqref{eq:approximatelargeDe}, is included in (c). \label{fig:Stresses_Undisturbed}}
\end{figure}

The deviatoric part of $\boldsymbol{\Pi}^{(0U)}$ is,
\begin{equation}
\hat{\boldsymbol{\Pi}}^{(0U)}=\hat{{\Pi}}^{(0U)}_{zz} \mathbf{E}.
\end{equation}
$\hat{{\Pi}}^{(0U)}_{zz}$ for three different ranges of $De$ is shown in figure \ref{fig:Stresses_Undisturbed}. For $De\le0.4$ (figure \ref{fig:Stresses_SmallDe_undisturbed}), a monotonic increase of $\hat{{\Pi}}^{(0U)}_{zz}$ with $De$ reflects the increase in polymer stretch with the applied extension rate. For large $L$, in this $De$ regime, the polymers are stretched much less than $L$, e.g. for $De=0.4$, $L=50$, $\sqrt{\text{tr}(\boldsymbol{\Lambda}^{(0U)})}=2.52\ll50$. Hence we observe an $L$ independence of $\hat{{\Pi}}^{(0U)}_{zz}$ for this regime, especially for $L\gtrsim 20$. This extends up to a value slightly less than $De=0.5$ for a finite $L\gtrsim 50$ (not shown).

The polymer stress $\hat{{\Pi}}^{(0U)}_{zz}$ in figure \ref{fig:Stresses_MediumDe_undisturbed} and \ref{fig:Stresses_LargeDe_undisturbed} is normalized with $L^2$. The rapid increase in $\hat{{\Pi}}^{(0U)}_{zz}/L^2$ with $De$ around $De=0.5$ is the aforementioned coil-stretch transition. $\hat{{\Pi}}^{(0U)}_{zz}\sim L^2$, for $L\gtrsim50$, and $0.5\lesssim De<0.6$, as the stretch $\sqrt{\text{tr}(\boldsymbol{\Lambda}^{(0U)})}\sim L$. For $De>0.6$ in figure \ref{fig:Stresses_LargeDe_undisturbed}, the $L^2$ scaling is valid for even lower $L$, as the curves for $10\le L\le500$ are indistinguishable. Throughout the $De$ range shown, a monotonic increase in stress with $De$ is due to the increasing polymer stretch. Using the approximations mentioned in \eqref{eq:largeDeApprox} and \eqref{eq:largeDeEstimate}, after the coil-stretch transition,
\begin{equation}\label{eq:approximatelargeDe}
\hat{{\Pi}}^{(0U)}_{zz}=\frac{4}{3}\Big(1-\frac{1}{2De} \Big)L^2.
\end{equation}
As shown in figure \ref{fig:Stresses_LargeDe_undisturbed}, this agrees closely with the full solutions.

\section{Kinematics of steady extensional flow around a sphere}\label{sec:Kinematics}
Insight into the kinematics of the flow around an isolated sphere provides an analogy between the polymer configuration around the sphere (equation \eqref{eq:u_field}) and the effect of the velocity on the fluid elements. In this section, we discuss the kinematics using velocity gradient, strain rate and Cauchy-Green strain tensors, before considering the forthcoming discussion about the polymer configuration in the next section.
\subsection{Local Kinematics: Velocity gradient and strain rate tensor}
\begin{figure}[h!]
	\centering	\subfloat{\includegraphics[width=0.33\textwidth]{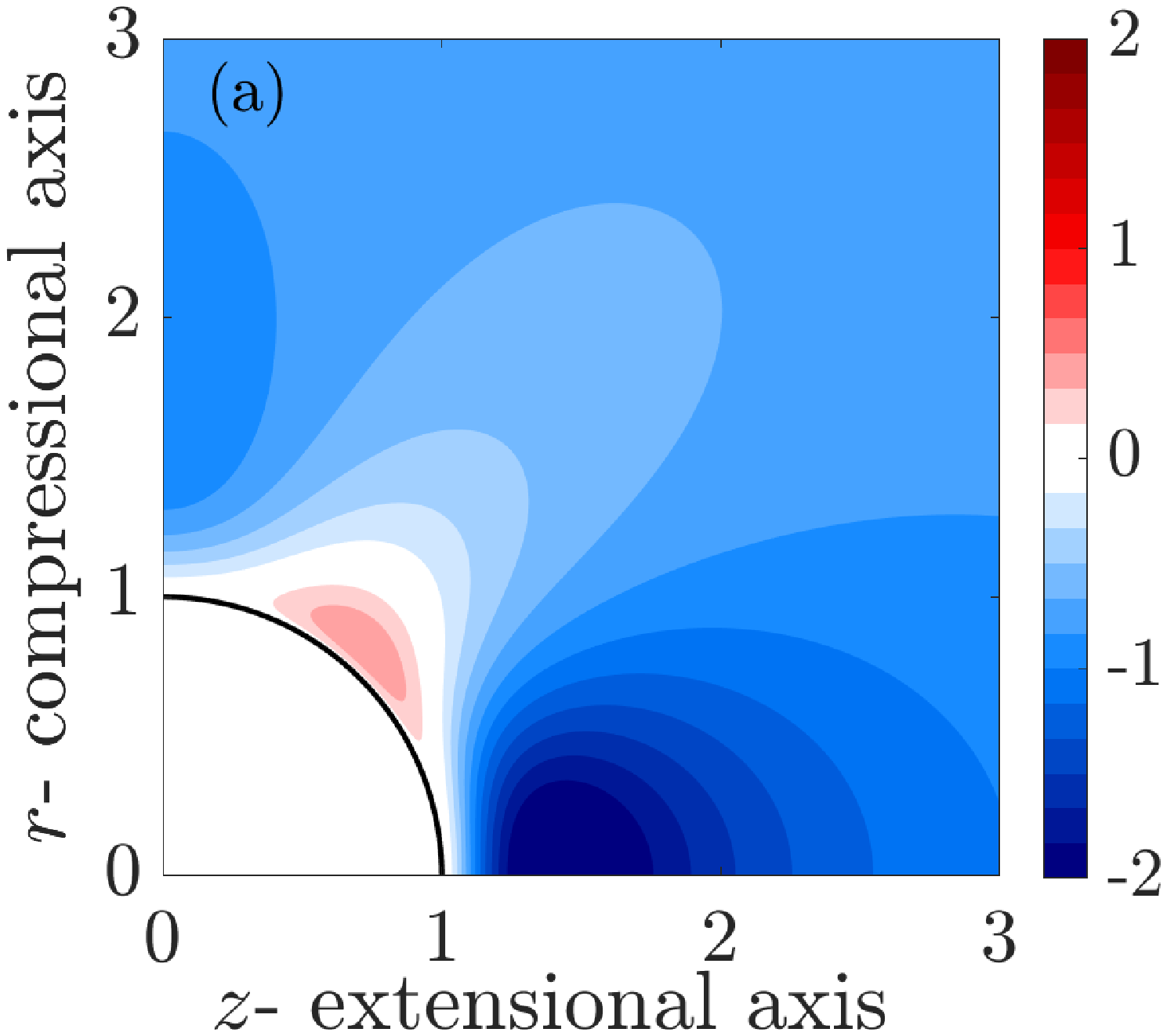}\label{fig:Q_extension.eps}}\hspace{0.2in}	\subfloat{\includegraphics[width=0.33\textwidth]{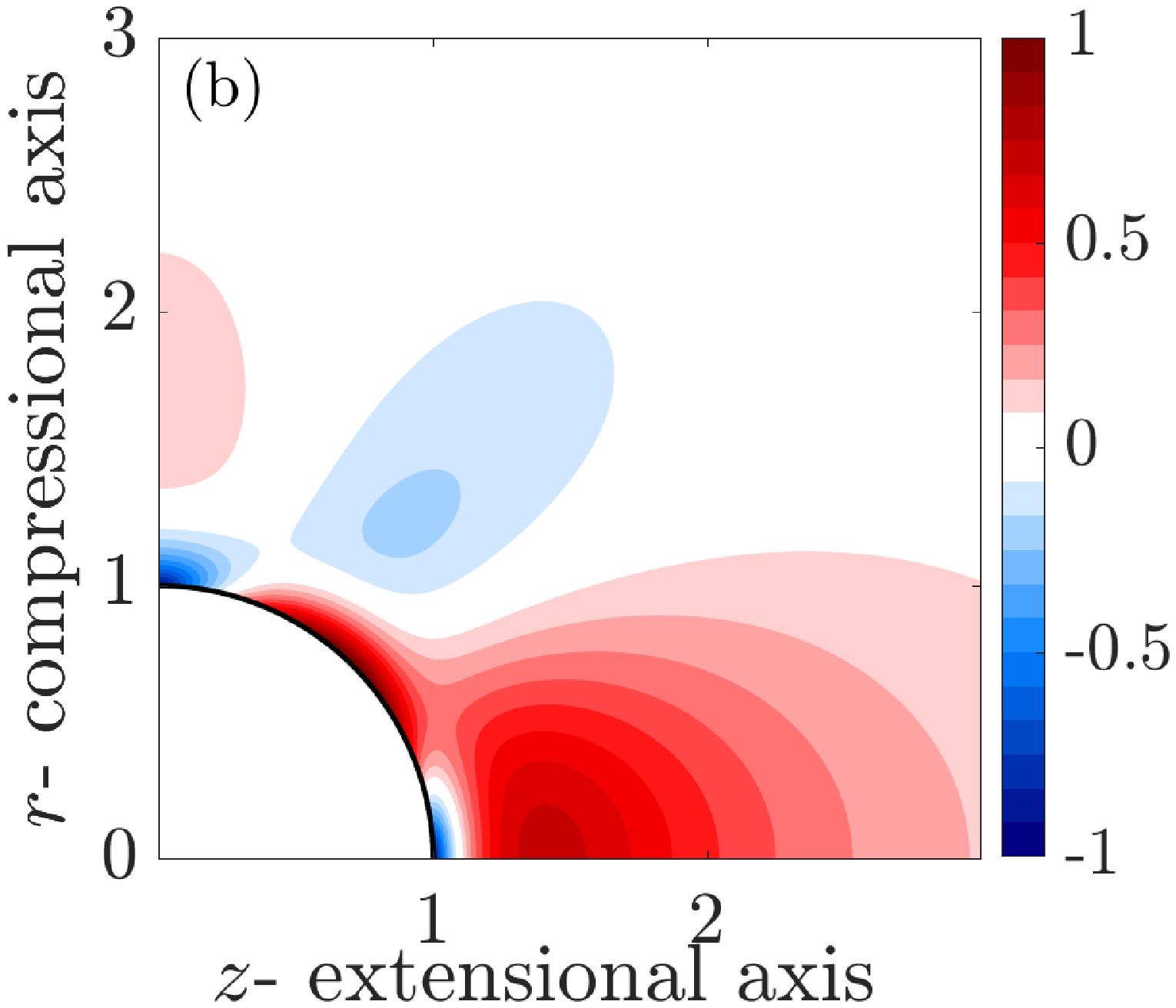}\label{fig:Local_De.eps}}
	\caption {Local kinematic diagnostic fields: (a) velocity gradient second invariant, $Q$ and, (b) fractional change in the local Deborah number field, $\Delta De_\text{local}$, due to a sphere in an imposed extensional flow.\label{fig:Kinematic_1}}
\end{figure}

Invariants of the velocity gradient and strain rate tensor have been extensively used to deduce the topology and dynamics of fluid flows \cite{chong1990general}. The second invariant of the characteristic equation for the eigenvalues of the velocity gradient tensor of an incompressible  flow with velocity, $\mathbf{u}$ is \cite{chong1990general},
\begin{equation}
Q=\frac{1}{2}(\text{tr}(\nabla\mathbf u)^2-\text{tr}((\nabla\mathbf u)^2))=-\frac{1}{2}\text{tr}((\nabla\mathbf u)^2)=\frac{1}{2}({\omega}_{ij}{\omega}_{ij}-{e}_{ij}{e}_{ij}),
\end{equation}
where $\boldsymbol{\omega}=(\nabla\mathbf{u}-(\nabla\mathbf{u})^\text{T})/2$ is the vorticity tensor and $\boldsymbol{e}= (\nabla\mathbf{u}+(\nabla\mathbf{u})^\text{T} )/2$ is the strain rate tensor.
$Q$ compares the rotation rate to the strain rate of the flow. A positive $Q$ indicates the dominance of enstrophy over strain, and a negative $Q$ indicates a weaker rotation. For undisturbed extensional flow with $\nabla\mathbf{u}=\mathbf{E}$ given by equation \eqref{eq:UndisturbedStrainRate}, $Q=-3/4$. Figure \ref{fig:Q_extension.eps} shows the $Q$ field for the extensional flow disturbed by an isolated sphere. We observe a rotation dominated region around 45$^\circ$ from the extensional axis, and a rotation deficient (implied strain dominance) region around the extensional axis. In the literature (such as  \cite{yang2018mechanism}), regions of negative $Q$ have been associated with high stretching rates. While it is true that negative $Q$ indicates regions where stretching exceeds rotation, it leads to false negatives in identifying regions of high absolute stretching rate for the present flow as discussed below.

The dissipation rate $\boldsymbol{e}:\boldsymbol{e}$ is a scalar estimate of the local rate of stretching experienced by an infinitesimal fluid element in the underlying velocity field. In an undisturbed uniaxial extensional flow $\boldsymbol{e}:\boldsymbol{e}=1.5$. The Deborah number, $De$, of the imposed flow defined in equation \eqref{eq:DeEquation} is based on the imposed extension rate. Using the local $\boldsymbol{e}:\boldsymbol{e}$, we define a scalar field termed the fractional change in the local Deborah number field, $\Delta De_\text{local}$, around the sphere in an imposed uniaxial extensional flow,
\begin{equation}
\Delta De_\text{local}=\sqrt{\frac{\boldsymbol{e}:\boldsymbol{e}}{1.5}}-1.
\end{equation}
It is dimensionally consistent and is zero in the far-field extensional flow implying no change in stretching by velocity gradients in the far-field. $\Delta De_\text{local}$ due to a spherical particle is shown in figure \ref{fig:Local_De.eps}. A $\Delta De_\text{local}>0$ region has more local stretching and $\Delta De_\text{local}<0$ region has less stretching in the presence of the particle. Compared to the $Q$ field of figure \ref{fig:Q_extension.eps}, this gives a relatively direct insight into the local stretching properties of the flow field around the sphere. It shows increased stretching regions in specific locations: near the surface of the sphere around 45$^\circ$ from the extensional axis (missed by the $Q$ field) and at axial positions between about 1.2 and 2 along the extensional axis. There is reduced stretching near the particle surface around the stagnation points on the extensional and compressional axis. The region near the particle surface around $45^\circ$ from the extensional axis has a large positive $Q$, indicating a rotation dominated region, but it also has increased local stretching due to a larger local strain rate. It is a region of relatively large local shear rate, which implies a high rotation rate but also a high strain rate. Both the $Q$ and $\Delta De_\text{local}$ fields only provide insight into the local stretching as they do not take account of the Lagrangian history.
\subsection{Kinematics with Lagrangian history: Finite time Lyapunov exponents}\label{sec:FTLE}
To characterize the stretching capability of the extensional flow around an isolated sphere accounting for the Lagrangian history, we modify a tool from non-linear dynamics, the finite time Lyapunov exponent (FTLE) \cite{haller2015lagrangian}. The evolution of an infinitesimal fluid element denoted by a vector $\boldsymbol{\xi}$, evolves due to the linearised local flow as,
\begin{equation}
{{\xi}(t)_i= \xi_j(0)\frac{\partial x_i (t;\boldsymbol{x}_0)}{\partial x_{0,j}}+\mathcal{O}({\xi}_i(0){\xi}_i(0)),}
\end{equation}
{where ${\partial x_i (t;\boldsymbol{x}_0)}/{\partial x_{0,j}}$ is the deformation gradient of the flow map,}
\begin{equation}
{x_i(t;\boldsymbol{x}_0)=x_i(0)+\int_{0}^{t}{v}_i(\boldsymbol{x}(\tau),\tau) \text{d}\tau,}
\end{equation} {that maps the initial position, $x_i(0)=x_{0,i}$, of a fluid particle to its later position $x_i$ at $t$. A measure of the relative stretch in time $t$ is}
\begin{equation}
{\frac{{\xi}_i(t){\xi}_i(t)}{\xi_j(0)\xi_j(0)}\approx\frac{\xi_k(0)\xi_i(0)C^t_{0,ik}(\boldsymbol{x}_0)}{\xi_j(0)\xi_j(0)},}
\end{equation}
where,
\begin{equation}
{C^t_{0,ik}(\boldsymbol{x}_0)=\frac{\partial x_j (t;\boldsymbol{x}_0)}{\partial x_{0,i}}\frac{\partial x_j (t;\boldsymbol{x}_0)}{\partial x_{0,k}},}
\end{equation}
is the Cauchy-Green strain tensor. It is a symmetric, positive definite tensor, with at least one eigenvalue less than 1 for an incompressible flow \cite{haller2015lagrangian}, in 3 dimensions
\begin{align}\begin{split}
&{C^t_{0,ik}(\boldsymbol{x}_0) {\eta}_k^{(l)}=\lambda^{(l)} {\eta}^{(l)}_i,\hspace{0.2in} \det C^t_{0,ik}(\boldsymbol{x}_0)=\lambda^{(1)}\lambda^{(2)}\lambda^{(3)}=1,}\\&{0<\lambda^{(1)}\le \lambda^{(2)} \le \lambda^{(3)},\hspace{0.2in} 0<\lambda^{(1)}\le 1\le \lambda^{(3)}.}
\end{split}\end{align}
{The largest possible deformation starting with all possible orientations at the initial location ${x}_i(0)={x}_{0,i}$, is used to detect the stretching regions within the fluid \cite{haller2015lagrangian},}
\begin{equation}
{\max_{\boldsymbol{\xi}(0)} \frac{{\xi}_i(t){\xi}_i(t)}{\xi_i(0)\xi_j(0)}\approx \max_{\boldsymbol{\xi}(0)}  \frac{\xi_k(0)\xi_i(0)C^t_{0,ik}(\boldsymbol{x}_0)}{\xi_j(0)\xi_j(0)}=\lambda^{(3)}(t;\boldsymbol{x}_0).}
\end{equation}
{A field of $\lambda^{(3)}(\boldsymbol{x}_0)$ is used to construct a scalar field,}
\begin{equation}
{\text{FTLE}(t;\boldsymbol{x}_0)=\frac{1}{2t}\ln [\lambda^{(3)}(t;\boldsymbol{x}_0)].}
\end{equation}
Regions of high FTLE are associated with stretching regions (see \cite{haller2015lagrangian} and references therein). The FTLE field constructed using the maximum deformation of fluid elements following the negative of the velocity field, i.e. backward in time, is known as the backward FTLE \cite{haller2015lagrangian}. FTLE identifies the locations that lead to maximum stretching (in forward or backward time); i.e. the elements starting from the regions of high FTLE undergo relatively large stretching.

\begin{figure}[h!]
	\centering
	\subfloat{\includegraphics[width=0.33\textwidth]{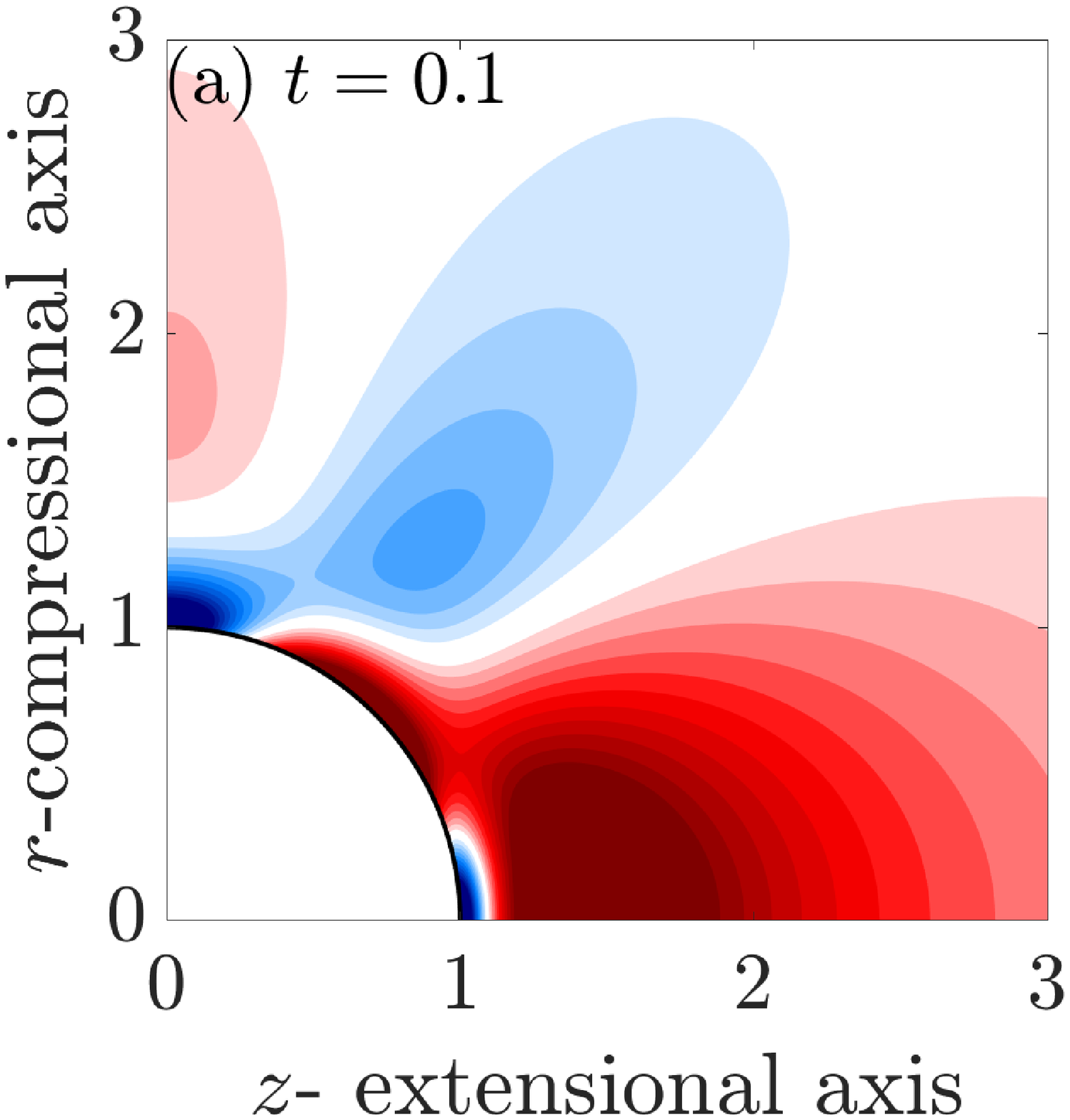}\label{fig:FTLE_pt1}}\hfill
	\subfloat{\includegraphics[width=0.33\textwidth]{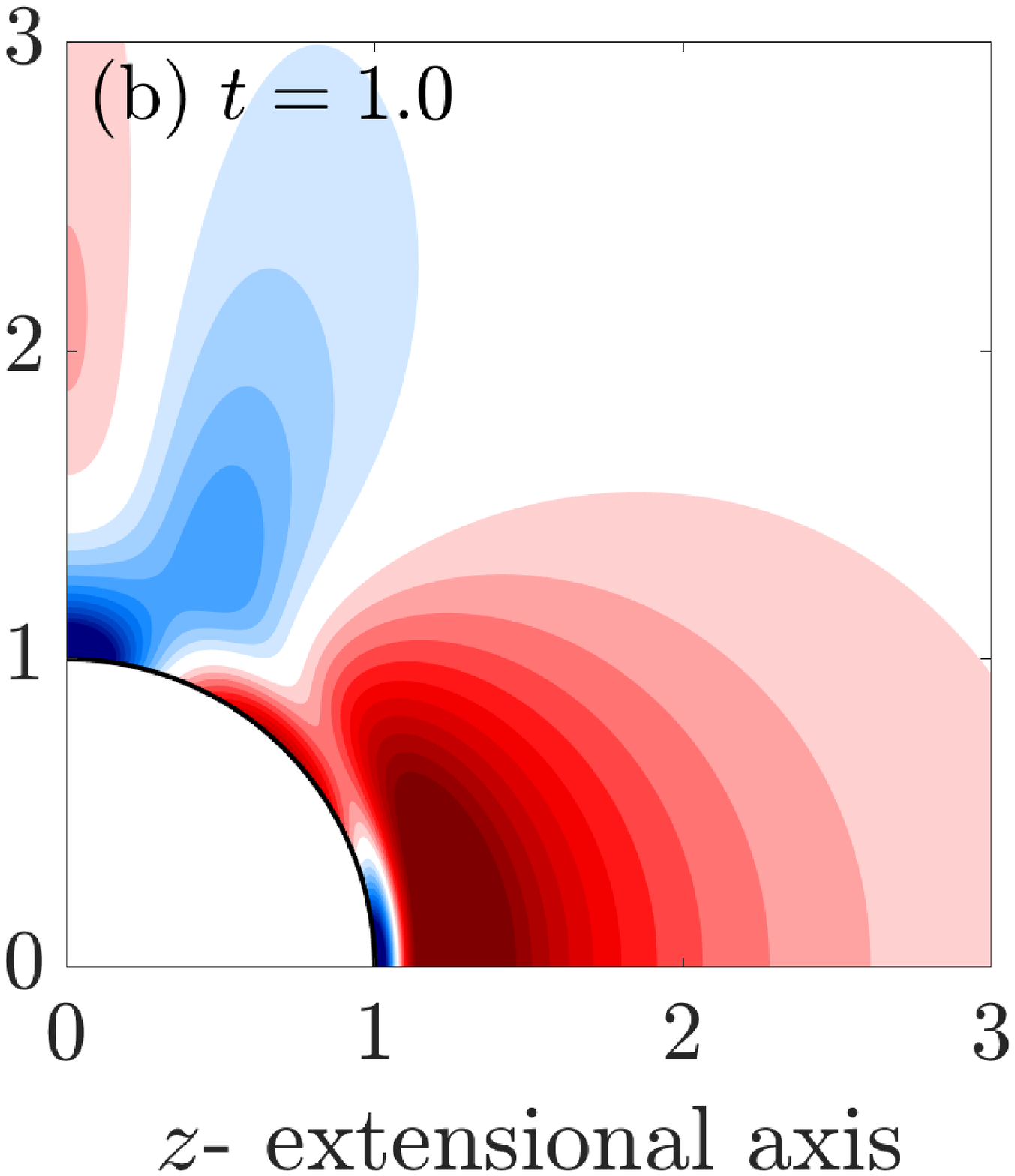}\label{fig:FTLE_1}}\hfill
	\subfloat{\includegraphics[width=0.33\textwidth]{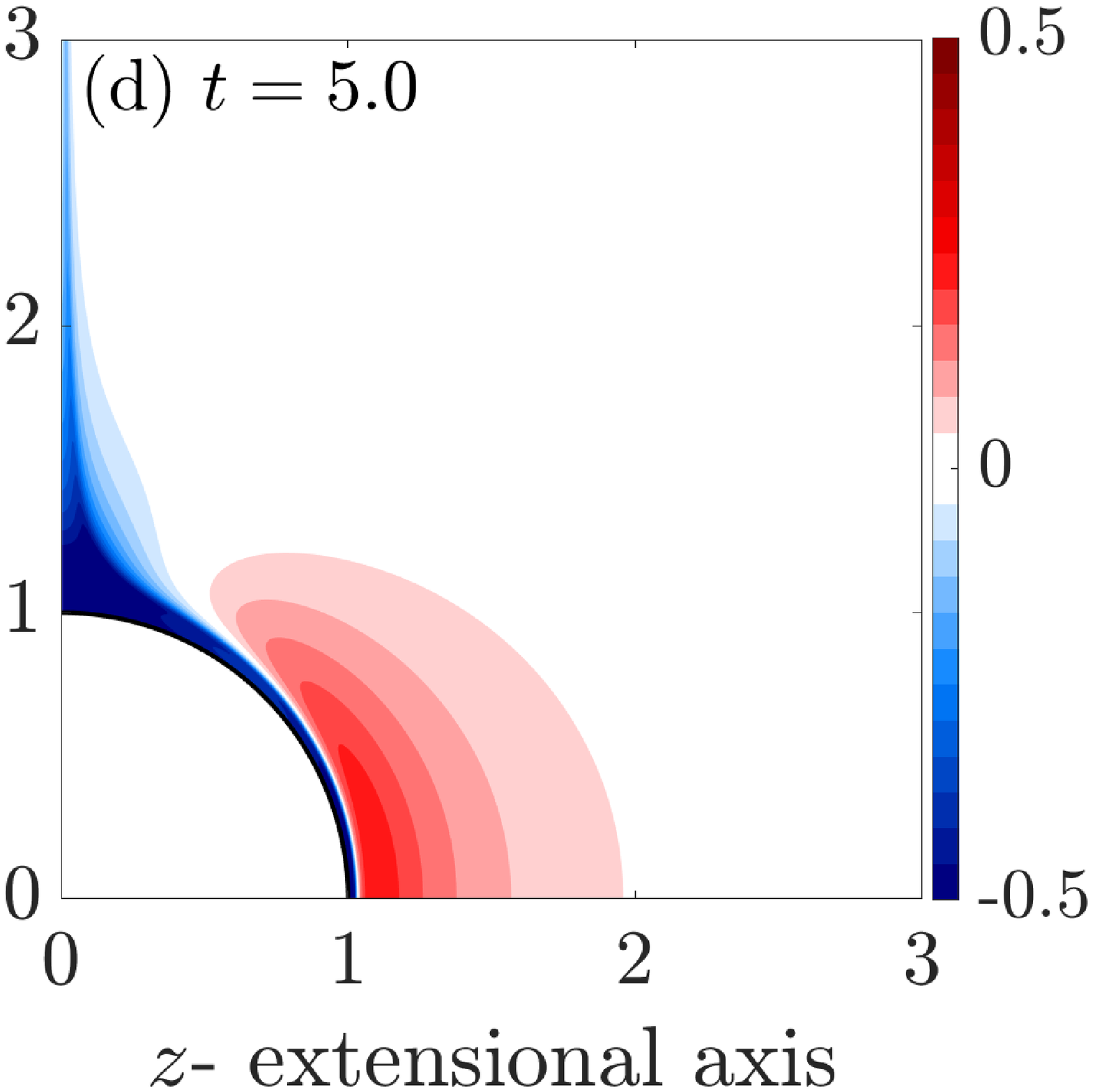}\label{fig:FTLE_5}}
	\caption {$\Delta\text{FTLE}(t;\boldsymbol{x}_0)$ due to the sphere in extensional flow for various $t$.\label{fig:FTLE}}
\end{figure}

\begin{figure}[h!]
	\centering
	\subfloat{\includegraphics[width=0.33\textwidth]{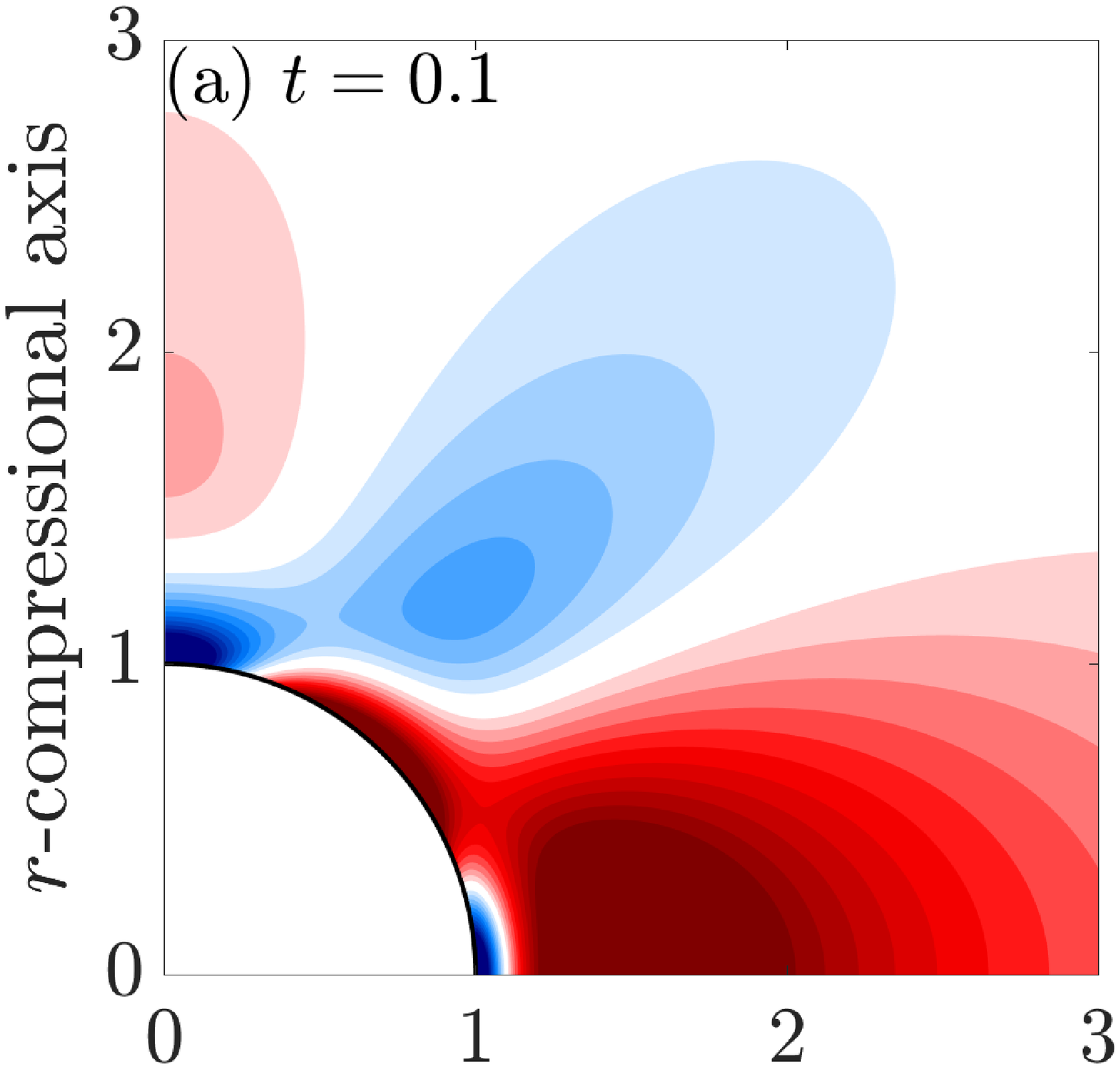}\label{fig:S_pt1}}\hfill
	\subfloat{\includegraphics[width=0.33\textwidth]{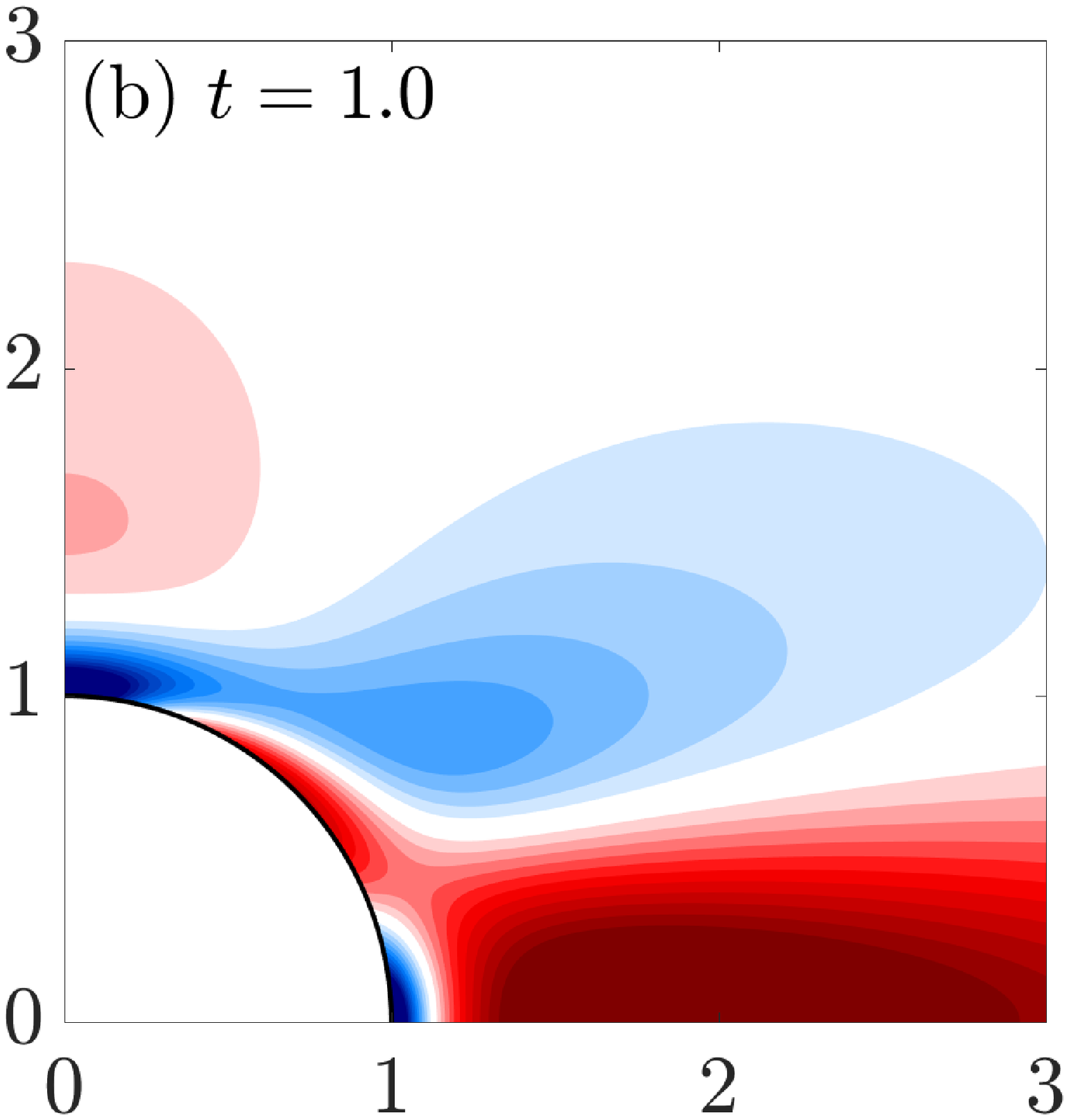}\label{fig:S_1}}\hfill
	\subfloat{\includegraphics[width=0.33\textwidth]{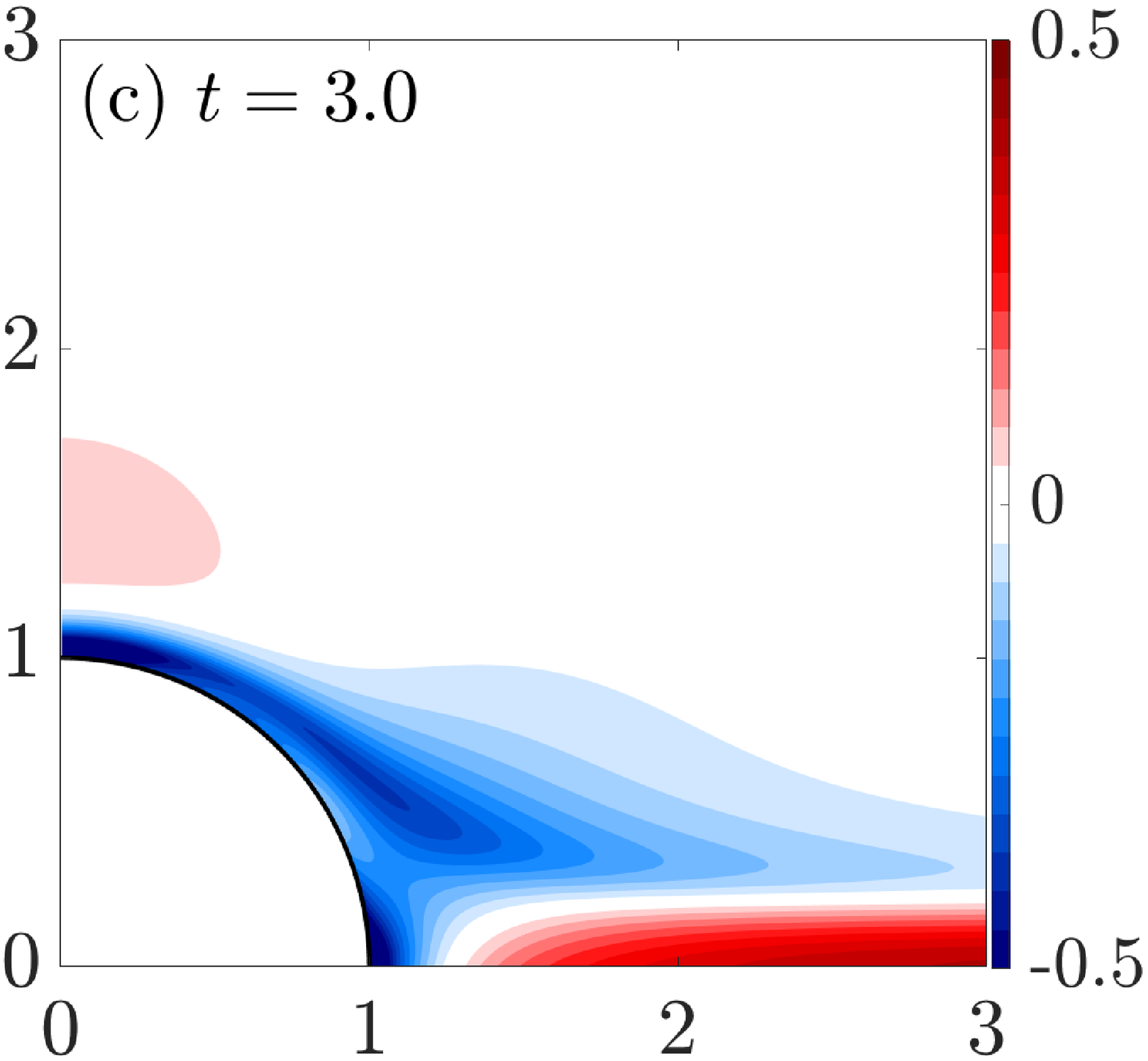}\label{fig:S_3}}\hfill
	\subfloat{\includegraphics[width=0.33\textwidth]{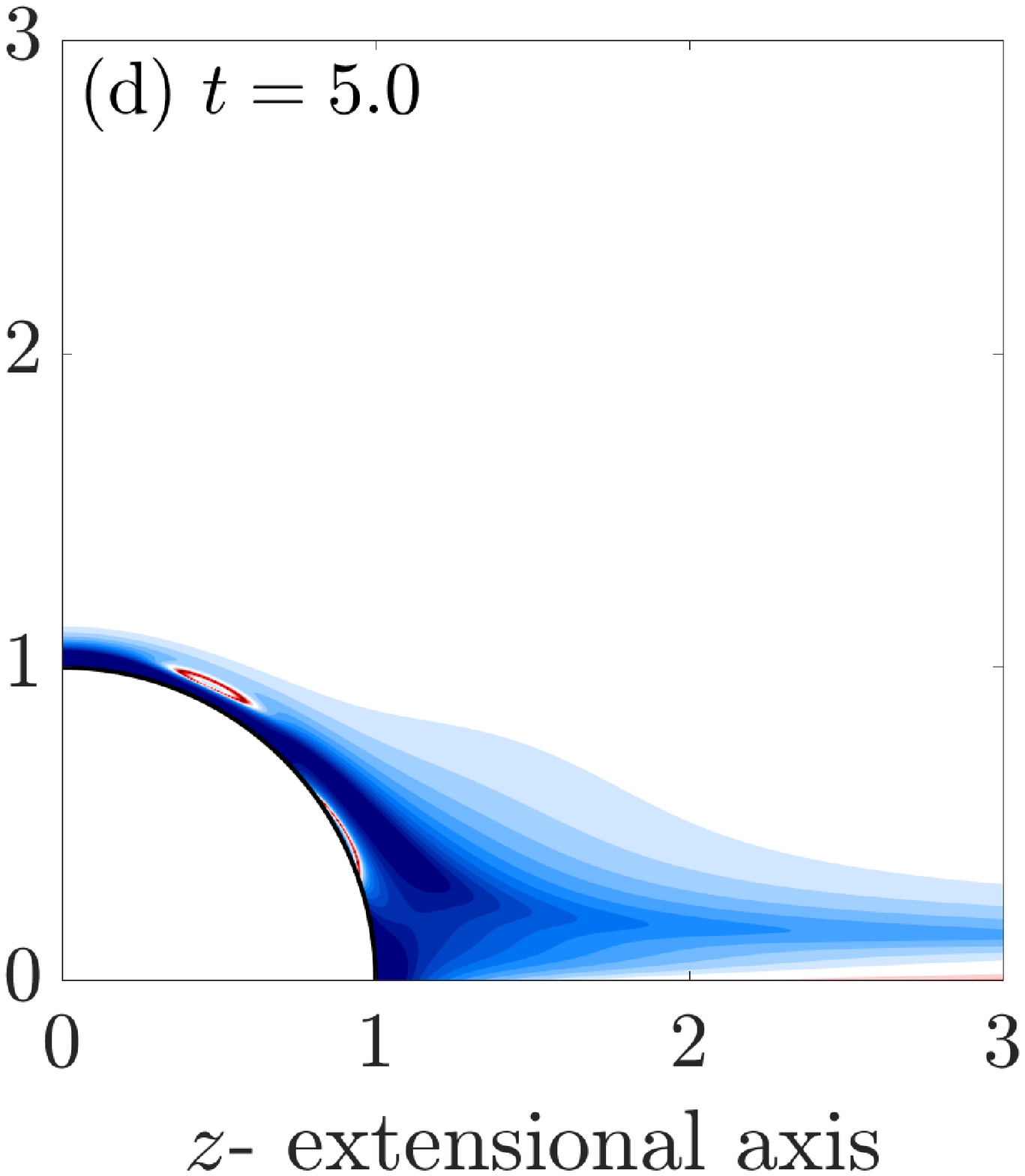}\label{fig:S_5}}\hfill
	\subfloat{\includegraphics[width=0.33\textwidth]{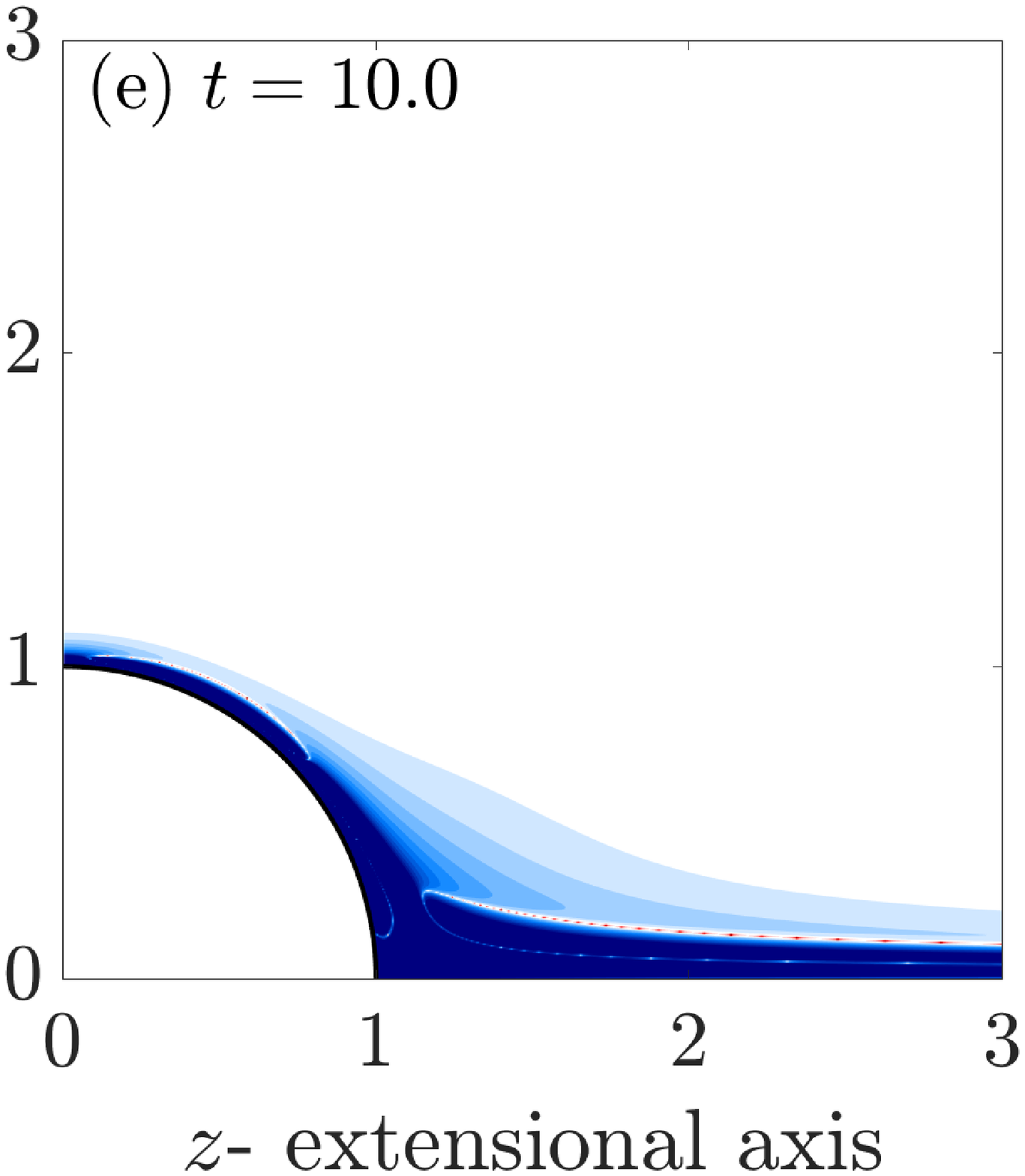}\label{fig:S_10}}\hfill
	\subfloat{\includegraphics[width=0.33\textwidth]{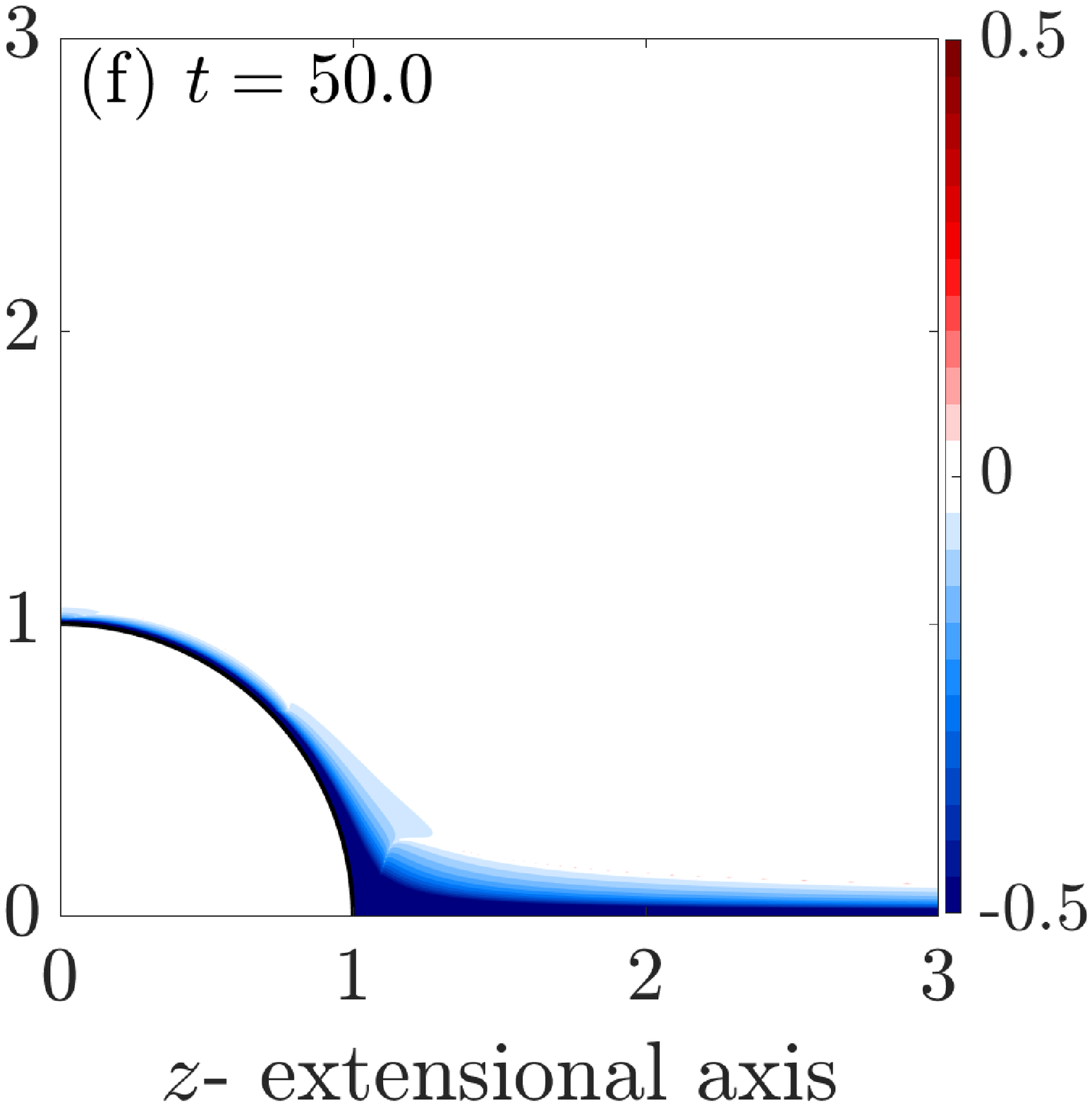}\label{fig:S_50}}
	\caption { $\Delta{\text{FTS}}(t;\boldsymbol{x}_0)$ due to the sphere in extensional flow for various $t$.\label{fig:S-field}}
\end{figure}

For our purpose, it is more useful to identify the locations where the most stretched fluid elements end up. To quantify this, we evaluate the maximum compression direction and rate in backward time. To this end, we use the backward flow map,
\begin{equation}
{{\tilde{x}}_i(t;\boldsymbol{x}_0)={x}_i(0)-\int_{0}^{t}{v}_i(\boldsymbol{\tilde{x}}(\tau),\tau) \text{d}\tau,}
\end{equation}
to construct the backward time deformation gradient and the corresponding backward Cauchy-Green tensor at each location in the domain,
\begin{equation}
{\tilde{C}^t_{0,ik}(\boldsymbol{x}_0)=\frac{\partial \tilde{x}_j (t;\boldsymbol{x}_0)}{\partial x_{0,i}}\frac{\partial \tilde{x}_j (t;\boldsymbol{x}_0)}{\partial x_{0,k}}.}
\end{equation}
{It has eigenvectors and eigenvalues, $\tilde{\lambda}^{(l)},l\in[1,3]$, satisfying,}
\begin{align}\begin{split}
&{\tilde{C}^t_{0,ik}(\boldsymbol{x}_0) {\eta}_k^{(l)}=\tilde{\lambda}^{(l)} {\eta}^{(l)}_i,\hspace{0.2in} \det \tilde{C}^t_{0,ik}(\boldsymbol{x}_0)=\tilde{\lambda}^{(1)}\tilde{\lambda}^{(2)}\tilde{\lambda}^{(3)}=1,}\\&{0<\tilde{\lambda}^{(1)}\le \tilde{\lambda}^{(2)} \le \tilde{\lambda}^{(3)},\hspace{0.2in} 0<\tilde{\lambda}^{(1)}\le 1\le \tilde{\lambda}^{(3)}.}
\end{split}\end{align}
{The minimum eigenvalue of this backward time Cauchy-Green tensor, $\tilde{\lambda}^{(1)}$ is used to estimate the maximum stretch of the fluid elements at the location $\boldsymbol{x}_0$, given they started as infinitesimal fluid elements at the appropriate location (defined by the flow map), at a time $t$ earlier. We define a finite time stretch field, FTS, as}
\begin{equation}
{\text{FTS}(t;\boldsymbol{x}_0)=\frac{1}{2t}\ln\Big(\frac{1}{\tilde{\lambda}^{(1)}(t;\boldsymbol{x}_0)}\Big).}
\end{equation}
The regions of large $\text{FTS}(t;\boldsymbol{x}_0)$ are the locations in the domain where a fluid element or non-diffusive line of dye released time $t$ ago is currently most stretched.
For the undisturbed extensional flow,
\begin{equation}
\text{FTLE}^{(0U)}(t;\boldsymbol{x}_0)=\text{FTS}^{(0U)}(t;\boldsymbol{x}_0)=1; \hspace{0.2in}\forall \boldsymbol{x}_0, t.
\end{equation}
Figure \ref{fig:FTLE} shows the change in the FTLE field, \begin{equation}
\Delta\text{FTLE}(t;\boldsymbol{x}_0)=\text{FTLE}(t;\boldsymbol{x}_0)-1,
\end{equation} and figure \ref{fig:S-field} shows the change in the $\text{FTS}$ field,
\begin{equation}
\Delta\text{FTS}(t;\boldsymbol{x}_0)=\text{FTS}(t;\boldsymbol{x}_0)-1,
\end{equation} due to the particle for various $t$. The $\Delta\text{FTLE}(t;\boldsymbol{x}_0)$ and $\Delta\text{FTS}(t;\boldsymbol{x}_0)$ fields capture the effect of the spherical particle on the stretching of the fluid elements or non-diffusing line of dye. Positive values indicate more stretching and negative values indicate less stretching in the presence of the particle. The topologies of the $\Delta\text{FTLE}(t;\boldsymbol{x}_0)$ and $ \Delta\text{FTS}(t;\boldsymbol{x}_0)$ with $t=0.1$ in figure \ref{fig:FTLE_pt1} and \ref{fig:S_pt1} are similar to the $\Delta De_\text{local}$ field shown in figure \ref{fig:Local_De.eps}. This is because all three capture the instantaneous ($t=0.1\ll 1$) stretching. The high stretch regions indicated by $\Delta\text{FTLE}$ shrink monotonically with $t$ (figure \ref{fig:FTLE}). $\Delta\text{FTLE}$ for large $t$ indicates a region very close to the particle surface such that a line of dye starting from there will be less stretched after time $t$ in the presence of the particle. There is a region just downstream of this less stretching region in which the starting elements will get more stretched as they are advected along the extensional axis.
However, the regions that possess highly stretched elements of dye, released time $t$ before, as indicated by positive $\Delta\text{FTS}$ in figure \ref{fig:S-field} for t$\gtrsim0.5$ are qualitatively different. For $t \lesssim 5$, there is a wake of highly stretched elements along the extensional axis which becomes thinner with $t$. There is a region of relatively less stretched elements around either stagnation point for $t=0.1$. With an increase in $t$, this region covers more of the particle surface and nearby region. It extends to lie over the highly stretched thin wake for $3\lesssim t \lesssim 5$. For even higher $t$, this region replaces the highly stretched wake, such that for $t\gtrsim 10$ there is instead a wake of relatively unstretched elements around the extensional axis and also over the entire particle surface. This wake of relatively less stretched elements for large $t$ also becomes thinner with $t$. At the end of the next section, we will show the analogy between the change in the steady-state polymer stretch due to the particle and the $\Delta\text{FTS}$ fields, indicating that polymers are stretched by the velocity gradient field in a similar way as the Lagrangian stretching of a line of fluid.

\section{Polymer Configuration around an isolated sphere}\label{sec:PolymerConfiguration}
Juxtaposition of the forthcoming discussion in this section and that from the study of Chilcott \& Rallison \cite{chilcott1988creeping} and Yang \& Shaqfeh \cite{yang2018mechanism} allows one to appreciate the different influence of changes in local velocity gradients created by a sphere in three different types of imposed flows on the steady state polymer configuration. Chilcott \& Rallison \cite{chilcott1988creeping} studied the uniform flow of a polymeric fluid past a rigid sphere. In their study, the polymers stretch just upstream of the front stagnation point and collapse on the stagnation point. Then they undergo a series of relaxation and stretching over the particle surface in response to the local velocity gradients. Beyond the rear stagnation point, a wake of stretched polymers develops. In a simple shear flow, Yang \& Shaqfeh \cite{yang2018mechanism} find the polymers to be most stretched within the closed streamlines around the sphere. 

In this section, we consider the change in polymer configuration, due to a spherical particle in an extensional flow, relative to the undisturbed configuration described in section \ref{sec:UndisturbedConfiguration}. We solve equation \eqref{eq:Configuration} for $\boldsymbol{\Lambda}=\boldsymbol{\Lambda}^{(0)}$ based on $\mathbf{u}=\mathbf{u}^{(0)}$ from equation \eqref{eq:u_field}. First, we consider the polymer configuration at the particle surface and along the extensional axis. Analytical progress is possible in both cases. Due to the continuity of the solutions in space, these give an idea of the polymer configuration in the region near the particle surface and the extensional axis. Then, we show the configuration change in the rest of the region around the sphere.

\subsection{Polymer stretch on the particle surface}\label{sec:Surfacestretch}
\begin{figure}[h!]
	\centering
	\subfloat{\includegraphics[width=0.4\textwidth]{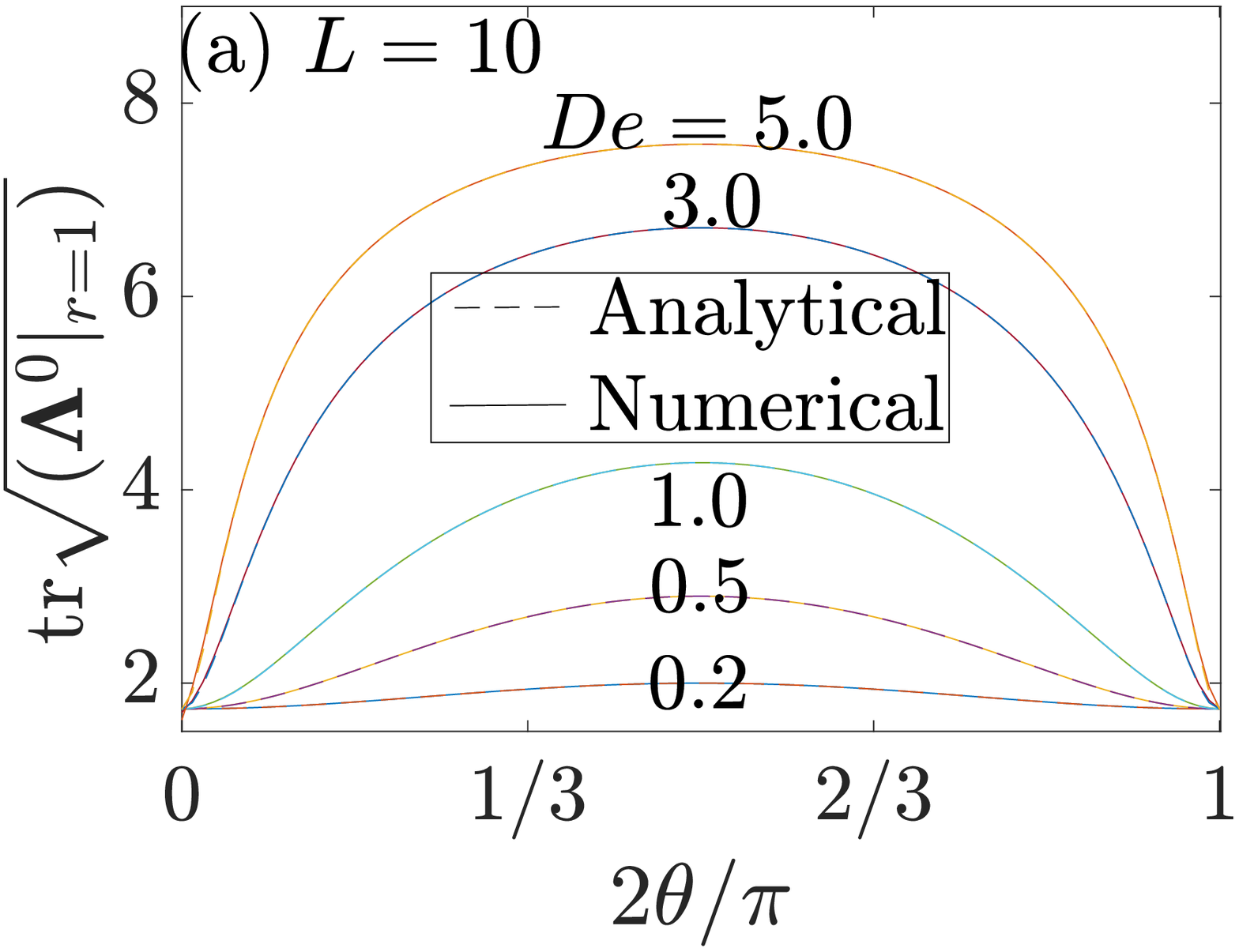}\label{fig:SmallLsurfacestretch}}\hspace{0.2in}
	\subfloat{\includegraphics[width=0.4\textwidth]{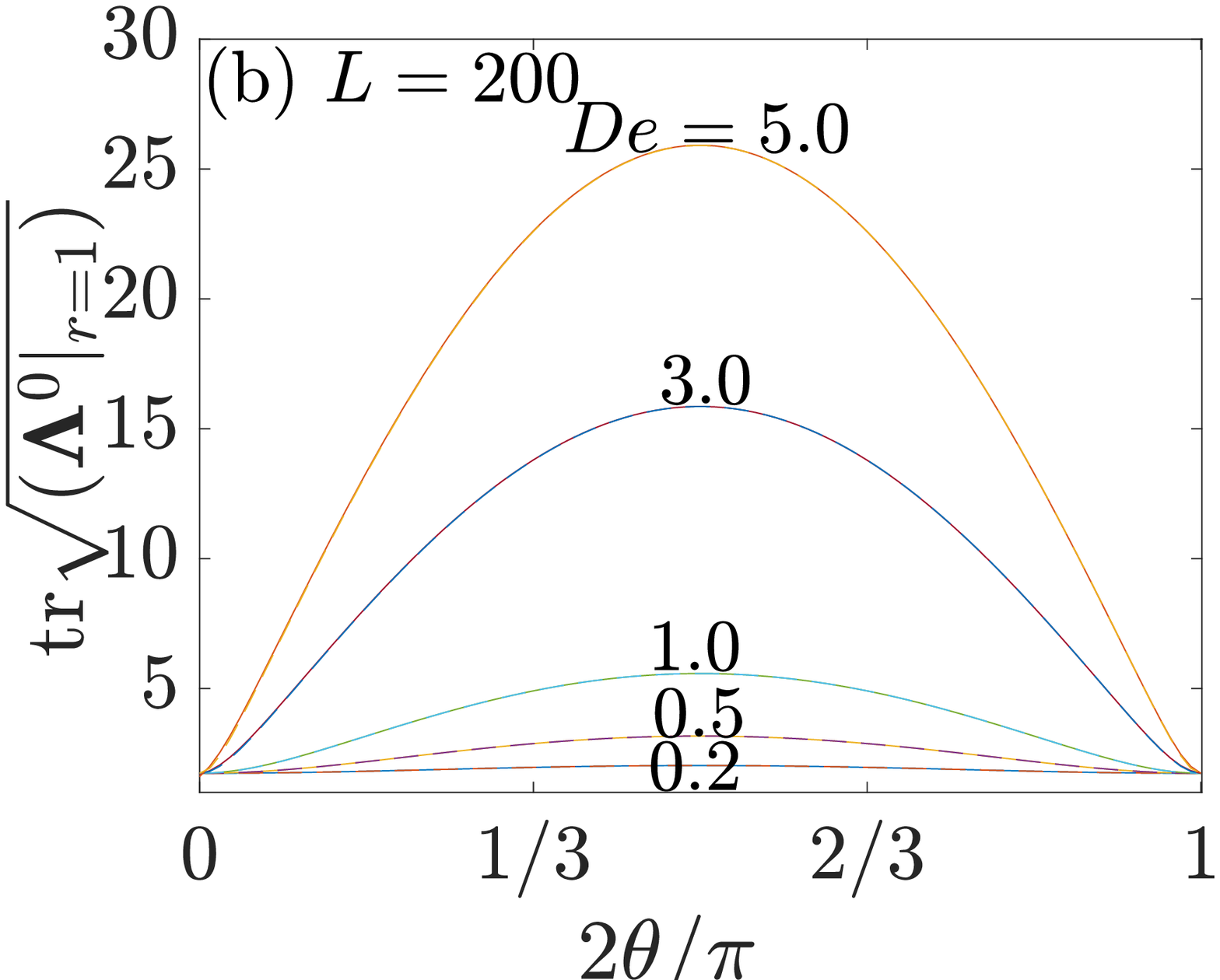}\label{fig:largeLsurfacestretch}}
	\caption {Polymer Stretch, $\sqrt{\text{tr}(\boldsymbol{\Lambda}^{(0)}|_{r=1})}$, at the particle surface, $r=1$ for various $De$ at (a) $L=10$, and (b) $L=200$ from the rear stagnation point on the extensional axis ($\theta=0$) to the front stagnation line ($\theta=\pi/2$) in the compressional plane.  The  $\sqrt{\text{tr}(\boldsymbol{\Lambda}^{(0)}|_{r=1})}$ distribution is symmetric about $\theta=\pi/2$ for $\theta\in[0,\pi]$, and here we show $\theta\in[0,\pi/2]$. Dashed lines represent the analytical solution and the solid lines the solution from numerical integration using the method of characteristics.\label{fig:surfaceConfiguration}}
\end{figure}

At the surface of the sphere, represented by  $r=1$, $\theta\in [0, \pi]$ and $\phi\in [0,2\pi]$, equation \eqref{eq:Configuration} leads to,
\begin{equation}\label{eq:analytical_surface_stress}
f|_{r=1}^3-\frac{L^2}{L^2-3}f|_{r=1}^2-\frac{225 De^2 }{2(L^2-3)}(\cos(\theta)^2-\cos(\theta)^4)=0,\hspace{0.2in} \theta\in [0, \pi].
\end{equation}
The individual tensor components of $\boldsymbol{\Lambda}^{(0)}$ are readily found and are combined to give the polymer stretch on the surface,
\begin{equation}\label{eq:exact_surface_configuration}
\sqrt{\text{tr}(\boldsymbol{\Lambda}^{(0)}|_{r=1})}(\theta)=\frac{L}{\sqrt{L^2-3}}\sqrt{\frac{3}{f|_{r=1,z}}+\frac{225De^2}{2f|_{r=1,z}^3}(\cos(\theta)^2-\cos(\theta)^4)}, \hspace{0.2in} \theta\in [0, \pi].
\end{equation}
The polymer stretch on the surface is only a function of the polar angle, $\theta$, from the extensional axis ($\theta=0$) due to axisymmetric flow and particle shape. The cubic equation \eqref{eq:analytical_surface_stress} is solved analytically for $f|_{r=1}$, but the expressions for the roots are unwieldy. A physical solution to the cubic equation must yield $f>1$ such that the polymer stretch, $\sqrt{\text{tr}(\boldsymbol{\Lambda}^{(0)}|_{r=1})}$ is positive and is limited by maximum polymer extensibility, $L$ (equation \eqref{eq:constitutive2}). We find that only one of the three roots is physical, at each location on the surface of the sphere, for a wide range of $De$ and $L$. The analytical results (obtained by using computer algebra) given by equation \eqref{eq:analytical_surface_stress} and \eqref{eq:exact_surface_configuration}  are shown along with the numerical results in figure \ref{fig:surfaceConfiguration}. The numerical and analytical curves are indistinguishable, which provides a first check for our numerical method. As mentioned earlier in section \ref{sec:Formulation} we use the method of characteristics to solve equation \eqref{eq:Configuration}. Since the velocity at the surface is zero, this cannot be done at the surface, where the numerical solution is instead extrapolated from the nearby non-stagnation streamlines. In general, the polymers are in the unstretched/equilibrium configuration at the stagnation points ($\boldsymbol{\Lambda}=\boldsymbol{\delta}$), and the stretch increases along the surface reaching a maximum at $\theta=\pi/4$ from the extensional axis. This complements the picture presented by $\Delta De_\text{local}$ (figure \ref{fig:Local_De.eps}), and  the $\Delta\text{FTLE}$ (figure \ref{fig:FTLE_pt1}) and $\Delta\text{FTS}$ (figure \ref{fig:S_pt1}) fields for $t=0.1$. As the polymer on the surface is not convected (zero velocity), it reacts to the local strain rate. As shown in figure \ref{fig:surfaceConfiguration}, at a given location on the surface, there is an increase in the stretch with extension rate ($De$). The effect of non-linearity of the spring force, used to model the polymer, is observed for the $De=3.0$ and 5.0 curves for $L=10$ (figure \ref{fig:SmallLsurfacestretch}), as increasing $De$ from 3.0 to 5.0 leads to a smaller increase in polymer stretch than for $L=200$ (figure \ref{fig:largeLsurfacestretch}). For $L=200$, at least up to $De=5$, the maximum stretch on the surface is very small compared to $L$. Polymer stretch on the surface for large $L$ is further examined. The maximum surface stretch is at $\theta=\pi/4$ (equation \eqref{eq:exact_surface_configuration}),
\begin{equation}
\max\Big(\sqrt{\text{tr}(\boldsymbol{\Lambda}^{(0)}|_{r=1})}(\theta)\Big)=\frac{L}{\sqrt{L^2-3}}\sqrt{\frac{3}{f|_{r=1,\theta=\pi/4}}\Big(1+\frac{75De^2}{8f|_{r=1,\theta=\pi/4}^2}\Big)}.\label{eq:maxsurfstretch}
\end{equation}
For $L$ large enough such that,
\begin{equation}\label{eq:stresslet_+approx_condition}
\frac{225 De^2 }{2 (L^2-3)}\ll 1,
\end{equation}
the physical solution to equation \eqref{eq:analytical_surface_stress} is simplified to
\begin{equation}f|_{r=1,z}\approx \frac{L^2}{L^2-3}\approx 1.\end{equation} and,
\begin{equation}
\max\Big(\sqrt{\text{tr}(\boldsymbol{\Lambda}^{(0)}|_{r=1})}(\theta)\Big)\approx\sqrt{3+\frac{225De^2}{8}}.\label{eq:maxsurfstretchLargeL}
\end{equation}
For $De=10$, this estimate leads to $\max\Big(\sqrt{\text{tr}(\boldsymbol{\Lambda}^{(0)}|_{r=1})}(\theta)\Big)\approx53$. Hence, for large $L$ the stretch at the surface of the sphere scales as $De$, and is very small compared to $L$ for $De\ll L$.

\subsection{Polymer configuration on the rear stagnation streamline}\label{sec:StagSteamline}
The mathematical analysis of the far-field polymer constitutive equations presented in this section on the streamline coinciding with the extensional axis, or the rear stagnation streamline, distinguishes two types of physical behavior of the polymers in the far-field:  at a given $De$ and $L$ polymers may be stretching to recover the far-field/ undisturbed configuration or they may be relaxing from their highly stretched state to approach the undisturbed configuration. Far-field analysis on the extensional axis is relevant because the deviation of the polymer configuration from its undisturbed state is expected to be most significant near the extensional axis. The numerical solution on the extensional axis provides useful physical information throughout the stagnation streamline starting from the particle's surface. The analytical and numerical solutions on the extensional axis match in the far-field.

\begin{figure}[h!]
	\centering
	\subfloat{\includegraphics[width=0.4\textwidth]{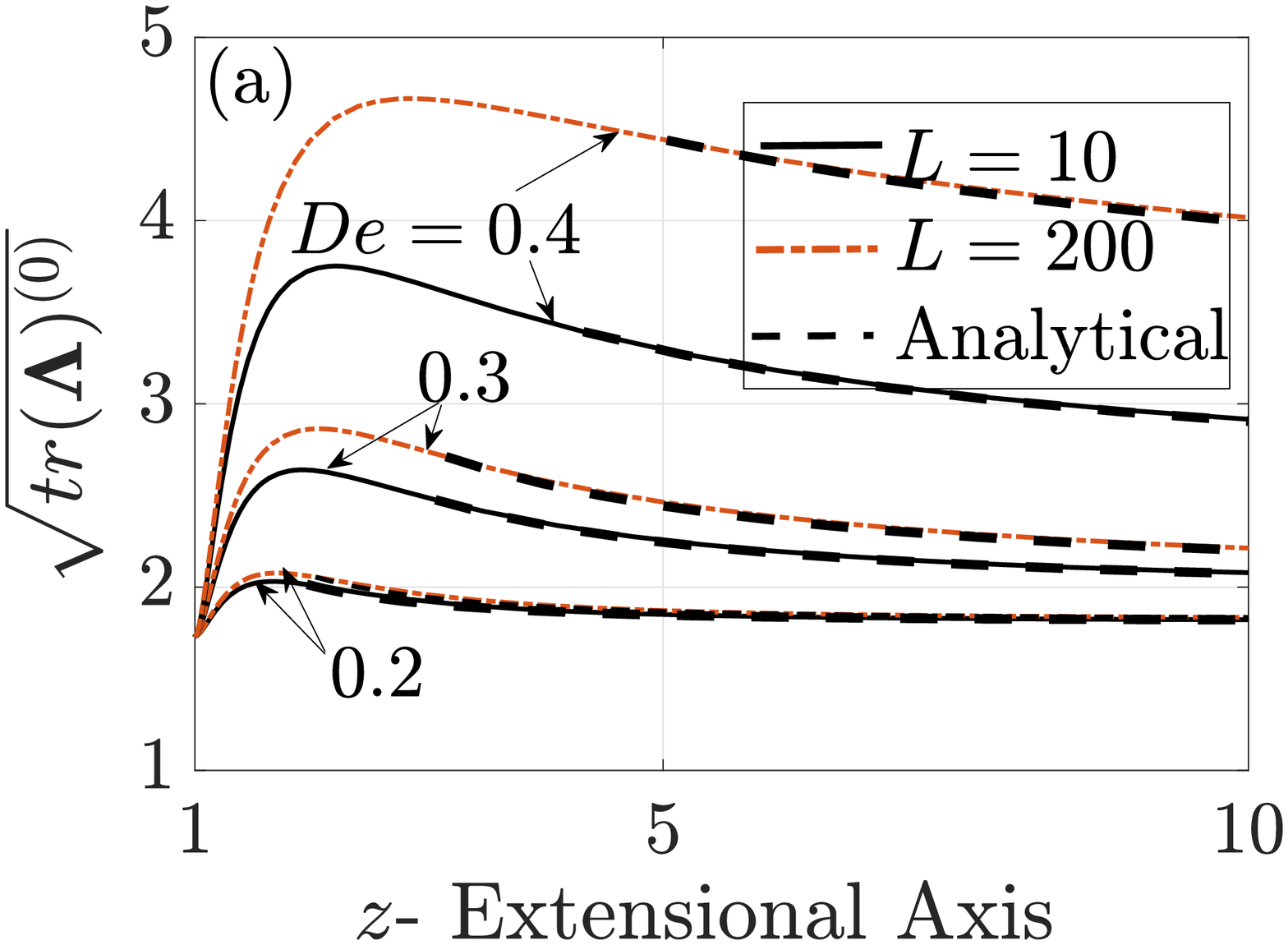}\label{fig:Stagnationstretch_smallDe}}\hspace{0.2in}
	\subfloat{\includegraphics[width=0.4\textwidth]{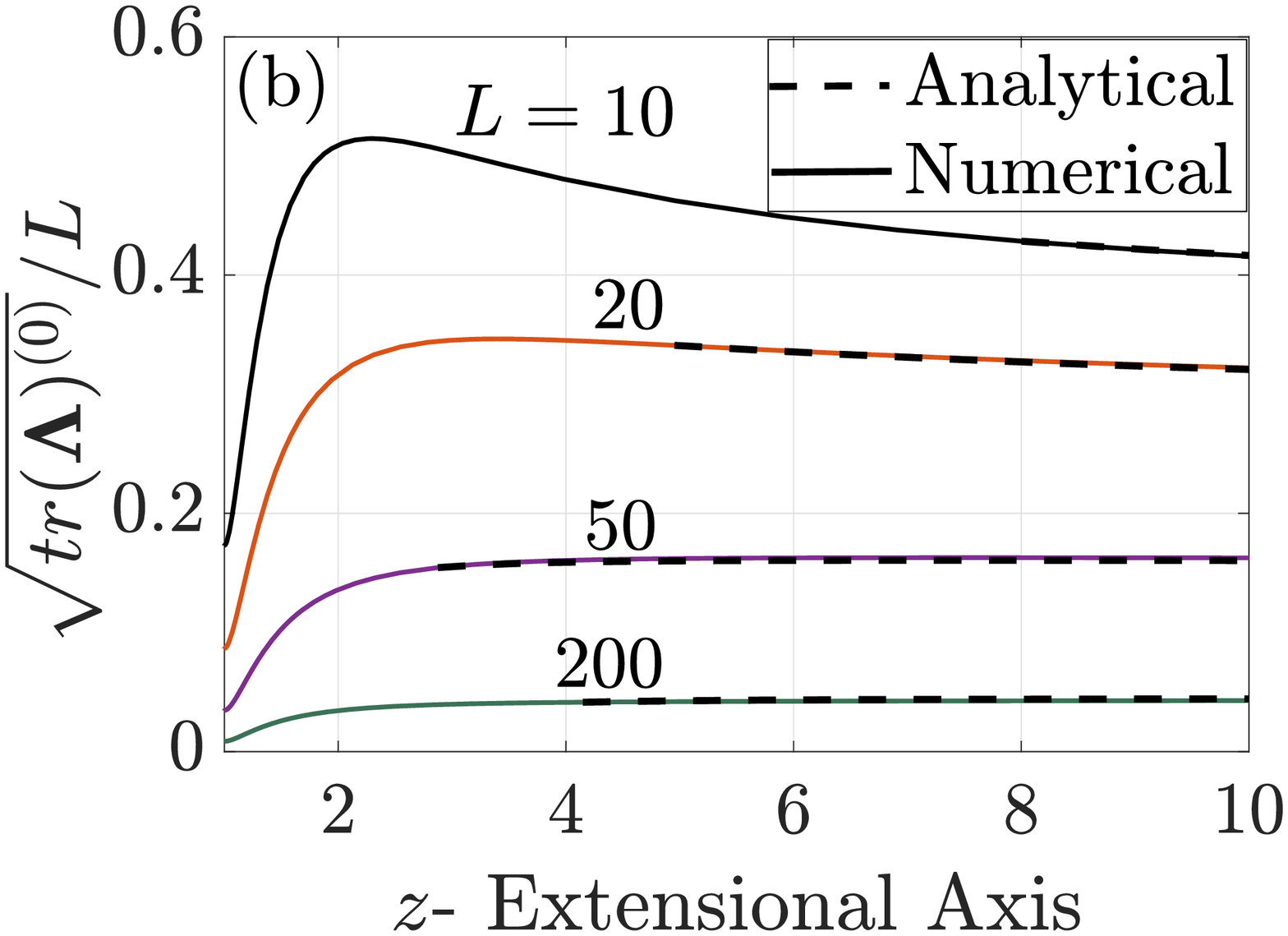}\label{fig:Stagnationstretch_Dept5}}\\	\subfloat{\includegraphics[width=0.4\textwidth]{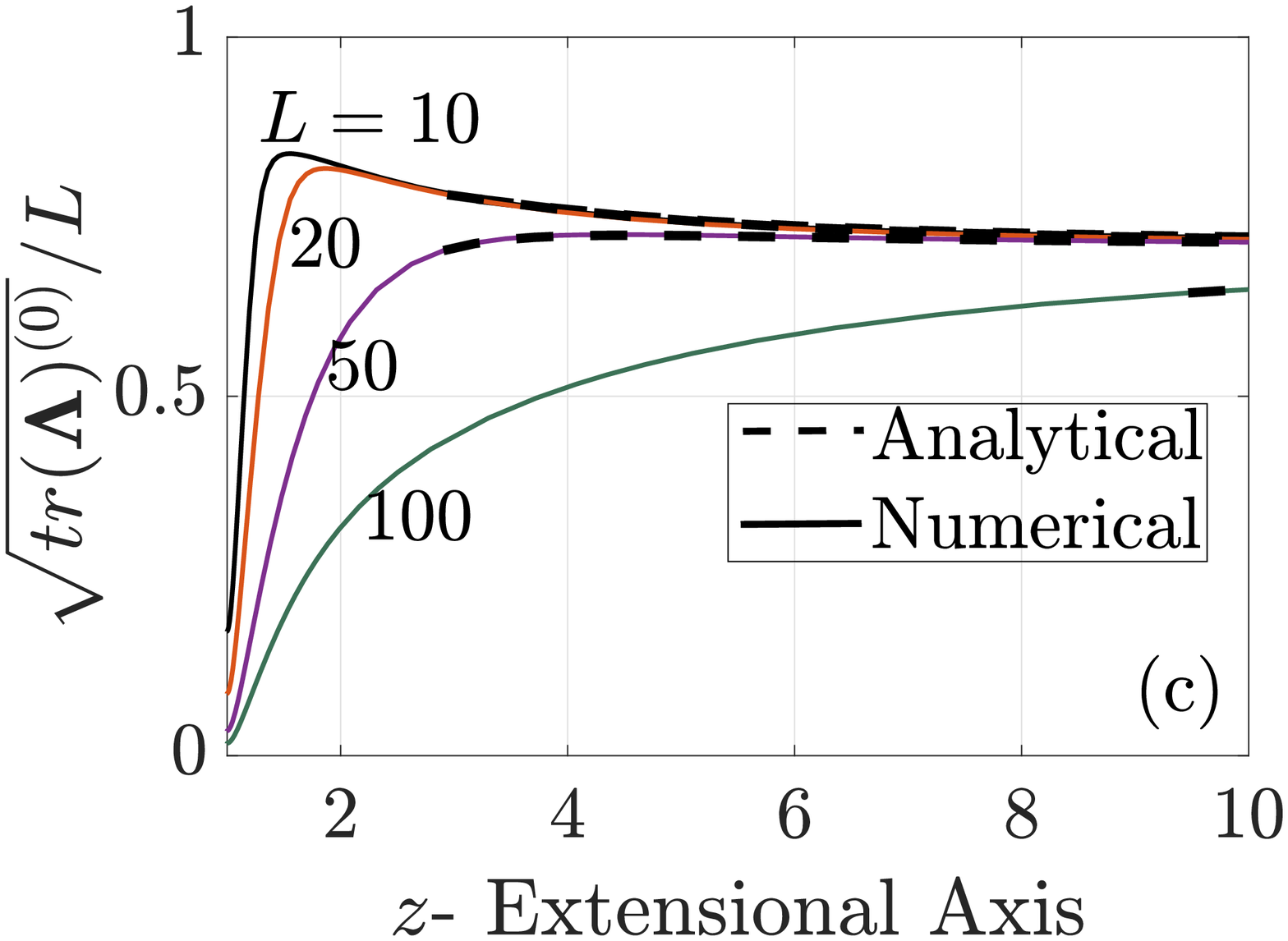}\label{fig:Stagnationstretch_De1}}\hspace{0.2in}
	\subfloat{\includegraphics[width=0.4\textwidth]{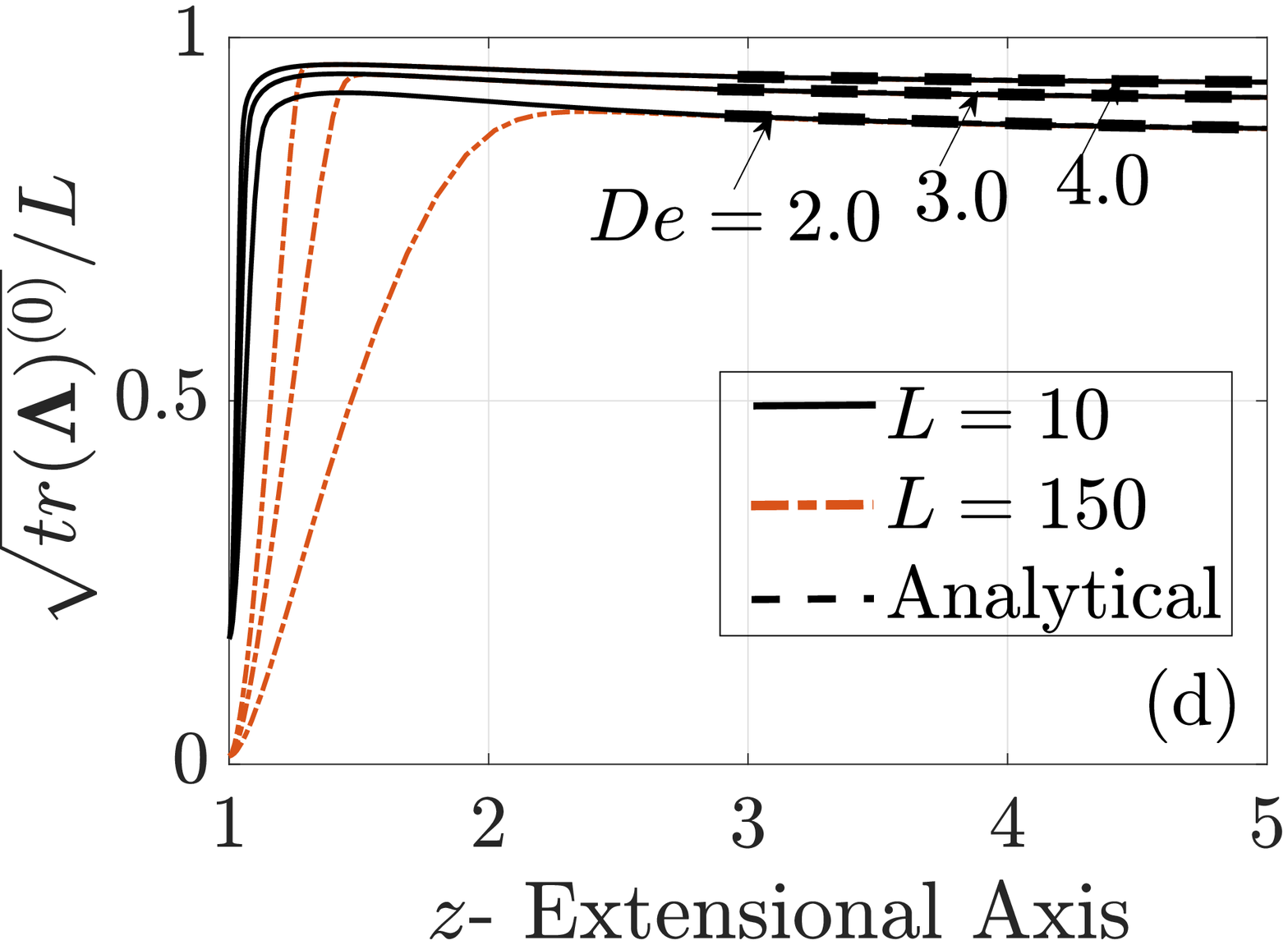}\label{fig:Stagnationstretch_largeDe}}
	\caption {Polymer stretch along the extensional axis for: (a) small De, (b) $De=0.5$, (c) $De=1.0$, and (d) large De. Stretch is normalized with $L$ for (b), (c) and (d).\label{fig:Stagnationstretch}}
\end{figure}

The only non-zero components of the velocity and its gradients on the stagnation streamline on the extensional axis ($r=0, z>1$), in the cylindrical coordinates $(r,\theta,z)$ are
\begin{eqnarray}
u_z=z+\frac{1.5}{z^4}-\frac{2.5}{z^2},\hspace{0.2in}\frac{\partial u_z}{\partial z}=-\frac{\partial u_r}{\partial r}=1-\frac{6}{z^5}+\frac{5}{z^3}.
\end{eqnarray}
Along the stagnation streamline, the streamwise velocity gradient ${\partial u_z}/{\partial z}$ starts from zero at the stagnation point, increases to a maximum value of about 1.7 at $z=\sqrt{2}$ and then decreases to the far-field value of 1. At low $De$, the polymers respond only to the local flow and the stretch of the polymer follows a similar qualitative pattern as ${\partial u_z}/{\partial z}$.
The equations governing the non-zero components of the configuration tensor in cylindrical coordinates are, \begin{equation}
u_z\frac{\text{d} \Lambda_{zz}}{\text{d} z}=2\frac{\partial u}{\partial z} \Lambda_{zz}+\frac{1}{De}(b-f\Lambda_{zz}),\hspace{0.05in}
u_z\frac{\text{d} \Lambda_{rr}}{\text{d} z}=-2\frac{\partial u}{\partial z} \Lambda_{rr}+\frac{1}{De}(b-f\Lambda_{rr}),\hspace{0.05in} \Lambda_{\theta\theta}=\Lambda_{rr}.
\end{equation}
At the stagnation point, $z=r=1$,
\begin{equation}
\Lambda_{rr}=\Lambda_{\theta\theta}=\Lambda_{zz}=1,
\end{equation}
leading to $f=b$ and zero polymer stress for all $De$ and $L$, i.e. the polymer at the stagnation point is in the equilibrium state due to the vanishing velocity gradient. A polymer traversing along this stagnation streamline starts to stretch from the nearly un-stretched state close to the stagnation point, and must obtain the stretch value corresponding to the one without the particle as $z\rightarrow \infty$. By examining the behavior of the polymer along the streamline, in the far field, we can determine whether this stretch is monotonic.
Splitting the configuration into the undisturbed ($\boldsymbol{\Lambda}^{(0U)}$) and the deviation from the undisturbed (${\boldsymbol{\Lambda}'}$) leads to the governing equations for the latter,
\begin{eqnarray}
\Lambda_{zz}={\Lambda}_{zz}'+\Lambda_{zz}^{(0U)},\hspace{0.2in}\Lambda_{rr}={\Lambda}_{rr}'+\Lambda_{rr}^{(0U)},\hspace{0.2in}f={f}'+f^{(0U)},\nonumber \\
u_z\frac{\text{d} {\Lambda}_{zz}'}{\text{d} z}=2\frac{\partial u_z'}{\partial z} {\Lambda}_{zz}^{(0U)}+2\frac{\partial u_z}{\partial z} {\Lambda}_{zz}'-\frac{1}{De}[{f}'\Lambda_{zz}^{(0U)}+{f}^{(0U)}{\Lambda}_{zz}'+{f}'{\Lambda}_{zz}'],\nonumber\\
u_z\frac{\text{d} {\Lambda}_{rr}'}{\text{d} z}=-\frac{\partial u_z'}{\partial z} {\Lambda}_{rr}^{(0U)}-\frac{\partial u_z}{\partial z} {\Lambda}_{rr}'-\frac{1}{De}[{f}'\Lambda_{rr}^{(0U)}+{f}^{(0U)}{\Lambda}_{rr}'+{f}'{\Lambda}_{rr}'],\label{eq:FarFieldDeviation}
\end{eqnarray}
where,
\begin{equation}
\frac{\partial u_z'}{\partial z}=-\frac{6}{z^5}+\frac{5}{z^3}.
\end{equation}
is the  deviation of the streamwise velocity gradient from the far field limit of 1. In the far field,
\begin{equation}
z\gg 1\hspace{0.2in}\rightarrow\hspace{0.2in}\frac{\partial u_z}{\partial z}\approx 1,\hspace{0.2in} u_z\approx z, \hspace{0.2in}\frac{\partial u_z'}{\partial z}\ll 1\label{eq:farfield_assumptions}.
\end{equation}
We do not know {\it a priori} the scaling of the different components of ${\boldsymbol{\Lambda}}'$ with $z$ in the far-field (large z). However, we assume that the far-field values of ${\Lambda}_{zz}'$ and ${\Lambda}_{rr}'={\Lambda}_{\theta\theta}'$ are smaller than their respective undisturbed polymer configuration components,
\begin{equation} {\Lambda}_{zz}'\ll\Lambda_{zz}^{(0U)},\hspace{0.2in} {\Lambda}_{rr}'\ll\Lambda_{rr}^{(0U)}\label{eq:farfield_assumptions2}.
\end{equation}
This is a valid approximation  because, in the far-field, the total polymer configuration approaches the undisturbed state. Thus, we linearize equations \eqref{eq:FarFieldDeviation} about the undisturbed values of the polymer configuration components, velocity and velocity gradients i.e. ignore the ${f}'{\Lambda}_{zz}'$ and ${f}'{\Lambda}_{rr}'$ terms and assume ${\Lambda}_{zz}'\gg{\Lambda}_{rr}' \rightarrow \text{tr}(\boldsymbol{{\Lambda}}')\approx{\Lambda}_{zz}'$. The latter assumption is valid for all but very small $De$ since the polymers on the extensional axis are aligned along the axis and we will see that the solution confirms the expectation. We obtain,
\begin{equation}
f=\frac{1}{1-\Lambda_{zz}/L^2}=f^{(0U)}\Bigg[1-\frac{f^{(0U)}}{L^2}{\Lambda}_{zz}'\Bigg]^{-1}\approx f^{(0U)}+\Bigg(\frac{f^{(0U)}}{L}\Bigg)^2{\Lambda}_{zz}'+\mathcal{O}\Big(\frac{(f^{(0U)})^3}{L^4}{\Lambda}_{zz}'^2\Big).
\end{equation}
Using this value of $f=f^{(0U)}+f'$ and computer algebra to integrate the linearized equations for ${\Lambda}_{zz}'$ and ${\Lambda}_{rr}'$ (ignoring ${f}'{\Lambda}_{zz}'$ and ${f}'{\Lambda}_{rr}'$ in corresponding equations from equation \eqref{eq:FarFieldDeviation})
from an arbitrary far-field location to $z\rightarrow\infty$ we obtain,
\begin{equation}
{\Lambda}_{zz}'=\frac{k_1}{z^\beta}+\frac{10\Lambda^{(0U)}_{zz}}{\beta-3}\frac{1}{z^3}+\frac{12\Lambda^{(0U)}_{zz}}{5-\beta}\frac{1}{z^5},\hspace{0.2in}
{\Lambda}_{rr}'={\Lambda}_{\theta\theta}'=\frac{k_2}{z^\gamma}+\frac{5\Lambda^{(0U)}_{rr}}{3-\gamma}\frac{1}{z^3}-\frac{6\Lambda^{(0U)}_{rr}}{5-\gamma}\frac{1}{z^5},\label{eq:approx_lambda}
\end{equation}
where,
\begin{equation}
\beta=\frac{1}{De}\Bigg\{\frac{(f^{(0U)})^2\Lambda^{(0U)}_{zz}}{L^2}+f^{(0U)}\Bigg\}-2,\hspace{0.2in}\gamma=\frac{1}{De}\Bigg\{\frac{(f^{(0U)})^2\Lambda^{(0U)}_{zz}}{L^2}+f^{(0U)}\Bigg\}+1, \label{eq:lambda_gamma}
\end{equation}
and $k_1$ and $k_2$ are constants to be obtained by matching these approximate solutions with the numerical solutions at a point beyond which the far-field is deemed to be applicable. The dominant term in the $z$- variation of ${\Lambda}_{zz}'$ is either $z^{-\beta}$ or $z^{-3}$ and in the variation of  ${\Lambda}_{rr}'$ ($={\Lambda}_{\theta\theta}'$) is either $z^{-\gamma}$ or $z^{-3}$. $f^{(0U)}$ can be approximated to a simple expression on either side of the coil-stretch transition. Equations \eqref{eq:smallDeEstimate} and \eqref{eq:largeDeEstimate} show that for $De<0.5$, $f^{(0U)}\approx1$ and $\Lambda^{(0U)}_{zz}\ll L^2$, and for $De>0.5$, $f^{(0U)}\approx2De$ and $\Lambda^{(0U)}_{zz}/L^2=1-1/2De$. Thus, from equation \eqref{eq:lambda_gamma}, $\gamma\ge3$ for every $De$ and the $z$- variation of the stretch depends on $\beta$. The result $\gamma\ge3, \forall De>0$ is compatible with our assumption that ${\Lambda}_{zz}'\gg{\Lambda}_{rr}' \rightarrow \text{tr}(\boldsymbol{{\Lambda}}')\approx{\Lambda}_{zz}'$. Also, using these estimates for $f^{(0U)}$,
\begin{equation}
\beta\approx\begin{cases}
\frac{1}{De}-2\hspace{0.2in}& De<0.5, L\gg 1,\\
4{De}-2\hspace{0.2in}& De>0.5.
\end{cases}\label{eq:beta_cases}
\end{equation}
When $\beta>3$, the dominant term in ${\Lambda}_{zz}'$ in the far-field is $z^{-3}$, and its coefficient in equation \eqref{eq:approx_lambda} is positive. This implies a larger than undisturbed stretch in the far-field. Hence, a non-monotonic variation of the stretch along the stagnation streamline is observed for these cases, as shown for $De=0.2$ in figure \ref{fig:Stagnationstretch_smallDe} and $De=3.0$, 4.0 and 5.0 in figure \ref{fig:Stagnationstretch_largeDe}. From equation \eqref{eq:beta_cases}, the condition $\beta>3$ is satisfied for $De\lesssim 0.2$ (strictly valid for $L\gg 1$) and $De\gtrsim1.25$.  For $De<0.2$, the undisturbed stretch is not very high and the polymer relaxation time is very small, i.e. polymers react immediately to the the local strain rate, and the extra strain rate created by the particle along the stagnation streamline causes the polymers to stretch more than the undisturbed value, before they contract to the latter as $z\rightarrow \infty$. For $De>1.25$, although the undisturbed stretch is very high, the increased local extension rate is enough to stretch the polymer more than the far-field value. The dominant variation for various $De$ regimes are
\begin{equation}
{\Lambda}_{zz}'\sim\begin{cases}
z^{2-\frac{1}{De}}\hspace{0.2in}& 0.2<De<0.5,\\
z^{2-4{De}}\hspace{0.2in}& 0.5<De<1.25,\\
z^{-3}\hspace{0.2in}& De<0.2, De>1.25.
\end{cases}\label{eq:farFieldfactor}
\end{equation}
For $0.2<De<1.25$, the fully analytical approach cannot ascertain far-field growth or decay, since the sign of the dominant term, $k_1$ in equation \ref{eq:approx_lambda}, is determined by matching with the full numerical solution.

The above analytical estimates are for the linearized (about undisturbed values of polymer configuration, velocity and velocity gradients) constitutive equations. They described the qualitative features observed in the numerical results. Incorporating quadratic non-linearities allow an almost exact match with the numerical solution in the far-field. While we do not show these unwieldy analytical expressions, they are incorporated in the plots (dashed black curves) shown in figure \ref{fig:Stagnationstretch}. A good match of these far-field analytical estimates and the actual numerical solution for a wide range of $De$ and $L$ is observed in figure \ref{fig:Stagnationstretch}. The analytical result for very small and very large $De$ mentioned above, i.e. an initial increase in the stretch which is larger than the undisturbed or far-field value, is confirmed from figures \ref{fig:Stagnationstretch_smallDe} and  \ref{fig:Stagnationstretch_largeDe}.

For $De<0.5$, the undisturbed stretch is independent of $L$, and the different curves for the same $De$ in figure \ref{fig:Stagnationstretch_smallDe} must asymptote to the same value. However, along the stagnation streamline, closer to the particle, the stretch at a given $De$ is larger for larger $L$. This difference increases with $De$, because the polymers start from the nearly unstretched state and the larger strain rate (than the far-field value) on the stagnation streamline close to the particle allows the polymer with larger $L$ to be extended more. For $0.2<De<0.5$, we are unable to analytically predict the extra stretch on the stagnation streamline, but the numerical evidence suggests this to be the case for all $L$.

For the $De\ge0.5$ plots in figure \ref{fig:Stagnationstretch_Dept5} to \ref{fig:Stagnationstretch_largeDe}, the stretch along the streamline is scaled with $L$, because the undisturbed stretch scales as $L$. For $De\ge3$ in figure \ref{fig:Stagnationstretch_largeDe}, the maximum stretch is reached much earlier for smaller $L$ due to the limited extensibility. For a particular $L$, there is an increase in stretch with $De$ (or imposed far field extension rate) at any given location on the stagnation streamline.

For the intermediate values of $De=0.5$ and 1.0 in figure \ref{fig:Stagnationstretch_Dept5} and \ref{fig:Stagnationstretch_De1}, the overshoot in the stretch due to the local increase in strain rate along the stagnation streamline occurs only for small $L$. For large $L$ the undisturbed stretch (which scales as $L$) is very high. The slow variation of the far-field stretch for $De=0.5$ in figure \ref{fig:Stagnationstretch_Dept5} for large $L$ matches the analytical prediction of equation \eqref{eq:farFieldfactor}. This is strong evidence for the need to remove the linear part of the polymer stress before ensemble-averaging the stress. This will be further discussed in section \ref{sec:RheologyTheory}.

{The polymer stretch field in the case of a uniform flow past a sphere investigated by Chilcott \& Rallison \cite{chilcott1988creeping} can be compared with that in a uniaxial extensional flow considered here. In a uniform flow, there are two stagnation points on the particle surface in the flow direction. In a uniaxial extensional flow, there are two stagnation points along the extensional axis and a stagnation line in the compressional plane. The neighboring stagnation points are 180$^\circ$ apart in uniform flow while they (on a plane including the extensional axis and passing through the center of the sphere) are 90$^\circ$ apart in uniaxial extensional flow. Around the front/ upstream stagnation point in uniform flow, fluid undergoes a biaxial extension, and around the rear/ downstream stagnation point a uniaxial extensional flow. Thus, the polymers undergo a larger stretch around the rear stagnation point than the front stagnation point. The polymer stretch is fore-aft symmetric in uniaxial extensional flow. In both flows, the polymers are in their equilibrium/ un-stretched configurations at the stagnation points as all the velocity gradient components are zero. Along the rear stagnation streamline in uniform flow, the polymers are first stretched and then advected far downstream of the particle surface before they relax to their equilibrium configuration. This is similar one type of behavior observed for extensional flow, where an overshoot in polymer stretch occurs close to the particle surface as a polymer translates along the extensional axis before the polymer relaxes to its undisturbed (but non-equilibrium) configuration.  The other type of behavior where the polymer stretches monotonically toward its The other scenario where the polymer stretch monotonically increases along the rear stagnation streamline up to the highly stretched far-field configuration is not found in the uniform flow as the undisturbed polymers are in equilibrium configuration. Similar to the uniaxial extensional flow (figure \ref{fig:surfaceConfiguration}), the polymer stretch along the particle surface is finite between the unstretched state at two stagnation points of the uniform flow.}

\subsection{Polymer configuration in the fluid surrounding the sphere}\label{sec:Fullpolymerconfiguration}
From section \ref{sec:UndisturbedConfiguration}, for large $L$, the undisturbed polymer stress, $\hat{\boldsymbol{\Pi}}^{(0U)}$, is independent of $L$ for $De\lesssim0.5$, and scales as $L^2$ for $De\gtrsim0.5$. Starting from the FENE-P equations, we show that for $L\gg1$ these scalings are valid for the polymer stress, ${\boldsymbol{\Pi}}$, in most of the fluid region, even in the presence of the particle. Before the coil-stretch transition, $||\boldsymbol{\Lambda}||_\text{max}\ll L^2$, for $L\gg1$, and an approximate form for the FENE-P configuration equation \eqref{eq:Configuration} follows,
\begin{equation}\label{eq:Configuration_SmallDe}
\frac{\partial \boldsymbol{\Lambda}}{\partial t}+\mathbf{u}\cdot \nabla \boldsymbol{\Lambda}=\nabla \mathbf{u}^\text{T}\cdot\boldsymbol{\Lambda}+\boldsymbol{\Lambda}\cdot\nabla\mathbf{u}+\frac{1}{De}(\boldsymbol{\delta}-\boldsymbol{\Lambda}).
\end{equation}
This is equivalent to the Oldroyd-B equation. After the coil-stretch transition, using $\text{tr}(\boldsymbol{\Lambda}) \gg 1$, for $L\gg 1$ ($b\approx 1$), the FENE-P configuration equation \eqref{eq:Configuration} is simplified to,
\begin{equation}\label{eq:Configuration_LargeDe}
\frac{\partial \tilde{\boldsymbol{\Lambda}}}{\partial t}+\mathbf{u}\cdot \nabla \tilde{\boldsymbol{\Lambda}}=\nabla \mathbf{u}^\text{T}\cdot\tilde{\boldsymbol{\Lambda}}+\tilde{\boldsymbol{\Lambda}}\cdot\nabla\mathbf{u}-\frac{\tilde{\boldsymbol{\Lambda}}}{De(1-\text{tr}(\tilde{\boldsymbol{\Lambda}}))},
\end{equation}
where $\tilde{\boldsymbol{\Lambda}}=\boldsymbol{\Lambda}/L^2$. Thus for $L\gg1$, we expect the dominant components of $\boldsymbol{\Lambda}$ to be independent of $L$ before the coil-stretch transition and scale with $L^2$ after. Using $f\approx b\approx 1$ for $De\lesssim0.5$, we simplify ${\boldsymbol{\Pi}}\approx(\boldsymbol{\Lambda}-\boldsymbol{\delta})/De$.  On the other hand, using $\boldsymbol{\Lambda}-\boldsymbol{\delta}\approx \boldsymbol{\Lambda}$, for $De\gtrsim 0.5$, we can simplify ${\boldsymbol{\Pi}}\approx(1/De) \boldsymbol{\Lambda}/(1-\text{tr}(\boldsymbol{\Lambda}/L^2))$. Hence, the polymer stress, ${\boldsymbol{\Pi}}$, also follows the $L$ independent and $L^2$ scaling, below and above $De=0.5$, respectively.

In certain regions very close to the sphere such as near the stagnation points, the polymers collapse to a nearly equilibrium state for every $L$ and $De$, due to the small velocity gradients and velocity (hence they spend enough time in these regions with small velocity gradients to collapse). Therefore, in these collapsed regions, the approximate form of the constitutive equation \eqref{eq:Configuration_LargeDe} for $De\gtrsim0.5$, and the $L^2$ dependence of ${\boldsymbol{\Pi}}$, based on $\text{tr}(\boldsymbol{\Lambda}) \gg 1$ is not valid. 
However, the approximate form of equation \eqref{eq:Configuration_SmallDe} for $De\lesssim0.5$, and the $L$ independence of ${\boldsymbol{\Pi}}$ is valid everywhere for large $L$.

For small polymer concentration, $c$, assuming the polymer to be in the undisturbed configuration far upstream of the sphere, we solve equation \eqref{eq:Configuration} under the steady-state assumption of equation \eqref{eq:steadystate} and velocity $\boldsymbol{u}=\boldsymbol{u}^{(0)}$ from equation \eqref{eq:u_field}. It is solved along a dense set of streamlines around the sphere for a wide range of $De$ and $L$ and the change in stretch due to the particle,
\begin{equation}
\Delta\mathcal{S}=\sqrt{\text{tr}(\boldsymbol{\Lambda}^{(0)})}-\sqrt{\text{tr}(\boldsymbol{\Lambda}^{(0U)})},
\end{equation}
 is analysed. We provided a first validation of the numerical calculations in figure \ref{fig:surfaceConfiguration}, where the numerically evaluated surface polymer stretch matches perfectly with that obtained by solving equation \eqref{eq:analytical_surface_stress} using computer algebra and using equation \eqref{eq:exact_surface_configuration}.  Further validation for the numerical calculations is presented later in table \ref{tab:Validation} in section \ref{sec:Rheology}, by comparing the rheological quantities at small $De$, to the ones availed theoretically \cite{koch2006stress,einarsson2018einstein}.

\begin{figure}[h!]
	\centering
	\subfloat{\includegraphics[width=0.22\textwidth]{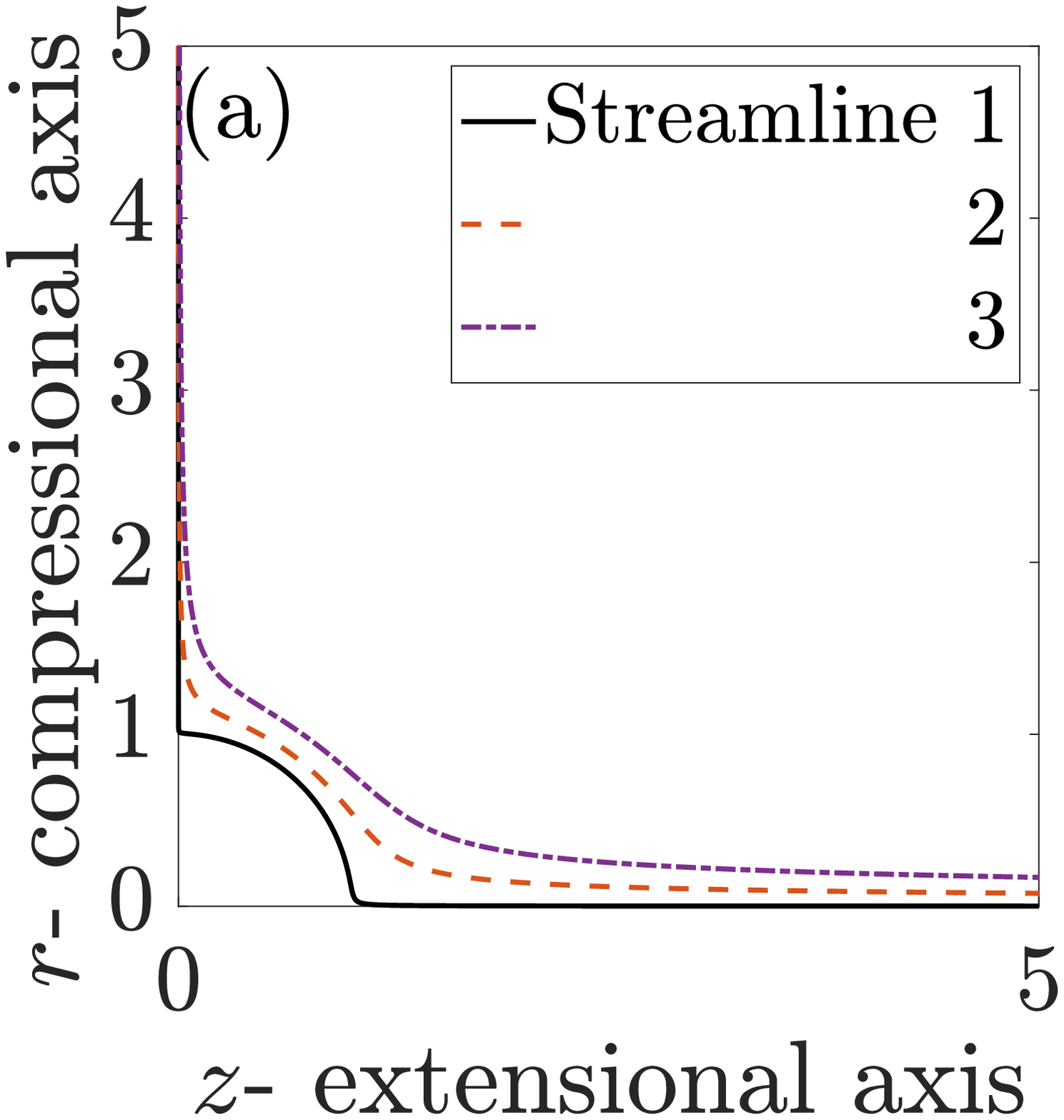}\label{fig:streamlines}}\hfill
	\subfloat{\includegraphics[width=0.33\textwidth]{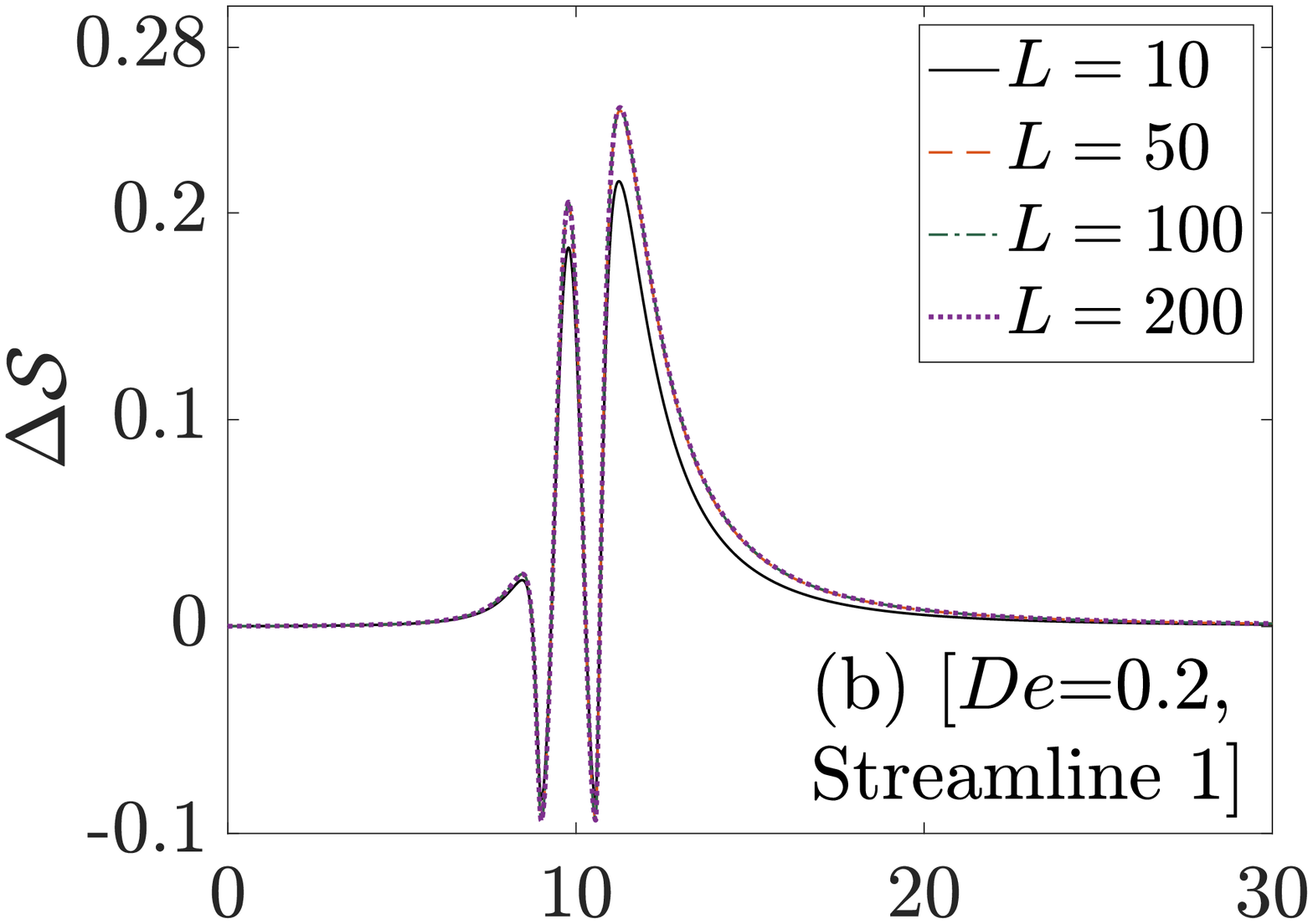}\label{fig:streamlinesLegend}}\hfill	\subfloat{\includegraphics[width=0.33\textwidth]{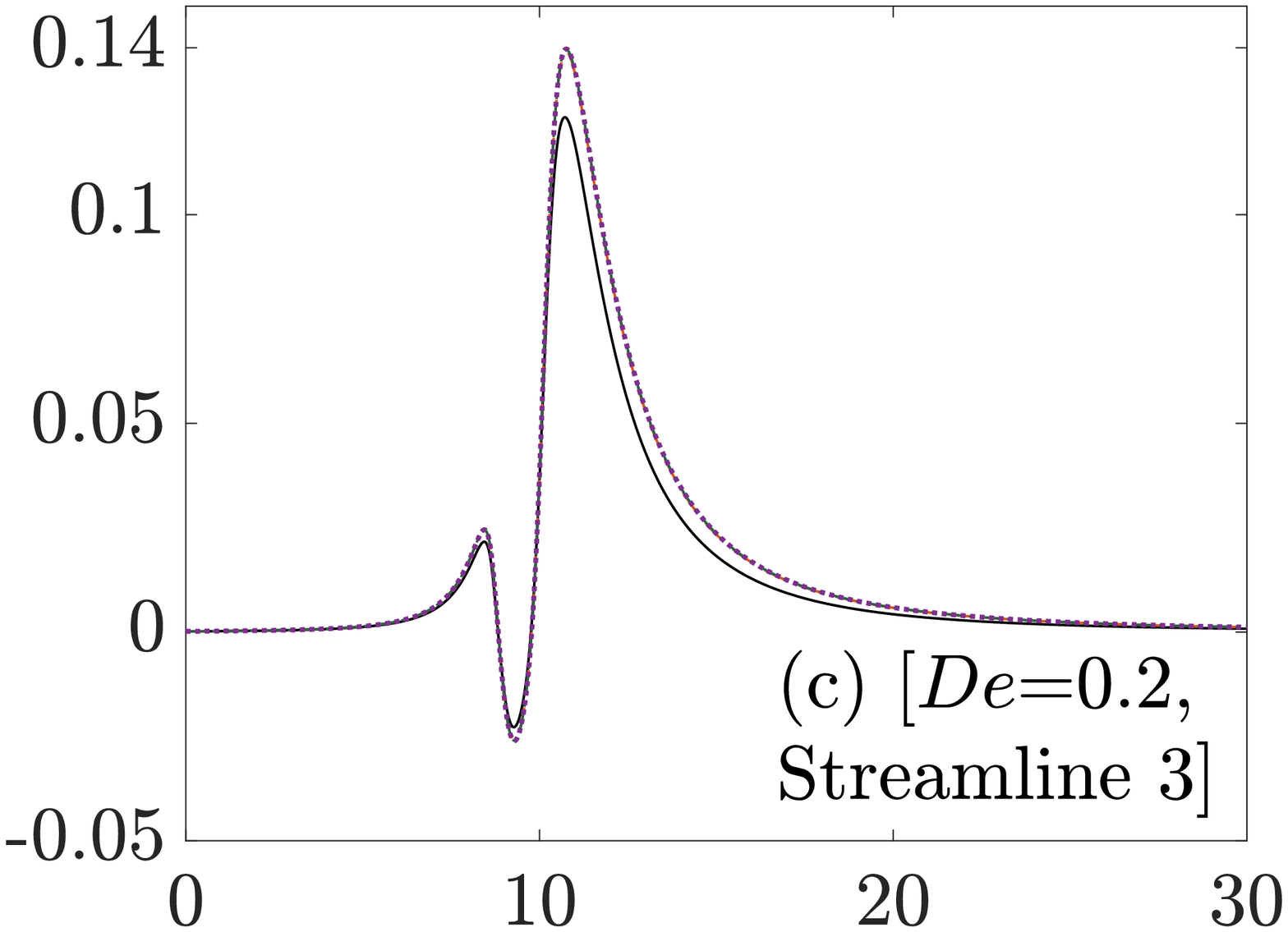}}\\\hspace{1.8in}
	\subfloat{\includegraphics[width=0.32\textwidth]{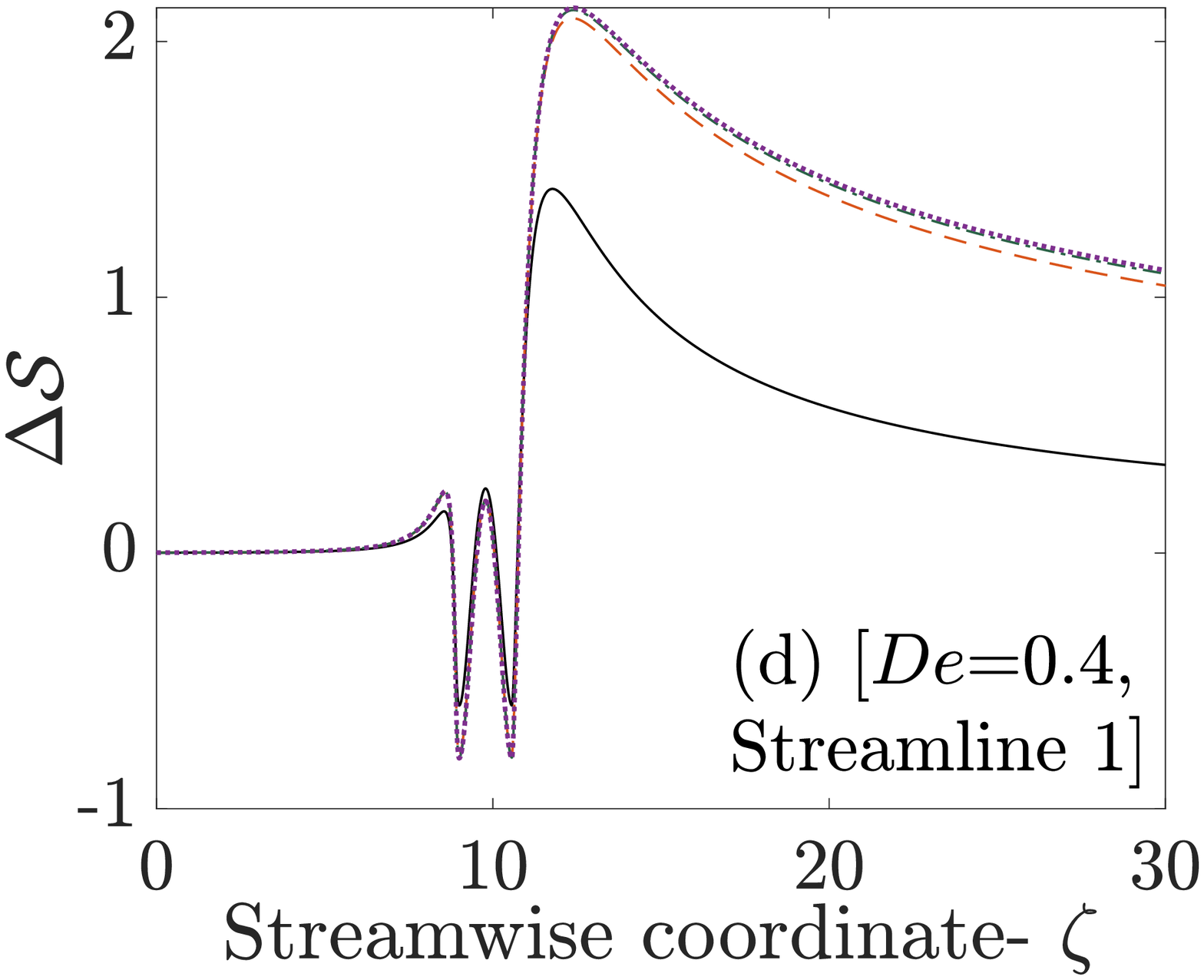}}\hfill	\subfloat{\includegraphics[width=0.315\textwidth]{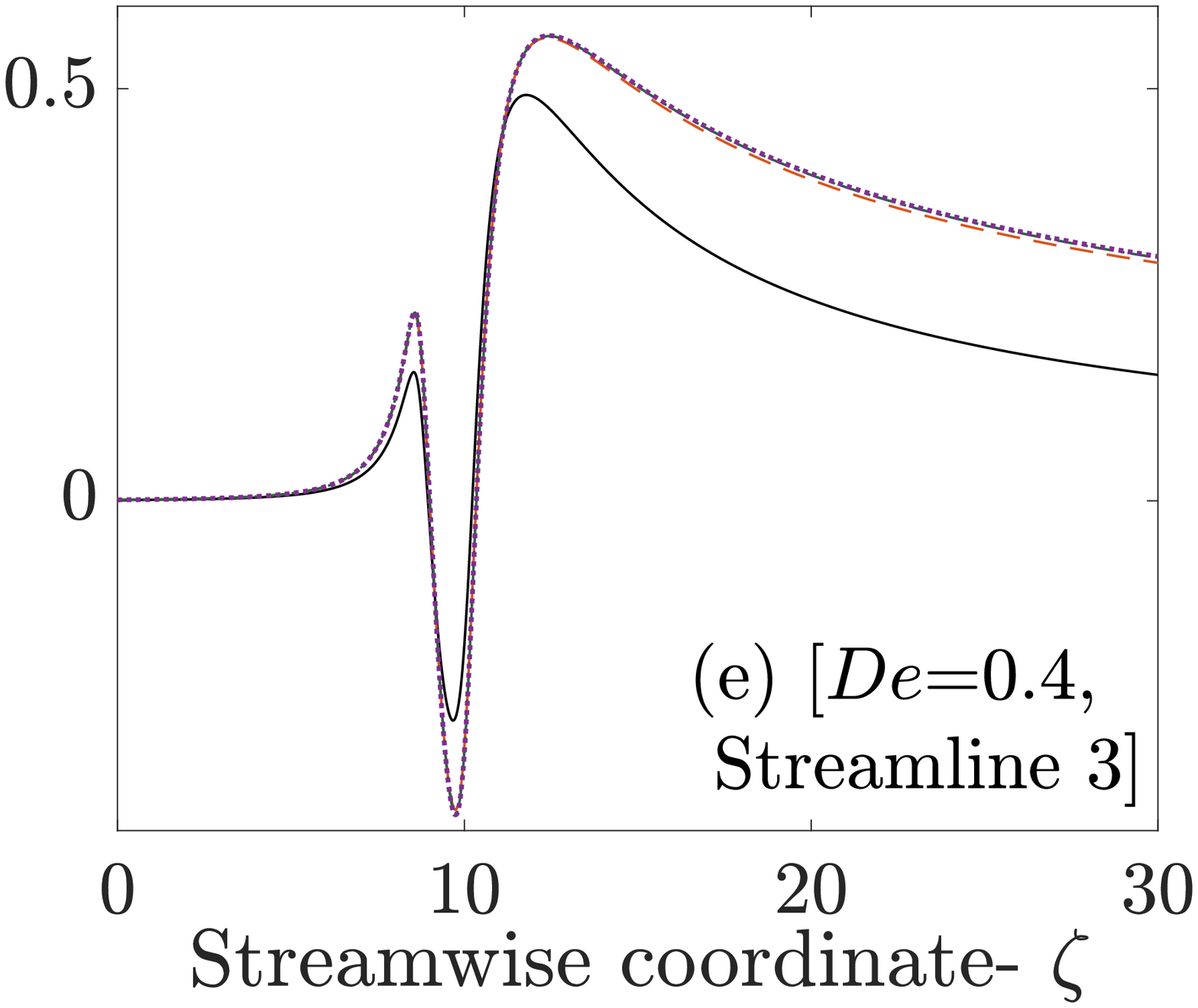}}
	\caption {$\Delta\mathcal{S}$ along streamlines indicated in figure (a). $\Delta\mathcal{S}$ for two different $De$ along streamlines 1 and 3 for various $L$. Figures (b) to (e) share the same legend. \label{fig:StreamlineStretch1}}
\end{figure}

\begin{figure}[h!]
	\centering
	\subfloat{\includegraphics[width=0.33\textwidth]{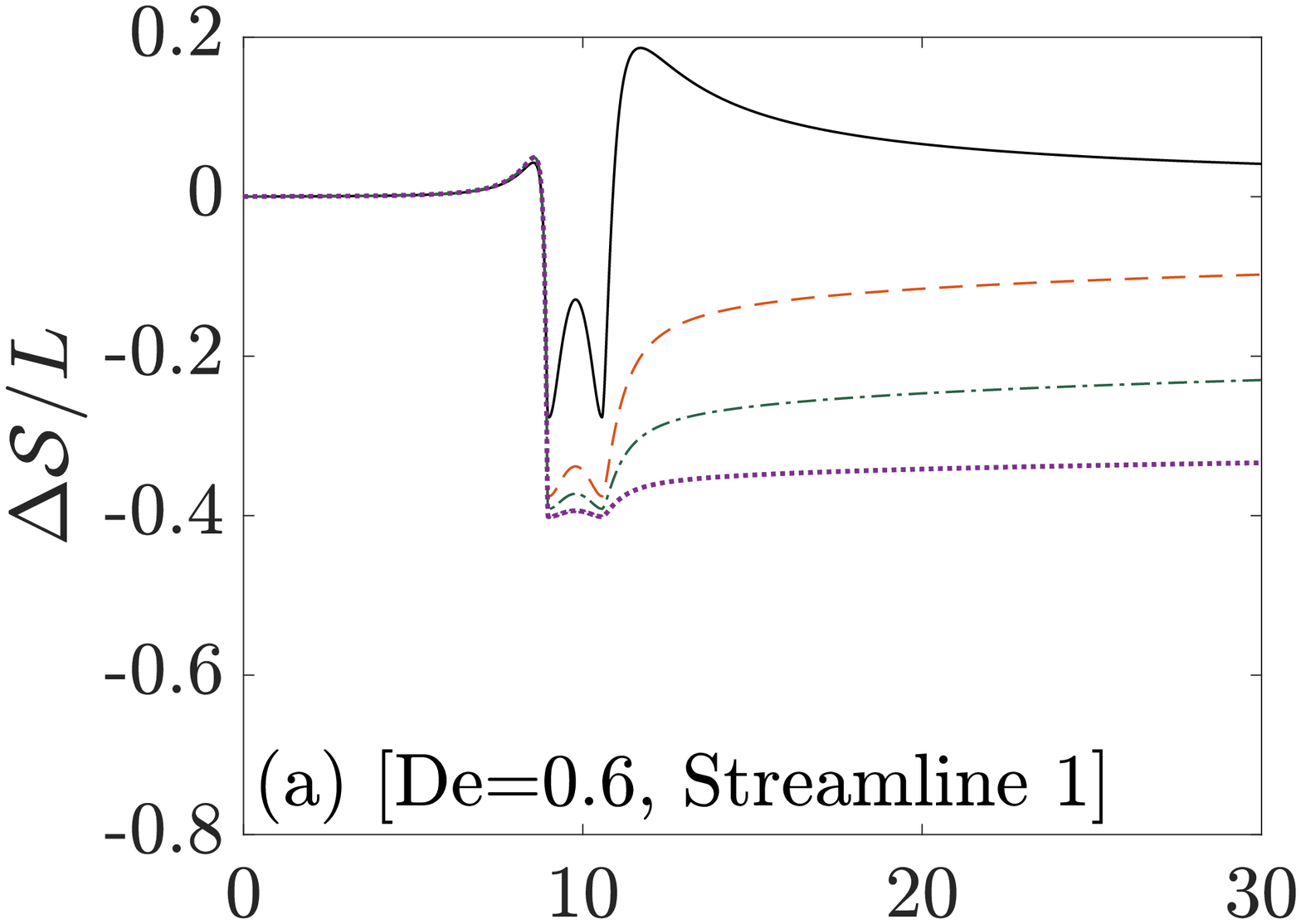}}\hfill
	\subfloat{\includegraphics[width=0.33\textwidth]{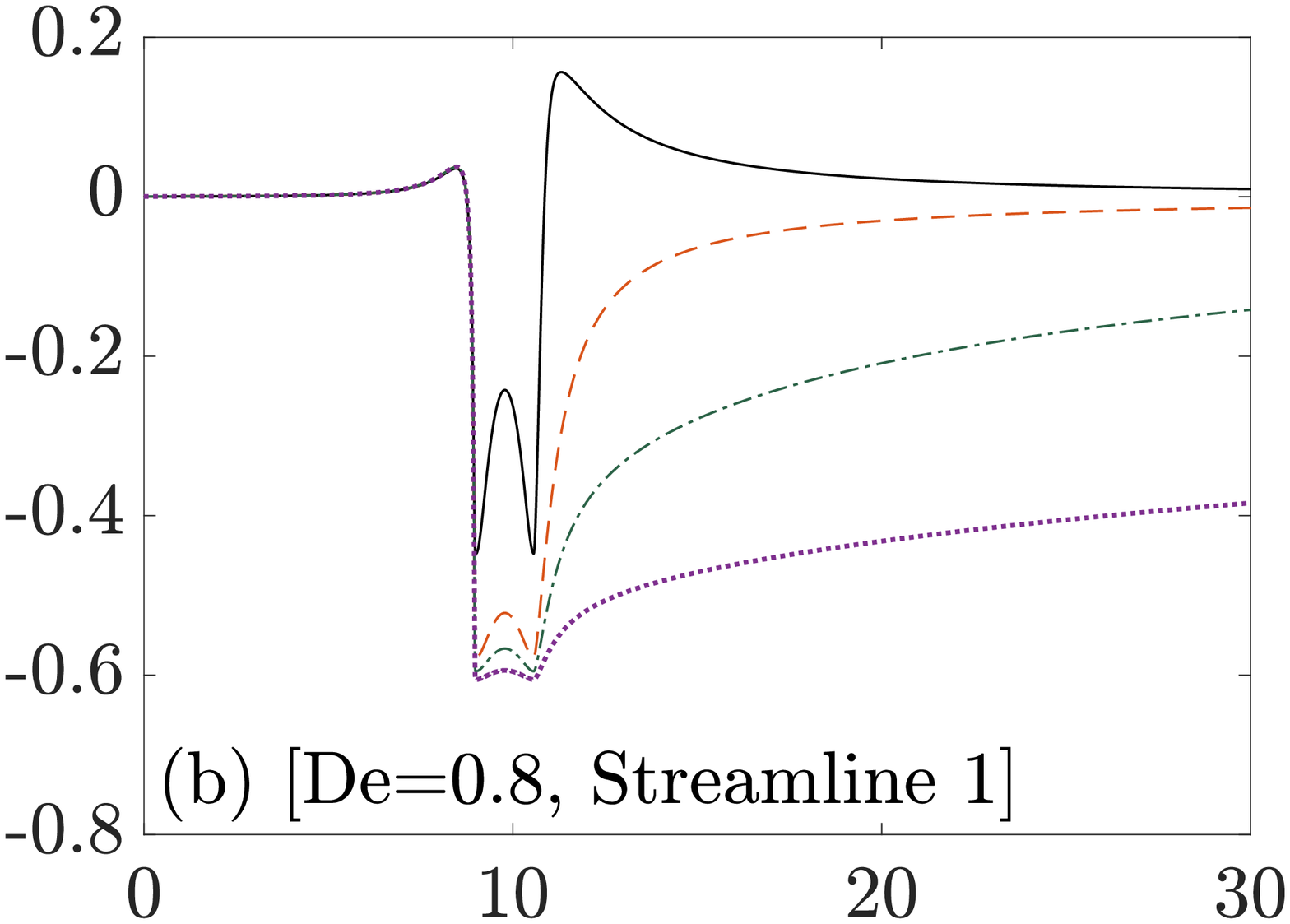}}\hfill
	\subfloat{\includegraphics[width=0.33\textwidth]{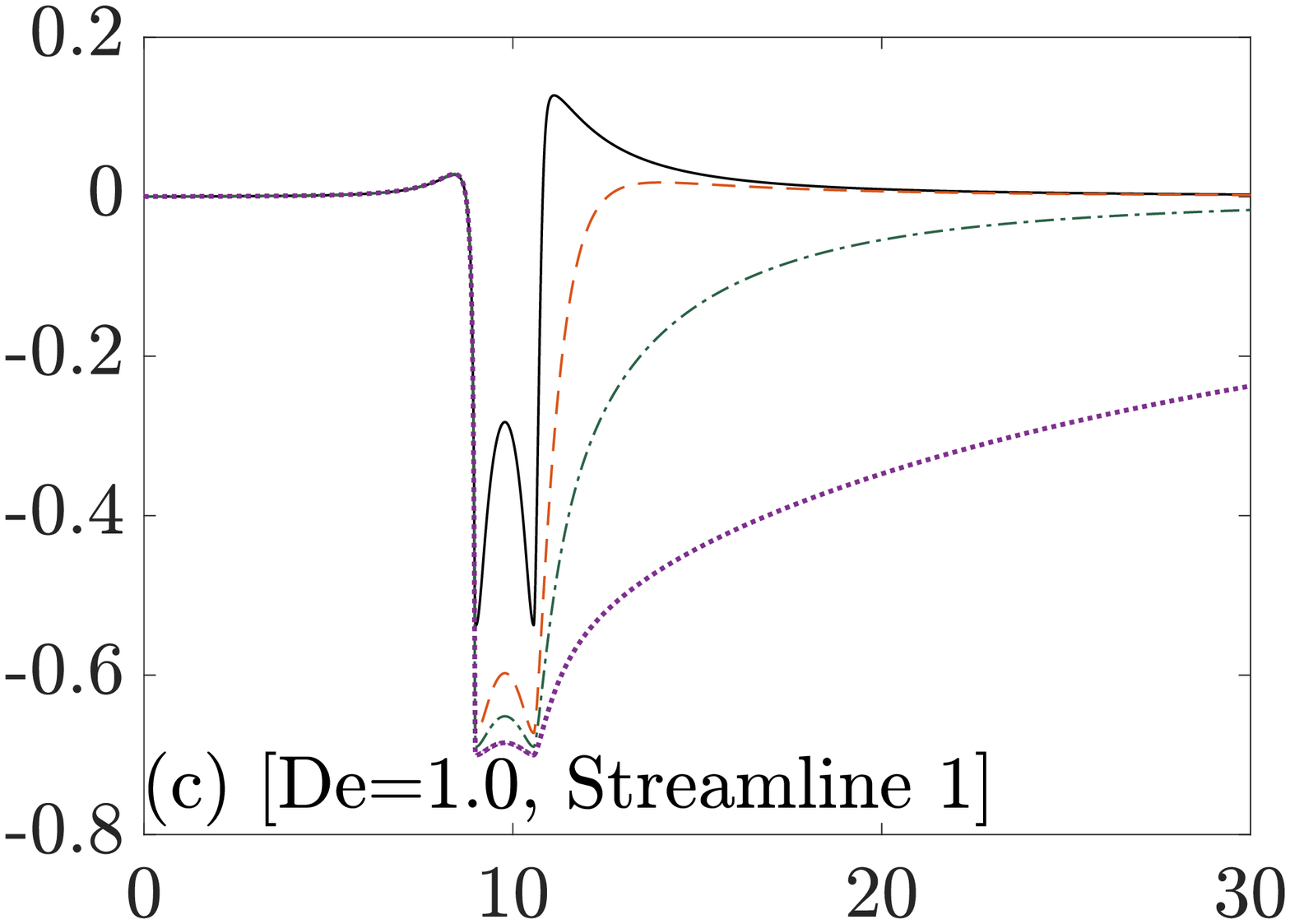}}\\
	\subfloat{\includegraphics[width=0.33\textwidth]{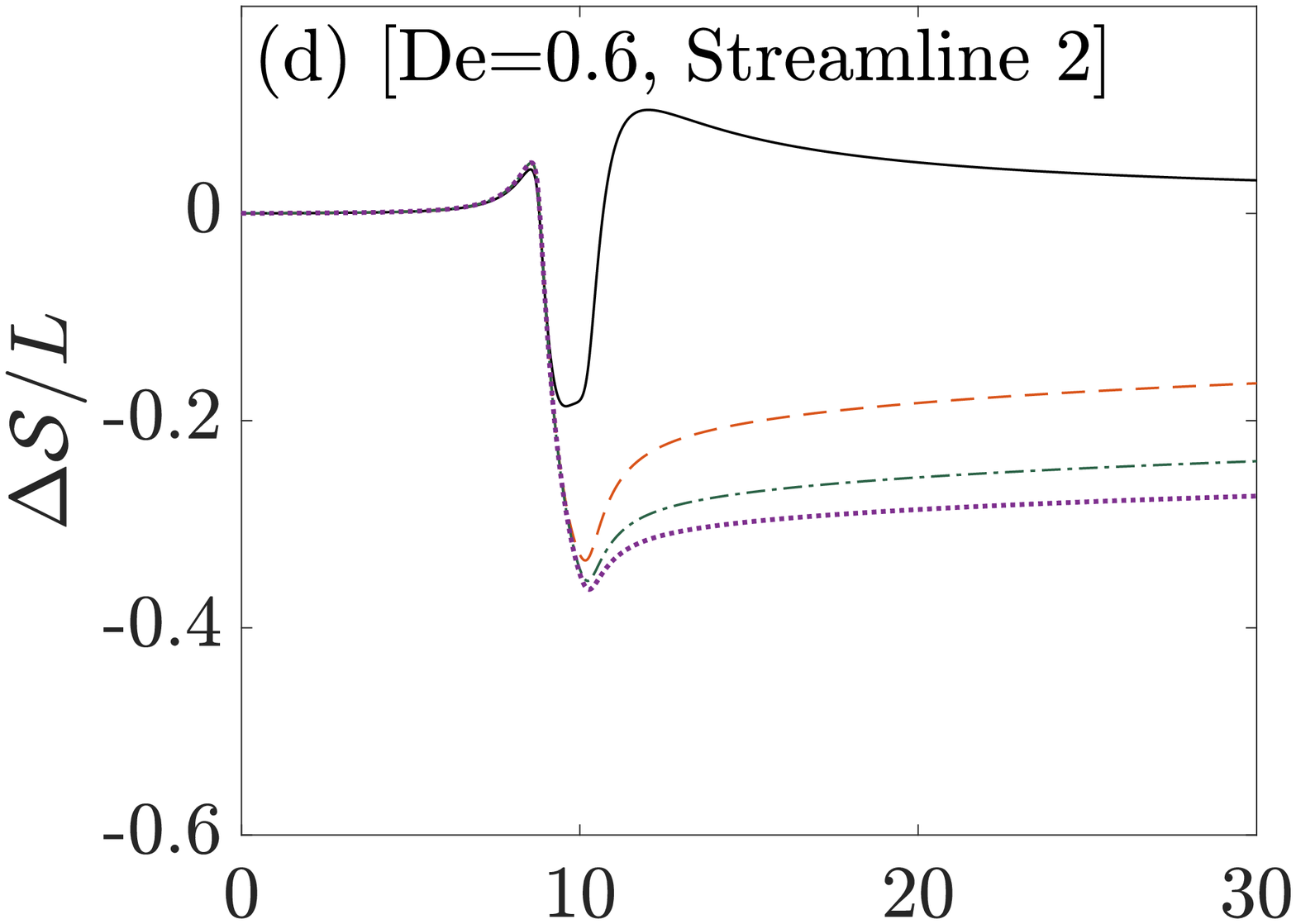}}\hfill
	\subfloat{\includegraphics[width=0.33\textwidth]{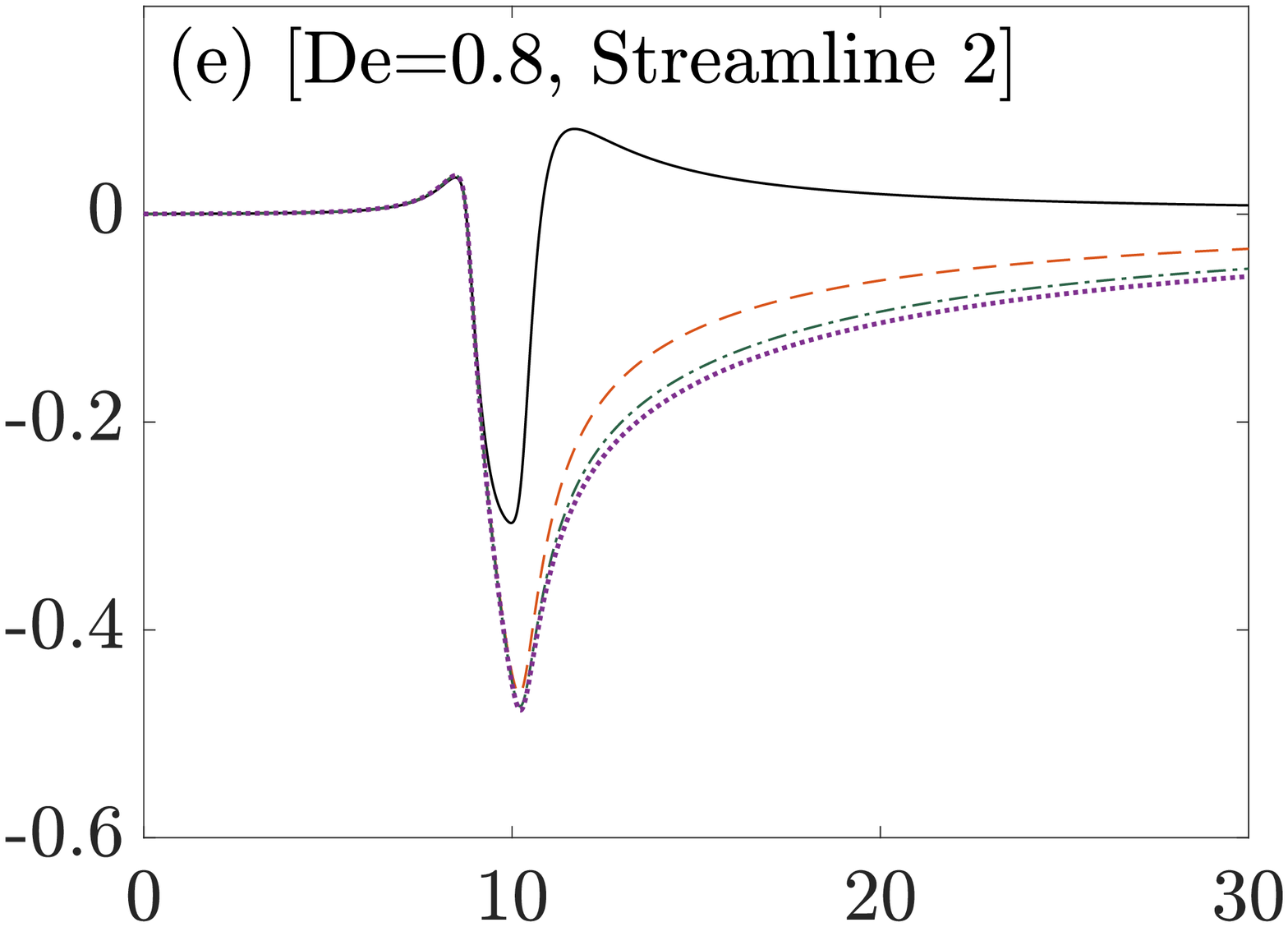}}\hfill
	\subfloat{\includegraphics[width=0.33\textwidth]{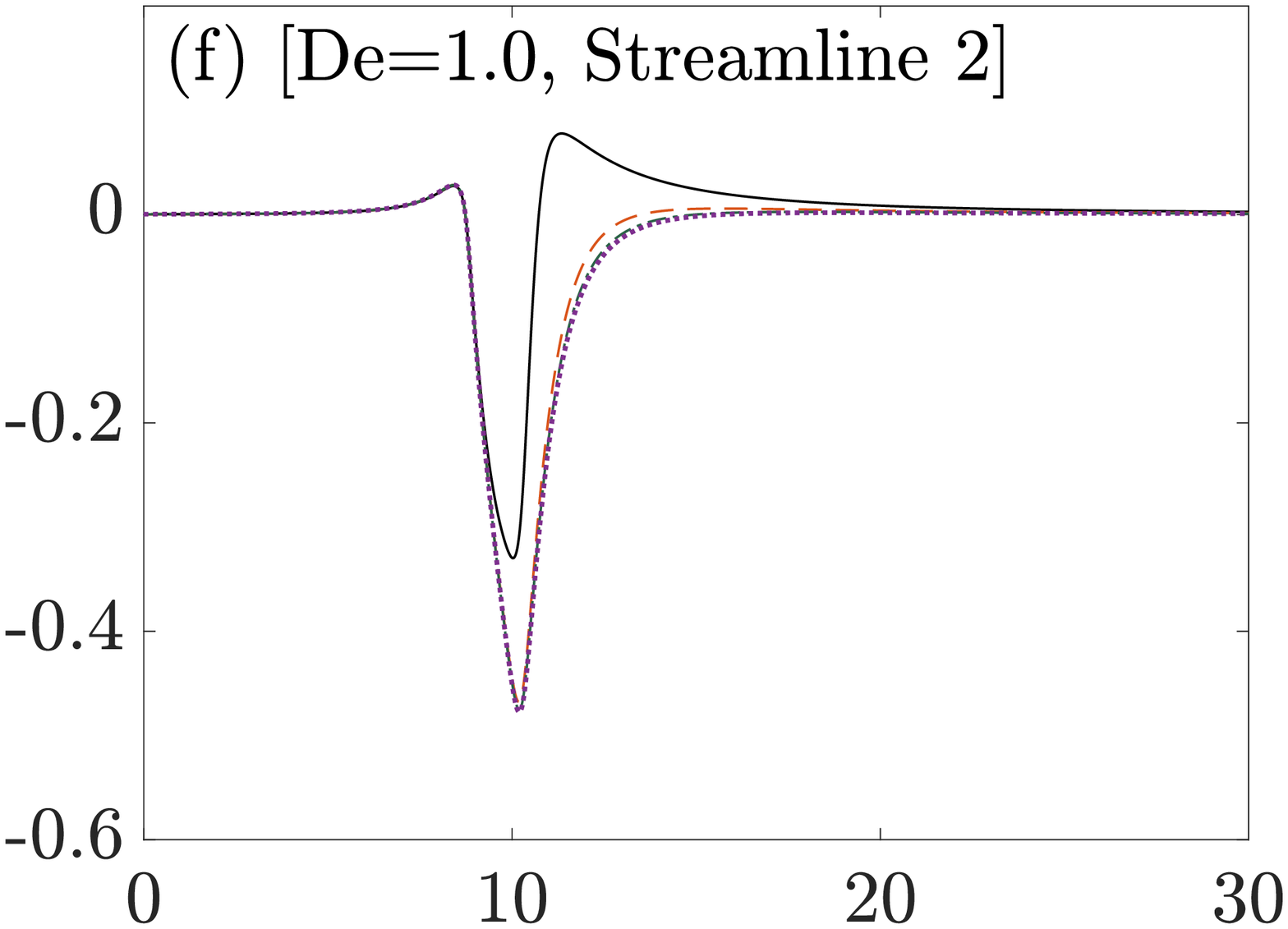}}\\
	\subfloat{\includegraphics[width=0.33\textwidth]{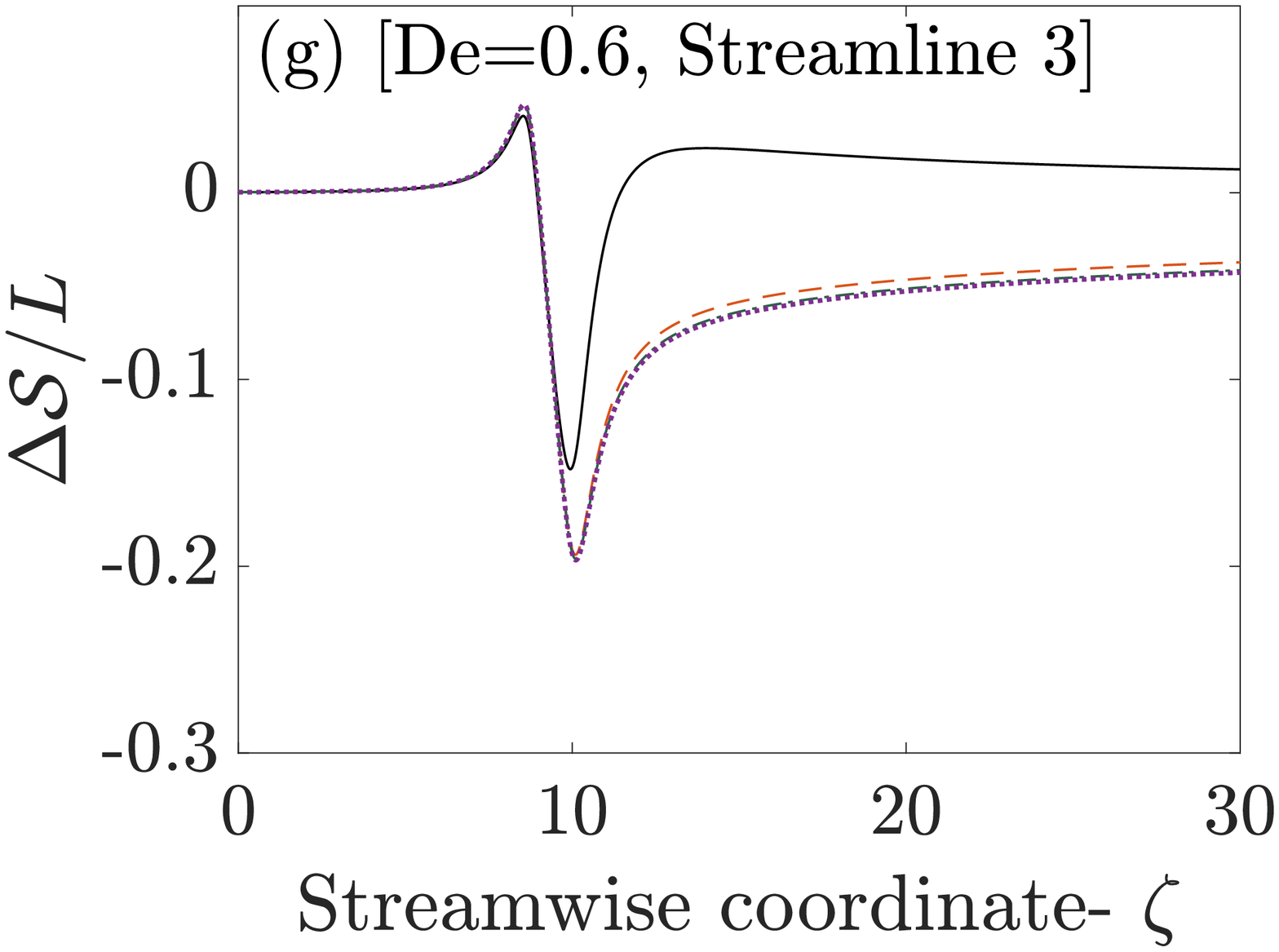}}\hfill
	\subfloat{\includegraphics[width=0.33\textwidth]{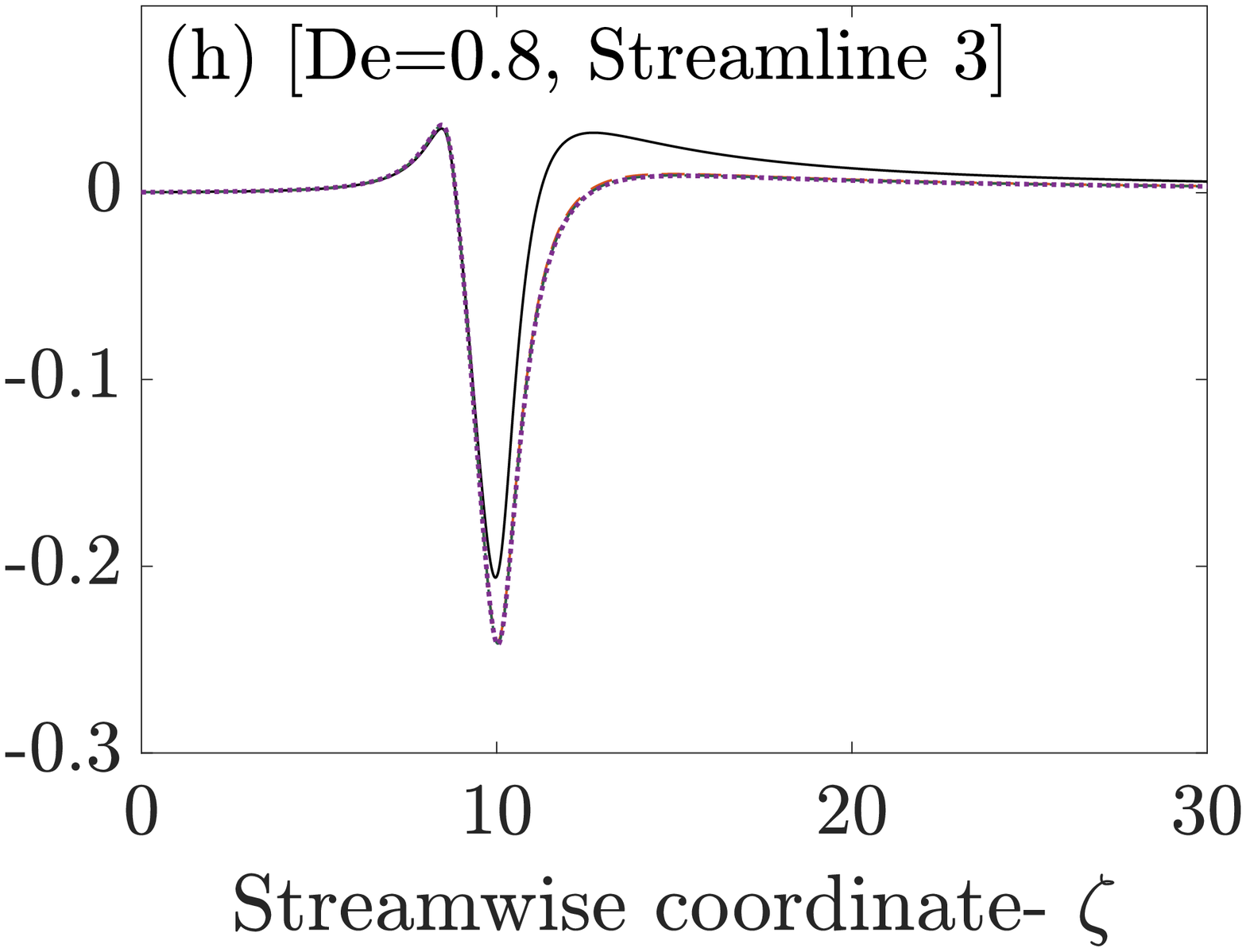}}\hfill
	\subfloat{\includegraphics[width=0.33\textwidth]{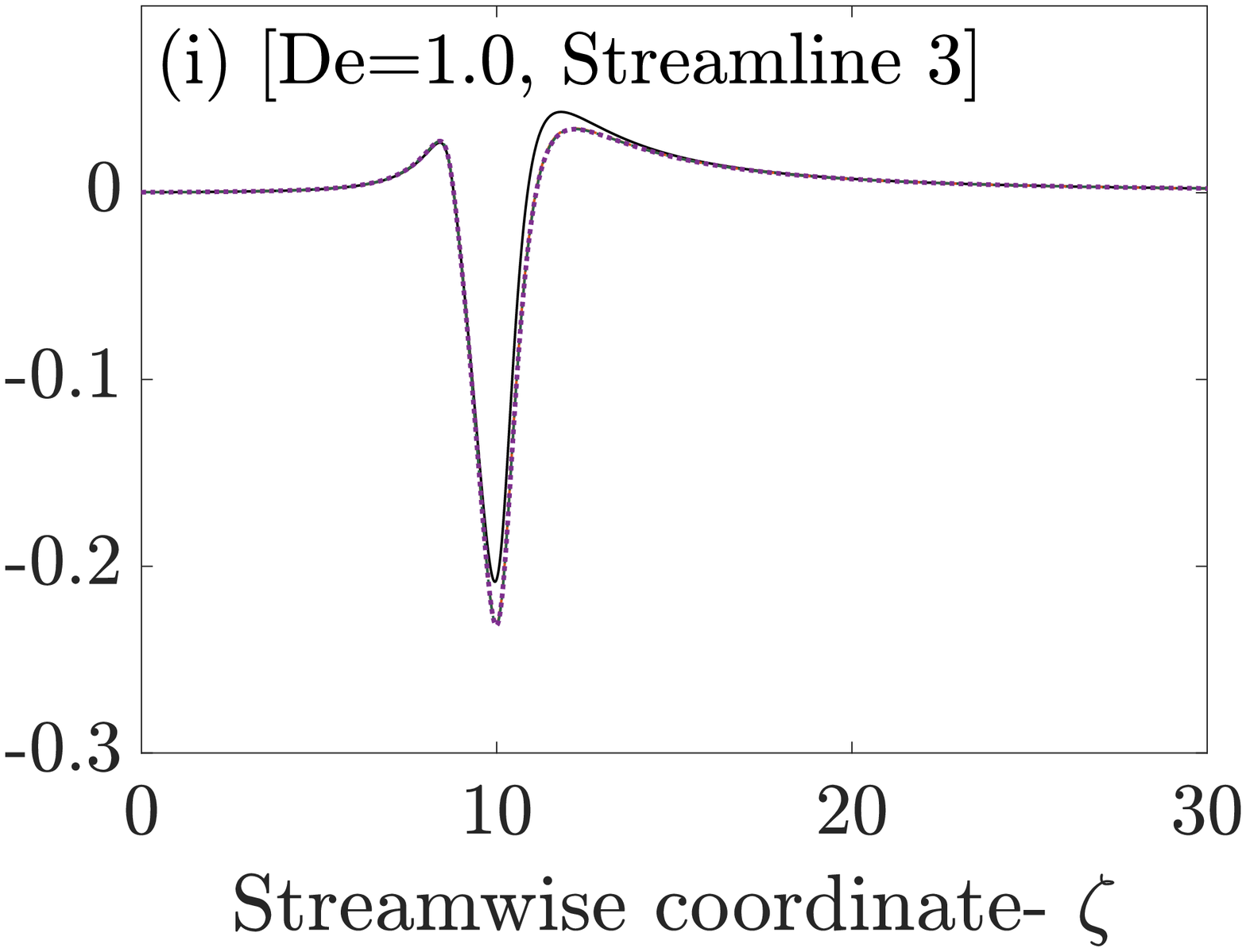}}
	\caption {$\Delta\mathcal{S}/L$ along the three streamlines indicated in figure \ref{fig:streamlines} for three different $De$ and various $L$. The legend is same as figure \ref{fig:streamlinesLegend}. \label{fig:StreamlineStretch2}}
\end{figure}

\begin{figure}[h!]
	\centering
	\subfloat{\includegraphics[width=0.33\textwidth]{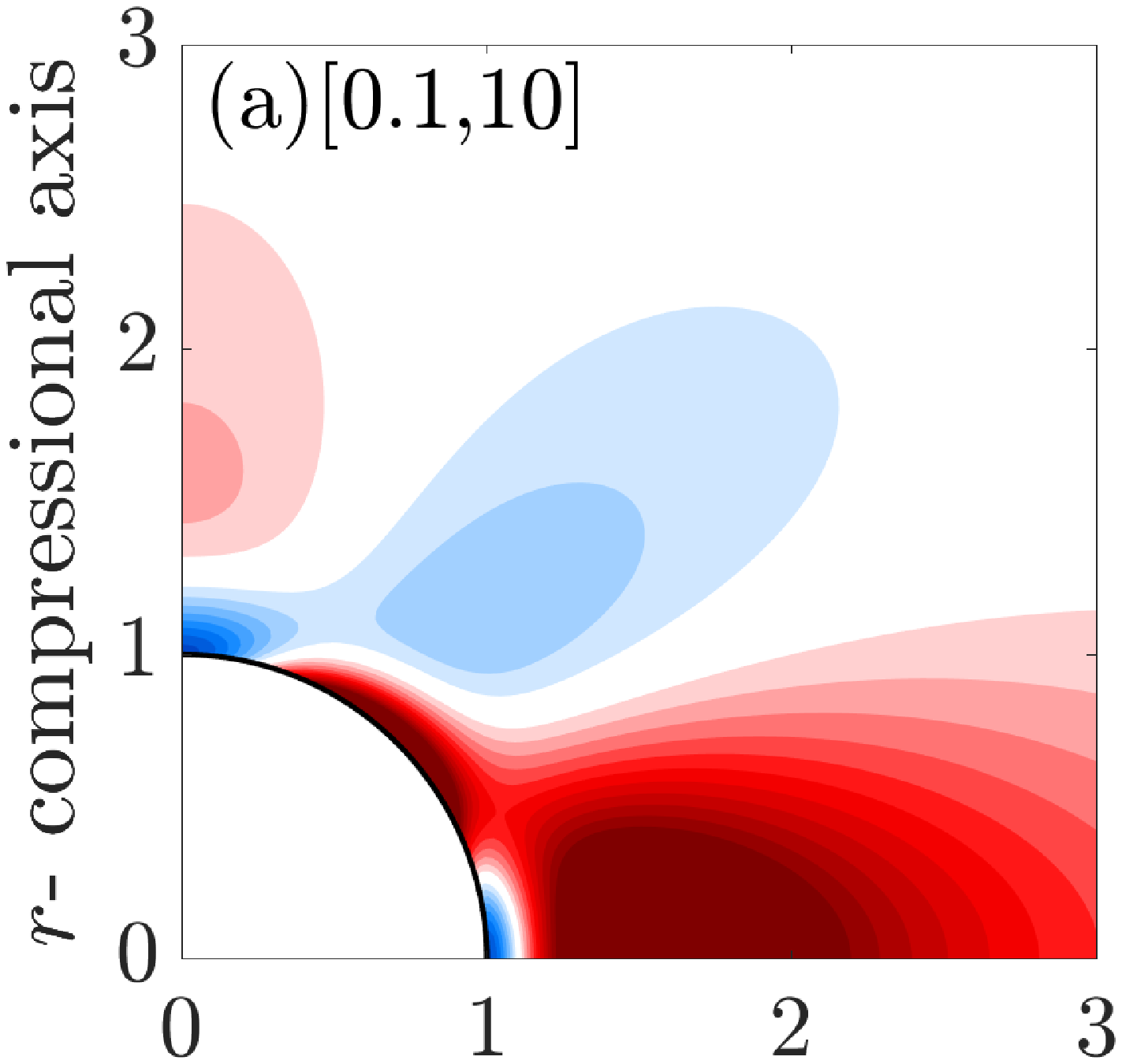}\label{fig:extraStretchDept1}}\hfill
	\subfloat{\includegraphics[width=0.33\textwidth]{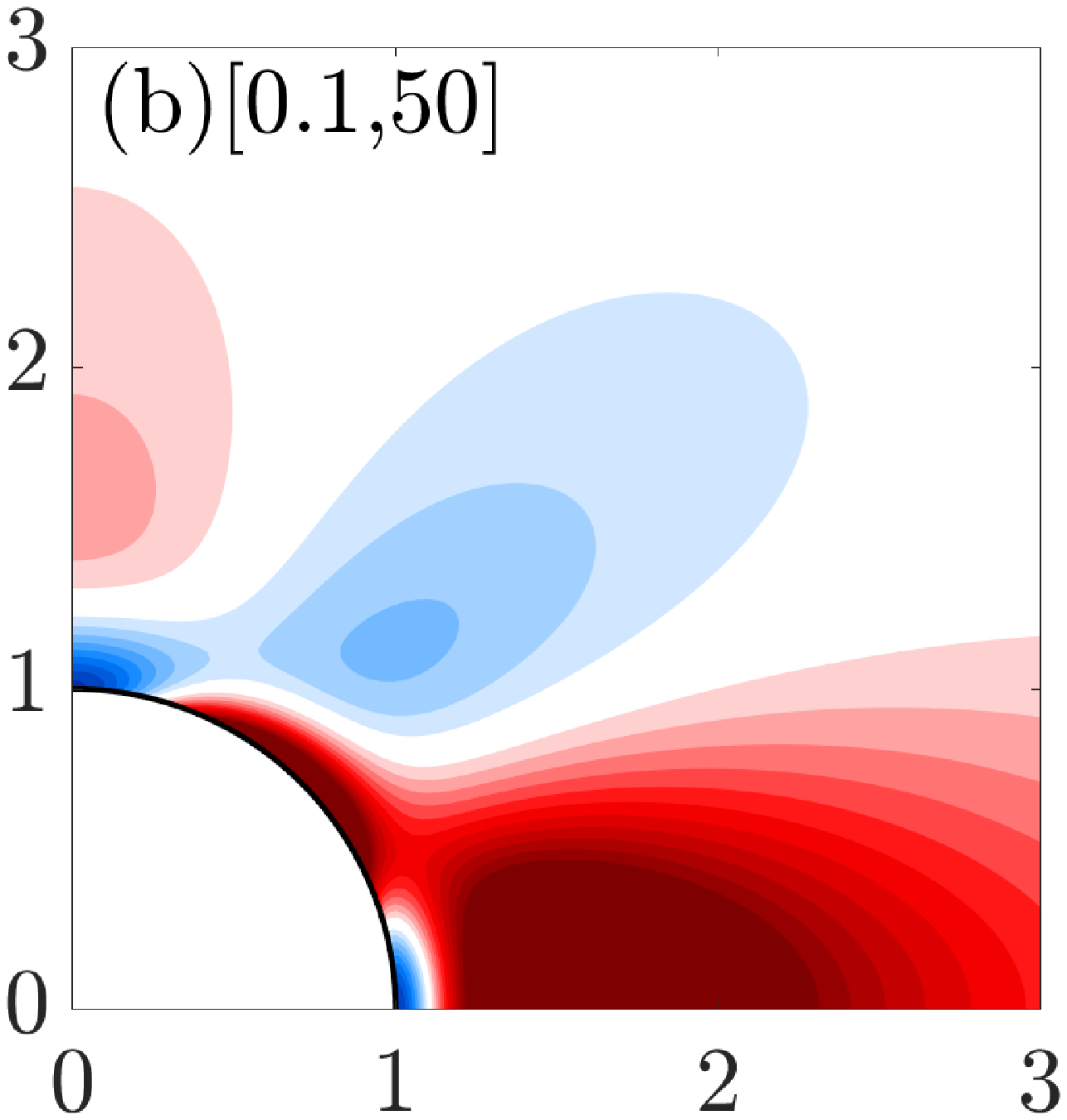}\label{fig:extraStretchDept1L50}}\hfill	\subfloat{\includegraphics[width=0.33\textwidth]{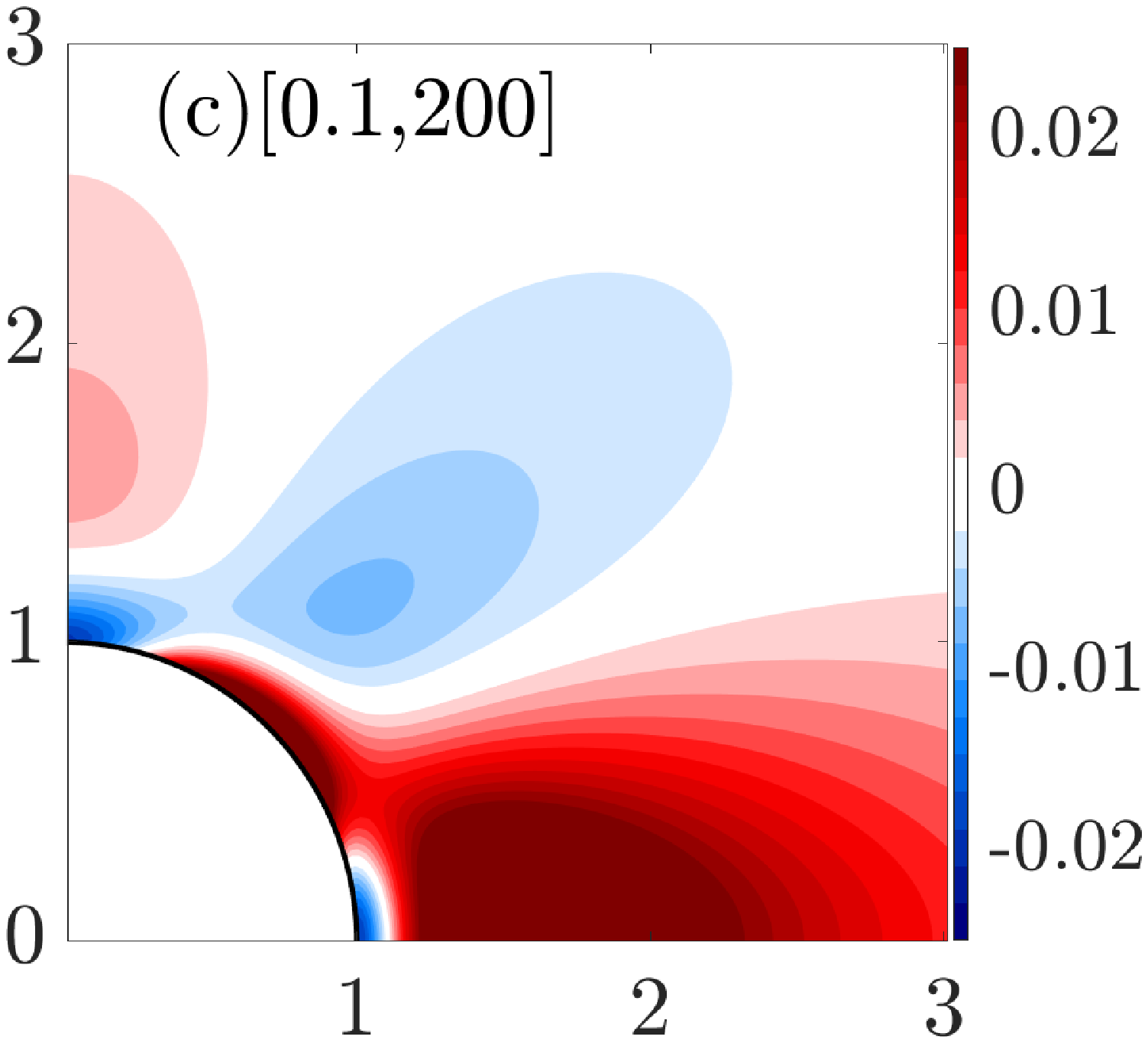}\label{fig:extraStretchDept1L200}}\hfill
	\subfloat{\includegraphics[width=0.33\textwidth]{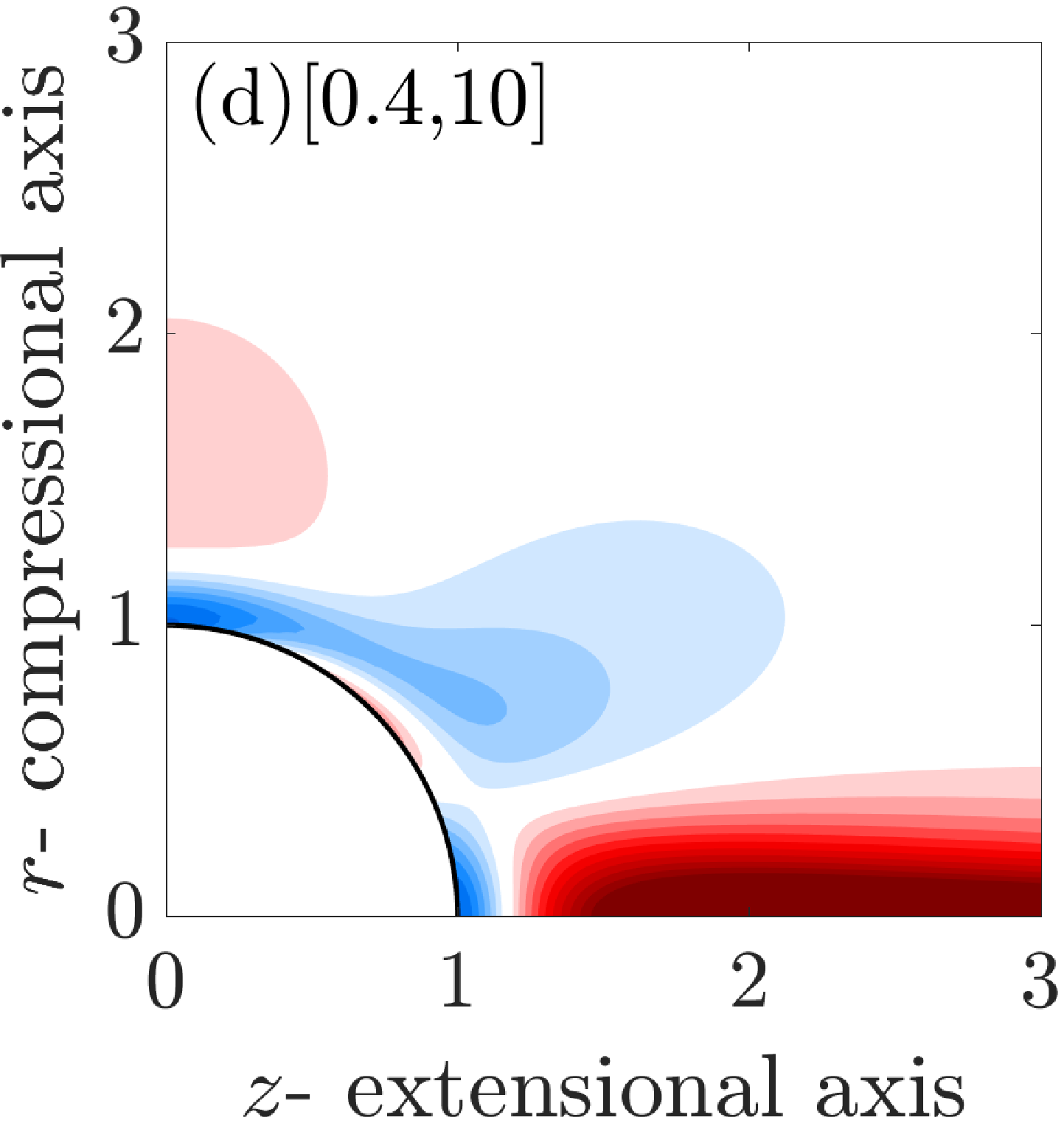}\label{fig:extraStretchDept4L10}}\hfill	\subfloat{\includegraphics[width=0.33\textwidth]{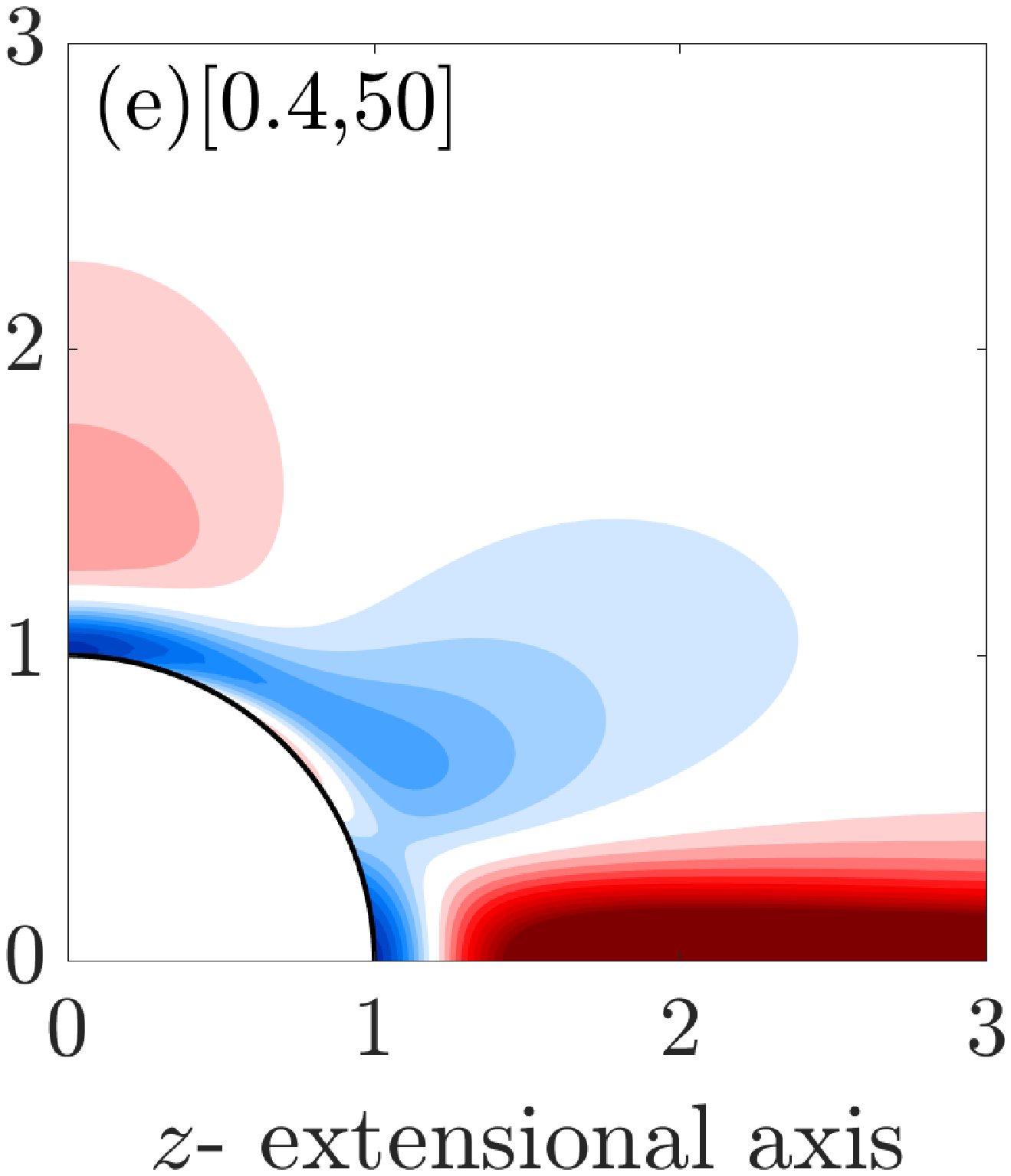}\label{fig:extraStretchDept4L50}}\hfill
	\subfloat{\includegraphics[width=0.33\textwidth]{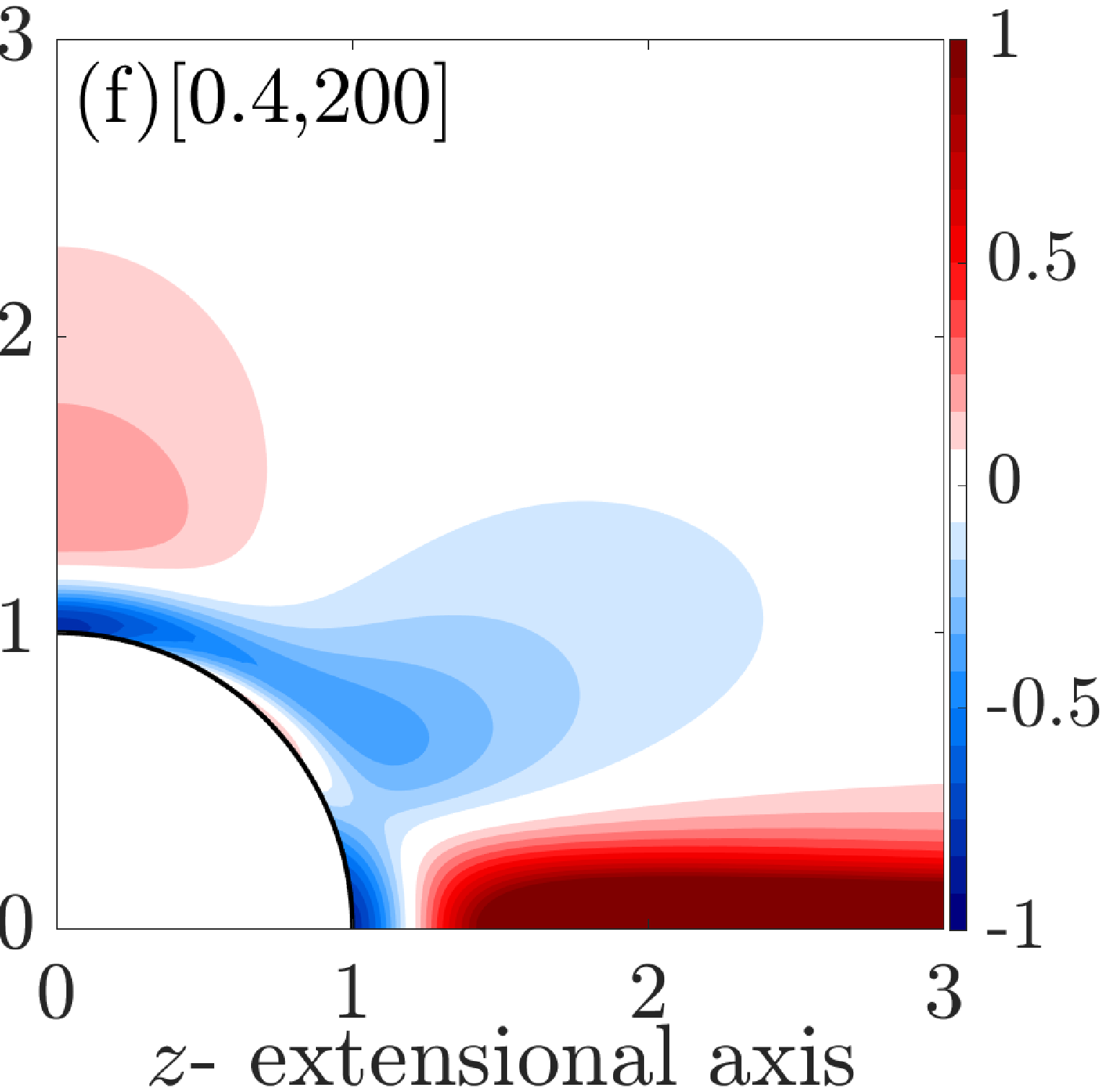}\label{fig:extraStretchDept4L200}}
	\caption {Contours of $\Delta\mathcal{S}$ for various $L$ and $De=0.1$ and 0.4. The parameters marked on each plot are [$De$, $L$]. The color axis is the same for all the plots in a given row and is indicated at the end. The most noticeable feature is the wake of highly stretched polymers (red regions around the extensional axis). \label{fig:polymerstretch_smallDe}}
\end{figure}

\begin{figure}[h!]
	\centering
	\subfloat{\includegraphics[width=0.33\textwidth]{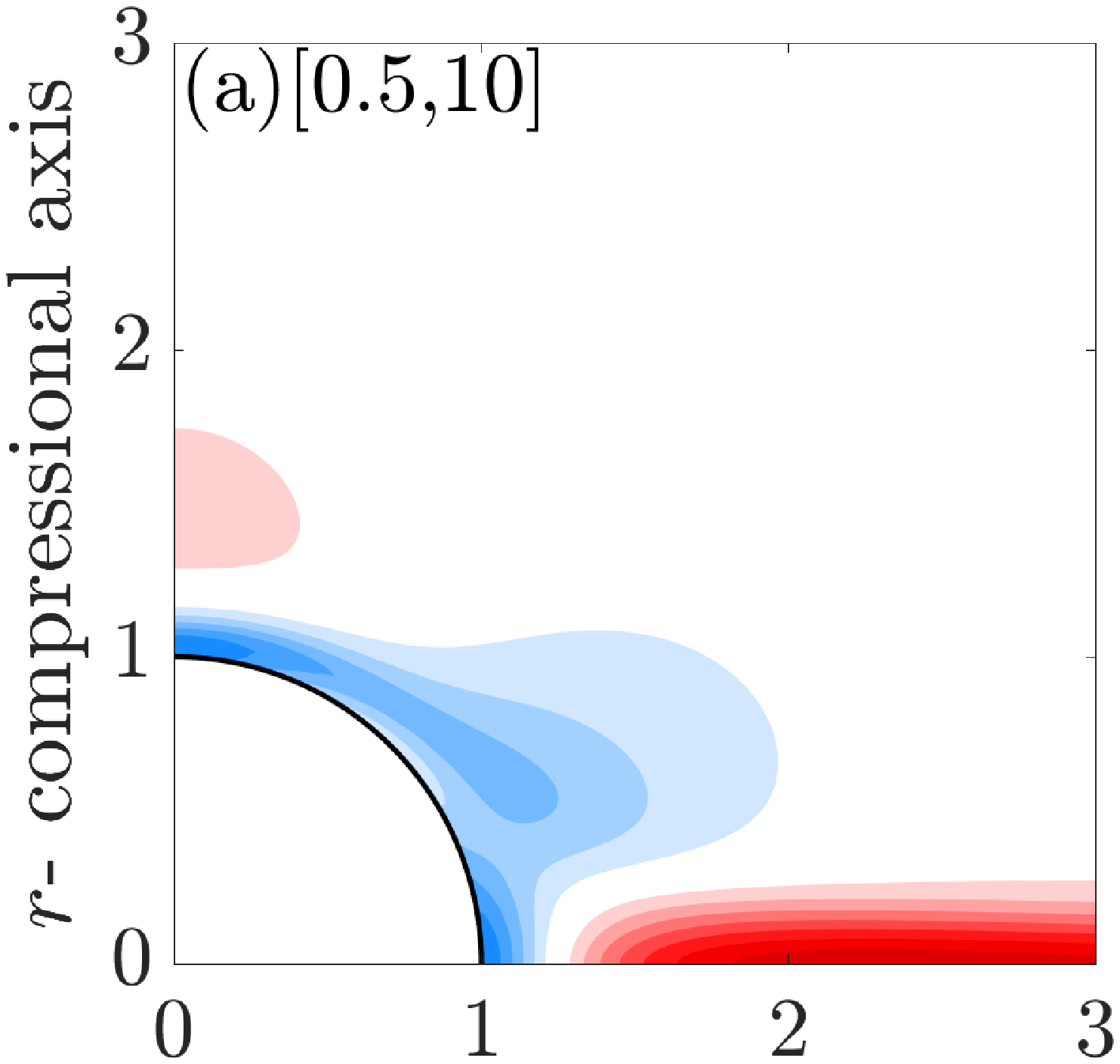}}\hfill
	\subfloat{\includegraphics[width=0.33\textwidth]{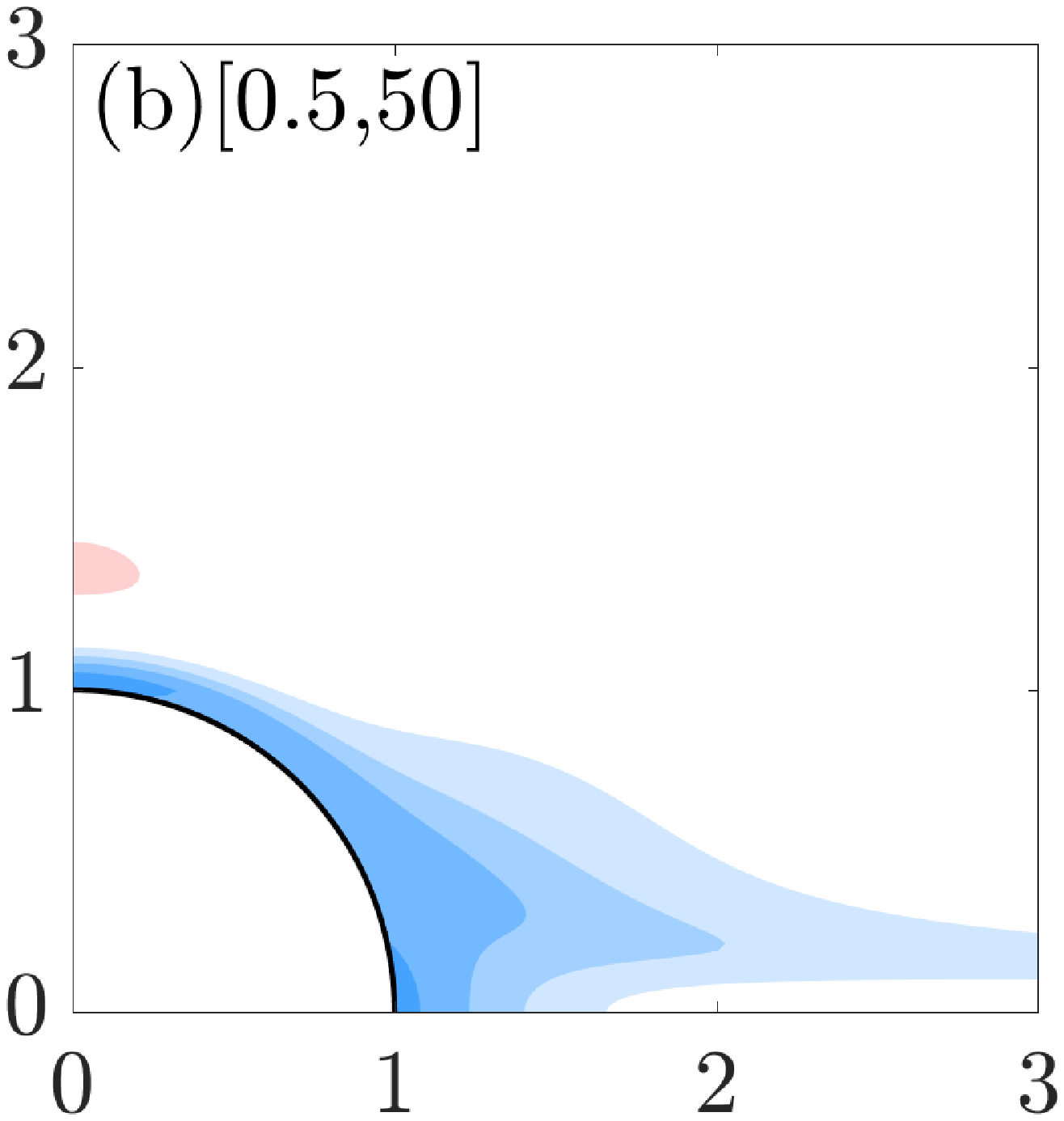}}\hfill
	\subfloat{\includegraphics[width=0.33\textwidth]{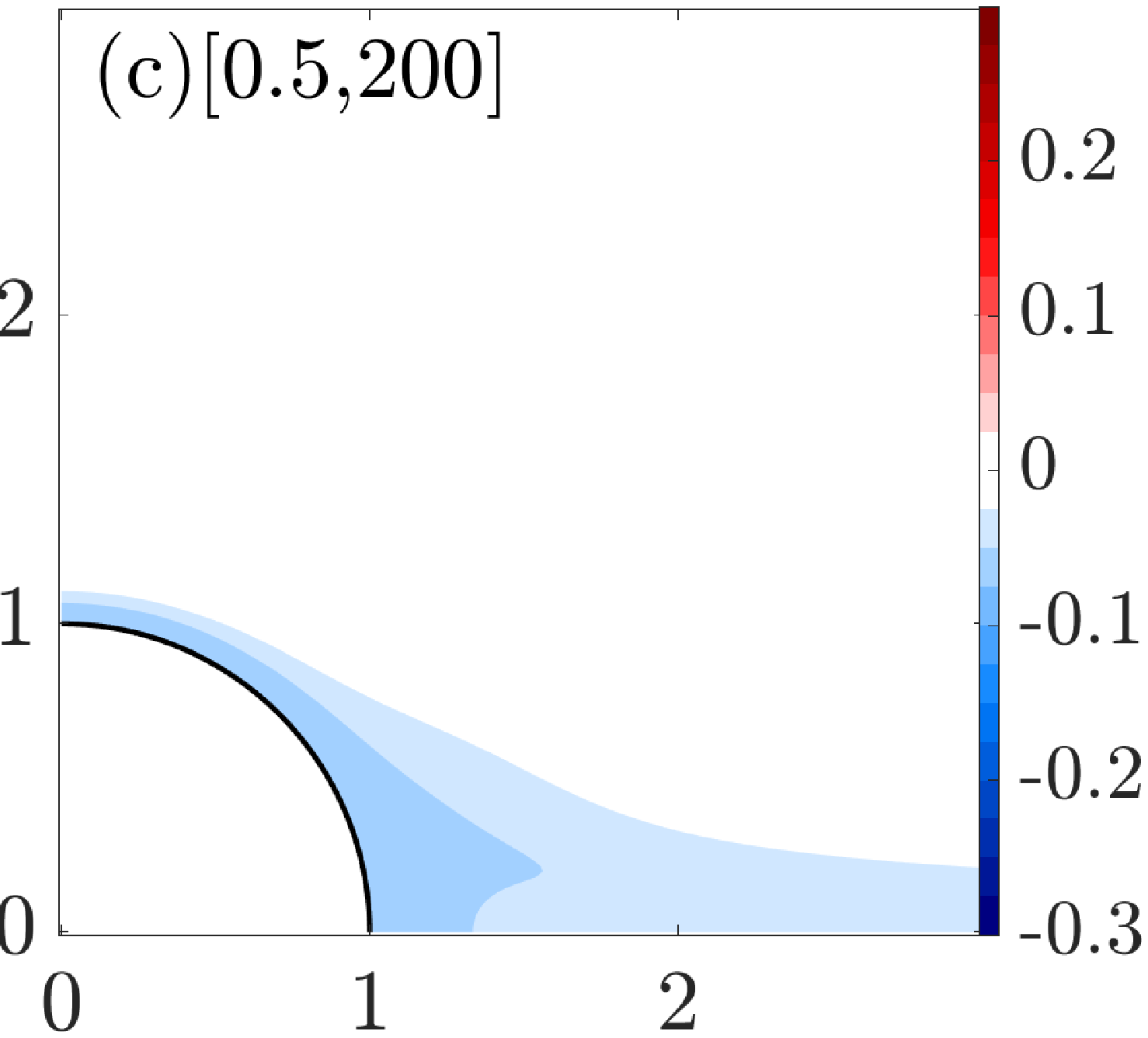}}\hfill
	\subfloat{\includegraphics[width=0.33\textwidth]{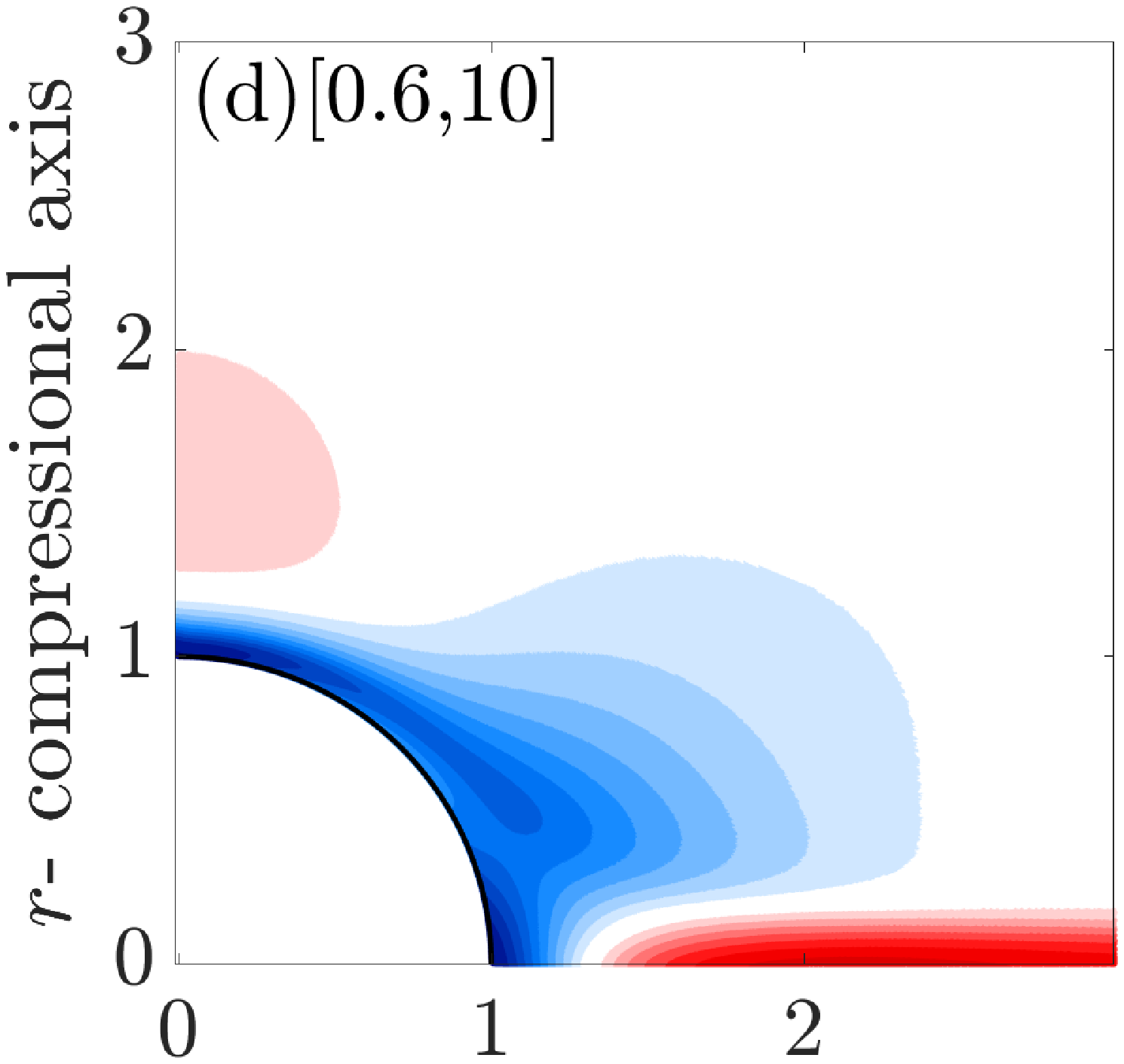}}\hfill
	\subfloat{\includegraphics[width=0.33\textwidth]{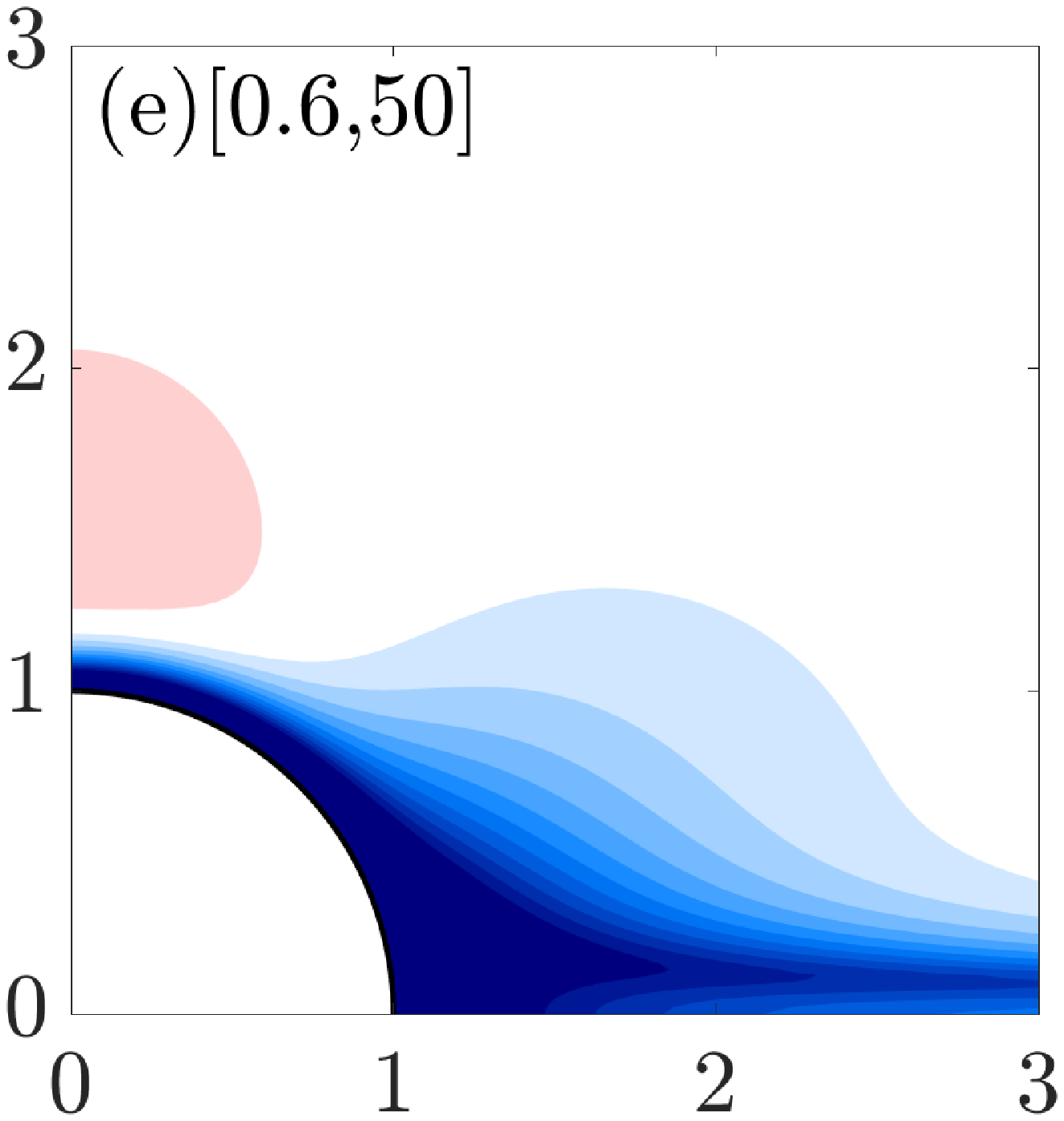}}\hfill
	\subfloat{\includegraphics[width=0.33\textwidth]{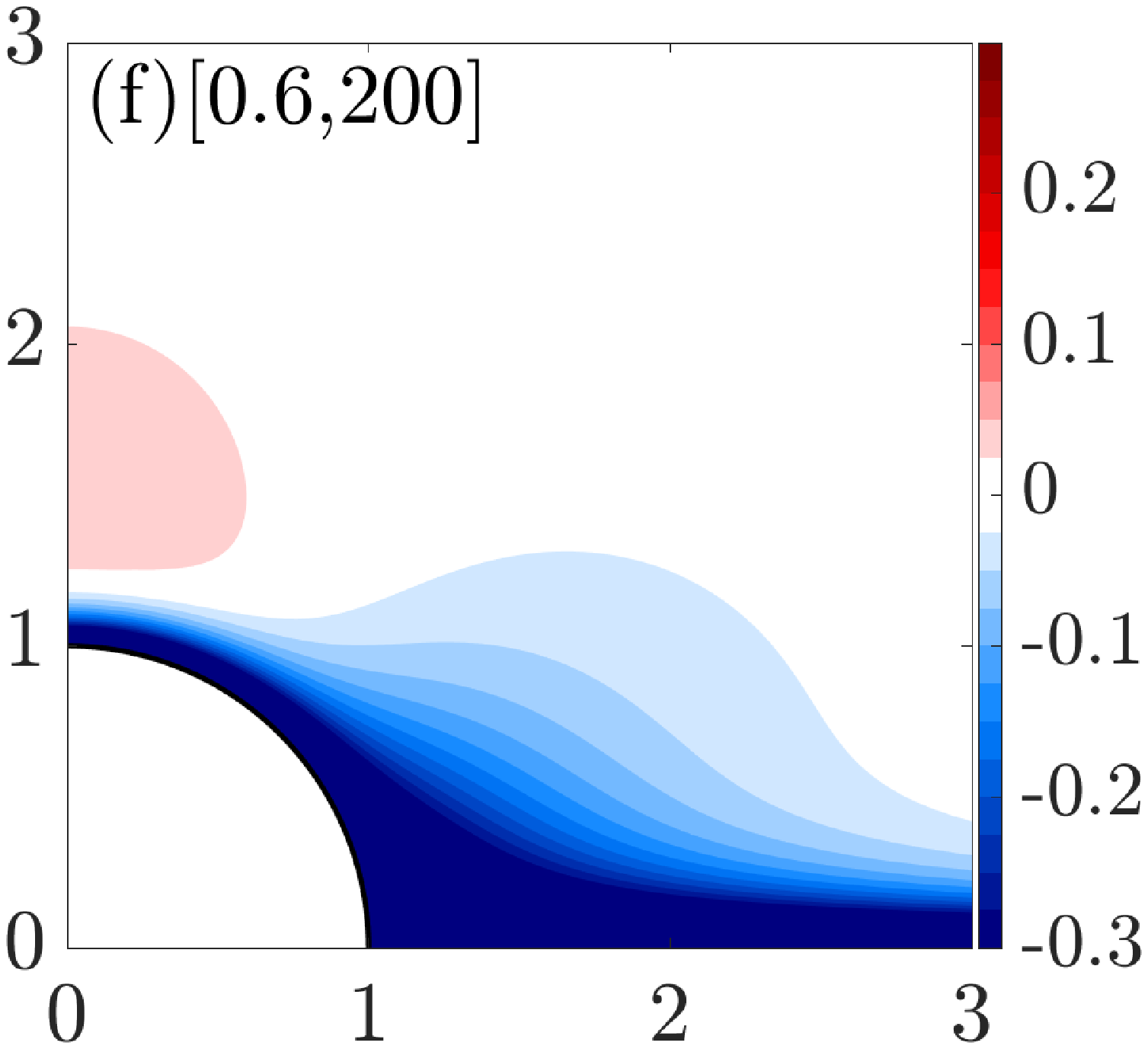}}\hfill
	\subfloat{\includegraphics[width=0.33\textwidth]{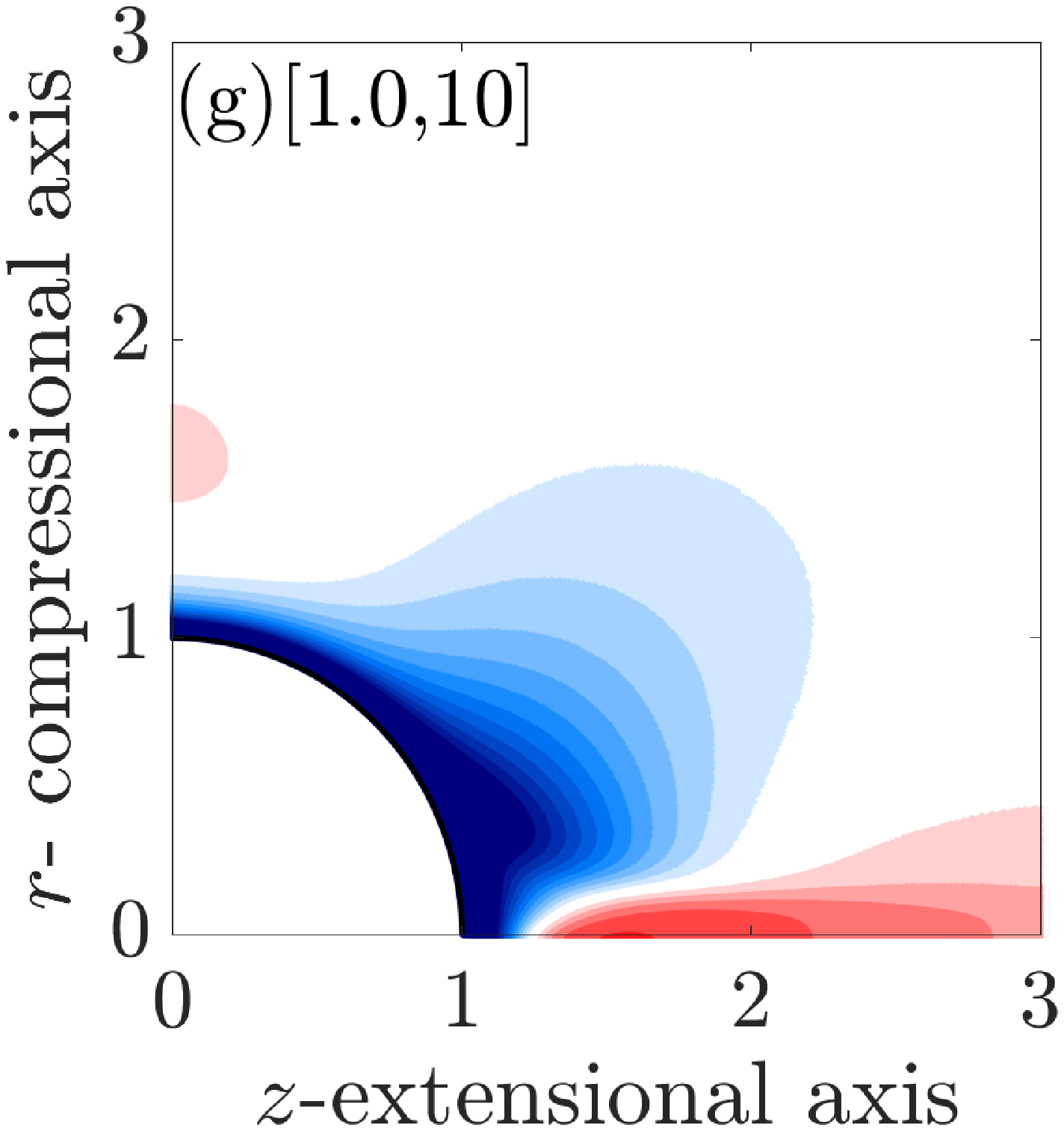}}\hfill
	\subfloat{\includegraphics[width=0.33\textwidth]{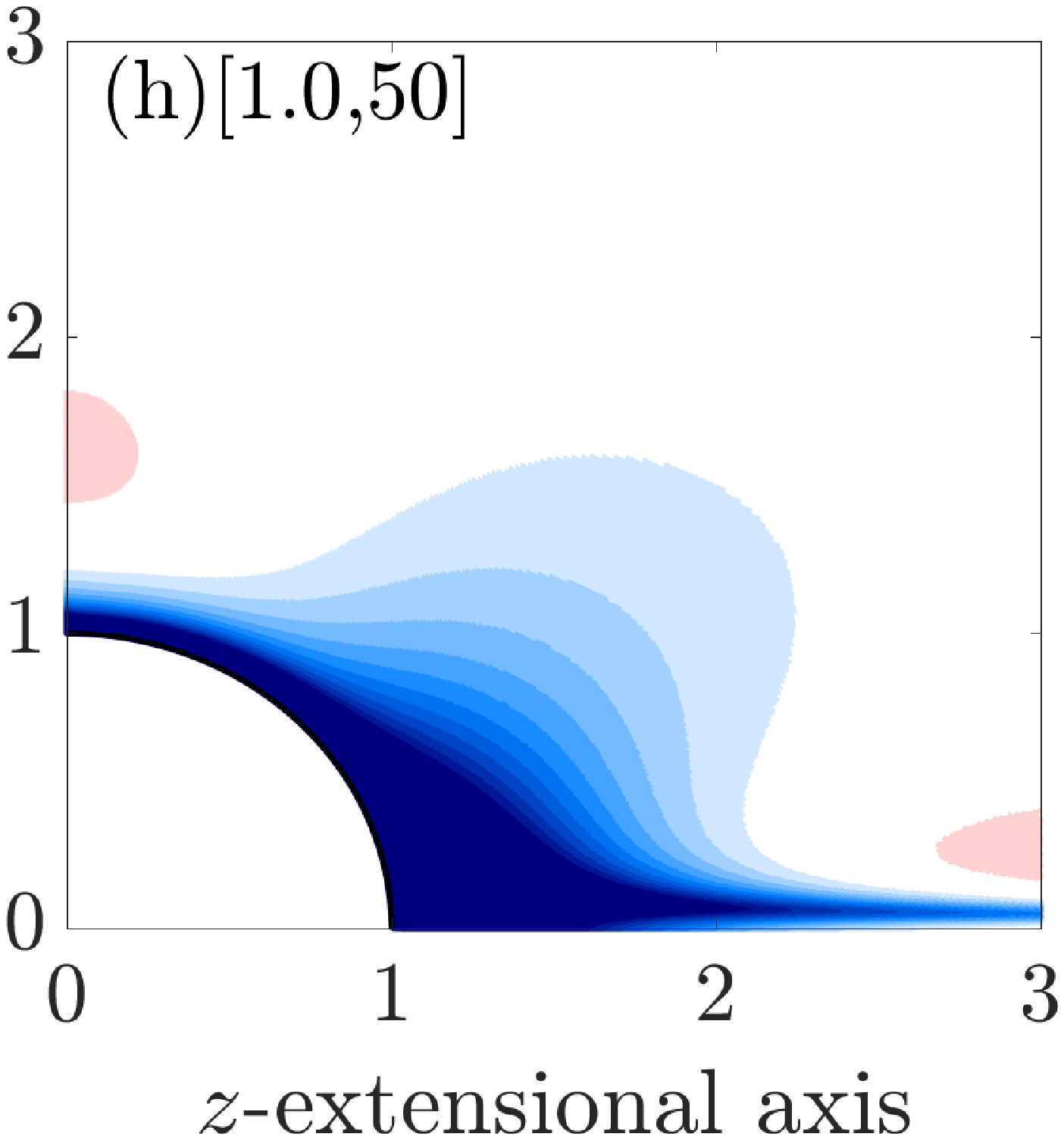}}\hfill
	\subfloat{\includegraphics[width=0.33\textwidth]{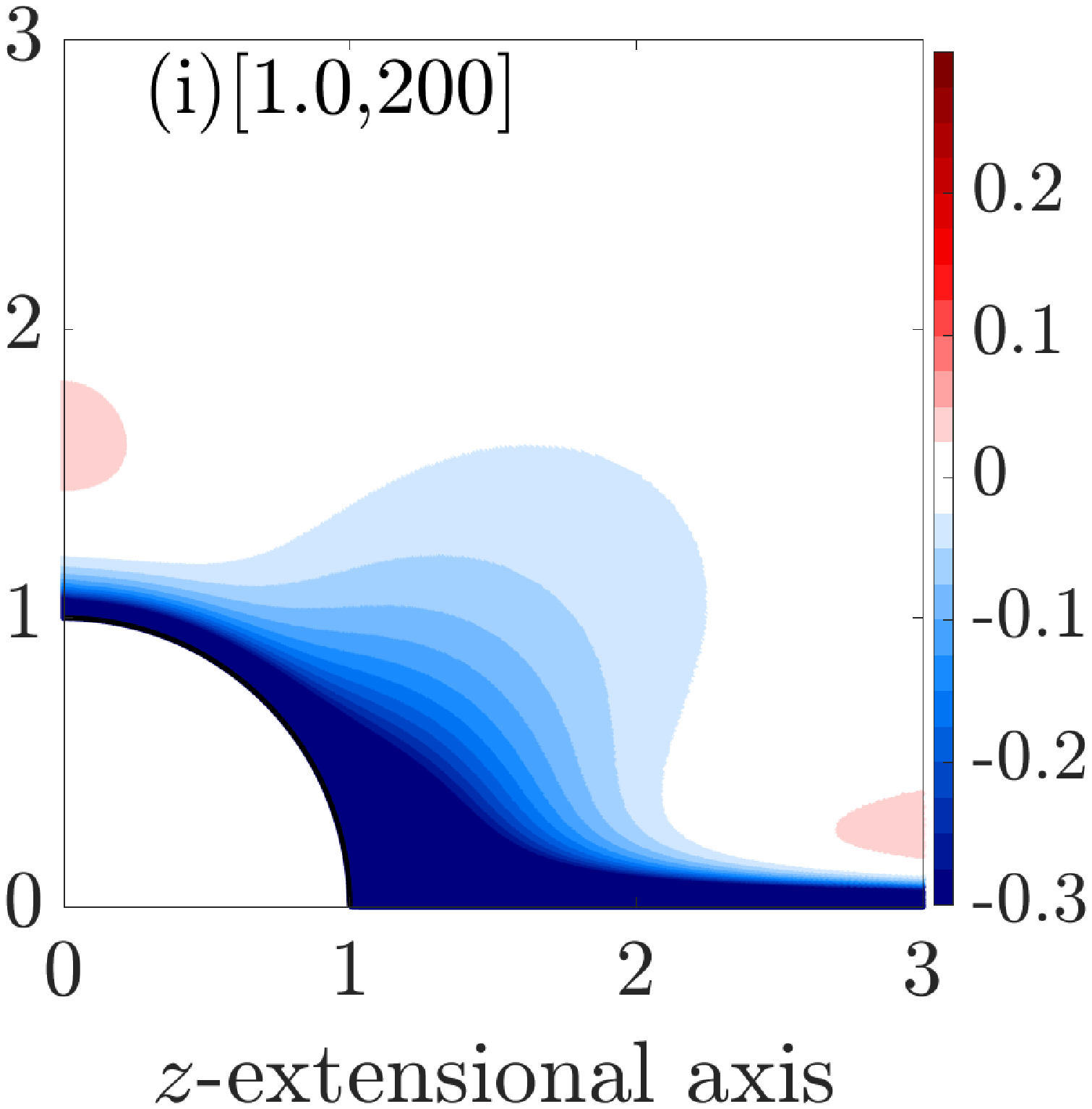}}\hfill
	\caption {Contours of $\Delta\mathcal{S}/L$ for various $L$ and $De$. The color axis is the same for all the plots and is indicated at the end of each row. The parameters marked on each plot are [$De$, $L$]. There is a region of collapsed/ unstretched polymers (blue regions) around the particle. It extends to form a wake for large $L$. At small $L$, there is a wake of highly stretched polymers (red regions) similar to that for the $De<0.5$ regime at all $L$. Similar contours of $\Delta\mathcal{S}/L$ are observed up to $De\approx 1.25$. \label{fig:polymerstretch_mediumDe}}
\end{figure}

\begin{figure}[h!]
	\centering
	\subfloat{\includegraphics[width=0.33\textwidth]{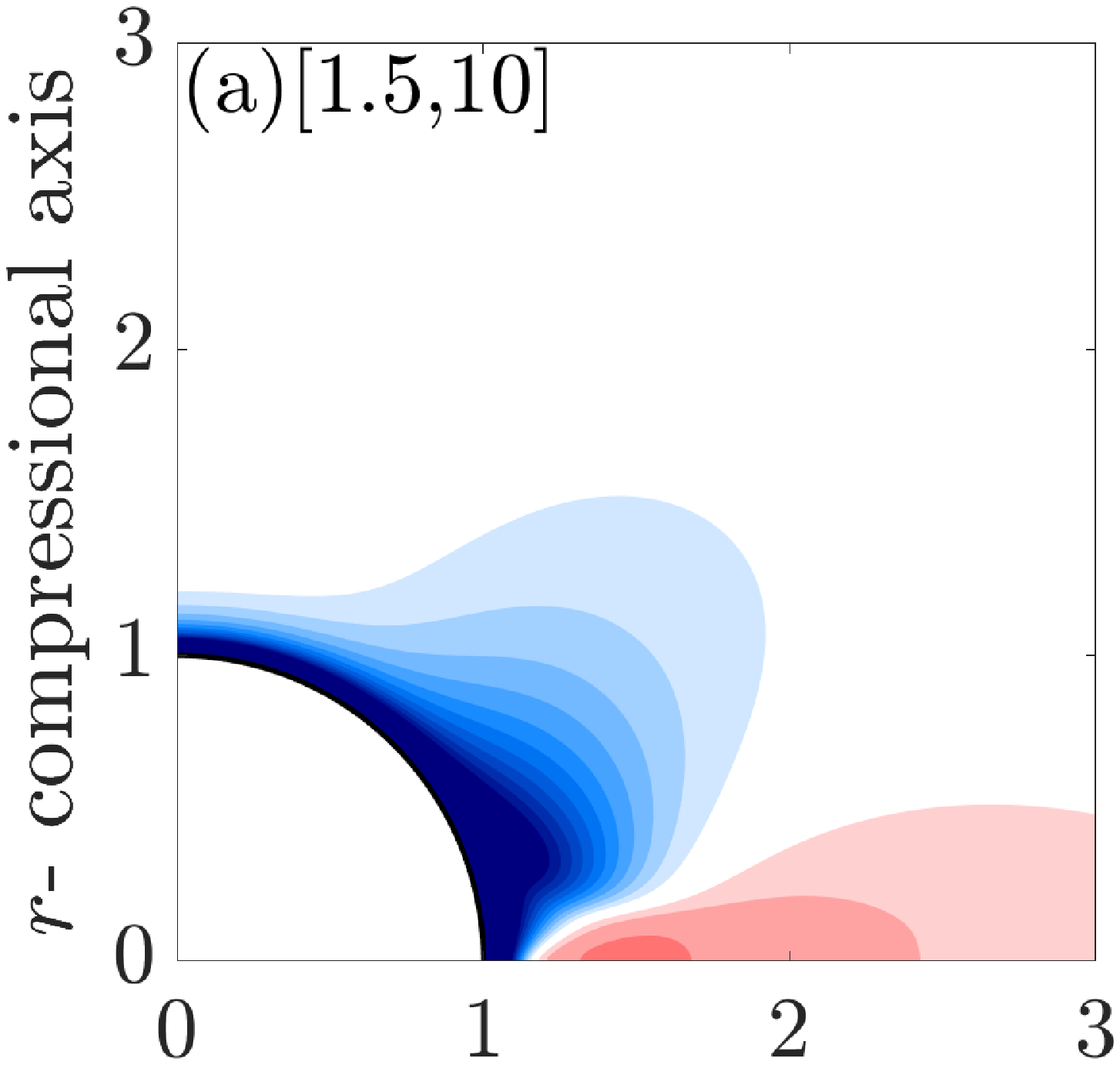}}\hfill
	\subfloat{\includegraphics[width=0.33\textwidth]{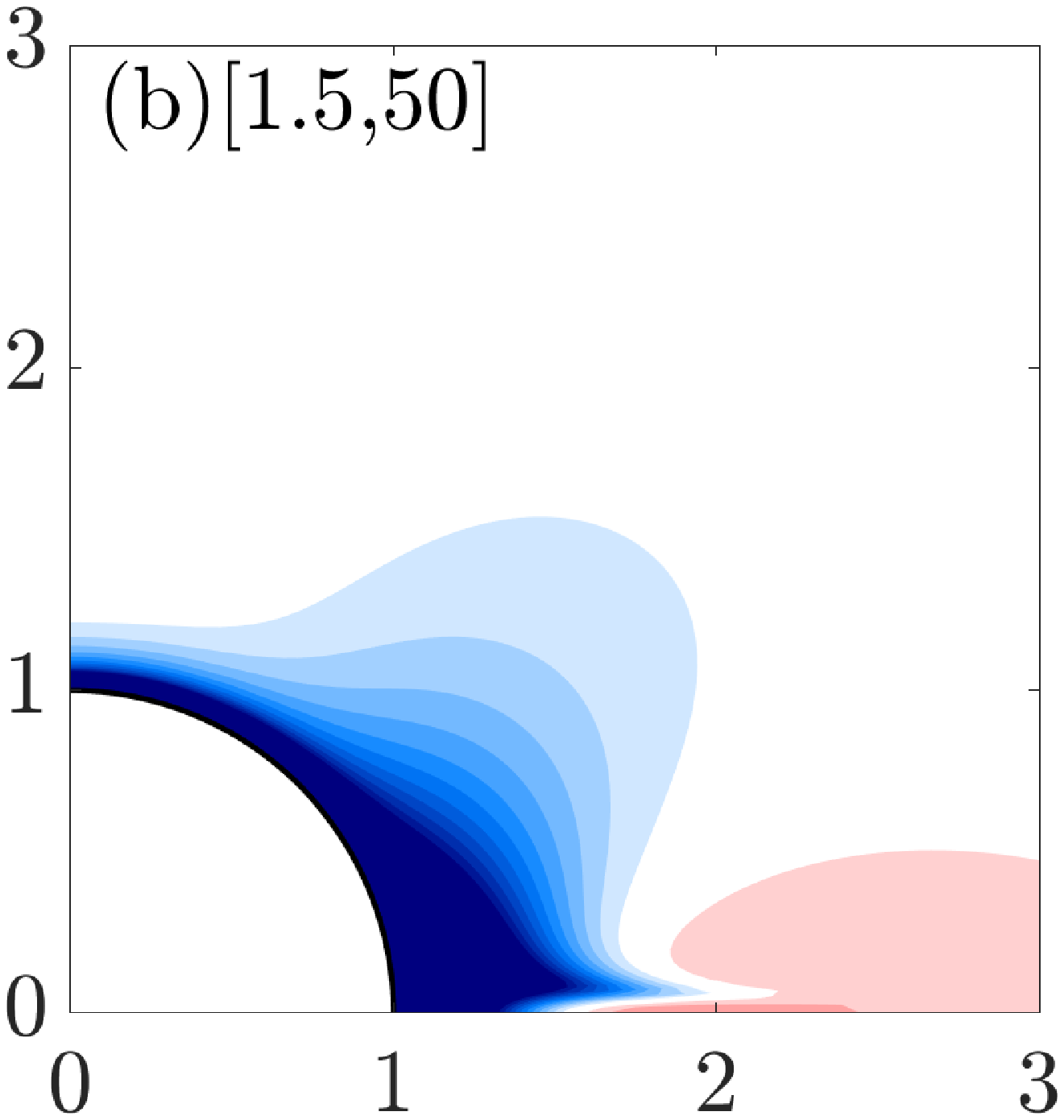}}\hfill
	\subfloat{\includegraphics[width=0.33\textwidth]{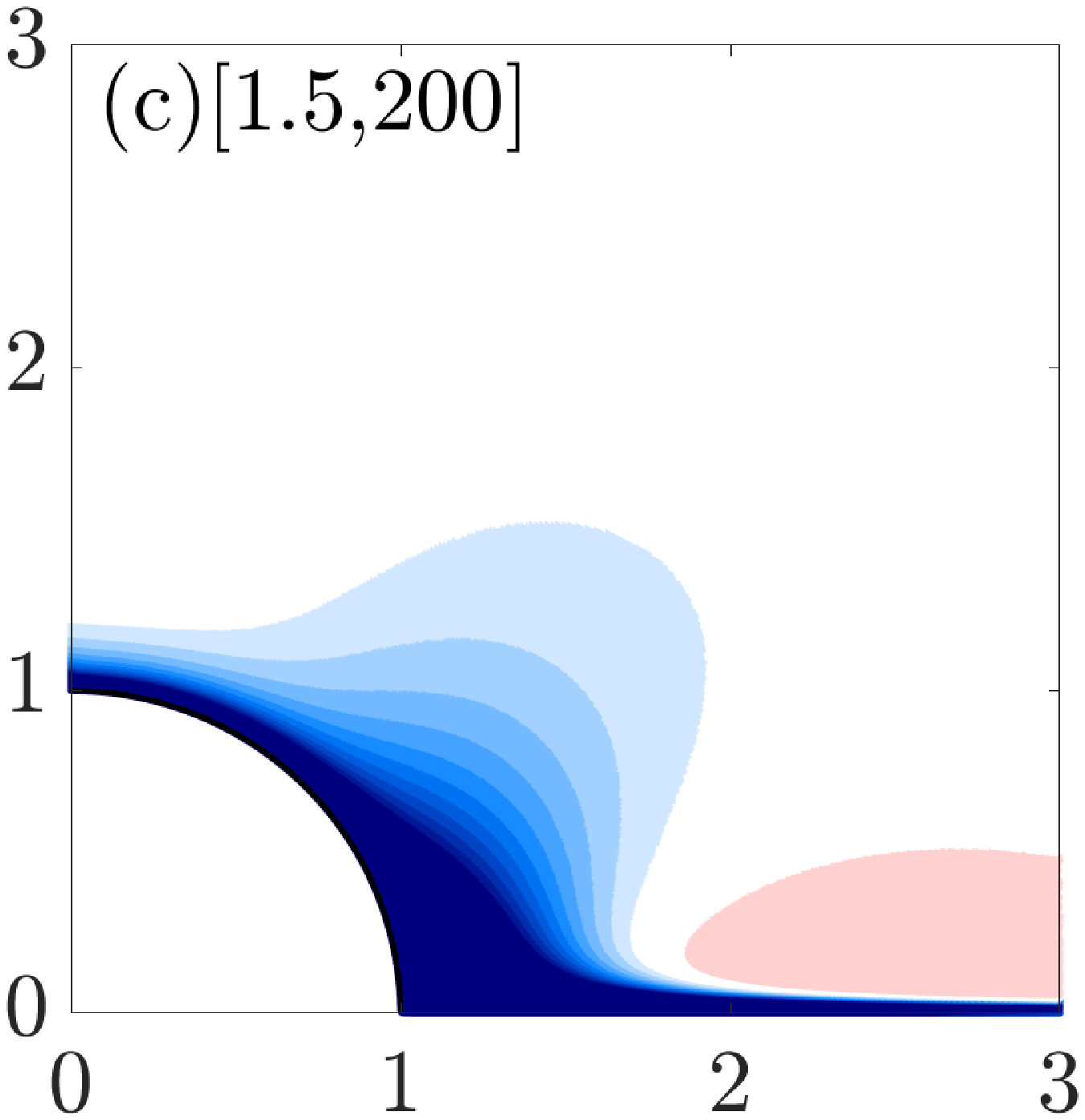}}\hfill
	\subfloat{\includegraphics[width=0.33\textwidth]{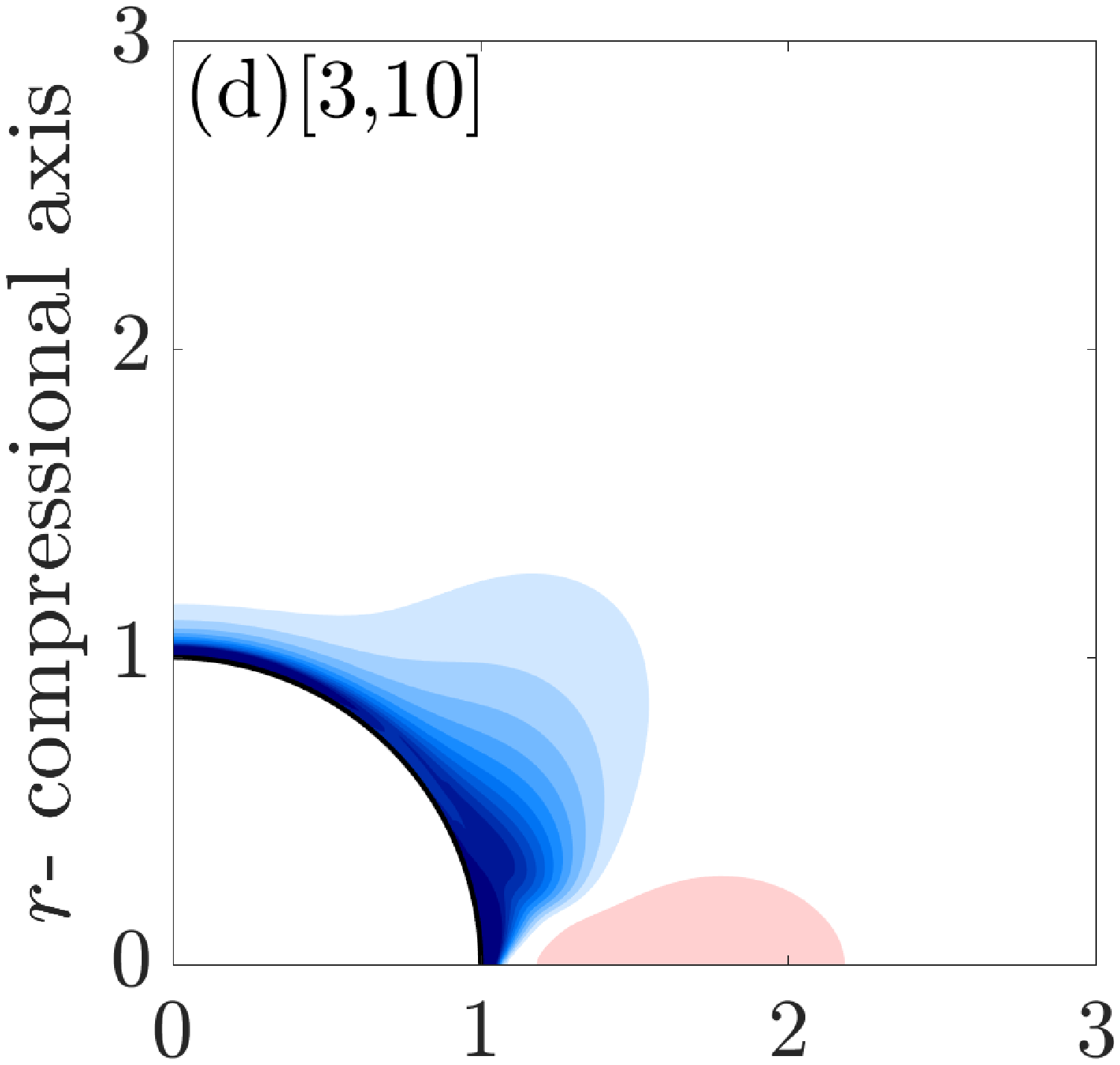}}\hfill	
	\subfloat{\includegraphics[width=0.33\textwidth]{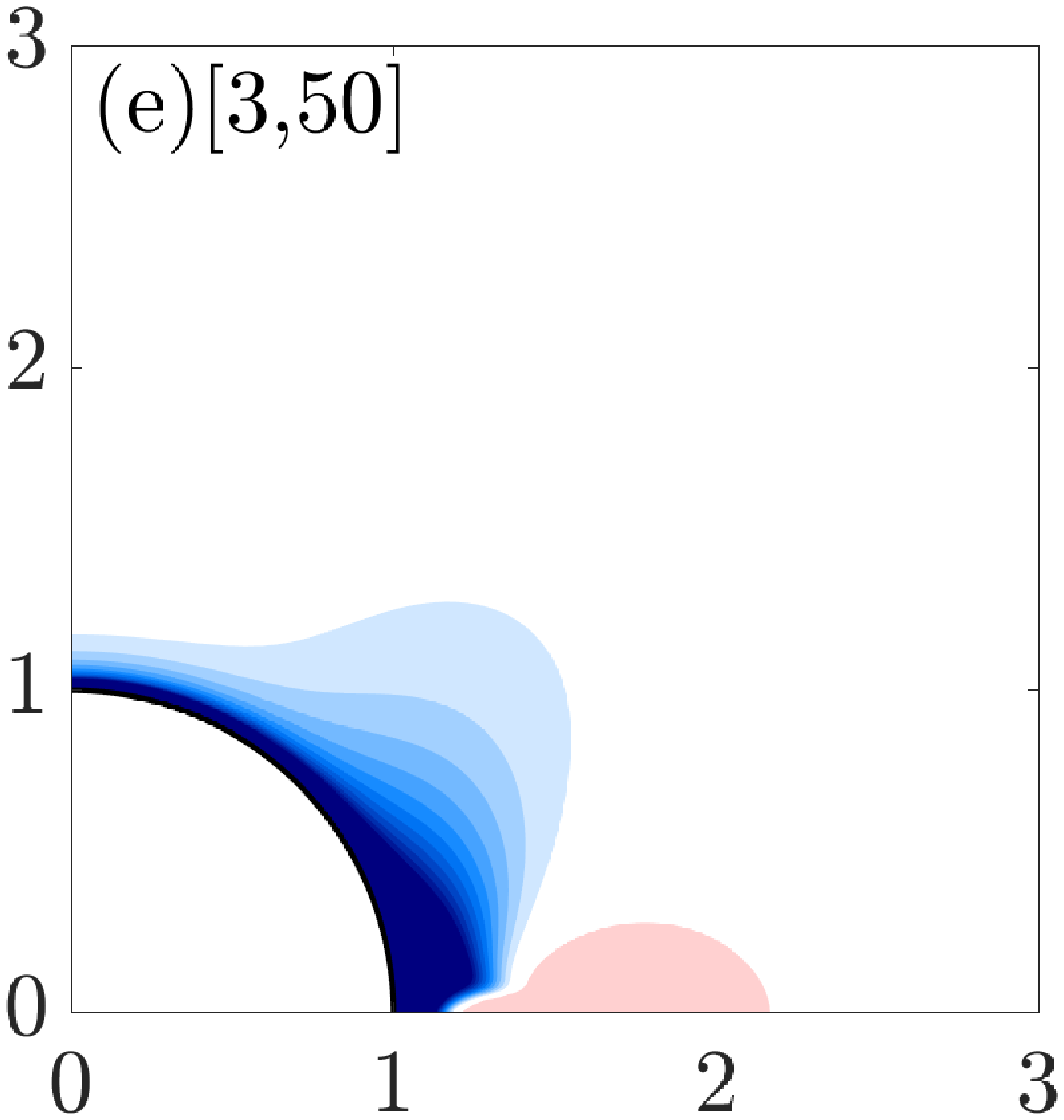}}\hfill
	\subfloat{\includegraphics[width=0.33\textwidth]{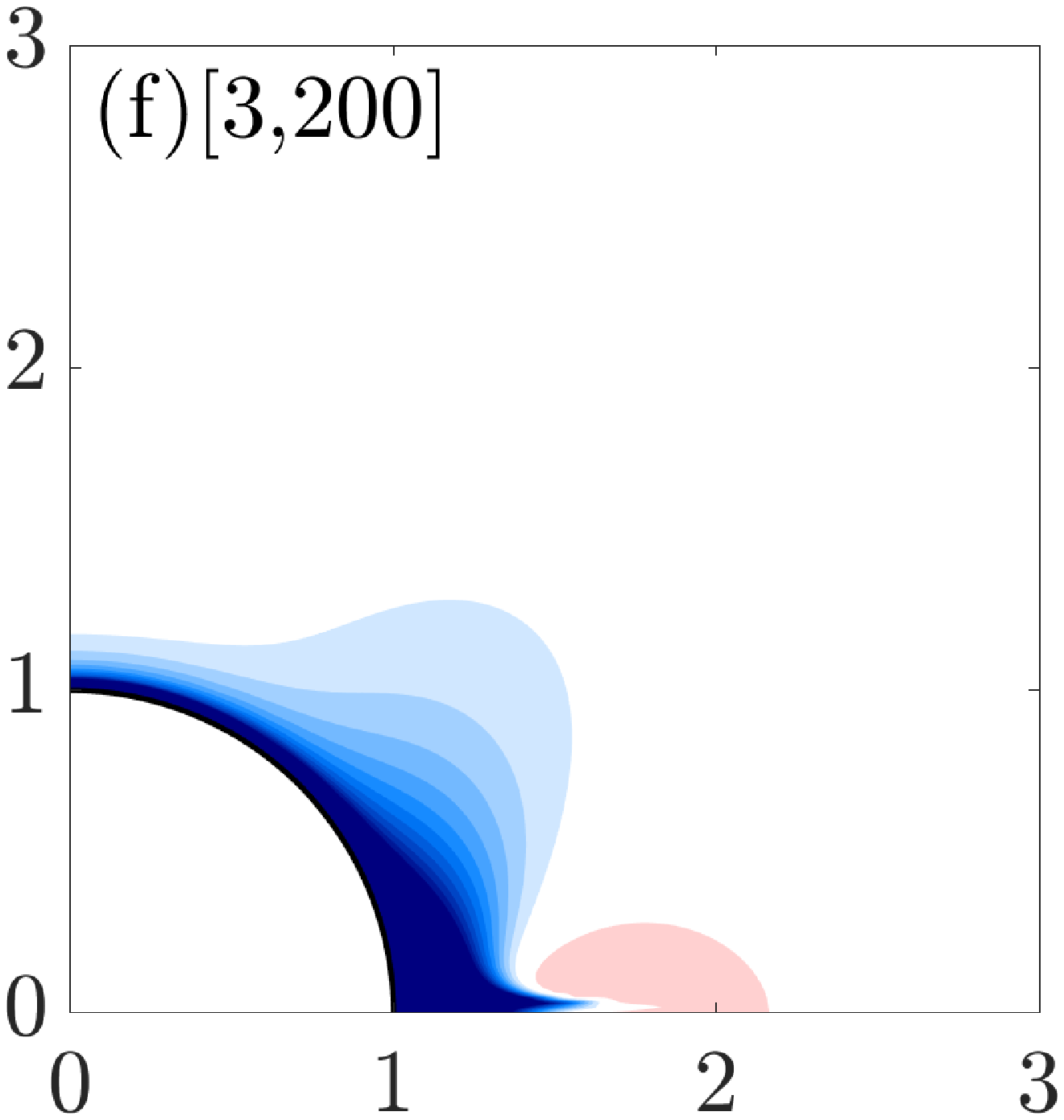}}\hfill
	\subfloat{\includegraphics[width=0.33\textwidth]{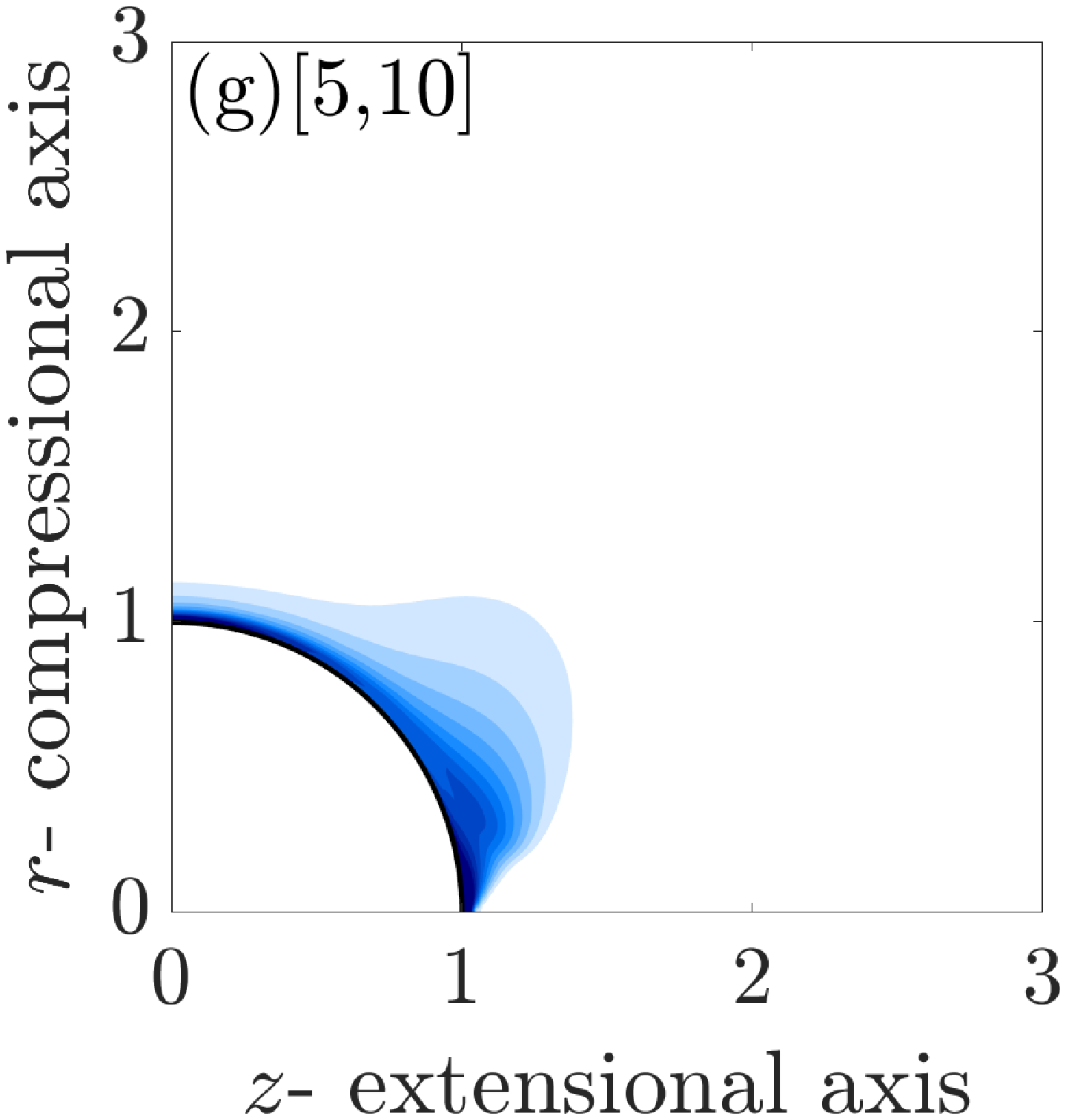}}\hfill
	\subfloat{\includegraphics[width=0.33\textwidth]{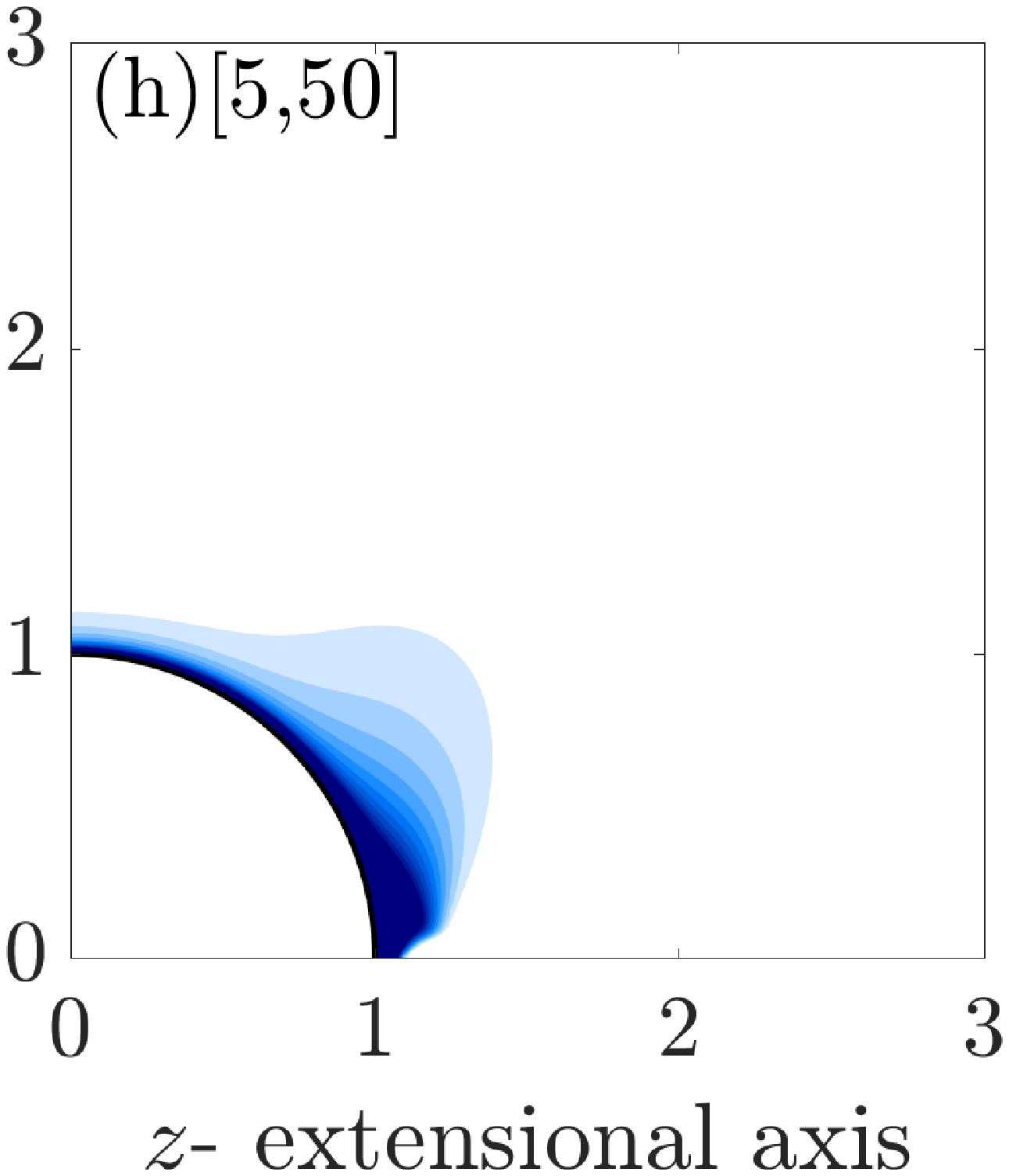}}\hfill	\subfloat{\includegraphics[width=0.33\textwidth]{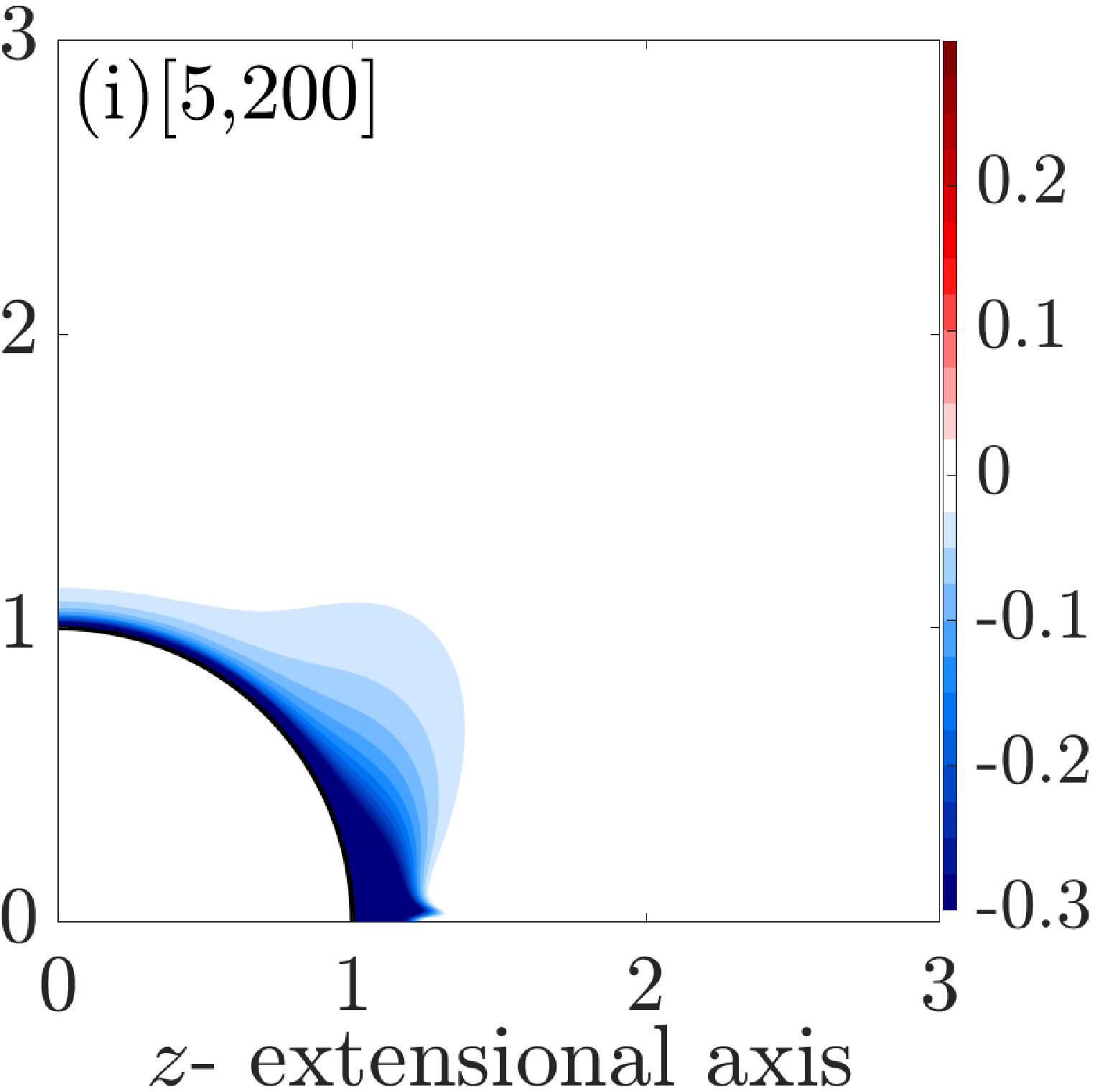}}
	\caption {Contours of $\Delta\mathcal{S}/L$ for various $L$ and $De\ge1.5$. The parameters marked on each plot are [$De$, $L$]. The color axis for all the plots is the same and is shown with the last figure. Most of the particle influence arises from a region of collapsed polymers near the particle surface (blue regions). The trends seen here extend to higher $De$. \label{fig:polymerstretch_largeDe}}
\end{figure}

Figures \ref{fig:StreamlineStretch1} and \ref{fig:StreamlineStretch2} show $\Delta\mathcal{S}$ along three streamlines and  figures \ref{fig:polymerstretch_smallDe} to \ref{fig:polymerstretch_largeDe} show the contours of  $\Delta\mathcal{S}$ in a region around the sphere. For $De\ge0.5$, $\Delta\mathcal{S}$ is presented after normalizing with $L$. Figure \ref{fig:streamlines} shows the position of the three streamlines considered relative to the particle. Streamline 1 comes close to the particle and traces it almost perfectly. Figure \ref{fig:StreamlineStretch1} shows that $\Delta\mathcal{S}$ is independent of $L$ for $L\gtrsim50$, along streamline 1 and 3. Compared to streamline 1, the magnitude of $\Delta\mathcal{S}$ is lower for streamline 3, which is further away from the sphere. The polymer stretch, $\sqrt{\text{tr}(\boldsymbol{\Lambda}^{(0)}})$, is thus independent of $L$, for $L\gtrsim50, De\lesssim0.5$, in the whole region around the sphere.

For $De=0.6$, 0.8 and 1.0, as presented in figure \ref{fig:StreamlineStretch2}, we see an expected breakdown of the $L^2$ scaling of $\boldsymbol{\Lambda}$ along streamline 1 (figure \ref{fig:StreamlineStretch2} first row), after it approaches the sphere and polymers collapse close to the equilibrium configuration. Along streamline 2 and 3 (figure \ref{fig:StreamlineStretch2} second and third row), the $L^2$ scaling is recovered for large $L$ as the imposed extension rate ($De$) is increased. On streamline 3, the $L^2$ scaling is observed for $De\gtrsim 0.6$ and $L\gtrsim 50$. Recovery happens at even lower $De$ and $L$ (i.e. for a wider parameter range) in the region outside streamline 3. The scaling is recovered for streamline 2, for $L\gtrsim 50$ and $De\gtrsim 1.0$. At a higher $L=100$, $De\gtrsim 0.8$ allows $L^2$ scaling on streamline 2. The region between the sphere and streamline 2 occupies a very small volume. Therefore, at large $L$, for a value of $De\propto 1/L$ beyond the coil-stretch transition, the $L^2$ scaling of the change in polymer stretch by the particle, and hence the polymer stress, is valid almost everywhere around the sphere. Even when the volume of the region where $L^2$ scaling breaks down is not negligibly small, we will find in section \ref{sec:RheologyResults} that the contribution from these regions to the suspension rheology is small as the extra stress in the suspension due to particle-polymer interaction still scales as $L^2$ for lower $De$ values at a given $L$ than indicated by the streamline analysis of polymer stretch discussed here.

The contours of $\Delta\mathcal{S}$ in a region around the particle are shown in figure \ref{fig:polymerstretch_smallDe} for three different $L$ at $De=0.1$ and 0.4. The aforementioned $L$ independence below the coil-stretch transition $De$ is confirmed as the contours in the plots in a given row are almost identical. A strong qualitative similarity occurs between $\Delta\mathcal{S}$ for $De=0.1$ (figure \ref{fig:extraStretchDept1} to \ref{fig:extraStretchDept1L200}), $\Delta De_\text{local}$ (figure \ref{fig:Local_De.eps}) and the $\Delta\text{FTLE}$ (figure \ref{fig:FTLE_pt1}) \& $\Delta\text{FTS}$ fields (figure \ref{fig:S_pt1}) associated with $t=0.1$. The three latter fields provide a good qualitative assessment of the polymer stretch, because for $De=0.1$ the polymer responds only to the local strain rate. $De=0.4$ plots in figure \ref{fig:extraStretchDept4L10} to \ref{fig:extraStretchDept4L200} show a wake of highly stretched polymers around the extensional axis. The increase in wake's intensity with $De$ is complemented by the analysis of the polymer stretch along the stagnation streamline for this regime performed in section \ref{sec:StagSteamline} (see figure \ref{fig:Stagnationstretch_smallDe}). Additionally, there is a collapsed region at the rear stagnation point and small highly stretched region on the surface, $45^\circ$ from the extensional axis. The region of highly stretched polymers around the extensional axis for $De=0.1$ is elongated to a wake for $De=0.4$ due to the finite relaxation time of the polymers. These observations are complemented by the earlier treatment of surface polymer stretch shown in figure \ref{fig:surfaceConfiguration}.

The polymer stretch observed here for $De$ below the coil-stretch transition point is mechanistically explained by considering the stretching effect of the velocity gradients, visualized through the $\Delta De_\text{local}$ field of figure \ref{fig:Local_De.eps}, on the polymers convecting past the sphere. As the coiled undisturbed polymers arrive at positive $\Delta De_\text{local}$ regions (red in figure \ref{fig:Local_De.eps}) around 45$^\circ$ from extension axis at the particle surface, they are stretched. These stretched polymers convect downstream, are collapsed by the low stretching region (blue on the particle surface near x-axis in figure \ref{fig:Local_De.eps}) but are again stretched in a high stretching region around the extensional axis (red downstream of the particle surface in figure \ref{fig:Local_De.eps}). As $De$ increases, the time taken for the polymers to relax to their undisturbed state increases and hence a wake of stretched polymers forms that persists for longer downstream distances from the particle at larger $De$ within this $De$ regime.

Before moving to the analysis of polymer stretch for $De\ge 0.5$, we briefly comment on the comparison of the polymer stretch around the sphere placed in a uniaxial extensional flow considered here and that in a uniform or a simple shear flow considered previously by Chilcott \& Rallison \cite{chilcott1988creeping} and Yang \& Shaqfeh \cite{yang2018mechanism} respectively. Local kinematics of the velocity field in the region just downstream of the sphere placed in a uniform flow is similar to that in a uniaxial extensional flow. Therefore, as found by Chilcott \& Rallison \cite{chilcott1988creeping} for a uniform flow case, just downstream of the particle, a large polymer stretch region exists. This is similar to that in a uniaxial extensional flow at low $De$ considered here. As we will see below, increasing $De$ beyond a certain $L$ dependent value changes the qualitative nature of the polymer stretch field in the uniaxial extensional flow, i.e., instead of a region of highly stretched polymers around the particle, there is rather a region of polymer collapse as compared to undisturbed polymers. However, in uniform flow considered by Chilcott \& Rallison \cite{chilcott1988creeping} the polymer stretch field remains qualitatively similar at all $De$ as the region of highly stretched polymers downstream of the particle becomes more intense and extends  further downstream  upon increasing $De$.

The topology of the streamlines around the sphere is drastically different for a simple shear flow than for the uniaxial extensional flow. As a result, the polymer stretches differently in the two flows. Unlike the uniaxial extensional flow described above, the simple shear flow induces a region of closed streamlines around the particle. As shown by Yang \& Shaqfeh \cite{yang2018mechanism}, a large polymer stretch region starts in the compressional quadrant of the imposed simple shear just inside the separatrix between the open and closed streamline regions. This region extends downstream into the extensional quadrant and goes through the separatrix. Increasing $De$ increases the intensity and downstream extent (perhaps due to the increasing polymer relaxation time) of the highly stretched polymer region.

Figure \ref{fig:polymerstretch_mediumDe} shows contours of $\Delta\mathcal{S}/L$ for three different $L$ at $De=$0.5, 0.6 and 1.0. In all cases the polymers on the particle surface are collapsed (blue region) relative to the undisturbed polymers. This collapse is related to the local extensional rate captured by $\Delta De_\text{local}$ in figure \ref{fig:Local_De.eps}. Unlike the low $De$ case the undisturbed polymers far from the particle have undergone a coil-stretch transition and are almost fully stretched near $L$. When these polymers from regions upstream of the particle arrive close to the surface while convecting along the compressional axis they first observe negative $\Delta De_\text{local}$ (blue region in figure \ref{fig:Local_De.eps} close to the particle on the y-axis) near the front stagnation point. In this region the velocity is small and the polymers spend a long time here to locally undergo a stretch-to-coil transition.  The positive $\Delta De_\text{local}$ (red region on particle surface in figure \ref{fig:Local_De.eps}) along the particle surface partially stretches them. This partial recovery can be observed from  surface stretch plots of figure \ref{fig:surfaceConfiguration}. From this figure and equation \eqref{eq:maxsurfstretchLargeL} it can also be observed that maximum surface stretch is much smaller than $L$. The polymers lose even this partial recovery as their stretch reduces when they arrive near the rear stagnation point. Therefore, relative to the undisturbed state polymers in the region close to the particle surface remain collapsed. They are fully collapsed to their equilibrium state on both stagnation points where velocity and its gradients are zero.

The effects of limited polymer extensibility, $L$, are clear for the cases in figure \ref{fig:polymerstretch_mediumDe} as the $L=10$ figures are qualitatively different from the larger $L$ cases of 50 and 200 in the region around the extensional axis. In this region we observe a wake of more stretched polymers for $L=10$ only (at all $De$). As the collapsed polymers near the particle surface convect along the extensional axis they undergo a coil-stretch transition. While the undisturbed polymers are close to fully stretched they can never be stretched at their maximum extensibility $L$ (in equation \eqref{eq:ConfigurationUndisturbed_simple} the polymer relaxation term is inversely proportional to $L^2-\text{tr}(\boldsymbol{\Lambda}^{(0U)})$ where $\sqrt{\text{tr}(\boldsymbol{\Lambda}^{(0U)})}$ is the undisturbed polymer stretch). If the local extension rate is large enough there is a room for slightly more extension. The local extension rate just downstream of the particle along the extensional axis is larger than the undisturbed extension rate. This is evident from the positive $\Delta De_\text{local}$ in that region in figure \ref{fig:Local_De.eps}. Therefore, a wake of polymers that are more stretched than the undisturbed polymers forms along the extensional axis for $L=10$ as shown in figure \ref{fig:polymerstretch_mediumDe}. This wake reduces in intensity with $De$ because the undisturbed polymer stretch increases.

Within the $De$ regime of figure \ref{fig:polymerstretch_mediumDe} discussed above if $L$ is large enough as exemplified by $L=50$ and 100, the collapsed polymers in the region close to the particle surface have a large amount of stretch to recover as they convect along the extensional axis. Therefore, the polymer stretch remains upper bounded by the large undisturbed value. As $L$ is increased at a given $De$ collapsed polymers near the particle surface must travel further downstream to reach the far-field stretch and hence we see a longer blue region around the extensional axis for $L=200$ than for $L$=50 at $De=0.5$ in figure \ref{fig:polymerstretch_mediumDe}. This also occurs for $De=0.6$ and $De=1.0$ at $L=50$ and $200$, but it is not visible in figure \ref{fig:polymerstretch_mediumDe} as the collapse is very intense over the spatial extent shown. The increase with $L$ of the downstream distance required to recover the polymer stretch can however be observed by comparing the $\sqrt{\text{tr}(\boldsymbol{\Lambda}^{(0)})}/L$ curves along the extensional axis for different $L$ at $De=1.0$ in figure \ref{fig:Stagnationstretch_De1}. For $De=1.0$ (and also $De=0.6$) $\Delta\mathcal{S}/L$ plots for $L=50$ and 200 in figure \ref{fig:polymerstretch_mediumDe} are very similar to each other. This indicates the $L^2$ scaling of polymer configuration, $\boldsymbol{\Lambda}^{(0)}$, in the fluid region even in the presence of the particle for $De>0.5$ and large $L$.

Increasing $De$ at any fixed $L$ in the moderate $De$ regime shown in figure \ref{fig:polymerstretch_mediumDe} increases the spatial extent of collapse (blue region) since the increase in undisturbed polymer stretch at higher $De$ is more than the increase in local extension rate around the extensional axis downstream of the particle. This increase in spatial extent can also be viewed as arising due to the increased relaxation time or longer memory of the polymers of their once collapsed state. Therefore, polymers take longer distances along the streamlines to recover from their undisturbed stretch upon increasing $De$ within this moderate $De$ regime.

As already indicated, the features around the extensional axis discussed above are consistent with the analysis of polymer conformation on the extensional axis made in section \ref{sec:StagSteamline}. The far-field stretch recovery at large $L$ (increase in polymer stretch towards undisturbed value) and relaxation at small $L$ (reduction of polymer stretch towards its undisturbed value) along the extensional axis for $De=0.5$ and 1.0 were shown in figure \ref{fig:Stagnationstretch_Dept5} and \ref{fig:Stagnationstretch_De1}. In other words, on the extensional axis beyond a certain distance downstream of the particle, polymers are stretched more than the undisturbed values for small $L$, while they remain collapsed compared to the undisturbed polymers for large $L$ at these moderate $De$ less than 1.25. Due to spatial continuity and smoothness of the polymer stretch, the behavior of polymers on the extensional axis extends to a finite region off the axis, as we previously observed (figure \ref{fig:polymerstretch_mediumDe}).

Figure \ref{fig:polymerstretch_largeDe} shows $\Delta\mathcal{S}/L$ for three different $L$ and three different large $De$. Some of the features in the plots in figure \ref{fig:polymerstretch_largeDe} at each $L$ can be understood by viewing them as resulting from a further increase in $De$  from the moderate $De$ values for that $L$ in figure \ref{fig:polymerstretch_mediumDe}. Due to the same mechanism as that discussed above for $0.5\le De\le 1.0$ polymers collapse in a region close to the particle surface for all $L$ and $De$ shown in figure \ref{fig:polymerstretch_largeDe}. According to this mechanism collapse occurs as fully stretched undisturbed polymers undergo a local stretch-to-coil transition when they arrive in the negative $\Delta De_\text{local}$ regions of figure \ref{fig:Local_De.eps} near the front stagnation point and remain collapsed until they leave the negative $\Delta De_\text{local}$ regions around the rear stagnation point. As $De$ or the imposed extension rate is increased, the undisturbed polymer stretch increases and hence the intensity of the wake of highly stretched polymers around the extensional axis for $L=10$ is reduced starting from the $De=0.5$ case of figure \ref{fig:polymerstretch_mediumDe} and proceeding to the $De=5.0$ case of figure \ref{fig:polymerstretch_largeDe}. The manner in which the large $De$ regime represented in figure \ref{fig:polymerstretch_largeDe}  differs from the moderate $De$ regime of figure \ref{fig:polymerstretch_mediumDe} is that for each $L$ upon increasing $De$ the spatial extent of the collapse (blue regions) around the particle surface reduces in the former while it increases in the latter. We already discussed the reasons for the moderate $De$ behavior. In the large $De$ regime, as the extension rate or $De$ is increased to very large values the collapsed polymers from the near surface region quickly recover to their undisturbed values. This is evidenced by the increasing white and reducing blue region upon increasing $De$ from 1.5 to 5.0 for a given $L$ in figure \ref{fig:polymerstretch_largeDe}. The collapse on the particle surface however becomes more intense as the undisturbed stress (and hence stretch) increases with $De$ for all $De$ as shown in figure \ref{fig:Stresses_Undisturbed}. For some $De$ values at the beginning of the large $De$ regime stronger recovery from collapse overwhelms the increase in undisturbed polymer stretch even for large $L$ and there is a region of slightly more polymer stretch. This is observed as a light red region in the $De=1.5$ and 3 plots at $L=50$ and 200 in figure \ref{fig:polymerstretch_largeDe}. It is consistent with the analysis of the stagnation streamline in section \ref{sec:StagSteamline}, where figure \ref{fig:Stagnationstretch_largeDe} indicates that for $De>1.25$, after a small distance from the particle, the polymer stretch increases very slightly above the undisturbed value.


As discussed in this section, the effect of a sphere on the polymer stretch in a uniaxial extensional flow is qualitatively different for small (a region of large polymer stretch downstream of the particle) and large (a region of polymer collapse around the particle) $De$. However, as noted earlier for the uniform and simple shear flow considered by Chilcott \& Rallison \cite{chilcott1988creeping} and Yang \& Shaqfeh \cite{yang2018mechanism} respectively, the polymer stretch does not change qualitatively with an increase in $De$. This occurs because the undisturbed or far-field polymers stretch differently at small and large $De$ in a uniaxial extensional flow, whereas, in a uniform and a simple shear flow, the undisturbed polymers behave similarly at all $De$. In a uniform flow, the undisturbed polymers remain in equilibrium (unstretched) configuration for all $De$. In a simple shear flow considered by Yang \& Shaqfeh \cite{yang2018mechanism}, while the undisturbed polymer stretch increases with $De$, it does not exhibit any drastic changes, such as a coil-stretch transition observed in a uniaxial extensional flow at $De=0.5$ (discussed in section \ref{sec:UndisturbedConfiguration}).

\subsubsection*{Polymers stretch like lines of dye released at previous times}
The $\Delta\mathcal{S}$ field at each of the $De$ and $L$ combinations shown in figures \ref{fig:polymerstretch_smallDe}, \ref{fig:polymerstretch_mediumDe} and \ref{fig:polymerstretch_largeDe} can be qualitatively matched to a $\Delta\text{FTS}$ field from $t=0.1$ to 50 shown in figure \ref{fig:S-field}. This suggests that the particle changes the steady-state polymer stretch ($\Delta\mathcal{S}$ field), for a given $De$ and $L$, and the stretch of a non-diffusive line of dye released in the flow a certain time, $t$ before, in a similar way. Considering $L=10$, we find that the changes in $\Delta\mathcal{S}$ observed by increasing $De$ from 0.1 through 5 in figures \ref{fig:polymerstretch_smallDe}, \ref{fig:polymerstretch_mediumDe} and \ref{fig:polymerstretch_largeDe} are similar to the changes in $\Delta\text{FTS}$ observed by increasing $t$ from 0.1 through 5 in figure \ref{fig:S-field}. For these values of $De$  and $t$ the polymers/ dye-elements are stretched more than their undisturbed counterparts in a region downstream of the particle and as $De$ or $t$ is increased the stretching region takes the form of a wake that first intensifies with increasing $De$ or $t$. Upon further increasing $De$ or $t$, this wake of stretched components becomes less intense, while a region of less stretched polymers/ dye-elements develops around the particle. At larger $L$, exemplified by $L=200$, changes in the $\Delta\mathcal{S}$ field observed by increasing $De$ from 0.1 through 1.5 in figures \ref{fig:polymerstretch_smallDe}, \ref{fig:polymerstretch_mediumDe} and \ref{fig:polymerstretch_largeDe} are analogous to changes in the $\Delta\text{FTS}$ field observed by increasing $t$ from 0.1 through 50 in figure \ref{fig:S-field}. Polymers with larger extensibility $L$ remember a longer history of previous stretching and this is reflected in the larger $t$ values corresponding to a given Deborah number when $L$ is larger.
 At $L=200$, after the coil-stretch transition at $De\ge0.5$, a region of collapsed polymers forms around the particle. Similarly, a significant region of less stretched dye-elements forms around the particle as observed in the $\Delta\text{FTS}$ field for $t\gtrsim3$ in figure \ref{fig:S-field}. Upon increasing $De$, the region of collapsed polymers intensifies, but becomes thinner, similar to the stretch of dye-elements from $t=5$ to 50 in the $\Delta\text{FTS}$ field of figure \ref{fig:S-field}. Beyond $De=1.5$, the polymer stretch ($\Delta\mathcal{S}$) is similar to the stretch of dye-elements ($\Delta\text{FTS}$) with large $t\approx 50$ in most of the volume around the sphere, except the extensional axis, which is slightly positive for $\Delta\mathcal{S}$ and slightly negative for $\Delta\text{FTS}$. Performing a similar analysis of the $\Delta\text{FTS}$ and $\Delta\mathcal{S}$ field may lead to useful insights in other relevant flows of polymeric fluid around a sphere such as those considered by Chilcott \& Rallison \cite{chilcott1988creeping} (uniform flow) and Yang \& Shaqfeh \cite{yang2018mechanism} (simple shear flow).

\subsubsection*{Non local effects}
A better match of $\Delta\mathcal{S}$ with the $\Delta\text{FTS}$ field, than with the $Q$ or $\Delta De_\text{local}$ field indicates the importance of non-local effects on the polymer configuration that arise due to polymer convection. In order to directly observe the non-local effect on the configuration from the FENE-P equation, we calculate the configuration after ignoring the convection term i.e. by solving the algebraic equation,
\begin{equation}\label{eq:ConfigurationLocal}
\nabla \mathbf{u}^\text{T}\cdot\boldsymbol{\Lambda}+\boldsymbol{\Lambda}\cdot\nabla\mathbf{u}+\frac{1}{De}(b\boldsymbol{\delta}-f\boldsymbol{\Lambda})=0.
\end{equation}
\begin{figure}[h!]
	\centering
	\subfloat{\includegraphics[width=0.25\textwidth]{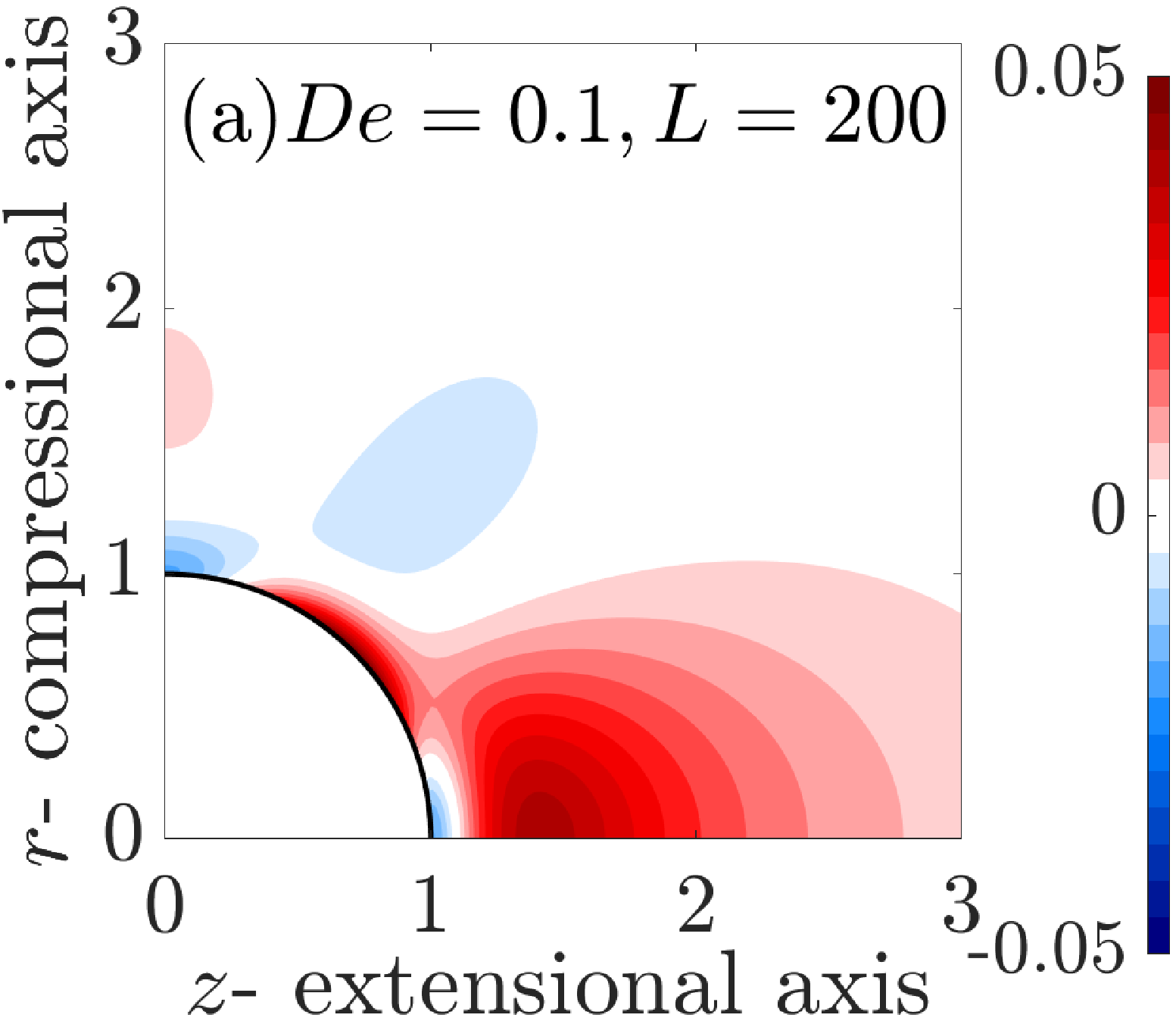}\label{fig:NonLocalDept1}}\hfill
	\subfloat{\includegraphics[width=0.25\textwidth]{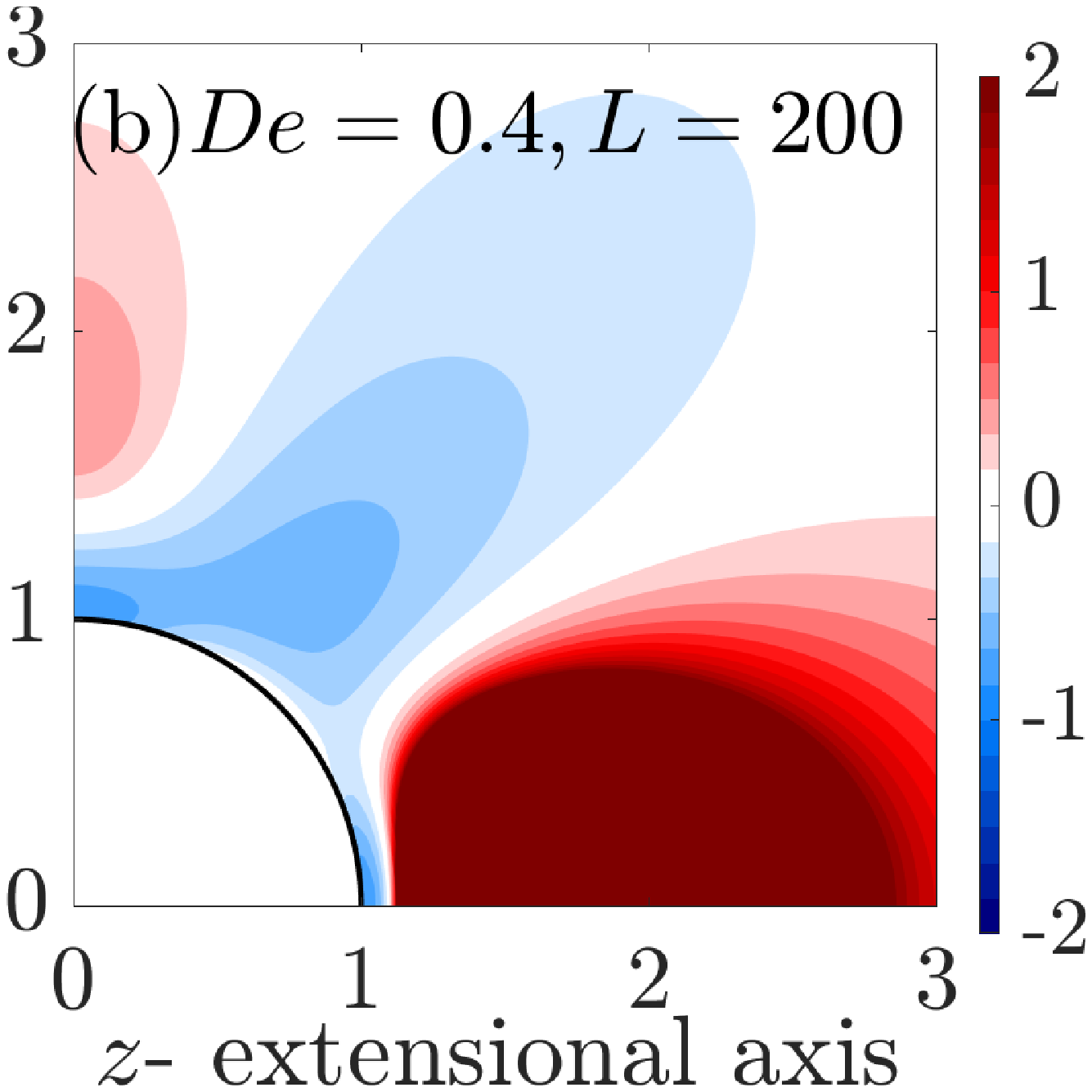}\label{fig:NonLocalDept4}}\hfill
	\subfloat{\includegraphics[width=0.25\textwidth]{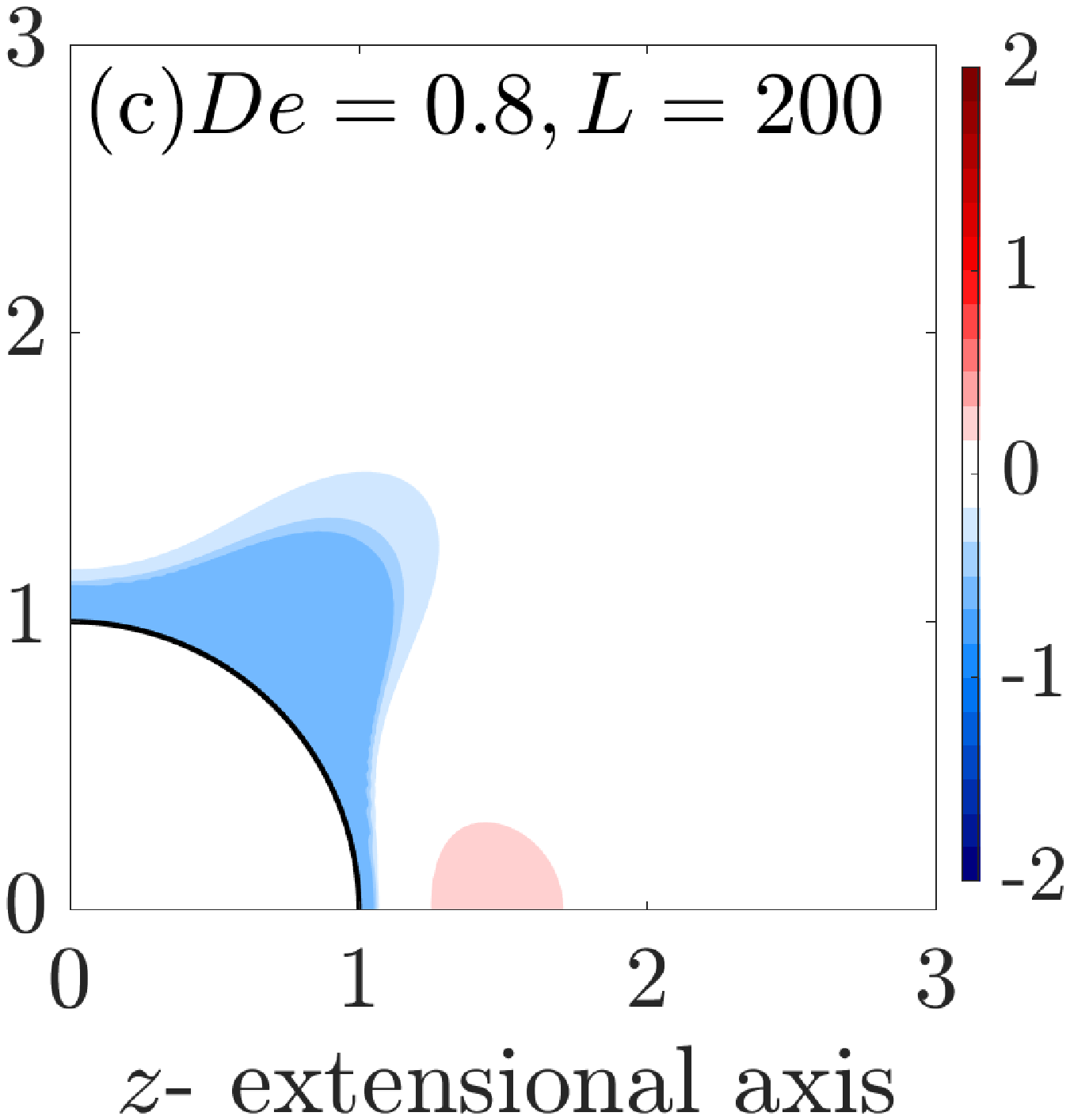}\label{fig:NonLocalDept8}}\hfill
	\subfloat{\includegraphics[width=0.25\textwidth]{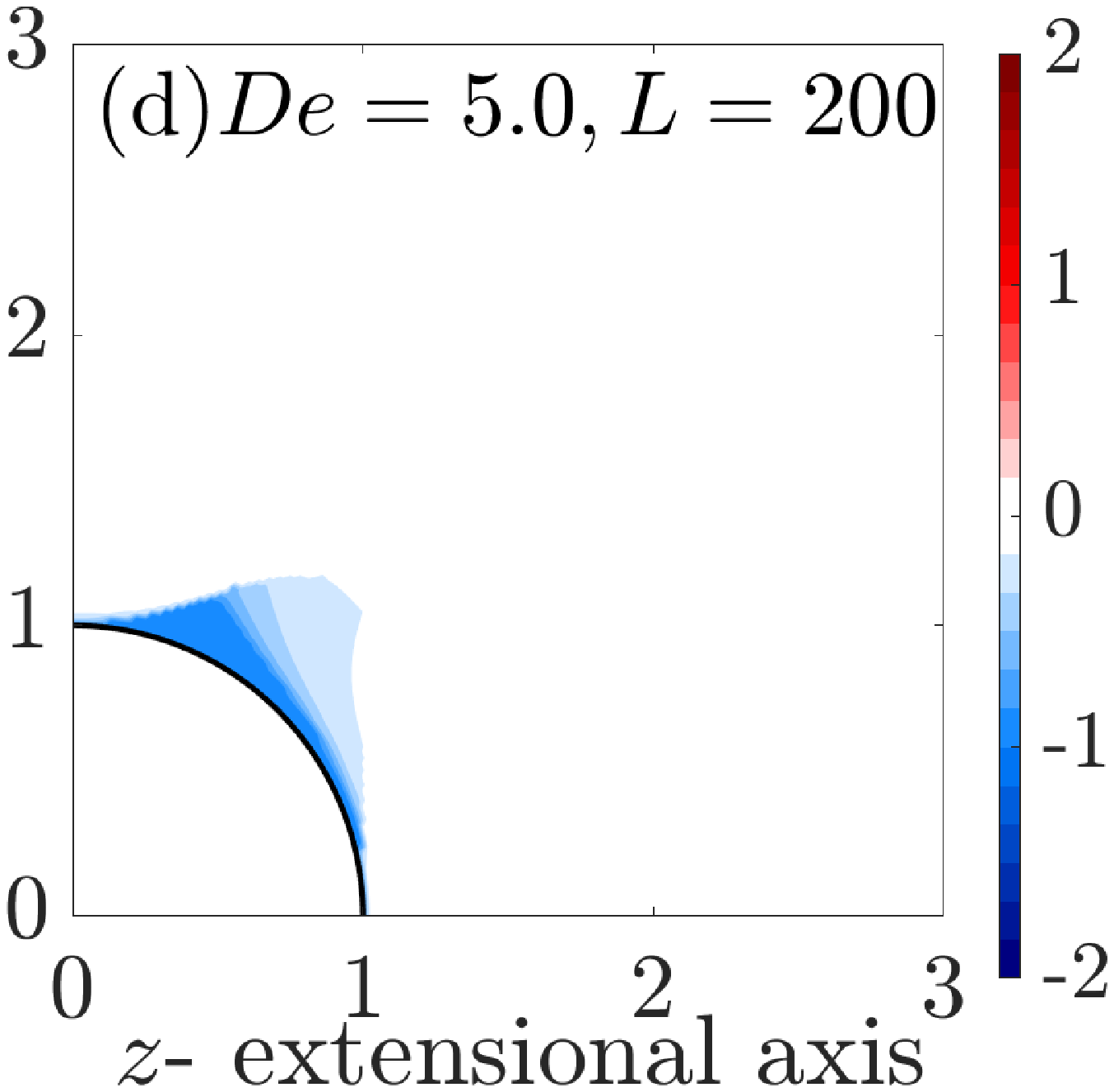}\label{fig:NonLocalDe5}}
	\caption {Change in polymer stretch from a local calculation: $\Delta\mathcal{S}$ for (a) $De=0.1$ and (b) 0.4 and $\Delta\mathcal{S}/L$  for (c) $De=0.8$ and (d) 5.0. $L=200$ for all cases.\label{fig:LocalConfig}}
\end{figure}
For $L=200$, we show the  $\Delta\mathcal{S}$ for $De=0.1$ and 0.4 and $\Delta\mathcal{S}/L$ for $De=0.8$ and 5.0 in  figure \ref{fig:LocalConfig}. The figure for $De=0.1$ is very similar to the actual polymer stretch in figure \ref{fig:extraStretchDept1L200} as at very low $De$ the non-local convective effects are negligible. However, at $De=0.4$ ignoring the non-local effects drastically changes the polymer stretch as observed by comparing figure \ref{fig:NonLocalDept4} with \ref{fig:extraStretchDept4L200}. The wake of highly stretched polymers in figure \ref{fig:NonLocalDept4} is very intense as the polymers are `fixed' in place in the high stretching region downstream of the particle. The wake is thicker for the non-convecting polymers in figure \ref{fig:NonLocalDept4} than that of the polymers in figure \ref{fig:NonLocalDept4} that are convected by the underlying velocity field. The stretch of the non-convecting polymer in figure \ref{fig:NonLocalDept4} is kinematically explained by the $Q$ and $\Delta De_\text{local}$ fields of figure \ref{fig:Kinematic_1}. Beyond the coil-stretch transition $De$, a non-convecting polymer collapses in the rotation dominated region of positive $Q$ as shown in figures 13c and 13d. These are not at all similar to the actual polymer stretch behavior described earlier. Therefore, when $De$ is not negligibly small, the non-local effects due to convection are very important in accurately determining the polymer stretch.

\section{Rheology}\label{sec:Rheology}
In this section we consider the rheology of the suspension. First, we describe the method for determining the mean stress.  We then validate our simulations at low $De$ with the theoretically available results for a suspension in a second order fluid. Finally, we present our results for finite $De$.
\subsection{Ensemble averaging formulation and generalized reciprocal theorem}\label{sec:RheologyTheory}
At any location in the suspension the stress is,
\begin{equation}\label{eq:constitutive3}
\boldsymbol{\sigma}=-p\boldsymbol{\delta}+2\boldsymbol{e}+\boldsymbol{\Pi}+\boldsymbol{\sigma}^\text{E},
\end{equation}
where $\boldsymbol{\sigma}^\text{E}$ is the extra stress inside a particle which is zero in the fluid region so equation \eqref{eq:constitutive3} reduces to \eqref{eq:constitutive1}.  As the isotropic part, $\text{tr}( \boldsymbol{\sigma})/3$, can be absorbed into the pressure, the  deviatoric stress is most relevant to the suspension rheology,
\begin{equation}
\hat{\boldsymbol{\sigma}}= \boldsymbol{\sigma}-\frac{1}{3}\text{tr}( \boldsymbol{\sigma})=2\boldsymbol{e}+ \hat{\boldsymbol{\Pi}} +\boldsymbol{\hat{\sigma}}^\text{E}.
\end{equation}
To study the rheology of a dilute suspension of particles we use the ensemble averaging technique \cite{koch2006stress,koch2016stress}. The ensemble average ($\langle . \rangle$) of the deviatoric stress in the suspension is,
\begin{equation}
\langle\hat{\boldsymbol{\sigma}}\rangle=2\langle\boldsymbol{e}\rangle+\langle \hat{\boldsymbol{\Pi}} \rangle+n\hat{\text{\textbf{S}}},
\end{equation}
where
\begin{equation}
n\hat{\text{\textbf{S}}}=\int_{|\mathbf{r}-\mathbf{r}_1|\le 1}\text{d}\mathbf{r}_1\langle \boldsymbol{\sigma}^\text{E}\rangle_1(\mathbf{r}|\mathbf{r}_1)P(\mathbf{r}_1),
\end{equation}
is the particle stresslet \cite{batchelor1970stress} and $P(\mathbf{r}_1)$ is the probability density of a particle being location at location $\mathbf{r}_1$. For a quantity $A$, $\langle A \rangle_1(\mathbf{r}|\mathbf{r}_1)$ represents the conditional ensemble average with one particle at $\mathbf{r}_1$ \cite{koch2016stress},
\begin{equation}
\langle A\rangle_1(\mathbf{r}|\mathbf{r}_1)=\int\text{d}\mathbf{r}_2\dots \mathbf{r}_N P(\mathbf{r}_2\dots \mathbf{r}_N|\mathbf{r}_1)A,
\end{equation}
where $P(\mathbf{r}_2\dots \mathbf{r}_N|\mathbf{r}_1)$ is the conditional probability density function within a suspension of $N$ particles.
The ensemble average of the rate of strain, $\langle\boldsymbol{e}\rangle$ is the imposed rate of strain, $\mathbf{E}$ (equation \eqref{eq:UndisturbedStrainRate}), determined by the motion of the suspension boundaries. For a dilute suspension with well separated particles, the hydrodynamic interaction between the particles is negligible. The conditional average stresses are then approximately the same as those around an isolated particle \cite{koch2016stress}. Hence, the conditional averaging symbols are removed, and the ensemble average of the deviatoric part of the stresslet \cite{batchelor1970stress} is,
\begin{eqnarray}
&\hat{\text{\textbf{S}}}(\boldsymbol{\sigma})=\int_{|\mathbf{r}-\mathbf{r}_1|= 1}\text{d}A\big\{\frac{1}{2}[{\text{\textbf{nn}}}\cdot  \boldsymbol{\sigma}+\text{\textbf{n}}\cdot  \boldsymbol{\sigma}\text{\textbf{n}} ]-\frac{1}{3}\boldsymbol{\delta}\text{\textbf{n}}\cdot  \boldsymbol{\sigma}\cdot \text{\textbf{n}}\big\}.
\end{eqnarray}
As we discuss later, unlike previous studies \cite{greco2007rheology,housiadas2009rheology}, $\langle {\boldsymbol{\Pi}} \rangle$ cannot be simply expressed as a volume average of $\boldsymbol{\Pi}$, for the integral would diverge logarithmically.
Similar to the other quantities (section \ref{sec:Formulation}), a regular perturbation in $c$ is used to expand the stresslet: $\hat{\text{\textbf{S}}}=\hat{\text{\textbf{S}}}^{(0)}+c\hat{\text{\textbf{S}}}^{(1)}+\mathcal{O}(c^2)$. The leading order deviatoric stresslet is the stresslet due to a unit sphere in a Newtonian fluid, given by Einstein \cite{einsteinoriginal},
\begin{equation}
\hat{\text{\textbf{S}}}^{(0)}= \hat{\text{\textbf{S}}}(\boldsymbol{\tau}^{(0)})=\frac{20 \pi}{3}\mathbf{E}.
\end{equation}

For the calculation of an isolated particle in an infinite expanse of polymeric fluid to be useful for evaluating dilute suspension rheology, the ensemble averaged polymer stress, $\langle{\boldsymbol{\Pi}} \rangle$ (or its deviatoric value $\langle \hat{\boldsymbol{\Pi}} \rangle$), needs to be related to a volume integral involving stresses in the vicinity of the particle under relevant assumptions. However, simplifying the ensemble average of the polymer stress requires a careful treatment that is forthcoming. The ensemble averaged deviatoric stress up to $\mathcal{O}(c)$ is,
\begin{equation}
\langle\hat{\boldsymbol{\sigma}}\rangle=2(1+2.5\phi)\mathbf{E}+c(\langle \hat{\boldsymbol{\Pi}}^{(0)} \rangle+n\hat{\text{\textbf{S}}}^{(1)}),
\end{equation}
where $\phi=4\pi n/3$ is the particle volume fraction of unit spheres.
The polymer stress, ${\boldsymbol{\Pi}}^{(0)}$, decays as $1/r^3$ at large distances from a particle. Therefore, if the stress in the dilute suspension due to the extra polymer stress in the presence of particle is approximated by the volume average of ${\boldsymbol{\Pi}}^{(0)}$ in an infinite expanse of fluid, as in \cite{greco2007rheology,housiadas2009rheology}, a logarithmic divergence occurs. Hence, it is important to carefully simplify the expression of ensemble averaging after identifying the source of the $1/r^3$ far-field scaling of  ${\boldsymbol{\Pi}}^{(0)}$. This is described for the Oldroyd-B equations by Koch et al. \cite{koch2016stress}. Here, we repeat that derivation for the FENE-P equations.

The velocity field can be decomposed into the imposed velocity and a perturbation caused by the particle,
\begin{equation}
\mathbf{u}^{(0)}=\mathbf{u}'+\mathbf{E}\cdot \boldsymbol{r},
\end{equation}
where $\mathbf{E}\cdot \boldsymbol{r}=\langle\mathbf{u}\rangle$. Decomposing the polymer configuration as the sum of undisturbed ${\boldsymbol{\Lambda}}^{(0U)}$, linear, ${\boldsymbol{\Lambda}}^{(0L)}$, and non-linear, ${\boldsymbol{\Lambda}}^{(0N)}$ components and linearizing the steady-state FENE-P equation \eqref{eq:constitutive2}, \eqref{eq:Configuration} and \eqref{eq:steadystate} leads to an equation for the linear polymer configuration that is forced by the perturbation in velocity gradients about the undisturbed value,
\begin{eqnarray}
&\begin{split}
\langle\mathbf{u}\rangle\cdot \nabla \boldsymbol{\Lambda}^{(0L)}-\nabla {\langle\mathbf{u}\rangle}^\text{T}\cdot\boldsymbol{\Lambda}^{(0L)}-\boldsymbol{\Lambda}^{(0L)}\cdot\nabla\langle\mathbf{u}\rangle+\frac{1}{De}\Big(f^{(0U)}\boldsymbol{\Lambda}^{(0L)}+&\frac{\text{tr}(\boldsymbol{\Lambda}^{(0L)})(f^{(0U)})^2}{L^2}\boldsymbol{\Lambda}^{(0U)}\Big)=\\&\nabla {\mathbf{u}'}^\text{T}\cdot\boldsymbol{\Lambda}^{(0U)}+\boldsymbol{\Lambda}^{(0U)}\cdot\nabla\mathbf{u}'.\label{eq:LinPolyConfig}
\end{split}\label{eq:LinearConfiguration}
\end{eqnarray}
The polymer stress is also decomposed into undisturbed, ${\boldsymbol{\Pi}}^{(0U)}$, linear, ${\boldsymbol{\Pi}}^{(0L)}$, and non-linear, ${\boldsymbol{\Pi}}^{(0N)}$, parts,
\begin{equation}
{\boldsymbol{\Pi}}^{(0)}={\boldsymbol{\Pi}}^{(0U)}+{\boldsymbol{\Pi}}^{(0L)}+{\boldsymbol{\Pi}}^{(0N)},\label{eq:Decomposetress}
\end{equation}
where ${\boldsymbol{\Pi}}^{(0U)}$ is defined in equation \eqref{eq:ConfigurationUndisturbed_simple},
\begin{equation}
\boldsymbol{\Pi}^{(0L)}=\frac{1}{De}\Big(f^{(0U)}\boldsymbol{\Lambda}^{(0L)}+\frac{\text{tr}(\boldsymbol{\Lambda}^{(0L)})(f^{(0U)})^2}{L^2}\boldsymbol{\Lambda}^{(0U)}\Big),\label{eq:LinStress}
\end{equation}
and ${\boldsymbol{\Pi}}^{(0N)}$ is defined as the difference between the total polymer stress, defined in equation \eqref{eq:constitutive2}, and the sum of undisturbed and linear polymer stresses. We solve equation \eqref{eq:LinearConfiguration} for ${\boldsymbol{\Lambda}}^{(0L)}$ using the method of characteristics, with the streamlines of the undisturbed velocity field, $\langle\mathbf{u}\rangle=\mathbf{E}\cdot \boldsymbol{r}$, acting as the characteristic curves and the far-field boundary condition,
\begin{equation}
\boldsymbol{\Lambda}^{(0L)}=0,\hspace{0.2in} r\rightarrow\infty. \label{eq:LinearBoundary}
\end{equation}
This solution combined with the solution for ${\boldsymbol{\Pi}}^{(0)}$ using the method of characteristics described earlier in section \ref{sec:Formulation} is used to obtain the nonlinear polymer stress field, ${\boldsymbol{\Pi}}^{(0N)}$.

By noting the $1/r^3$  far field scaling of $\nabla {\mathbf{u}'}$, we identify that $1/r^3$ far-field scaling of  ${\boldsymbol{\Pi}}^{(0)}$ arises from $\boldsymbol{\Lambda}^{(0L)}$. The ensemble averages of velocity and velocity gradients are their respective undisturbed values. Therefore, the ensemble averages of their disturbance about the undisturbed states are zero and the ensemble average of \eqref{eq:LinearConfiguration} is,
\begin{equation}
\langle\mathbf{u}\rangle\cdot \nabla \langle\boldsymbol{\Lambda}^{(0L)}\rangle-\nabla {\langle\mathbf{u}\rangle}^\text{T}\cdot\langle\boldsymbol{\Lambda}^{(0L)}\rangle-\langle\boldsymbol{\Lambda}^{(0L)}\rangle\cdot\nabla\langle\mathbf{u}\rangle+\frac{1}{De}\Big(f^{(0U)}\langle\boldsymbol{\Lambda}^{(0L)}\rangle+\frac{\text{tr}(\langle\boldsymbol{\Lambda}^{(0L)}\rangle)(f^{(0U)})^3}{L^2}\Big)=0.\label{eq:LinearConfigurationAverage}
\end{equation}
The far-field boundary condition for $\langle\boldsymbol{\Lambda}^{(0L)}\rangle$ is zero from the ensemble average of equation \eqref{eq:LinearBoundary}. The solution to equation \eqref{eq:LinearConfigurationAverage} subject to zero boundary conditions yields the ensemble average of linear polymer stress, $\langle{\boldsymbol{\Lambda}}^{(0L)}\rangle=0$  and hence,
\begin{equation}
\langle\hat{\boldsymbol{\Pi}}^{(0L)}\rangle =0.
\end{equation}
Therefore, the volume average of the ensemble average of the deviatoric polymer stress is,
\begin{equation}
\langle\hat{\boldsymbol{\Pi}}^{(0)}\rangle=\hat{\boldsymbol{\Pi}}^{(0U)}+\langle\hat{\boldsymbol{\Pi}}^{(0N)}\rangle
\end{equation}
While the deviatoric polymer stress inside the particle vanishes, $\hat{\boldsymbol{\Pi}}^{(0)}=0$, the non-linear stress $\hat{\boldsymbol{\Pi}}^{(0N)}$ is finite and is obtained as,
\begin{equation}
\hat{\boldsymbol{\Pi}}^{(0N)}=-\hat{\boldsymbol{\Pi}}^{(0U)}-\hat{\boldsymbol{\Pi}}^{(0L)},\hspace{0.2in}r<1.\label{eq:NLParticle}
\end{equation}

Thus, we have shown that the linearized polymer stress does not contribute to the ensemble average stress.  This observation implies that there is no need to integrate the slowly decaying component $\hat{\boldsymbol{\Pi}}^{(0L)}\sim 1/r^3$ in the far field that would have otherwise led to a non-convergent integral for the particle-induced polymer stress.  The dilute particle assumption which allows one to remove the conditional averaging and approximate the ensemble average of the polymer stress as the volume integral of a quantity in an infinite fluid around an isolated particle is only applied to the non-linear polymer stress, $\hat{\boldsymbol{\Pi}}^{(0N)}$.

The $\mathcal{O}(c)$ stresslet, $\hat{\text{\textbf{S}}}^{(1)}$, can be decomposed into the contribution due to the first order perturbation in the solvent stress, $\hat{\text{\textbf{S}}}^{1\boldsymbol{\tau}}$, and that due to the polymer stress, $\hat{\text{\textbf{S}}}^{1\boldsymbol{\Pi}}$, acting on particle surface, i.e.
\begin{eqnarray}
&\hat{\text{\textbf{S}}}^{(1)}=\hat{\text{\textbf{S}}}^{1\boldsymbol{\tau}}+\hat{\text{\textbf{S}}}^{1\boldsymbol{\Pi}},\label{eq:stresslet_order1}\\
&\hat{\text{\textbf{S}}}^{1\boldsymbol{\tau}}=\hat{\text{\textbf{S}}}(\boldsymbol{\tau}^{(1)})=\int_{|\mathbf{r}-\mathbf{r}_1|= 1}\text{d}A\big\{\frac{1}{2}[{\text{\textbf{nn}}}\cdot \boldsymbol{\tau}^{(1)}+\text{\textbf{n}}\cdot \boldsymbol{\tau}^{(1)}\text{\textbf{n}} ]-\frac{1}{3}\boldsymbol{\delta}\text{\textbf{n}}\cdot  \boldsymbol{\tau}^{(1)}\cdot \text{\textbf{n}}\big\}, \label{eq:S1_tau}\\
&\hat{\text{\textbf{S}}}^{1\boldsymbol{\Pi}}=\hat{\text{\textbf{S}}}(\boldsymbol{\Pi}^{(0)})=\int_{|\mathbf{r}-\mathbf{r}_1|= 1}\text{d}A\hspace{0.1in}\big\{\frac{1}{2}[{\text{\textbf{nn}}}\cdot \boldsymbol{\Pi}^{(0)}+\text{\textbf{n}}\cdot \boldsymbol{\Pi}^{(0)}\text{\textbf{n}} ]-\frac{1}{3}\boldsymbol{\delta}\text{\textbf{n}}\cdot  \boldsymbol{\Pi}^{(0)}\cdot \text{\textbf{n}}\big\}\label{eq:S1_pi}.
\end{eqnarray}
Once the polymer configuration, $\boldsymbol{\Pi}^{(0)}$, is determined, $\hat{\text{\textbf{S}}}^{1\boldsymbol{\Pi}}$ can be calculated. However, $\hat{\text{\textbf{S}}}(\boldsymbol{\tau}^{(1)})=\hat{\text{\textbf{S}}}(\boldsymbol{\tau}(\boldsymbol{u}^{(1)}))$ depends upon the $\mathcal{O}(c)$ velocity $\boldsymbol{u}^{(1)}$. $\boldsymbol{u}^{(1)}$ is driven by $\nabla\cdot{\boldsymbol{\Lambda}}^{(0)}$ via the $\mathcal{O}(c)$ momentum equation.  Thus, $\hat{\text{\textbf{S}}}(\boldsymbol{\tau}^{(1)})$ indirectly depends on the polymer configuration, ${\boldsymbol{\Lambda}}^{(0)}$.

Using a generalized reciprocal theorem and the divergence theorem, Koch et al. \cite{koch2016stress} provide a mathematical framework to obtain $\hat{\text{\textbf{S}}}^{1\boldsymbol{\tau}}$ directly from $\boldsymbol{\Pi}^{(0)}$  and $\boldsymbol{\Pi}^{(0U)}$, thus avoiding the need to numerically evaluate $\boldsymbol{u}^{(1)}$ from the $\mathcal{O}(c)$ momentum conservation. An equivalent expression for $\hat{\text{\textbf{S}}}^{1\boldsymbol{\tau}}$ to the one given in Koch et al. \cite{koch2016stress} is,
\begin{align}\label{eq:S1_tau2}\begin{split}
\hat{\text{\textbf{S}}}^{1\boldsymbol{\tau}}=&-\int_{r=1} dA \text{\textbf{ n}}\cdot[\boldsymbol{\Pi}^{(0)}-\boldsymbol{\Pi}^{(0U)}]\cdot \mathbf{v} + \int_{r\rightarrow\infty} dA \text{\textbf{ n}}\cdot[\boldsymbol{\Pi}^{(0)}-\boldsymbol{\Pi}^{(0U)}]\cdot \mathbf{v}\\&-\int_{V_f}dV\text{  } [\boldsymbol{\Pi}^{(0)}-\boldsymbol{\Pi}^{(0U)}]:\nabla\mathbf{v},
\end{split}\end{align}
where $\mathbf{v}$ is the auxiliary velocity field used in the reciprocal theorem. The divergence-less Stokes auxiliary velocity field, $\mathbf{v}$, is chosen such that it undergoes extensional deformation at the particle surface and decays to zero far from the particle \cite{koch2016stress}. Hence, its expression depends on the particle shape and for a spherical particle,
\begin{equation}
v_{jkl}=\frac{5}{2}\Big(\frac{1}{r^5}-\frac{1}{r^7}\Big)r_jr_kr_l+\frac{1}{2r^5}(r_k\delta_{jl}+r_l\delta_{jk})+\Big(\frac{1}{2r^5}-\frac{5}{6r^3}\Big)r_j\delta_{kl}.\end{equation}
The second term in equation \eqref{eq:S1_tau2} is zero because the integrand decays as $r^{-5}$ when $r\rightarrow \infty$ ($[\boldsymbol{\Pi}^{(0)}-\boldsymbol{\Pi}^{(0U)}]\sim r^{-3}$ and $\mathbf{v}\sim r^{-2}$). Therefore,
\begin{equation}\label{eq:S1_tau3}
\hat{\text{\textbf{S}}}^{1\boldsymbol{\tau}}=-\int_{r=1} dA \text{\textbf{ n}}\cdot[\boldsymbol{\Pi}^{(0)}-\boldsymbol{\Pi}^{(0U)}]\cdot \mathbf{v} -\int_{V_f}dV\text{  } [\boldsymbol{\Pi}^{(0)}-\boldsymbol{\Pi}^{(0U)}]:\nabla\mathbf{v}, 
\end{equation}
Thus, in equation \eqref{eq:stresslet_order1} we have expressed the $\mathcal{O}(c)$ stresslet, $\hat{\text{\textbf{S}}}^{(1)}$, as the sum of $\hat{\text{\textbf{S}}}^{1\boldsymbol{\tau}}$ and $\hat{\text{\textbf{S}}}^{1\boldsymbol{\Pi}}$ which can be computed from the $\mathcal{O}(1)$ polymer stress field using \eqref{eq:S1_tau3} and \eqref{eq:S1_pi}, respectively.  This decomposition is based on the physical origins of the stress (Newtonian solvent and polymeric stress).

Next, we will derive a second decomposition of the $\mathcal{O}(c)$ stresslet.  We start with the observation that for any tensor stress field, $\boldsymbol{B}$ we find,
\begin{equation}\label{eq:generalResult}
\hat{\text{\textbf{S}}}(\boldsymbol{B})=\int_{r=1} dA \text{\textbf{ n}}\cdot\boldsymbol{B}\cdot \mathbf{v}.\end{equation}
Therefore,
\begin{equation}
\int_{r=1} dA \text{\textbf{ n}}\cdot[\boldsymbol{\Pi}^{(0)}-\boldsymbol{\Pi}^{(0U)}]\cdot \mathbf{v}=\hat{\text{\textbf{S}}}(\boldsymbol{\Pi}^{(0)})-\hat{\text{\textbf{S}}}(\boldsymbol{\Pi}^{(0U)})=\hat{\text{\textbf{S}}}^{1\boldsymbol{\Pi}}-\hat{\text{\textbf{S}}}^{\boldsymbol{\Pi} 0U},\label{eq:S1_tau_4}
\end{equation}
and substituting equation \eqref{eq:S1_tau_4} into \eqref{eq:S1_tau3} leads to
\begin{equation}
\hat{\text{\textbf{S}}}^{1\boldsymbol{\tau}}=\hat{\text{\textbf{S}}}^{\boldsymbol{\Pi} 0U}-\hat{\text{\textbf{S}}}^{1\boldsymbol{\Pi}}+\hat{\text{\textbf{S}}}^{1,\text{volume}}\label{eq:S1_tau4},
\end{equation}
where,
\begin{equation}
\hat{\text{\textbf{S}}}^{\boldsymbol{\Pi} 0U}=\hat{\text{\textbf{S}}}(\boldsymbol{\Pi}^{(0U)}),\hspace{0.2in}
\hat{\text{\textbf{S}}}^{1,\text{volume}}=-\int_{V_f}dV\text{  } [\boldsymbol{\Pi}^{(0)}-\boldsymbol{\Pi}^{(0U)}]:\nabla\mathbf{v}.\label{eq:StressletDecomp2}
\end{equation}
 Here, we can decomposition the $\mathcal{O}(c)$ stresslet as,
\begin{equation}
\hat{\text{\textbf{S}}}^{(1)}=\hat{\text{\textbf{S}}}^{\boldsymbol{\Pi} 0U}+\hat{\text{\textbf{S}}}^{1,\text{volume}}\label{eq:stresslet_order2},
\end{equation}
where $\hat{\text{\textbf{S}}}^{\boldsymbol{\Pi} 0U}$ is the stresslet on a unit fluid in the far field, and $\hat{\text{\textbf{S}}}^{1,\text{volume}}$ is the contribution due to the difference between the actual and undisturbed polymer stress in the fluid volume around the particle. We refer to the former as the undisturbed stresslet and the latter as the volumetric stresslet. For a spherical volume the undisturbed stresslet, $\hat{\text{\textbf{S}}}^{\boldsymbol{\Pi} 0U}$, is determined analytically as the following function of $\boldsymbol{\Pi}^{(0U)}$,
\begin{equation}\label{eq:stresslet_new}
\hat{\text{\textbf{S}}}^{\boldsymbol{\Pi} 0U}=\frac{4\pi}{3}\hat{\boldsymbol{\Pi}}^{(0U)}=\frac{4\pi}{3}\hat{\Pi}_{zz}^{(0U)}\mathbf{E}.
\end{equation}

To summarize,  the ensemble averaged deviatoric stress for polymer concentration $c$ and particle volume fraction $\phi$, is
\begin{equation}
\langle\hat{\boldsymbol{\sigma}}\rangle=(2+5\phi)\mathbf{E}+c\hat{\boldsymbol{\Pi}}^{(0U)}+c\phi\frac{3}{4\pi}(\hat{\text{\textbf{S}}}^{(1)}+\hat{\boldsymbol{\Pi}}^{\text{PP}}).\label{eq:averagestress}
\end{equation}
The symmetry of the imposed uniaxial extensional flow is maintained in $\langle\hat{\boldsymbol{\sigma}}\rangle$, so that,
\begin{equation}
\langle\hat{\boldsymbol{\sigma}}\rangle=[2+5\phi+c\hat{\Pi}^{(0U)}_{zz}+c\phi\frac{3}{4\pi}(\hat{\text{S}}^{(1)}_{zz}+\hat{\Pi}^{\text{PP}}_{zz})]\mathbf{E}.\label{eq:ExtVisc}
\end{equation}
 The two possible decompositions of the interaction stresslet contribution $\hat{\text{S}}^{(1)}_{zz}$ are,
\begin{equation}
\hat{\text{S}}^{(1)}_{zz}=\hat{\text{S}}^{1\boldsymbol{\tau}}_{zz}+\hat{\text{S}}^{1\boldsymbol{\Pi}}_{zz}=\hat{\text{{S}}}^{\boldsymbol{\Pi} 0U}_{zz}+\hat{\text{S}}^{1,\text{volume}}_{zz}, \label{eq:StressletContributions}
\end{equation}
and the particle-induced polymer stress contribution is,
\begin{equation}
\hat{\Pi}^{\text{PP}}_{zz}=\int_{V_f+V_p} dV\text{  } \hat{{\Pi}}^{(0N)}_{zz}.
\end{equation}
$\hat{\Pi}^{\text{PP}}_{zz}$ is further decomposed into the contribution from the fluid and particle volume, \label{eq:fluidstressTotal}
\begin{equation}
\hat{\Pi}^{\text{PP,fluid}}_{zz}=\int_{V_f} dV\text{  } \hat{{\Pi}}^{(0N)}_{zz},\hspace{0.2in} \hat{\Pi}^{\text{PP,particle}}_{zz}=\int_{V_p} dV\text{  } \hat{{\Pi}}^{(0N)}_{zz}. \label{eq:fluidstress}
\end{equation}

We can define the extensional viscosity of the suspension from the various components of the suspension stress discussed above. 
Equation \eqref{eq:ExtVisc} can be expressed as
\begin{equation}
\langle\hat{\boldsymbol{\sigma}}\rangle=2\mu\mathbf{E},\label{eq:ExtVisc2}
\end{equation} where,
\begin{equation}
\mu=1+2.5\phi+0.5c\hat{\Pi}^{(0U)}_{zz}+c\phi\frac{3}{8\pi}(\hat{\text{S}}^{(1)}_{zz}+\hat{\Pi}^{\text{PP}}_{zz})=1+\mu^\text{poly}+\mu^\text{part}\phi,\label{eq:ExtensionalViscosity}
\end{equation}
is the extensional viscosity of the suspension.
\begin{equation}
\mu^\text{poly}=0.5c\hat{\Pi}^{(0U)}_{zz}\label{eq:ExtenPolymers}
\end{equation}
is the polymer contribution to extensional viscosity in a particle-free polymeric fluid, and,
\begin{equation}
\mu^\text{part}=[2.5+c\frac{3}{8\pi}(\hat{\text{S}}^{(1)}_{zz}+\hat{\Pi}^{\text{PP}}_{zz})]\phi=2.5\phi+\mu^\text{intr}.\label{eq:ExtensionalViscosityParticles}
\end{equation}
is the extensional viscosity due to the presence of the particles. Within $\mu^\text{part}$, $2.5\phi$ is the Einstein \cite{einarsson2018einstein} viscosity that arises due to the stress on the particle surface in a Newtonian fluid and
\begin{equation}
\mu^\text{intr}=\frac{3}{8\pi}c\phi(\hat{\text{S}}^{(1)}_{zz}+\hat{\Pi}^{\text{PP}}_{zz})\label{eq:Interactionviscosity}
\end{equation}
is the extensional viscosity due to the particle-polymer interaction stress.
This completes the mathematical formulation of the suspension rheology. In the remaining part of this subsection we derive estimates for some quantities that will aid the forthcoming discussion of the results in section \ref{sec:RheologyResults}.

\subsubsection{Estimates of some of the components of the interaction stresslet}\label{sec:SomeStressletRelations}
\begin{figure}
	\centering
	\includegraphics[width=0.4\textwidth]{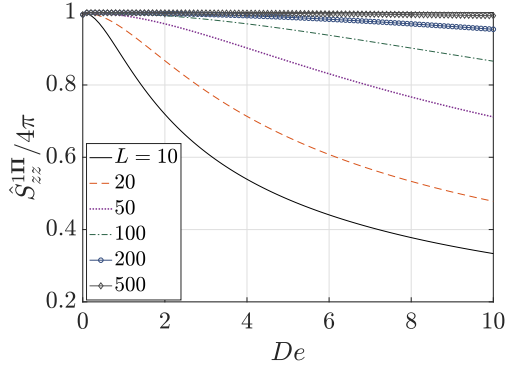}
	\caption {Stresslet due to the polymer stress, $\hat{\text{{S}}}^{1\boldsymbol{\Pi}}_{zz}$ as a function of $De$.\label{fig:polymerstresslet}}
\end{figure}

The estimates and discussion of this section will aid in our forthcoming discussion of the interaction stresslet, $\hat{\text{S}}^{(1)}_{zz}$ , and its sub-components.
Substituting equation \eqref{eq:approximatelargeDe} into \eqref{eq:stresslet_new} we obtain an analytical estimate the expression for the undisturbed stresslet or the stresslet on a unit fluid in the far field, $\hat{\text{{S}}}^{\boldsymbol{\Pi} 0U}_{zz}$, in the $De>0.5$ regime,
\begin{equation}
\hat{\text{{S}}}^{\boldsymbol{\Pi} 0U}_{zz}\approx \frac{16\pi}{9}\Big(1-\frac{1}{2De}\Big)L^2=\frac{4\pi}{3}\hat{{\Pi}}^{(0U)}_{zz},\hspace{0.2in}De>0.5.\label{eq:far_field_fluid_Stresslet_largeDe}
\end{equation}
$\hat{\text{S}}^{1\boldsymbol{\Pi}}_{zz}$ only depends upon the configuration tensor at the surface of the sphere, $\boldsymbol{\Lambda}^{(0)}|_{r=1}$, and is analytically determined once $\boldsymbol{\Lambda}^{(0)}|_{r=1}$ is known. Calculation of $\boldsymbol{\Lambda}^{(0)}|_{r=1}(z)$ requires computer algebra since it involves solving the cubic equation \eqref{eq:analytical_surface_stress}. However, under the assumption of equation \eqref{eq:stresslet_+approx_condition},
$$\frac{225 De^2 }{2 (L^2-3)}\ll 1,$$
valid for $L\gg 10De$, the stresslet contribution from the polymer stress is approximately, \begin{equation}
\hat{\text{{S}}}^{1\boldsymbol{\Pi}}_{zz}\approx 4\pi.\label{eq:approximate_polymer_stresslet}
\end{equation}
For example, for  $L\approx 100$, the assumption \eqref{eq:stresslet_+approx_condition}  and hence $\hat{\text{{S}}}^{1\boldsymbol{\Pi}}_{zz}$ in \eqref{eq:approximate_polymer_stresslet} is valid only for $De\lesssim 3$, as shown in figure \ref{fig:polymerstresslet}. At high $De$ the magnitude of $\hat{\text{{S}}}^{1\boldsymbol{\Pi}}_{zz}$ decreases, but remains positive. In section \ref{sec:RheologyResults}, we will find that whenever $\hat{\text{{S}}}^{1\boldsymbol{\Pi}}_{zz}$ contributes significantly to the fluid rheology ($De<0.5$), equation \eqref{eq:approximate_polymer_stresslet} remains a good estimate.

\subsection{Validation}
The particle-polymer interaction contributions for very small $De$  can be compared with the theoretical results for second-order-fluid suspensions from Koch and Subramanian \cite{koch2006stress} and Einarsson et al. \cite{einarsson2018einstein}. As $De$ approaches zero both the Oldroyd-B and FENE-P constitutive relations can be used to model a second order fluid via an asymptotic expansion in $De$. The expressions for second order fluid properties of a FENE-P fluid in the small $De$ limit are long and unwieldy, but differ from those for an Oldroyd-B fluid only by $\mathcal{O}(1/L^2)$. Therefore, in this sub-section, we invoke an $L\gg1$ assumption in the FENE-P equations and use the properties of an Oldroyd-B fluid in the second order fluid limit from Koch and Subramanian \cite{koch2006stress} to compare with our numerical estimates.

At small $De$, apart from a boundary layer thickness of $\mathcal{O}( De)$ \cite{koch2016stress} near the particle surface, the non-linear component of the stress within the particle is constant,
\begin{equation}
\hat{\boldsymbol{\Pi}}^{(0N)}\approx 2De\mathbf{E}, \hspace{0.2in} De\ll1,
\end{equation}
This result is obtained by expanding equation \eqref{eq:LinearConfiguration} in $De$ (with $L\gg1$) and using equations \eqref{eq:Decomposetress} and \eqref{eq:LinStress} in the limit of small $De$.
The contribution to the particle-induced polymer stress arising from the particle region is,
\begin{equation}\label{eq:particle_region_PIFS}
\hat{\Pi}^{\text{PP,particle}}_{zz}\approx \frac{8\pi}{3} De.
\end{equation}
In our simulations for $0.001<De<0.005$ with $L=500$, we find $2De\mathbf{E}\lesssim\hat{\boldsymbol{\Pi}}^{(0N)}\lesssim 2.04 De\mathbf{E}$ within the particle (apart from a thin boundary layer close to the particle surface). 

\begin{figure}[h!]
	\centering
	\subfloat{\includegraphics[width=0.33\textwidth]{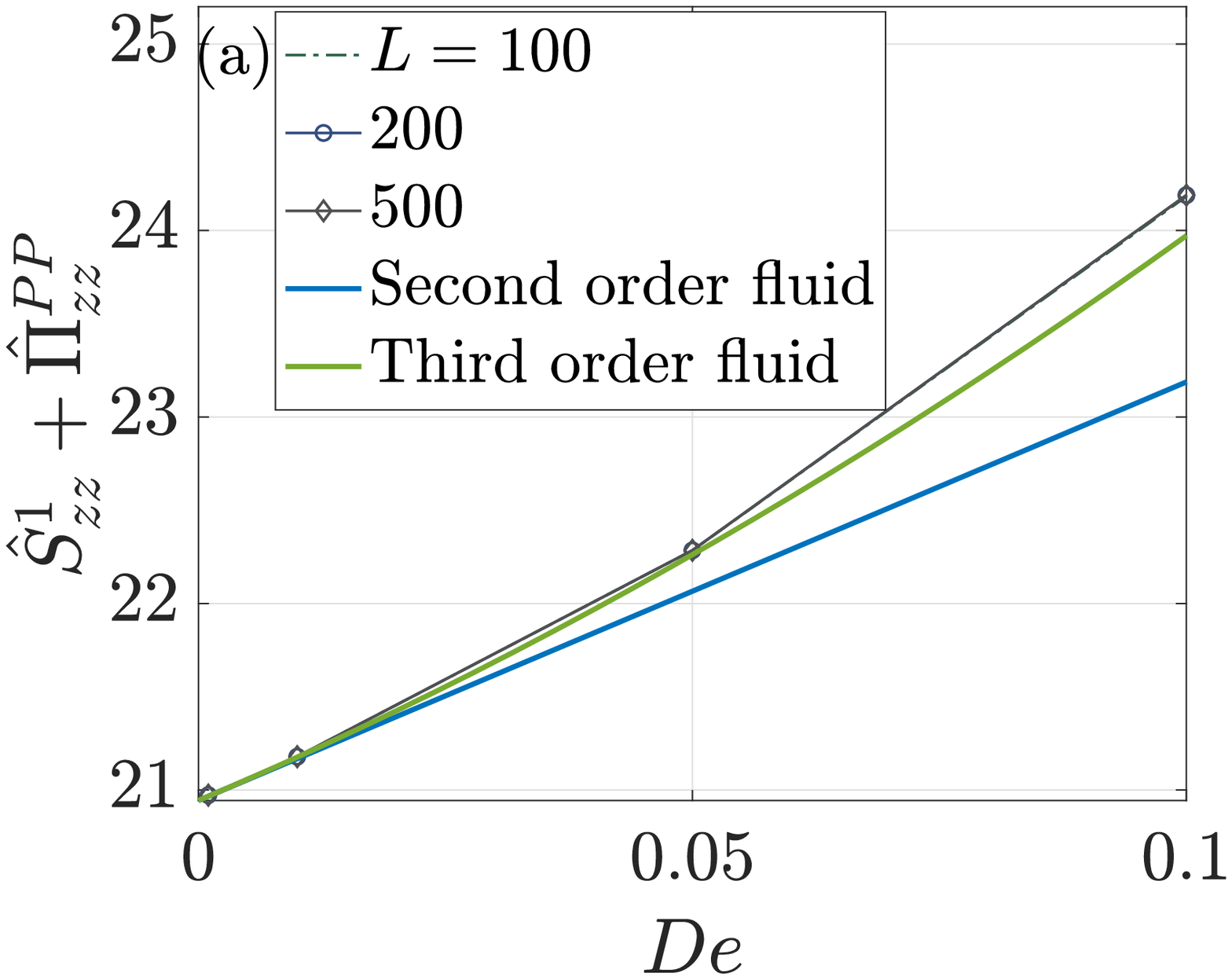}\label{fig:TotalInteractionSmallDeValidation}}\hfill
	\subfloat{\includegraphics[width=0.33\textwidth]{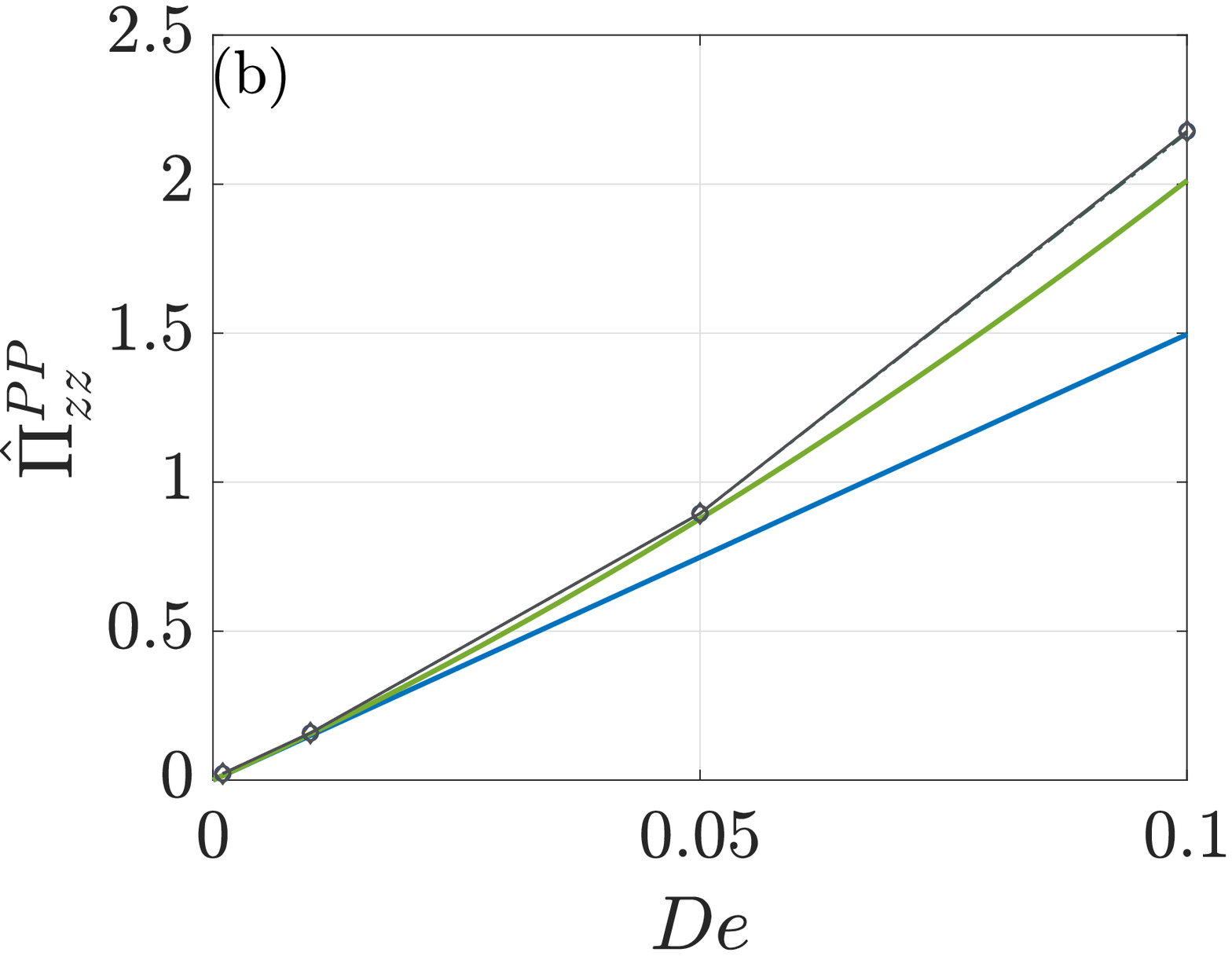}\label{fig:TotalPIPSSmallDeValidation}}\hfill
	\subfloat{\includegraphics[width=0.33\textwidth]{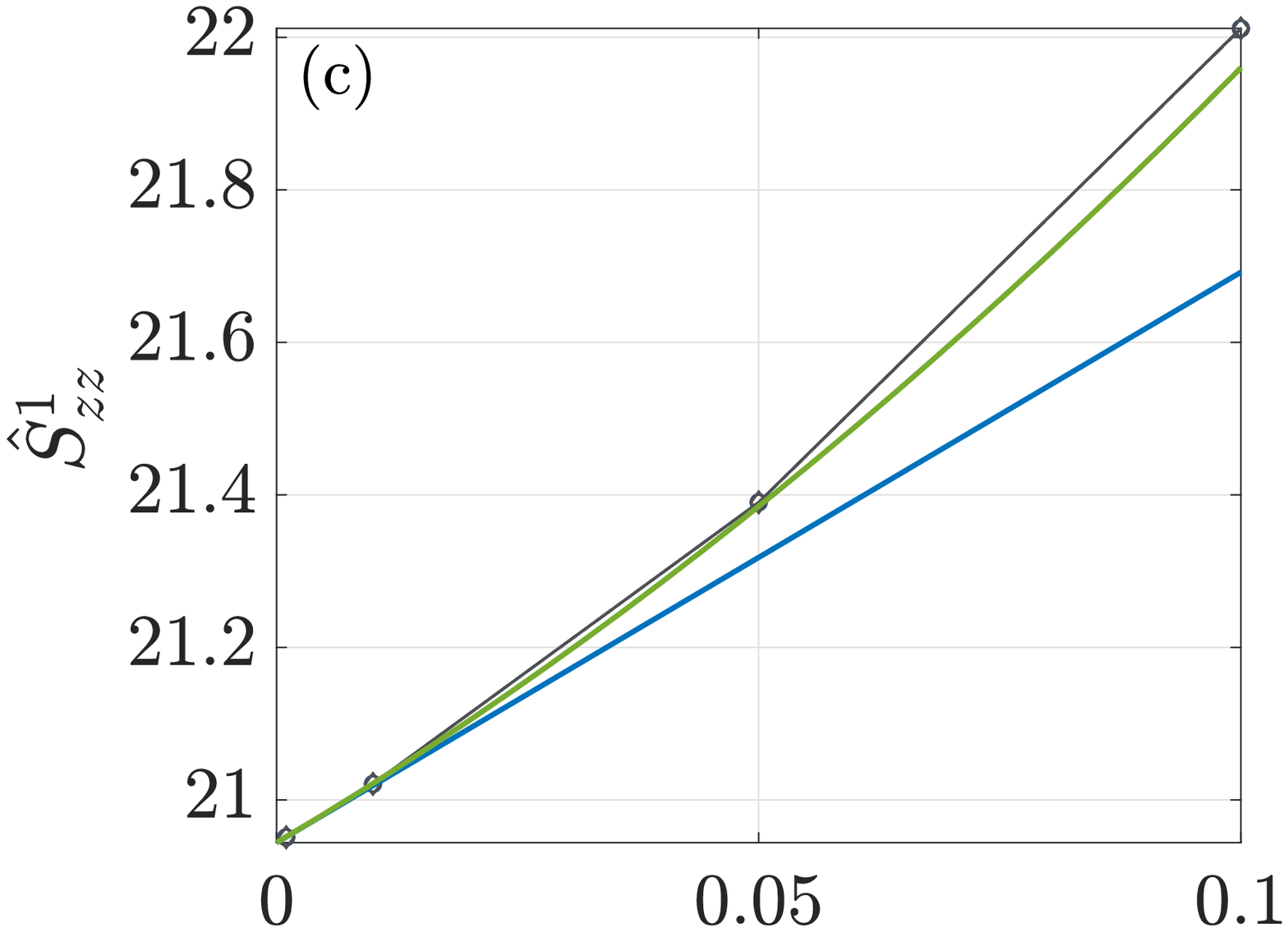}\label{fig:TotalStressletSmallDeValidation}}\hfill
	\caption {Validation of our methodology for large polymer extensibility, $L=100$, 200 and 500. The total interaction stress (a) and its decomposition into the particle induced polymer stress ($\hat{\Pi}^{\text{PP}}_{zz}$ or PIPS) in (b), and the interaction stresslet ($\hat{\text{S}}^{(1)}_{zz}$) in (c) are compared with results for a second order \cite{koch2016stress} and third order \cite{einarsson2018einstein} fluid at low $De$. These graphs are shown for a larger $De$ range (beyond the validity of lower order fluids) in figures \ref{fig:Stresses_SmallDe_disturbed}, \ref{fig:Total_extra_polymer_small_de} and \ref{fig:TotalStressletSmallDe}. All three plots share the same legend as (a). \label{fig:Stresses_Validation}}
\end{figure}

The comparison between the numerical and the analytical stress components at small $De$  and $L=500$ is shown in table \ref{tab:Validation}. As mentioned in section  \ref{sec:RheologyTheory}, $\hat{\text{S}}^{1\boldsymbol{\Pi}}_{zz}$ is obtained analytically for any $De$ and $L$, using computer algebra, but the value reported in table \ref{tab:Validation} is obtained numerically, using the method of characteristics.
\begin{table}[h!]
	\centering
	\caption{Comparison of various stress components at small $De$ ($0.001\le De\le 0.005$) and large $L=500$, evaluated numerically and theoretically. Theoretical values of $\hat{\text{S}}^{(1)}_{zz}$ and $\hat{\Pi}^{\text{PP}}_{zz}$ are mentioned in \cite{einarsson2018einstein}. Further decomposition into $\hat{\text{S}}^{1\boldsymbol{\tau}}_{zz}$,  $\hat{\text{{S}}}^{1\boldsymbol{\Pi}}_{zz}$, $\hat{\Pi}^{\text{PP,fluid}}_{zz}$ and $\hat{\Pi}^{\text{PP,particle}}_{zz}$ are provided in \cite{koch2006stress}.  \label{tab:Validation}}
	\begin{tabular}{c||c|c}
		Interaction Stress component &   Theoretical & Numerical\\ \hline \hline
		$\hat{\text{{S}}}^{1\boldsymbol{\Pi}}_{zz}$&${4\pi+\mathcal{O}(De^2)} \approx 12.57+\mathcal{O}(De^2)$&$12.56-0.002De$ \\
		$\hat{\text{S}}^{1\boldsymbol{\tau}}_{zz}$&${8\pi}/{3}+{50\pi}/{21}De\approx 8.38+7.48De$ &$8.38+7.49 De$ \\	
		$\hat{\Pi}^{\text{PP,fluid}}_{zz}$&${44\pi}/{21} De\approx 6.58De$&$0.00+6.76De$\\			
		$\hat{\Pi}^{\text{PP,particle}}_{zz}$&${8\pi}/{3} De\approx 8.38 De$&$0.00+8.54De$\\			
	\end{tabular}
\end{table}
The errors in the values presented are generally low. The most erroneous terms are the ones requiring volume integrals in their computation i.e. $\hat{\text{S}}^{1\boldsymbol{\tau}}_{zz}$ and $\hat{\Pi}^{\text{PP}}_{zz}$. We also present the comparison of our large $L$ results (where $L=100$ and 200 are shown in addition to $L=500$) at small $De\le0.1$ with  second \cite{koch2016stress} and third \cite{einarsson2018einstein} order fluid results in figure \ref{fig:Stresses_Validation}. The quadratic increase of the interaction stress ($\hat{\text{S}}^{(1)}_{zz}+\hat{\Pi}^{\text{PP}}_{zz}$) in figure \ref{fig:TotalInteractionSmallDeValidation} and its components i.e. the particle induced polymer stress ($\hat{\Pi}^{\text{PP}}_{zz}$ or PIPS) in figure \ref{fig:TotalPIPSSmallDeValidation} and the interaction stresslet ($\hat{\text{S}}^{(1)}_{zz}$) in figure \ref{fig:TotalStressletSmallDeValidation} with $De$ at small $De$ matches well with the third order fluid results of Einarsson et al. \cite{einarsson2018einstein} up to $De\approx0.05$. Beyond this value the magnitude of stresses in the FENE-P fluid is more than that in the third order fluid. 

\subsection{Results and Discussion}\label{sec:RheologyResults}
\begin{figure}[h!]
	\centering
	\subfloat{\includegraphics[width=0.33\textwidth]{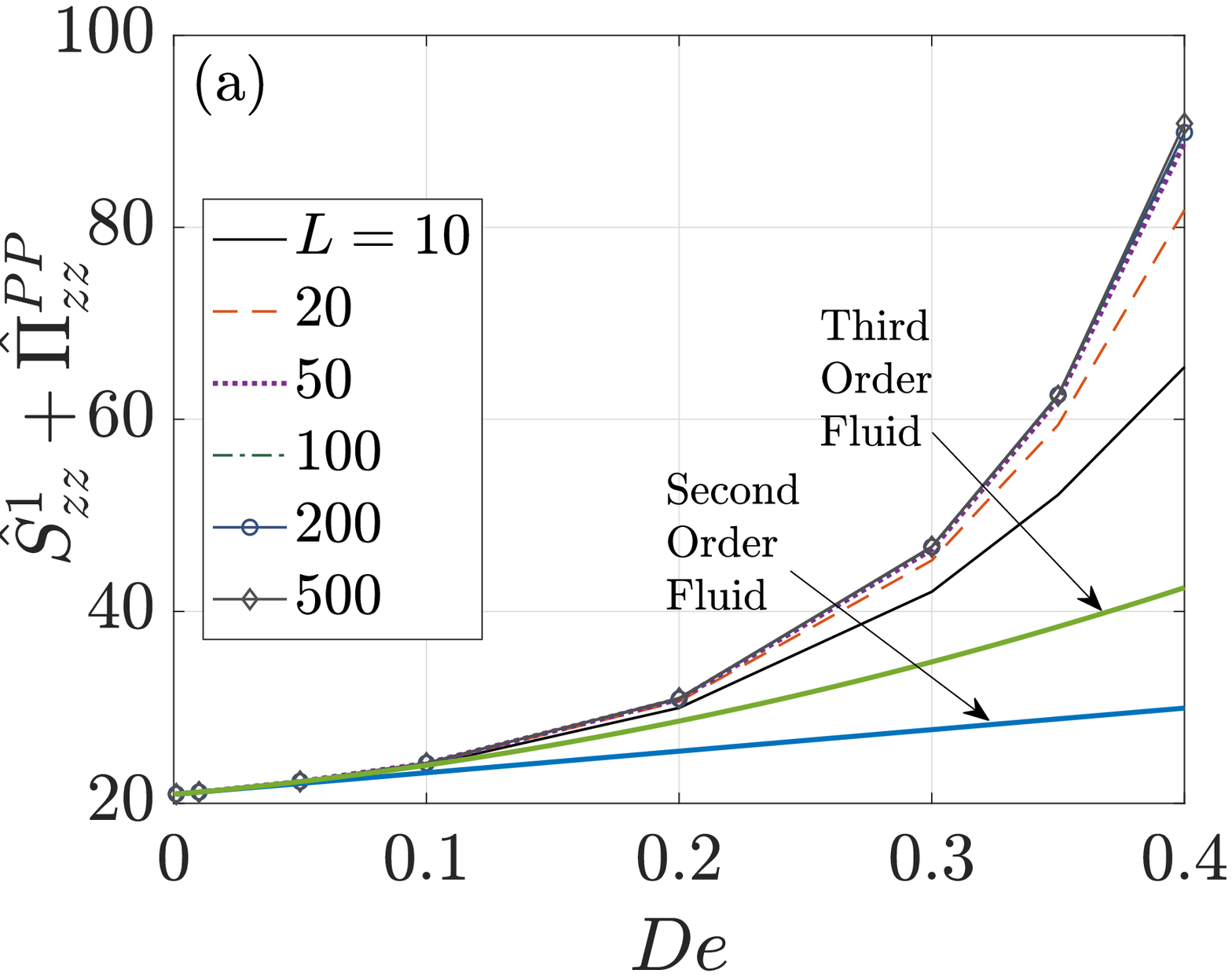}\label{fig:Stresses_SmallDe_disturbed}}\hfill
	\subfloat{\includegraphics[width=0.33\textwidth]{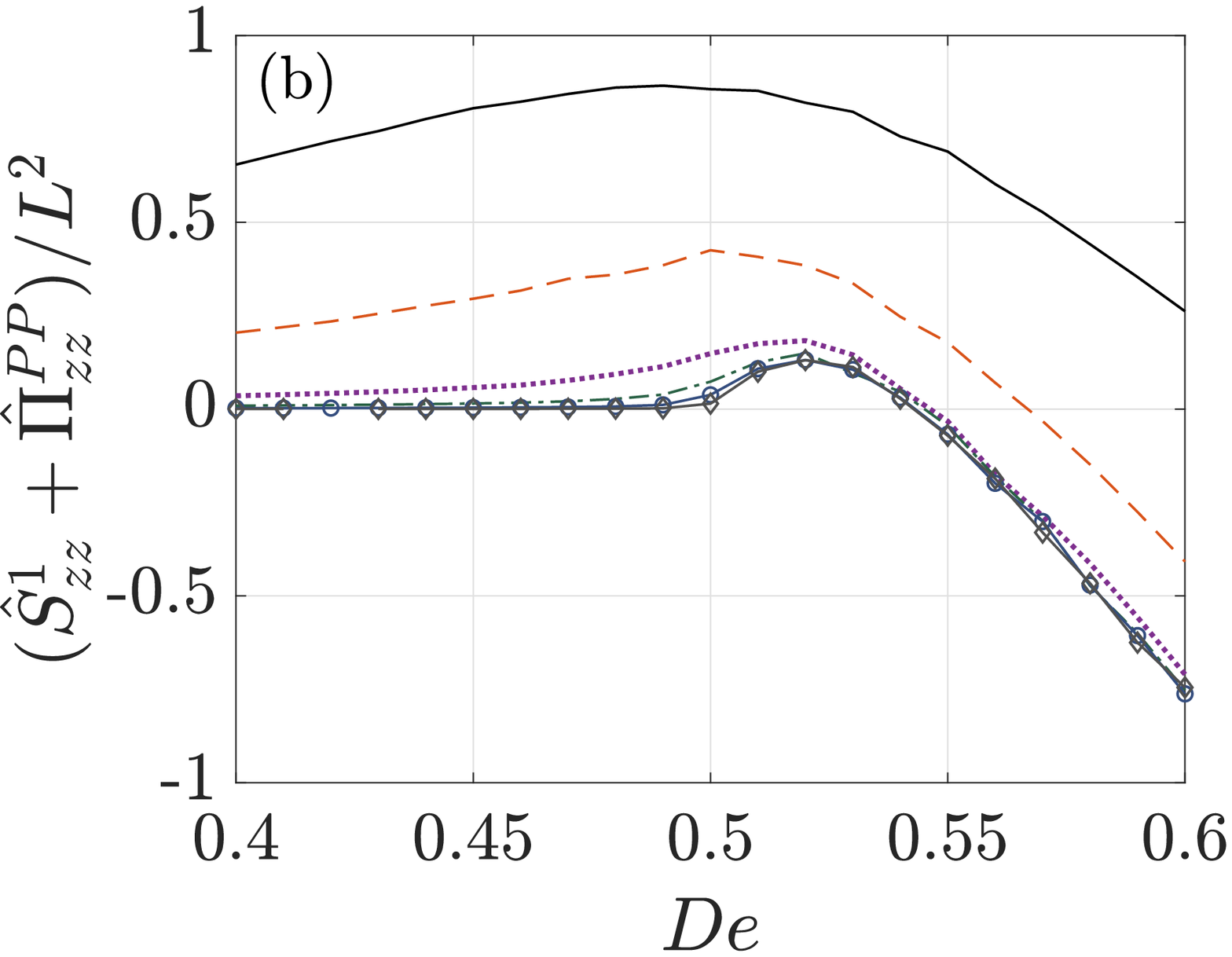}\label{fig:Particle_extra_medium_de}}\hfill
	\subfloat{\includegraphics[width=0.33\textwidth]{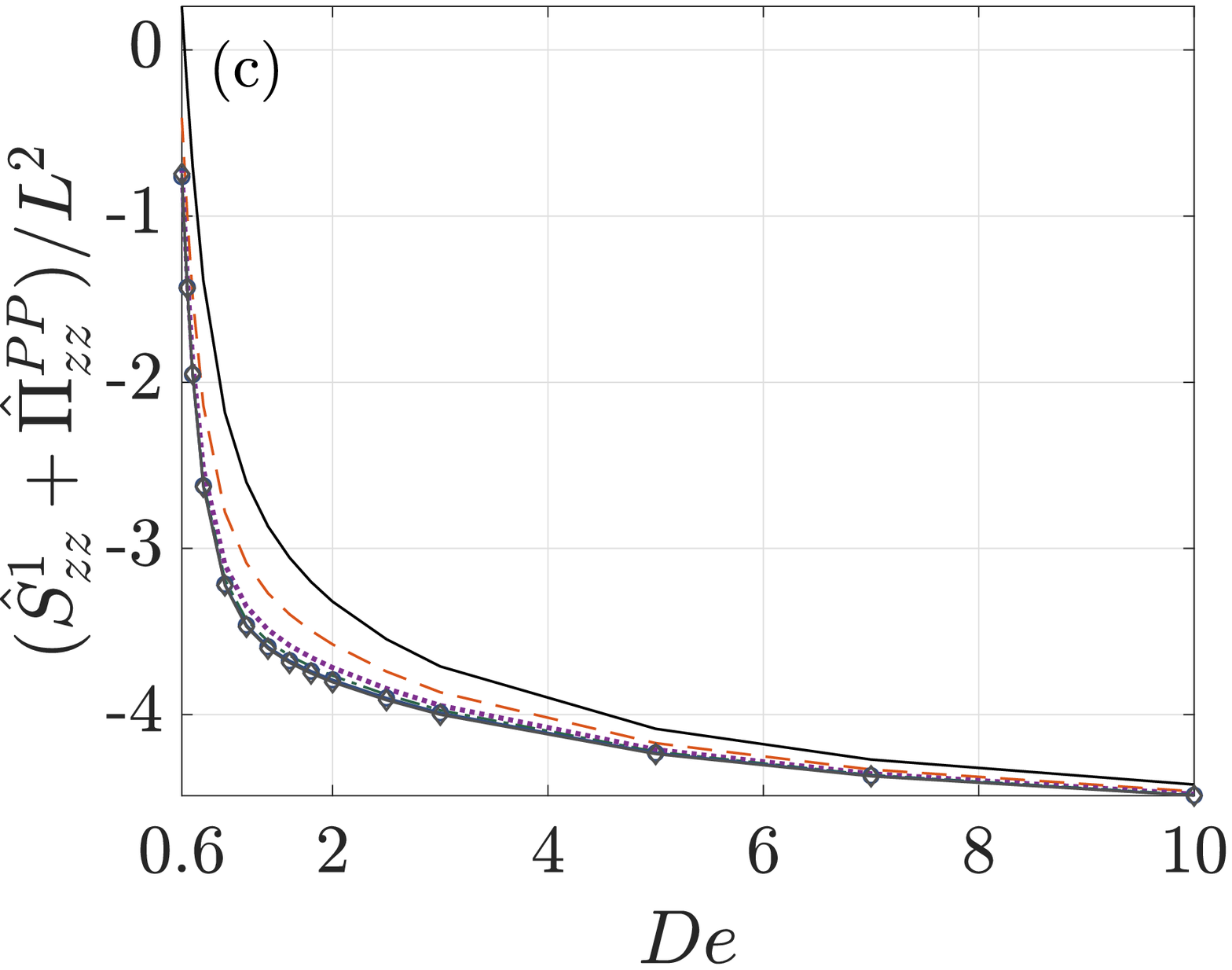}\label{fig:Total_Fluid_extra_polymer_large_de}}\hfill
	\caption {Total particle-polymer interaction stress, $\hat{\text{S}}^{(1)}_{zz}+\hat{\Pi}^{\text{PP}}_{zz}$, in a dilute suspension of spheres in a dilute polymeric liquid at 6 different $L$ in $10\le L\le500$ for (a) $De<0.4$, (b) $0.4<De<0.6$ and (c) $De>0.6$. All figures share the legend shown in (a). The stresses in (b) and (c) are scaled with $L^2$ in view of the coil-stretch transition at $De=0.5$. In the small $De\le0.4$ regime of (a) the interaction stresses for a second order fluid from \cite{koch2006stress,einarsson2018einstein} and a third order fluid from \cite{einarsson2018einstein} are also shown. \label{fig:Stresses_Total}}
\end{figure}

We begin with a summary of the overall effect of particle-polymer interactions on the suspension rheology.
Figure \ref{fig:Stresses_Total} shows the variation of the extensional component of the total deviatoric particle-polymer interaction stress, $\hat{\text{S}}^{(1)}_{zz}+\hat{\Pi}^{\text{PP}}_{zz}$, for three different $De$ regimes: $De\le0.4$, $0.4\le De\le0.6$ and $De\ge0.6$ that cover the entire $De$ range and six $L$ in the range $10\le L\le 500$. Since the coil- stretch transition is expected to influence the latter two $De$ regimes, the stresses in these regimes are plotted after normalizing with $L^2$. From equation \eqref{eq:Interactionviscosity}, the extensional viscosity due to this interaction, $\mu^\text{intr}$, is simply $3c\phi/8$ times the interaction stress displayed in figure \ref{fig:Stresses_Total}. In the plot for $De\le0.4$ in figure \ref{fig:Stresses_SmallDe_disturbed} we also show the theoretical curves of second- \cite{koch2006stress,einarsson2018einstein} and third-order \cite{einarsson2018einstein} Oldroyd-B fluids, which both under-predict  the magnitude of interaction stress when $De\gtrapprox0.1$. From figure \ref{fig:Stresses_SmallDe_disturbed} we find that the interaction stress is positive and increases with $De$ for $De\le0.4$. In this regime, the total change in the extensional viscosity due to adding particles ($\mu^\text{part}$ from equation \ref{eq:ExtensionalViscosityParticles}) is positive and is larger than that in a Newtonian fluid. Interestingly, upon further increase in $De$, within the $0.4\le De\le0.6$ regime of figure \ref{fig:Particle_extra_medium_de} the interaction stress stops increasing with $De$ and instead starts to decrease. The overall particle-polymer interaction stress changes from positive to negative around $De=0.52$, for high $L$ values. For $L=10$ and 20, it happens later and more gradually. Therefore, depending upon $L$, there is a $De$ slightly greater than 0.5 beyond which the extensional viscosity due to the particles, $\mu^\text{part}$ from equation \ref{eq:ExtensionalViscosityParticles}, is reduced by the particle-polymer interaction in contrast to the enhancement by the same mechanism at lower $De$. Further increase in $De$ leads to more negative particle-polymer interaction stress as shown in figure \ref{fig:Particle_extra_medium_de} and for $De\ge0.6$ in figure \ref{fig:Total_Fluid_extra_polymer_large_de} (with a large magnitude as the values in the corresponding figure are normalized with $L^2$). We will later show that there are combinations of $c$ and $De$ such that the net extensional viscosity due to the particles, $\mu^\text{part}$ from equation \ref{eq:ExtensionalViscosityParticles} is negative or in other words adding particles leads to a reduction in suspension stress. The positive interaction stress at smaller $De$ is due to the wake of highly stretched polymers shown in figure \ref{fig:polymerstretch_smallDe} and the negative interaction stress at larger $De$ is due to the region of collapsed polymers shown in figure \ref{fig:polymerstretch_mediumDe} and \ref{fig:polymerstretch_largeDe}. To justify this claim and understand finer features of the suspension rheology in figure \ref{fig:Stresses_Total} we consider the decompositions of the components of interaction stress $\hat{\text{S}}^{(1)}_{zz}+\hat{\Pi}^{\text{PP}}_{zz}$ while revisiting the discussion and figures of section \ref{sec:PolymerConfiguration} in light of the rheological observations.

In section \ref{sec:Fullpolymerconfiguration}, for large $L$, we discussed the $L$ independence and $L^2$ scaling of the polymer configuration (and hence polymer stress) for small and $De$ respectively. For large $L\gtrapprox50$, the total interaction stress is also independent of $L$ in the regime $De\le 0.4$ (figure \ref{fig:Stresses_SmallDe_disturbed}) and scales as $L^2$ in the regime $De\ge 0.6$ (figure \ref{fig:Total_Fluid_extra_polymer_large_de}). The validity of the $L^2$ scaling justifies  the claim in section \ref{sec:Fullpolymerconfiguration}, that the contribution of the region to suspension rheology where $L^2$ scaling in the change in polymer stretch due to the particle and hence the polymer stress breaks down is very small. We have also confirmed by numerical integration over this volume near the extensional axis that its contributions are too small to affect the scaling of the averaged quantities. Various components of the interaction stress discussed in the rest of this section also follow the $L$ independent (for $De\le 0.4$) and $L^2$ (for $De\ge 0.6$) scalings below and above the coil-stretch transition respectively.

\begin{figure}[h!]
	\centering
	\subfloat{\includegraphics[width=0.33\textwidth]{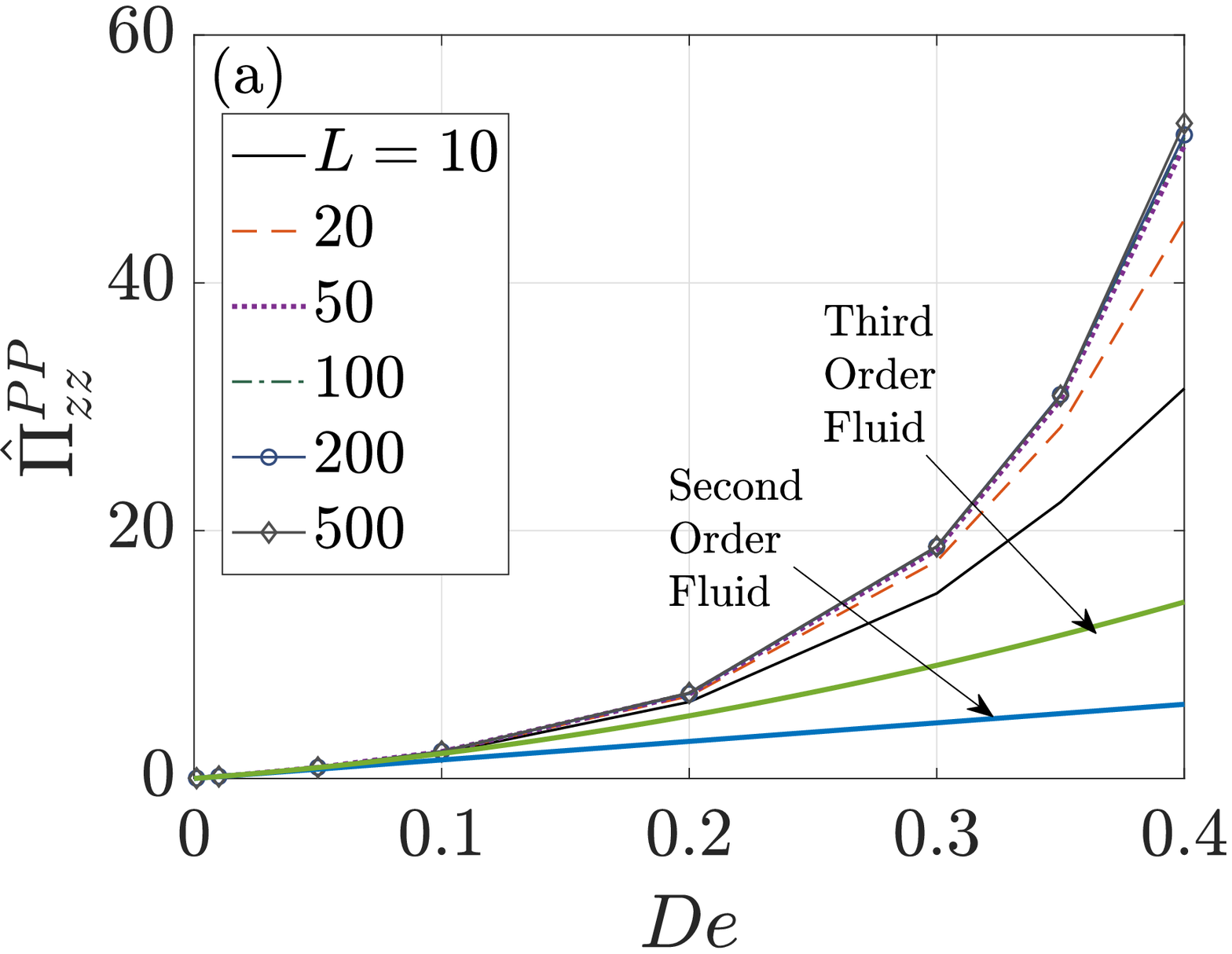}\label{fig:Total_extra_polymer_small_de}}\hfill
	\subfloat{\includegraphics[width=0.33\textwidth]{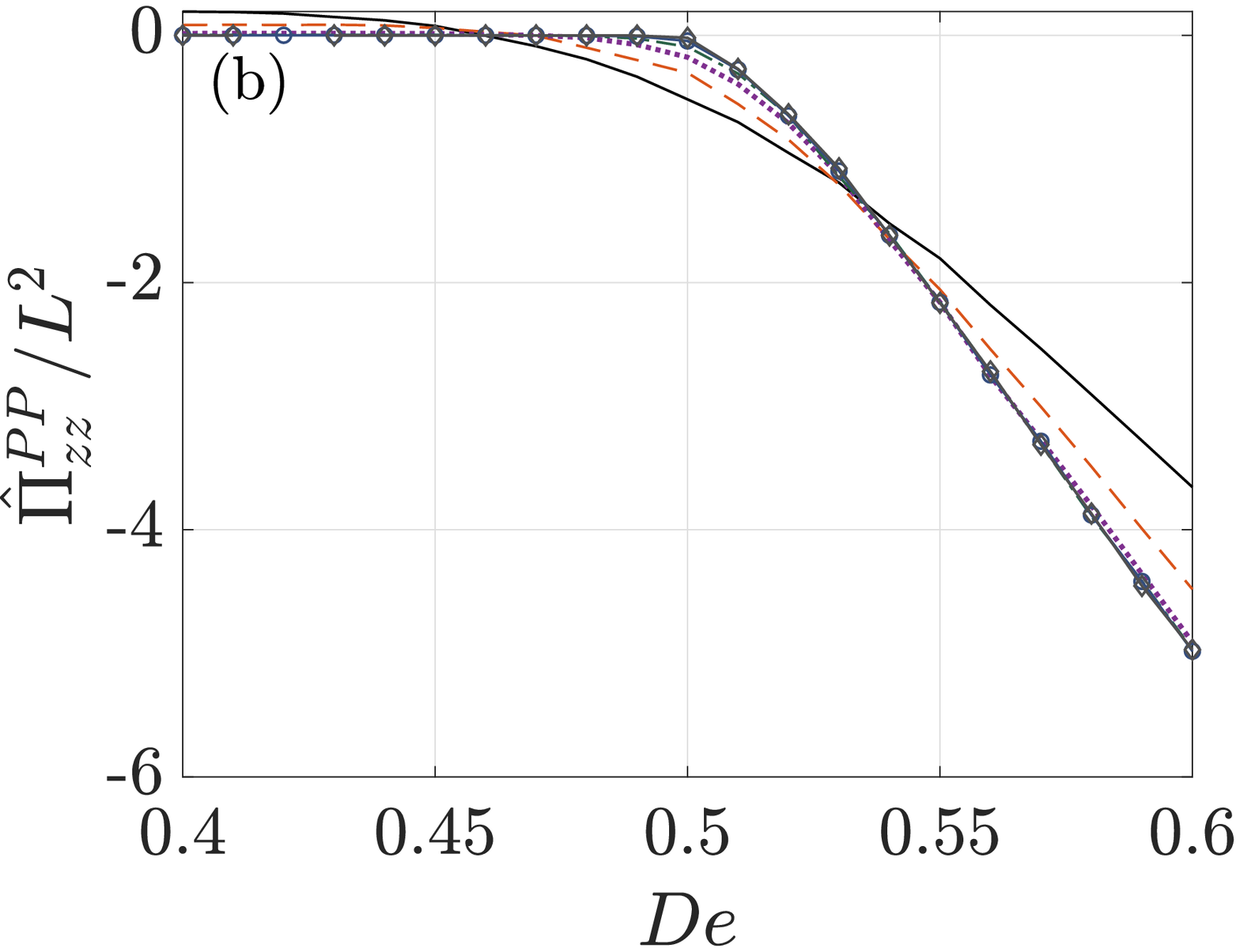}\label{fig:Total_extra_polymer_medium_de}}\hfill
	\subfloat{\includegraphics[width=0.33\textwidth]{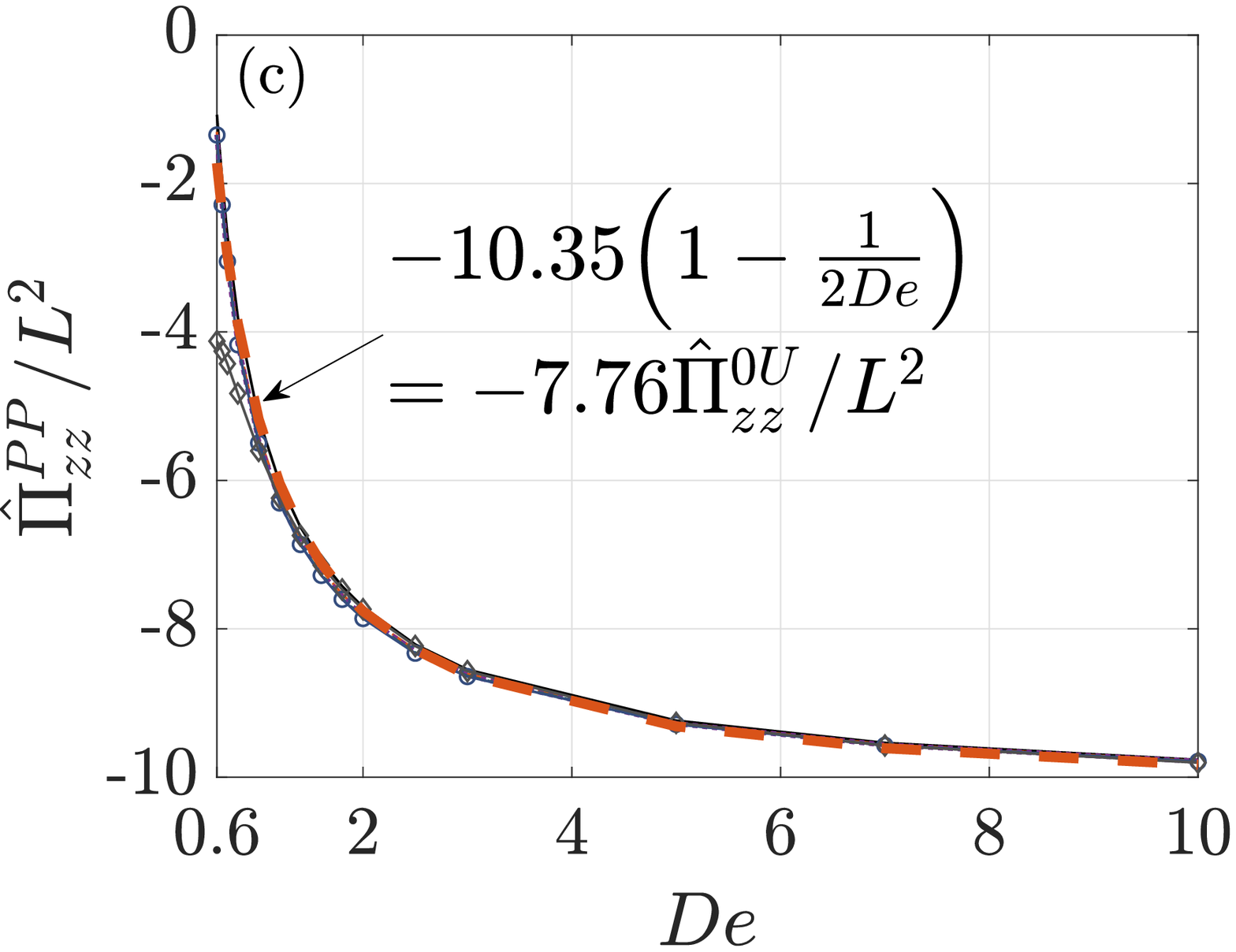}\label{fig:Total_extra_polymer_large_de}}\hfill
	\subfloat{\includegraphics[width=0.33\textwidth]{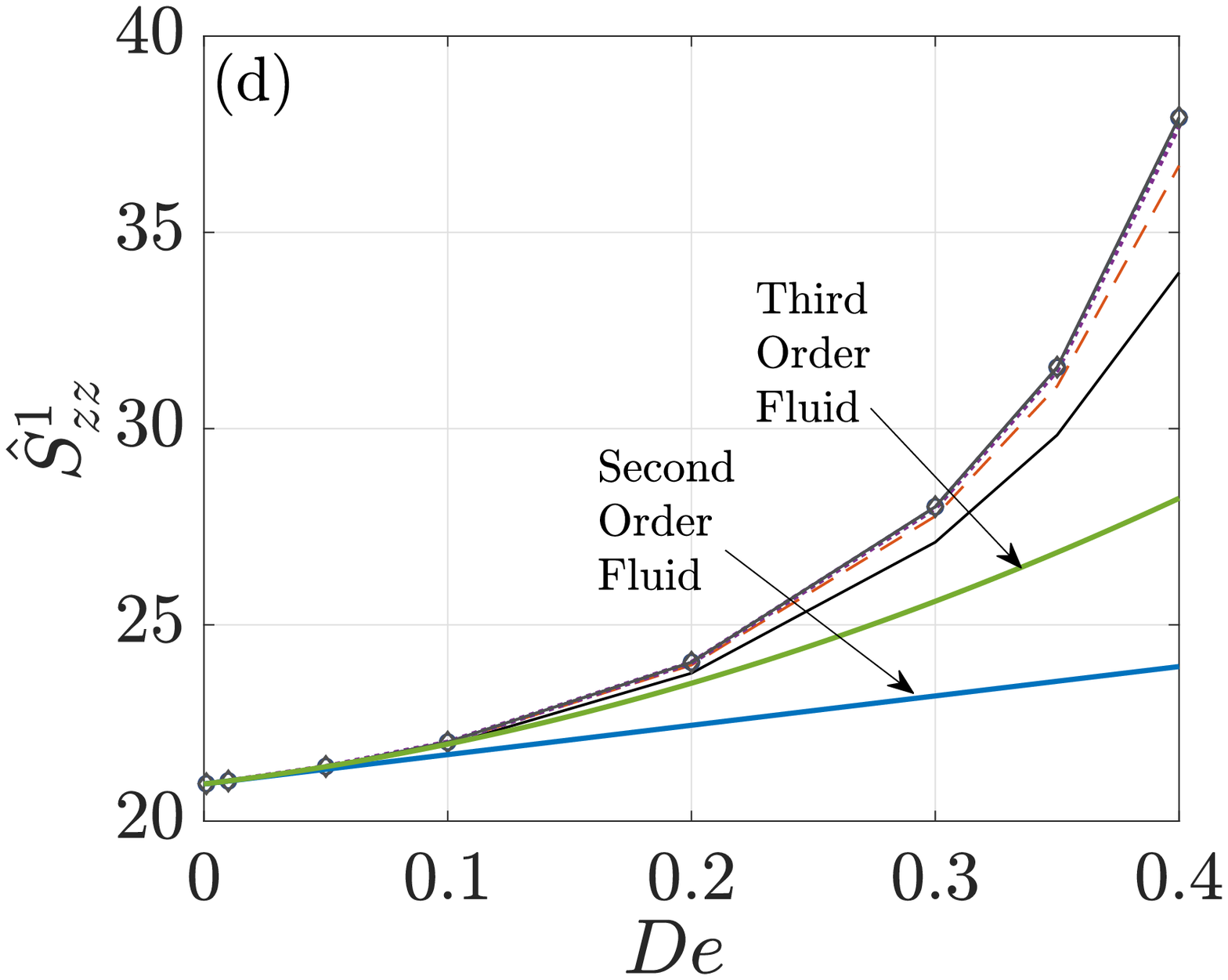}\label{fig:TotalStressletSmallDe}}\hfill
	\subfloat{\includegraphics[width=0.33\textwidth]{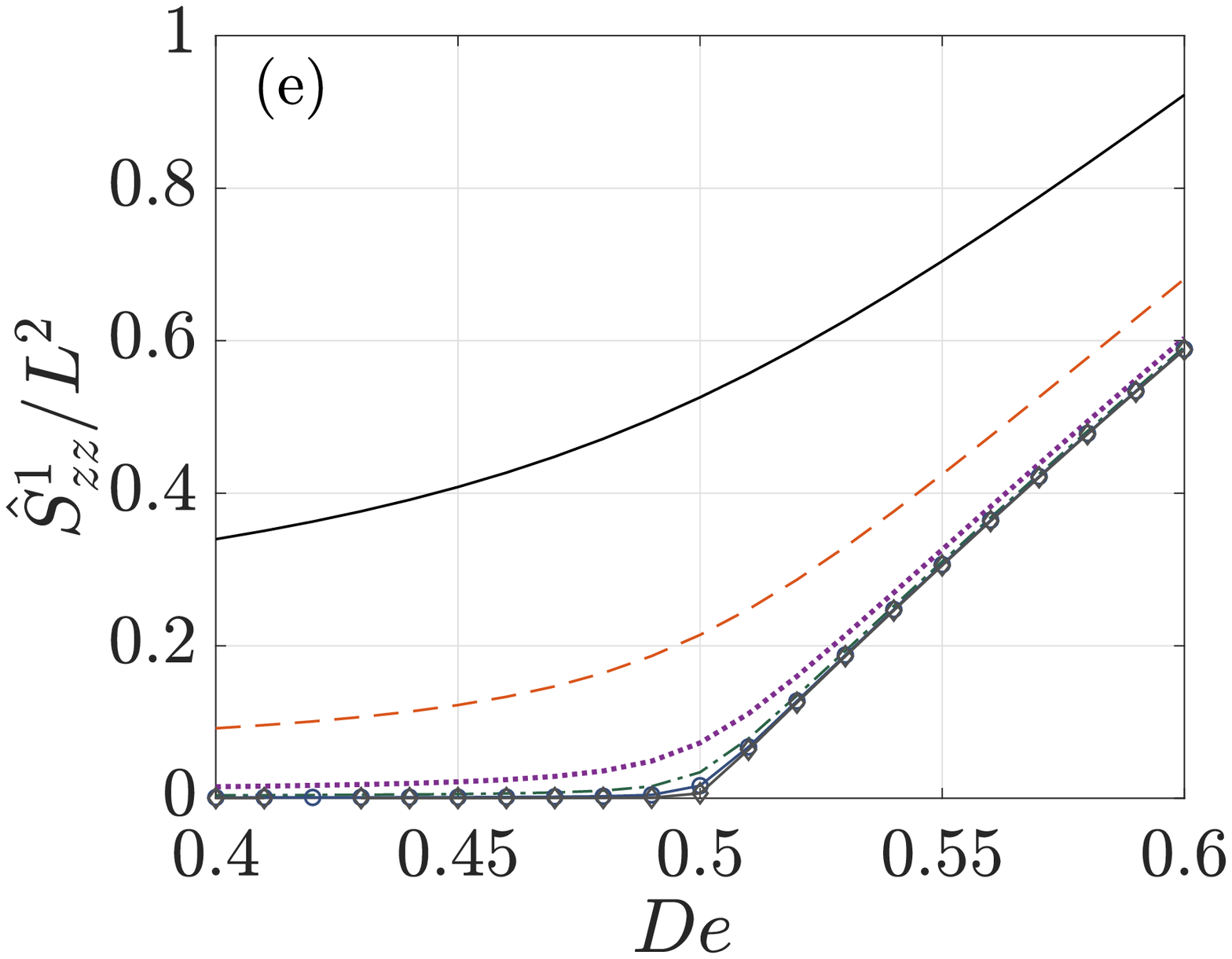}\label{fig:TotalStressletMediumDe}}\hfill
	\subfloat{\includegraphics[width=0.33\textwidth]{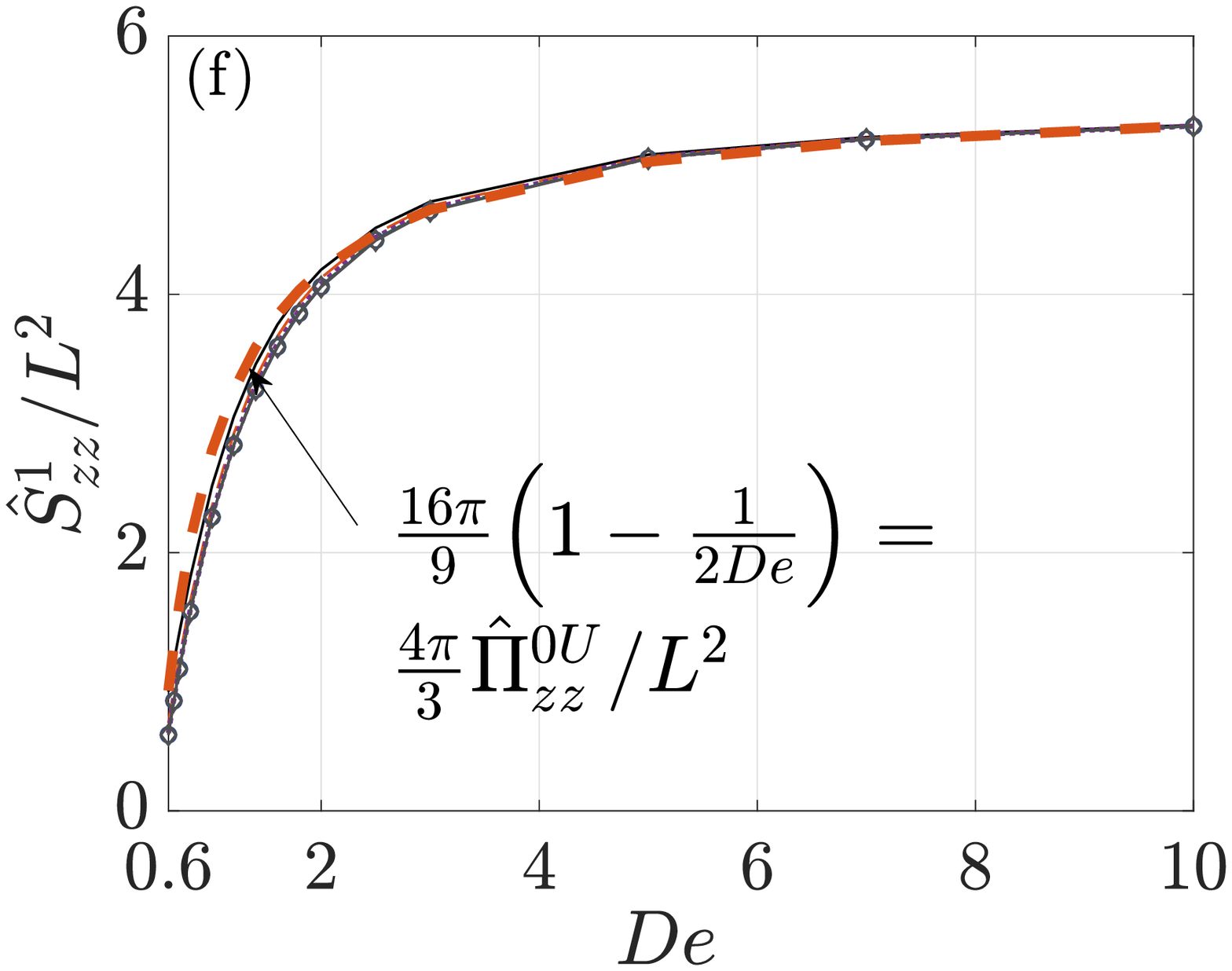}\label{fig:TotalStressletLargeDe}}\hfill
	\caption {Decomposition of the total particle polymer interaction stress into the particle induced polymer stress ($\hat{\Pi}^{\text{PP}}_{zz}$ or PIPS) in figures (a)-(c) and the interaction stresslet ($\hat{\text{S}}^{(1)}_{zz}$) in figures (d)-(f) for 6 different $L$ in $10\le L\le500$ for $De\le0.4$ ((a),(d)), $0.4< De<0.6$ ((b),(e)) and $De\ge0.6$ ((c),(f)). For the latter two $De$ regimes the stresses are scaled with $L^2$. All figures share the legend shown in (a). In figures (c) and (f) for the large $De\ge0.6$ regime, the approximate fits $-10.35(1-1/(2De))$ ($=-7.76\hat{{\Pi}}^{(0U)}_{zz}/L^2$ from equation \eqref{eq:approximatelargeDe})  and $16\pi/9(1-1/(2De))$ ($=4\pi/3\hat{{\Pi}}^{(0U)}_{zz}/L^2$ from equation \eqref{eq:approximatelargeDe}) are also shown. \label{fig:Stresses_Decomposition_Stresslet_PIPS}}
\end{figure}

The primary decomposition of the interaction stress into the interaction stresslet,  $\hat{\text{S}}^{(1)}_{zz}$, and the particle induced polymer stress, $\hat{\Pi}^{\text{PP}}_{zz}$ or the PIPS, is shown in figure \ref{fig:Stresses_Decomposition_Stresslet_PIPS}.
We find that the interaction stresslet, $\hat{\text{S}}^{(1)}_{zz}$, shown in plots (d)-(f) of figure \ref{fig:Stresses_Decomposition_Stresslet_PIPS} is positive, increases monotonically with $De$ and undergoes a rapid increase near the coil-stretch transition point at $De=0.5$. Qualitatively the interaction stresslet behaves similar to the undisturbed polymer stress shown in figure \ref{fig:Stresses_Undisturbed} within and across the $De$ regimes. We will later note that the interaction stresslet is proportional to the undisturbed stress in certain $De$ regimes.

\begin{figure}[h!]
	\centering
	\subfloat{\includegraphics[width=0.33\textwidth]{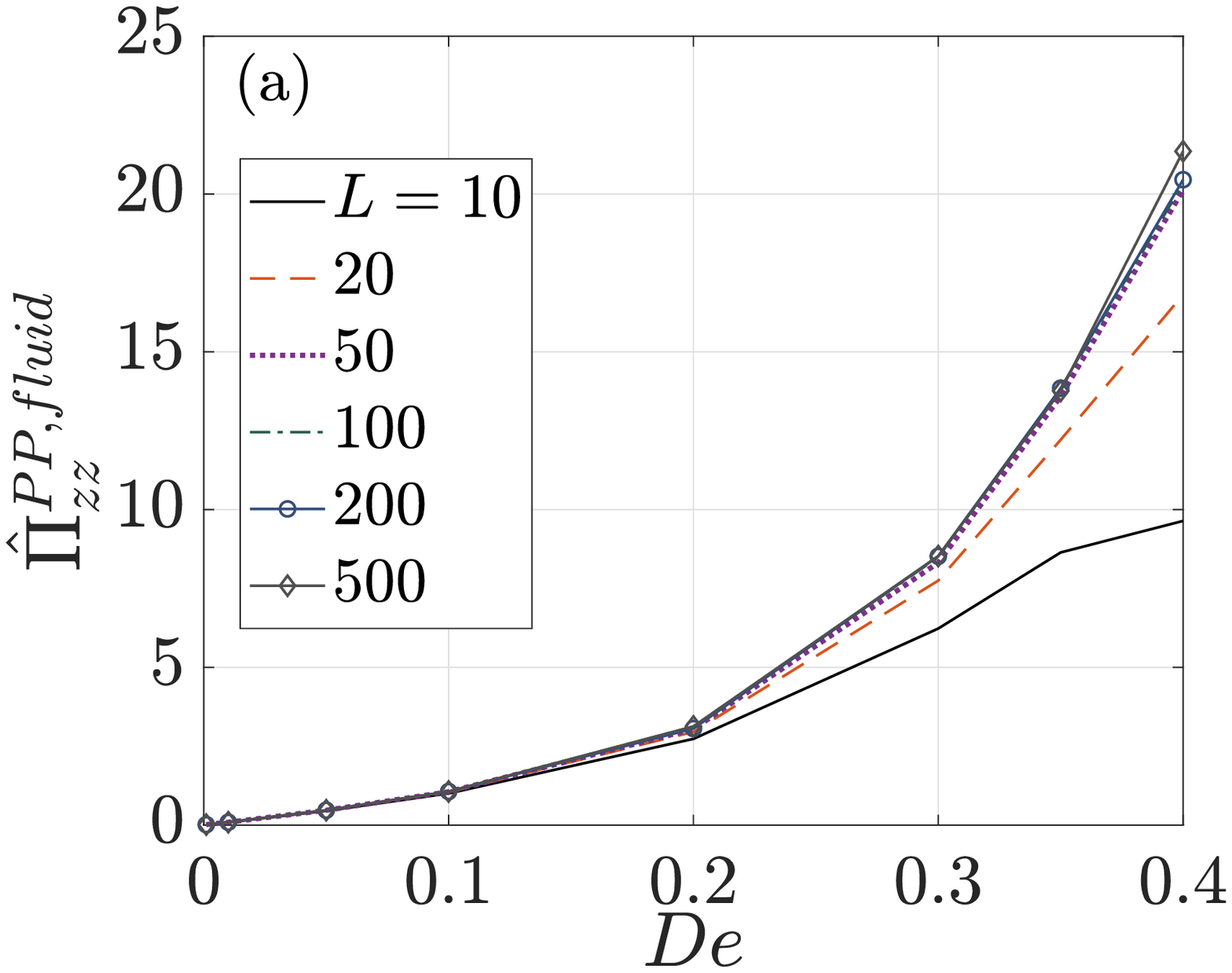}\label{fig:FluidPIPSSmallDe}}\hfill
	\subfloat{\includegraphics[width=0.33\textwidth]{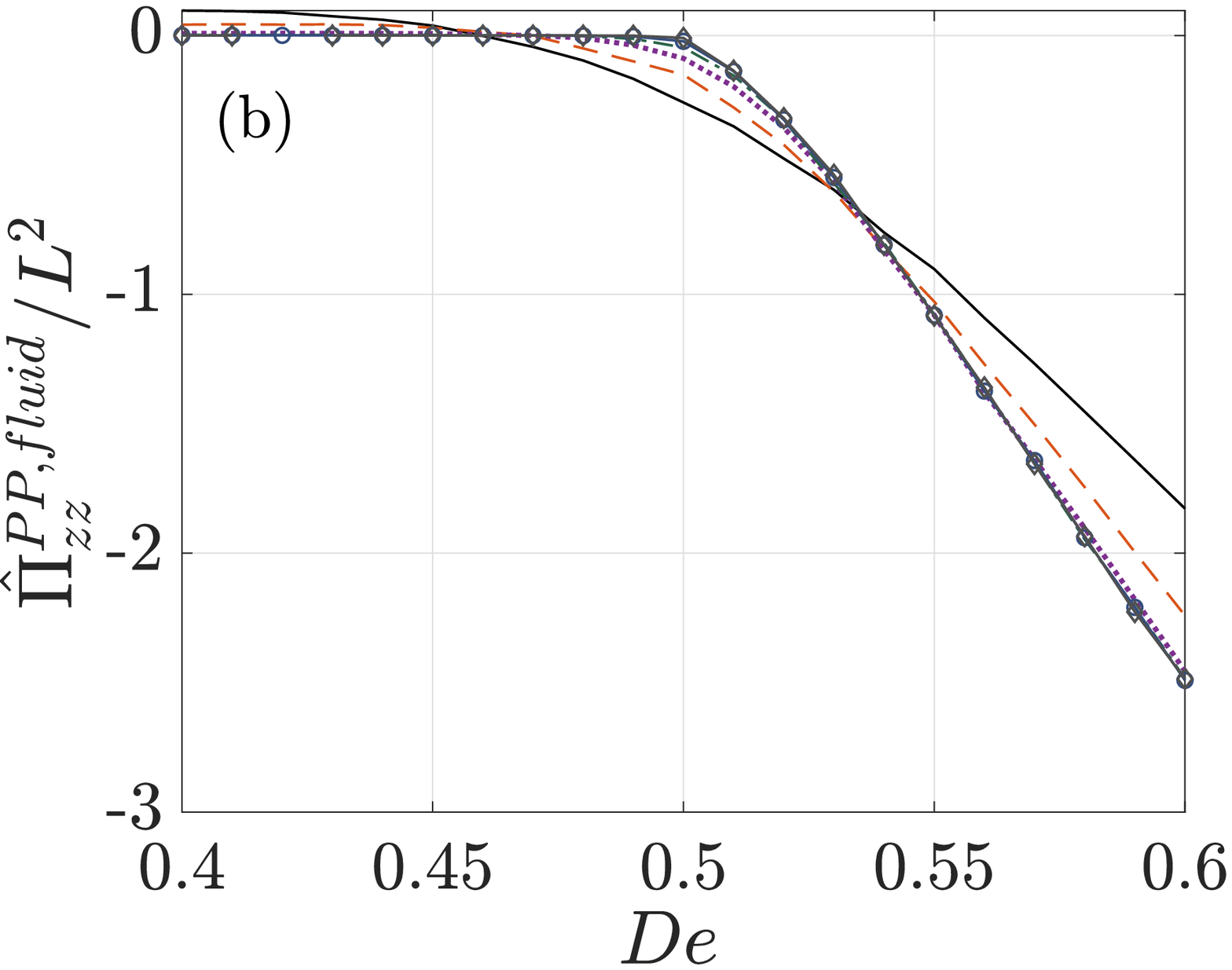}\label{fig:FluidPIPSMediumDe}}\hfill
	\subfloat{\includegraphics[width=0.33\textwidth]{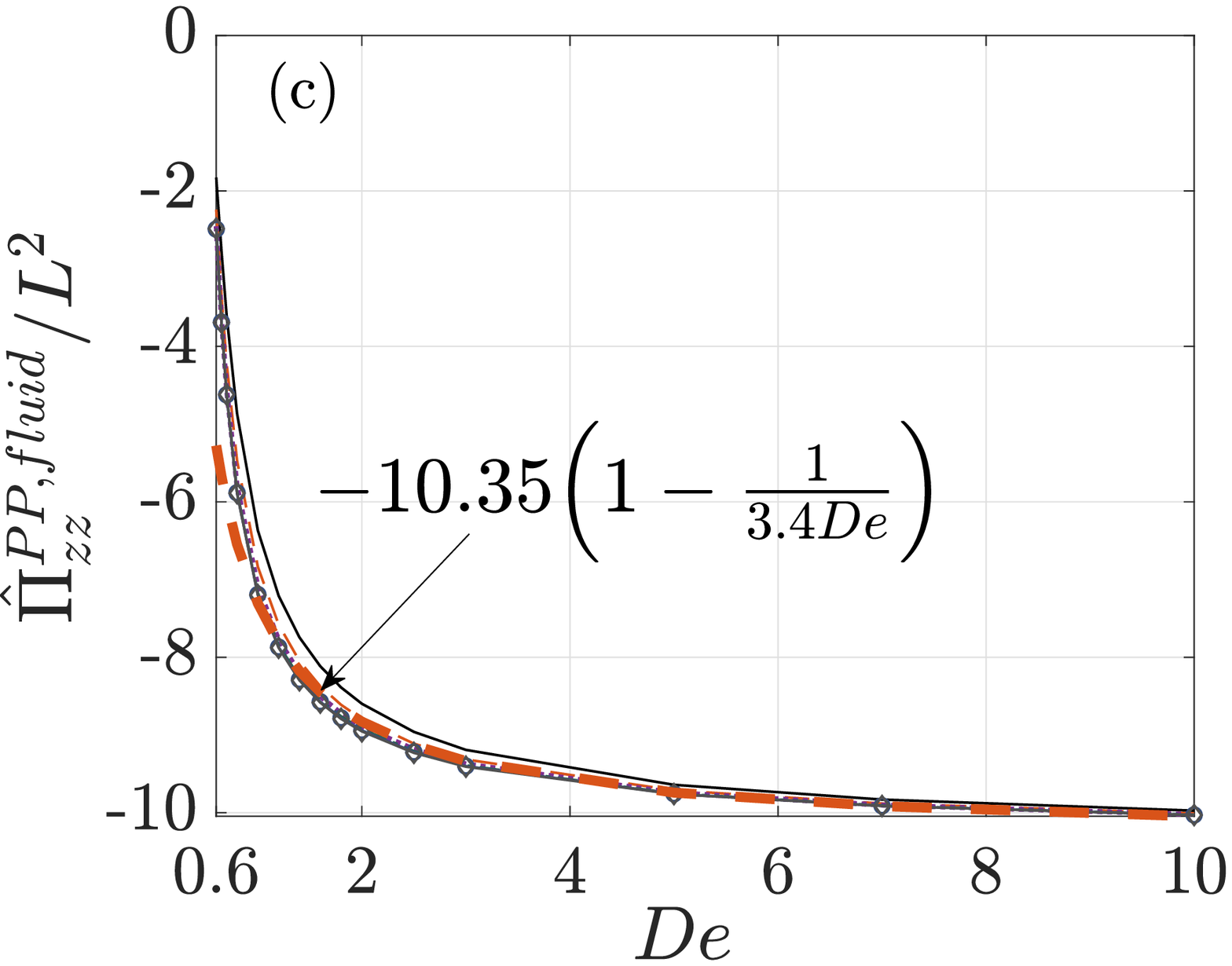}\label{fig:FluidPIPSLargeDe}}\hfill
	\subfloat{\includegraphics[width=0.33\textwidth]{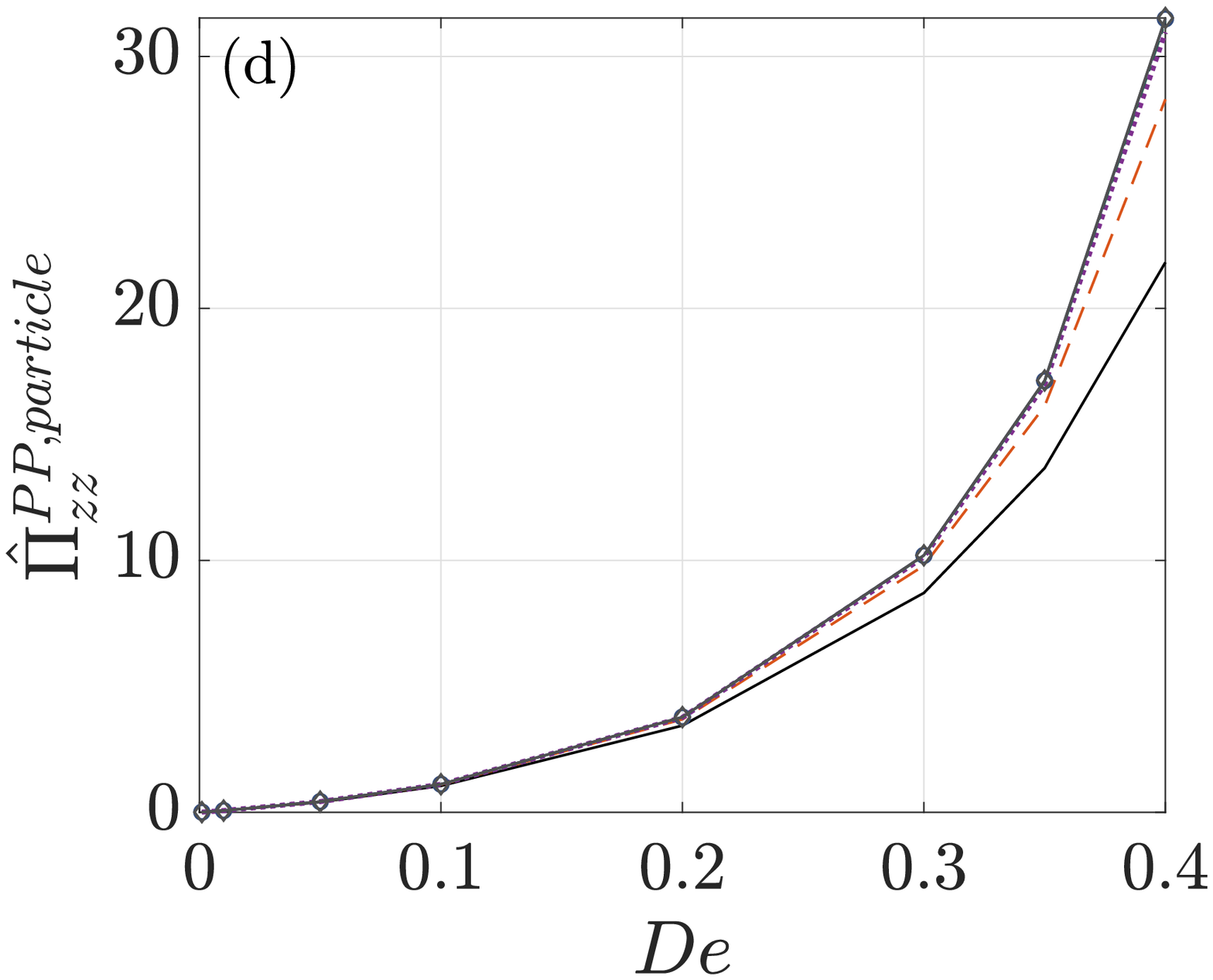}\label{fig:ParticlePIPSSmallDe}}\hfill
	\subfloat{\includegraphics[width=0.33\textwidth]{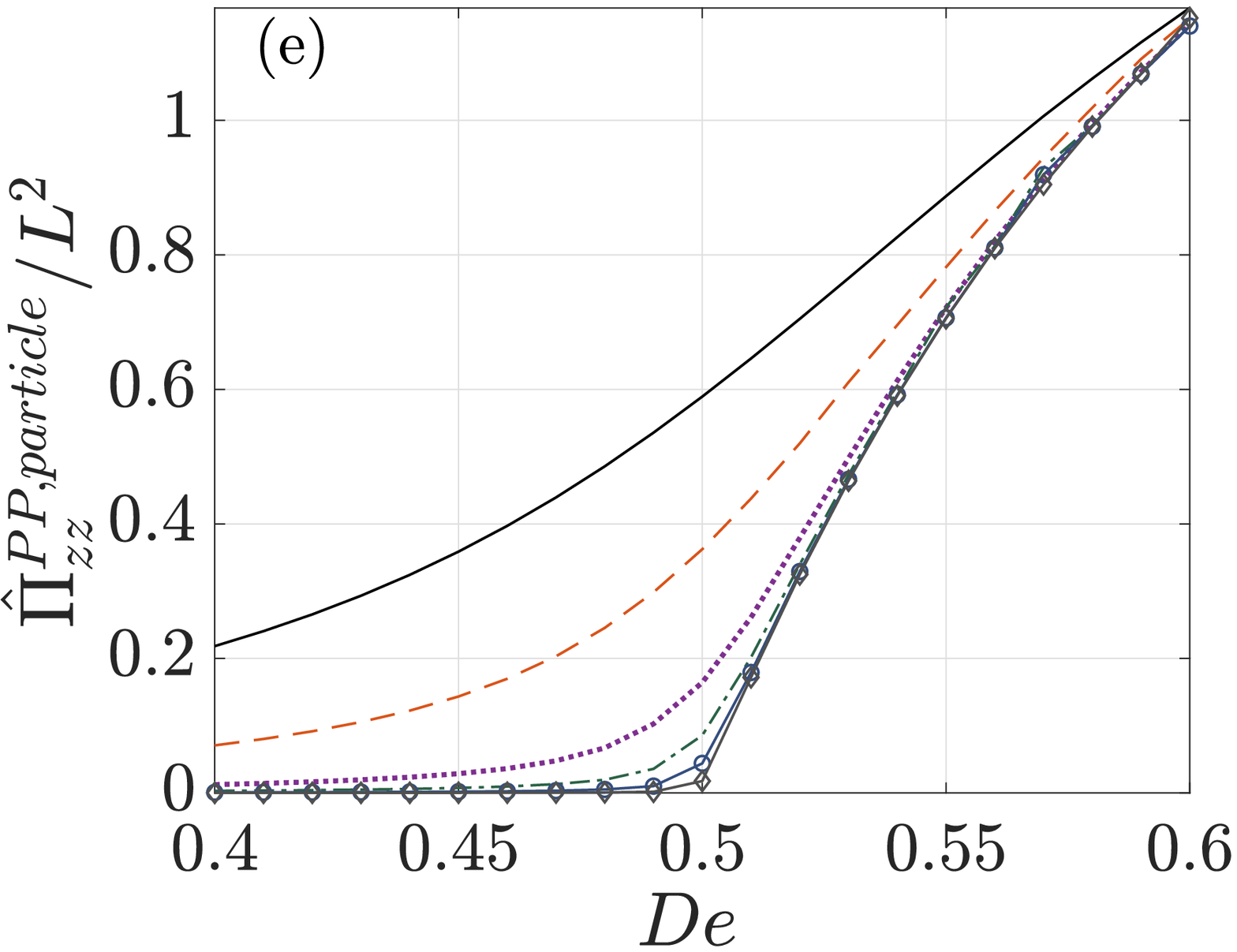}\label{fig:ParticlePIPSMediumDe}}\hfill
	\subfloat{\includegraphics[width=0.33\textwidth]{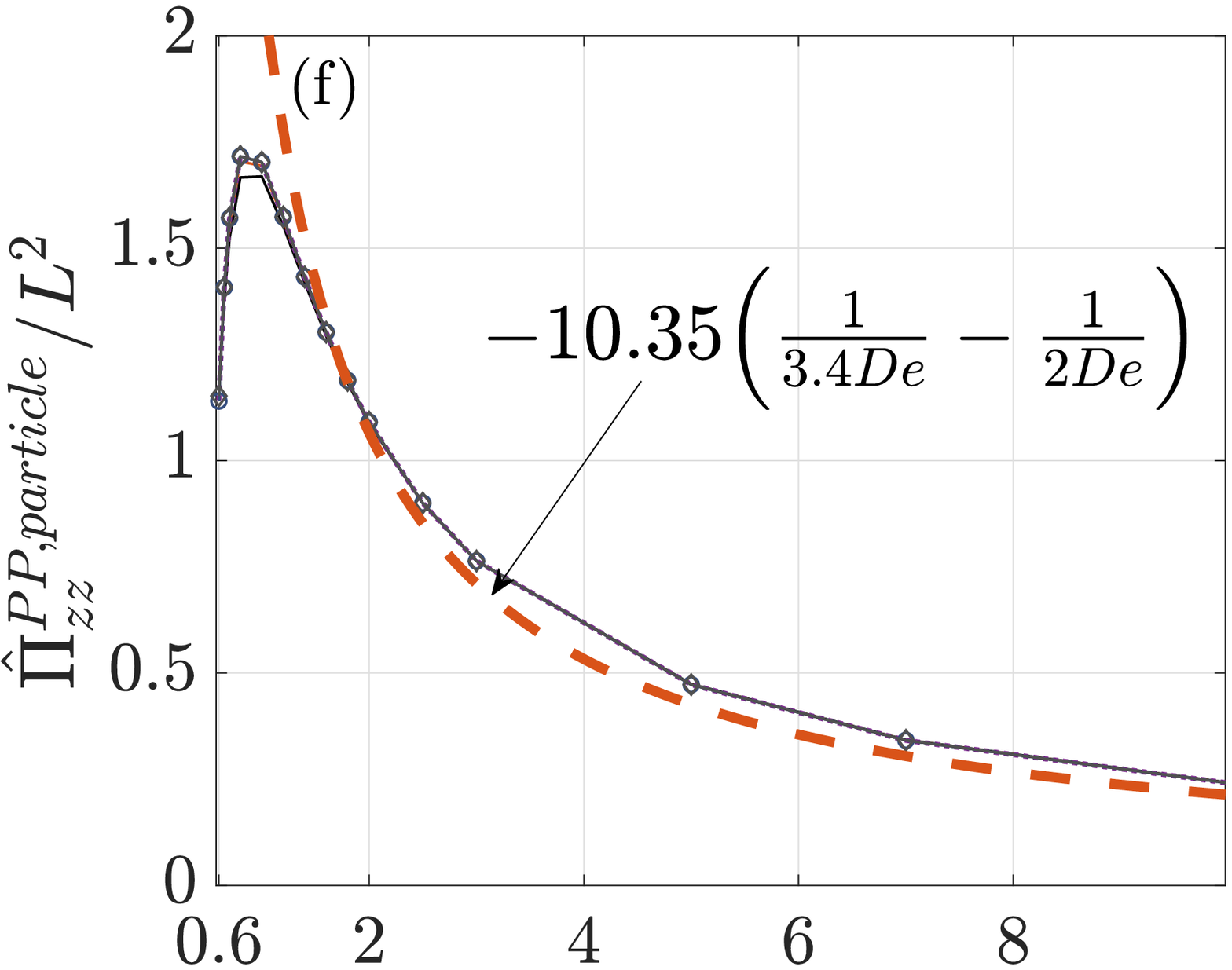}\label{fig:ParticlePIPSLargeDe}}
	\caption {Further decomposition of the particle induced polymer stress, $\hat{\Pi}^{\text{PP}}_{zz}$ or PIPS, into the contribution from the fluid region, $\hat{\Pi}^{\text{PP,fluid}}_{zz}$ ((a) to (c)) and the particle region $\hat{\Pi}^{\text{PP,particle}}_{zz}$ ((d) to (e)) for 6 different $L$ in $10\le L\le500$ for $De\le0.4$ ((a),(d)), $0.4< De<0.6$ ((b),(e)) and $De\ge0.6$ ((c),(f)). For the latter two $De$ regimes the stresses are scaled with $L^2$. All figures share the legend shown in (a). Similar to figure \ref{fig:Total_extra_polymer_large_de} individual components of PIPS (normalized with $L^2$) also follow a scaling with $De$ indicated on figures (c) and (f) here. \label{fig:Stresses_Decomposition_PIPS_split}}
\end{figure}

But first we discuss the PIPS because the non-monotonic variation of the interaction stress with $De$ arises from PIPS as shown in plots (a)-(c) of figure \ref{fig:Stresses_Decomposition_Stresslet_PIPS}. Compare the sub-figures corresponding to each PIPS for $De$ regime in figures \ref{fig:Stresses_Decomposition_Stresslet_PIPS} and the total interaction stress in figure \ref{fig:Stresses_Total} to observe the qualitative similarity. Further decomposition of the PIPS into the contribution from the fluid, $\hat{\Pi}^{\text{PP,fluid}}_{zz}$, and the particle, $\hat{\Pi}^{\text{PP,particle}}_{zz}$, region in figure \ref{fig:Stresses_Decomposition_PIPS_split} indicates the observed qualitative trend in the total interaction stress (figure \ref{fig:Stresses_Total}) and the PIPS (plots (a)-(c) of figure \ref{fig:Stresses_Decomposition_Stresslet_PIPS}) arises from the PIPS in the fluid region ($\hat{\Pi}^{\text{PP,fluid}}_{zz}$). The PIPS from the particle region also behaves non-monotonically but in a different fashion to the net PIPS or interaction stress around $De=0.5$. The particle PIPS or $\hat{\Pi}^{\text{PP,particle}}_{zz}$ is positive at all $De$, and undergoes a coil-stretch transition at $De=0.5$. But it decays in magnitude with $De$ for $De\gtrapprox0.8$ towards small values at large $De$. 

The fluid PIPS or $\hat{\Pi}^{\text{PP,fluid}}_{zz}$ is directly related to the behavior of polymers in the fluid region around the particle. In particular the change in polymer configuration from the far-field value is important. In the $De\le0.4$ regime, the fluid PIPS is positive due to the wake of highly stretched polymers downstream of the particle, represented by the red regions in the $\Delta\mathcal{S}/L$ plots of figure \ref{fig:polymerstretch_smallDe}. As discussed in section \ref{sec:Fullpolymerconfiguration}, this wake is a consequence of stretching of the polymers, that are coiled far upstream of the particle, by the high stretching region (large velocity gradients) around the extensional axis. At larger $De=\lambda \dot{\epsilon}$ (equation \ref{eq:DeEquation}) within this regime, the larger imposed extension rate, $\dot{\epsilon}$, causes the wake to become more intense as discussed in section \ref{sec:Fullpolymerconfiguration}. Therefore, the fluid PIPS or $\hat{\Pi}^{\text{PP,fluid}}_{zz}$ in figure \ref{fig:FluidPIPSSmallDe} increases with $De$ for $De\le 0.4$. The intensification of the wake of highly stretched polymers (red region) is accompanied by the appearance of a region of collapsed polymers (blue region) that first appears at $De\approx0.4$ as shown in figure \ref{fig:polymerstretch_smallDe}. As $De$ is  further increased, the region of collapsed polymers overwhelms the highly stretched polymer wake as shown in the plots of figure \ref{fig:polymerstretch_mediumDe}. This occurs because the far-field polymers get highly stretched upon increasing $De$, and when they arrive close to the particle, low stretching (small velocity and its gradients) regions around the stagnation points on the particle's surface collapse them to a coiled state: making them undergo a stretch-to-coil transition. This manifests as a rapid decrease in the fluid PIPS to negative values starting  at $0.45\lessapprox De\lessapprox 0.5$ in figure \ref{fig:FluidPIPSMediumDe}. Upon further increase in $De$, the far-field polymers become more stretched. On the particle surface, for large $L$ and $De>0.5$, the polymers collapse to an almost equilibrium configuration as shown in figure \ref{fig:largeLsurfacestretch} where the surface polymer stretch, $\sqrt{\text{tr}(\boldsymbol{\Lambda}^{(0)}|_{r=1})}$ is small compared with $L$. Figure \ref{fig:polymerstretch_largeDe} shows that the thin collapsed layer around the particle surface becomes thinner as $De$ is increased, while the intensity of collapse increases as discussed in section \ref{sec:Fullpolymerconfiguration} and evidenced by the increasingly negative fluid PIPS with $De$ in figure \ref{fig:FluidPIPSLargeDe}. For moderate $De$, $De \lesssim 1$, the collapsed region is near the particle surface and the neighboring region around 45$^\circ$ from the extensional axis ($L=10$ plots of figure \ref{fig:polymerstretch_mediumDe}). In this $De$ regime, at $L\gtrsim50$ the collapsed region extends downstream of the particle into a wake of unstretched polymers ($L=50$ and 100 plots of figure \ref{fig:polymerstretch_mediumDe}). Therefore, increasing $L$ leads to a slightly more negative contribution from fluid PIPS in figure \ref{fig:FluidPIPSLargeDe} (also in figure \ref{fig:FluidPIPSMediumDe} for larger $De$). For larger $De$, $De\gtrsim1.5$, the negative contribution mainly arises from the collapsed polymers near the particle (figure \ref{fig:polymerstretch_largeDe}). Hence, a very small volume contributes to a very large change in the stresses for $De\gtrsim 1.5$.

\begin{figure}[h!]
	\centering
	\subfloat{\includegraphics[width=0.33\textwidth]{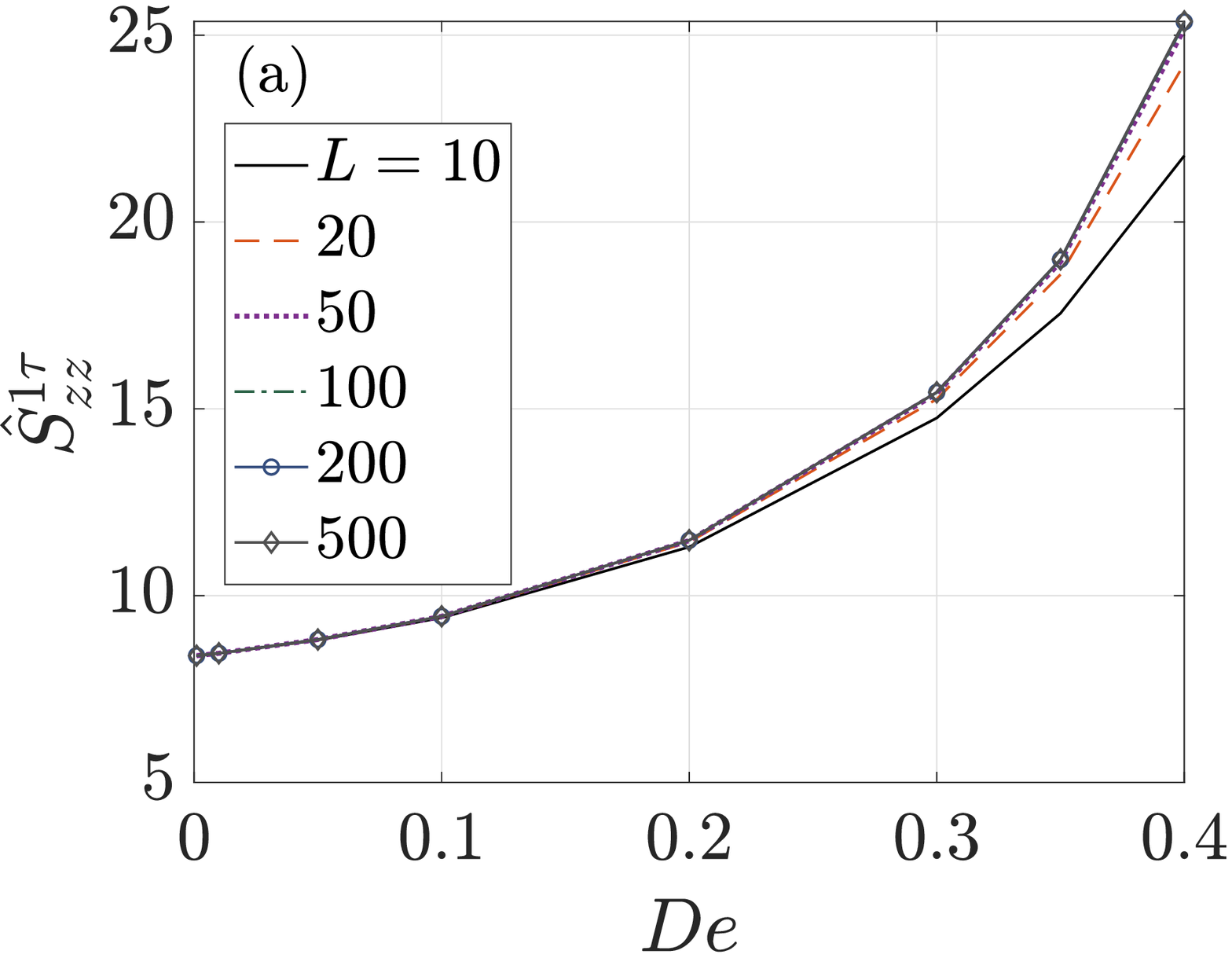}\label{fig:S1TauSmallDe}}\hfill
	\subfloat{\includegraphics[width=0.33\textwidth]{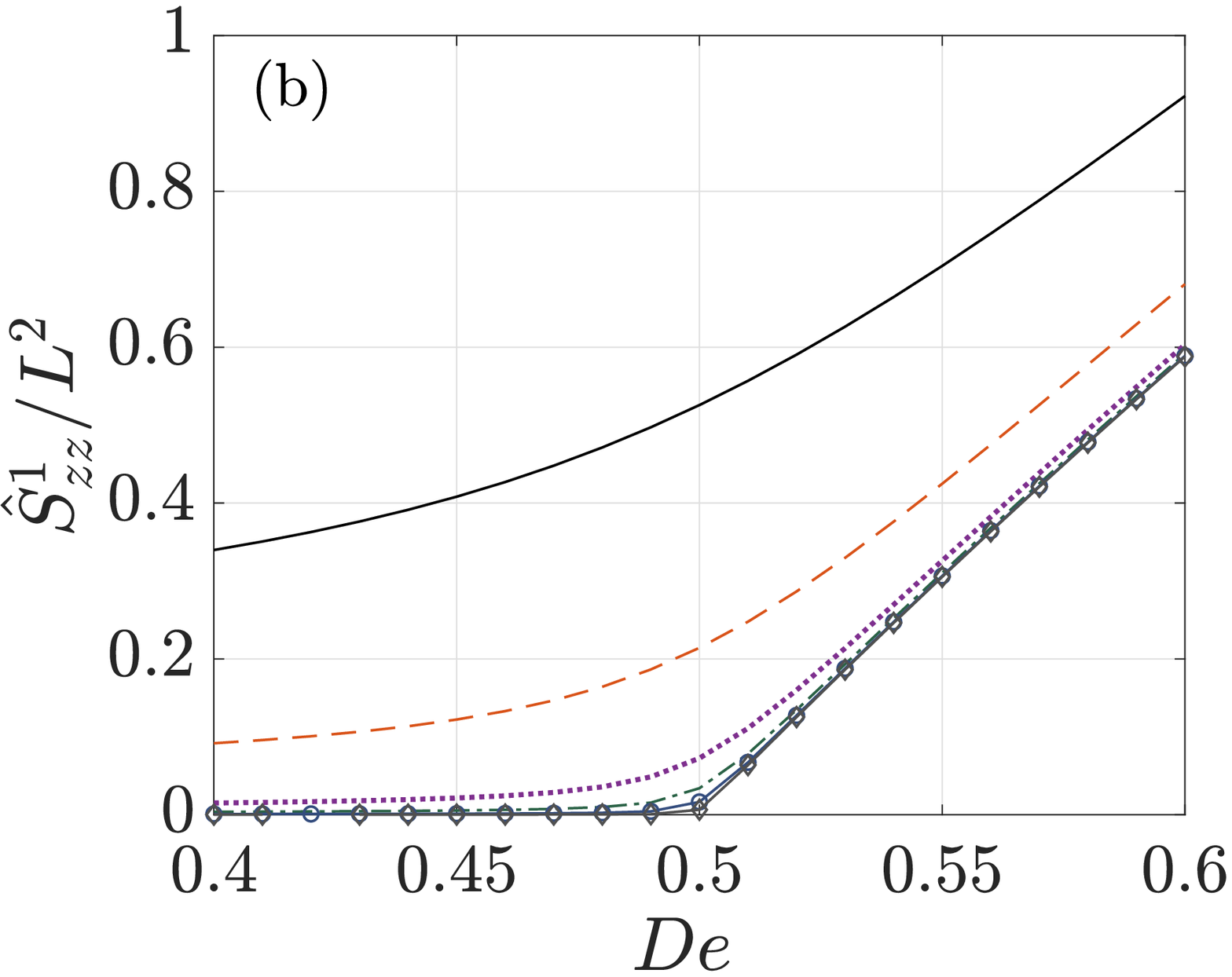}\label{fig:S1TauMediumDe}}\hfill
	\subfloat{\includegraphics[width=0.33\textwidth]{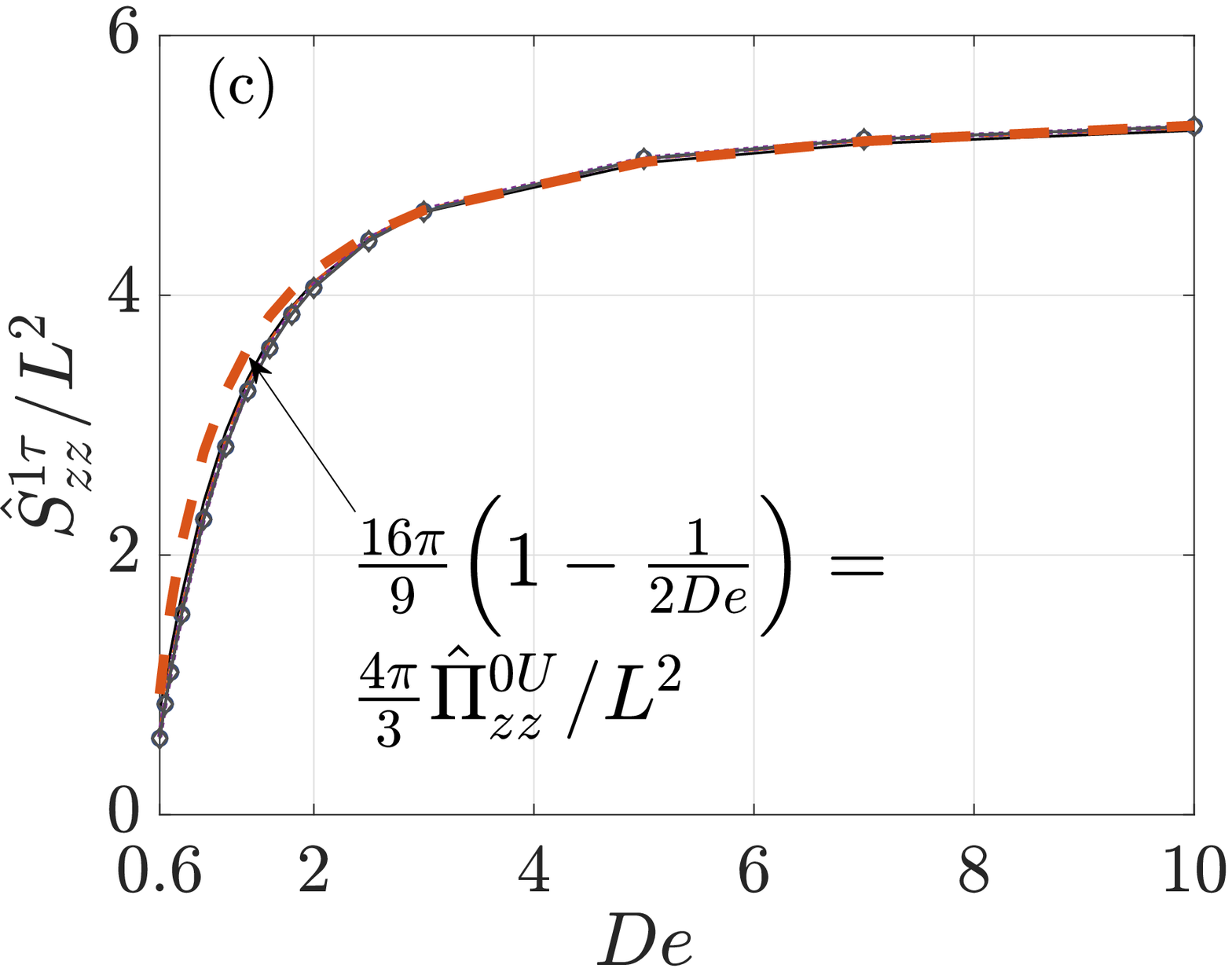}\label{fig:S1TauLargeDe}}\hfill
	\subfloat{\includegraphics[width=0.33\textwidth]{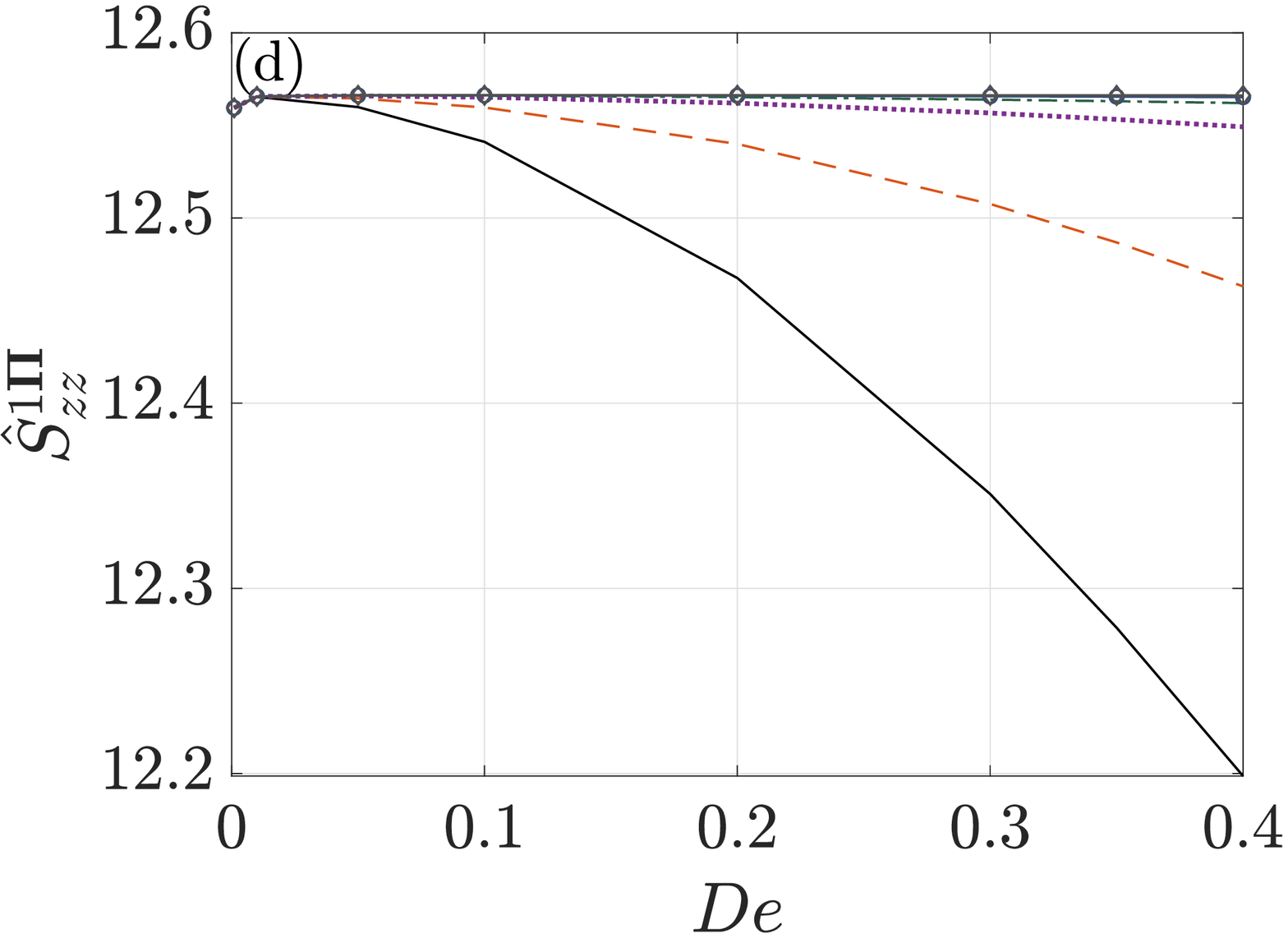}\label{fig:S1PiSmallDeIPSSmallDe}}\hfill
	\subfloat{\includegraphics[width=0.33\textwidth]{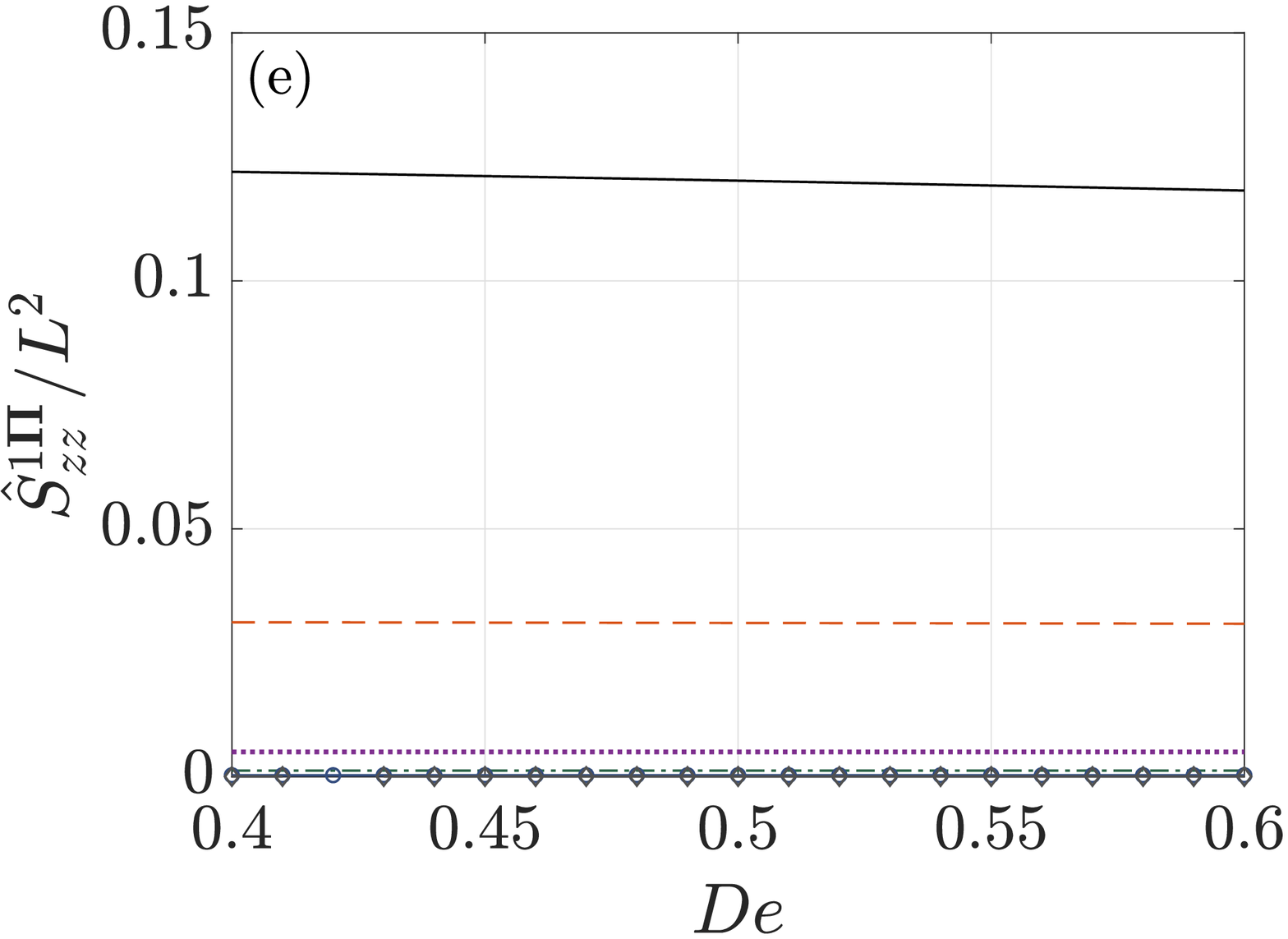}\label{fig:S1PiMediumDe}}\hfill
	\subfloat{\includegraphics[width=0.33\textwidth]{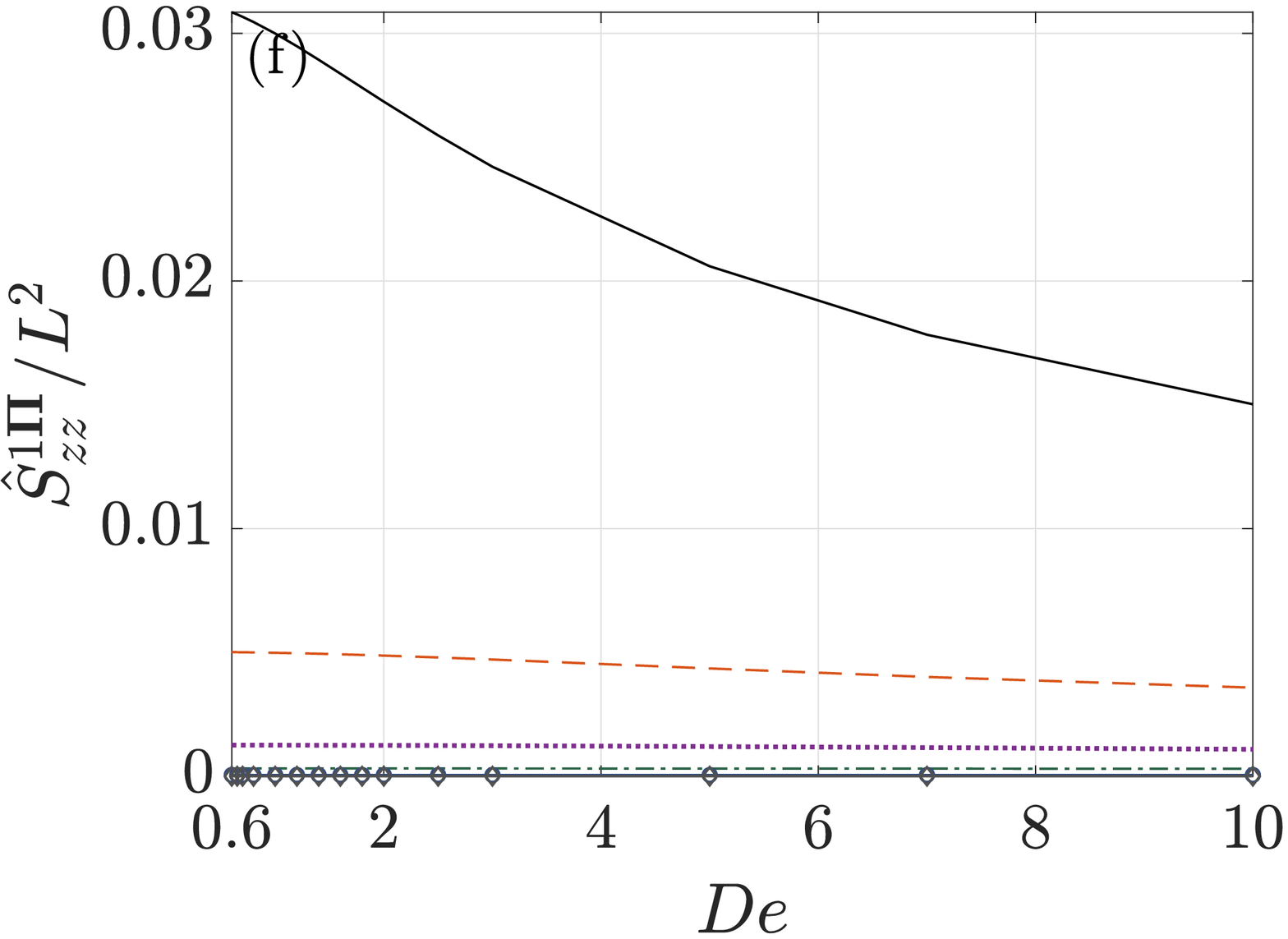}\label{fig:S1PiLargeDe}}
	\caption {Decomposition 1 of the the interaction stresslet $\hat{\text{S}}^{(1)}_{zz}$ into the contribution due to the $\mathcal{O}(c)$ perturbation in the solvent stress, $\hat{\text{S}}^{1\boldsymbol{\tau}}_{zz}$ ((a) to (c)) and the polymeric stress, $\hat{\text{S}}^{1\boldsymbol{\Pi}}_{zz}$ ((d) to (e)) for 6 different $L$ in $10\le L\le500$ for $De\le0.4$ ((a),(d)), $0.4< De<0.6$ ((b),(e)) and $De\ge0.6$ ((c),(f)). For the latter two $De$ regimes the stresses are scaled with $L^2$. All figures share the legend shown in (a).\label{fig:Stresses_Decomposition_Stresslet1}}
\end{figure}
\begin{figure}[h!]
	\centering
	\subfloat{\includegraphics[width=0.33\textwidth]{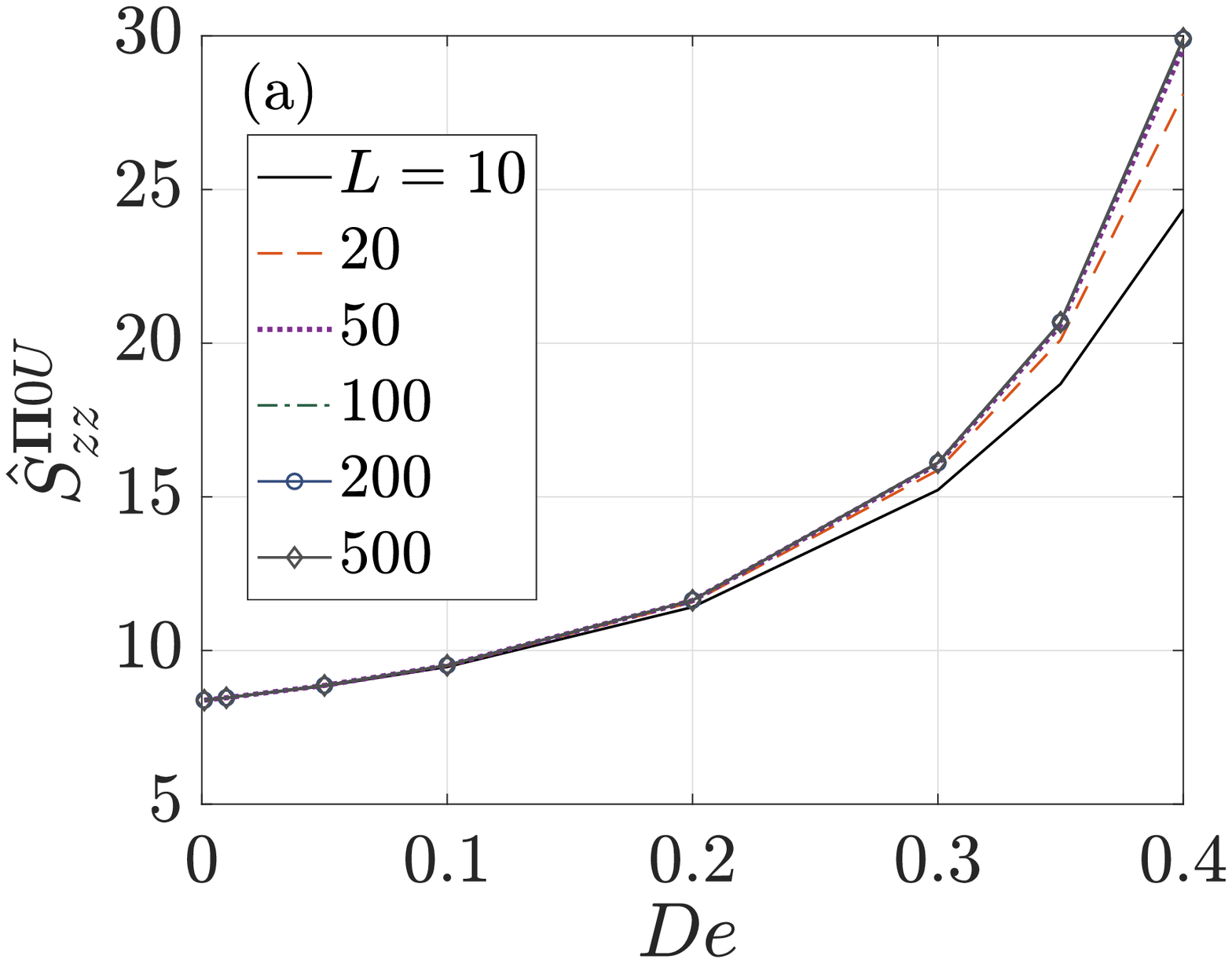}\label{fig:S1Pi0USmallDe}}\hfill
	\subfloat{\includegraphics[width=0.33\textwidth]{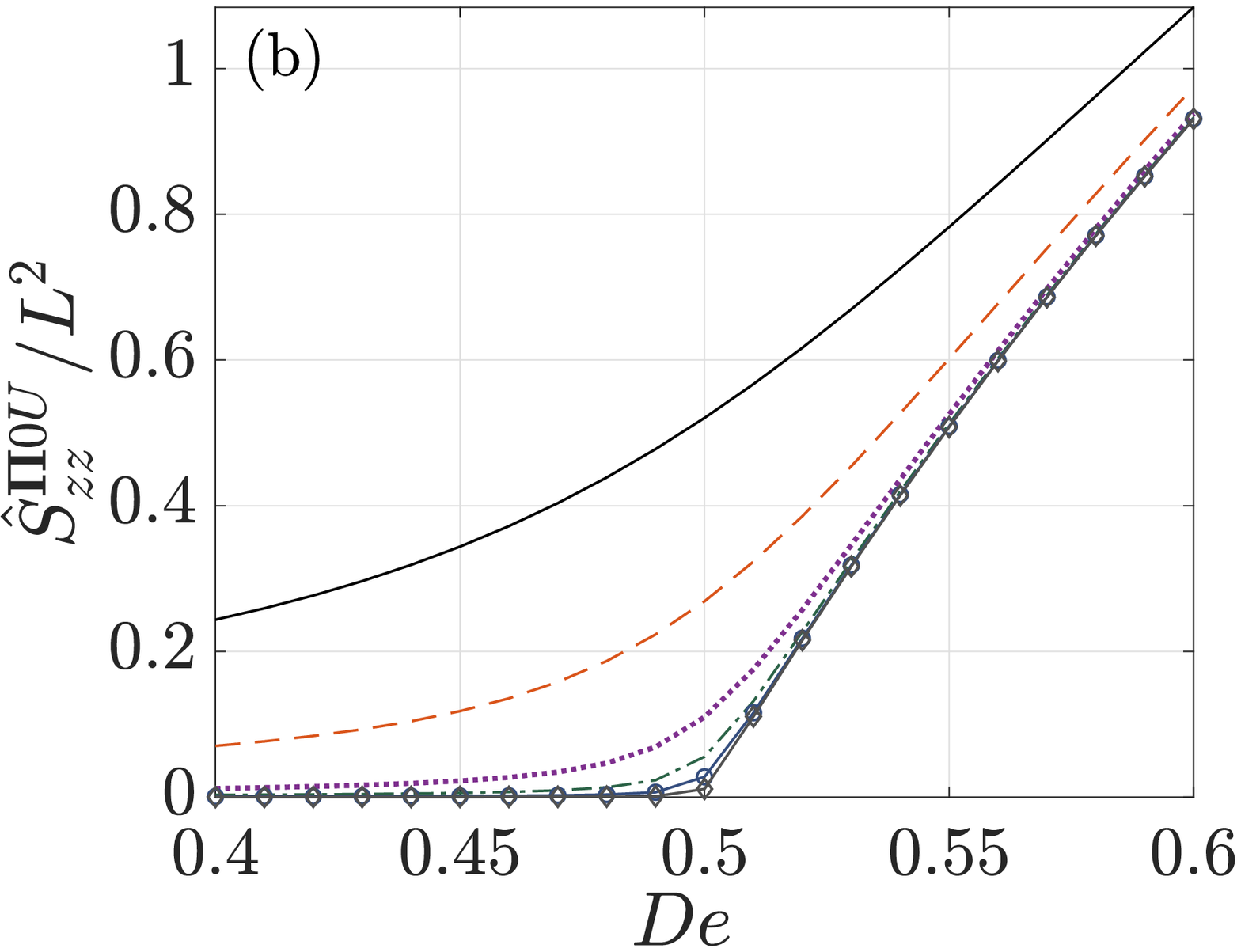}\label{fig:S1Pi0UMediumDe}}\hfill
	\subfloat{\includegraphics[width=0.33\textwidth]{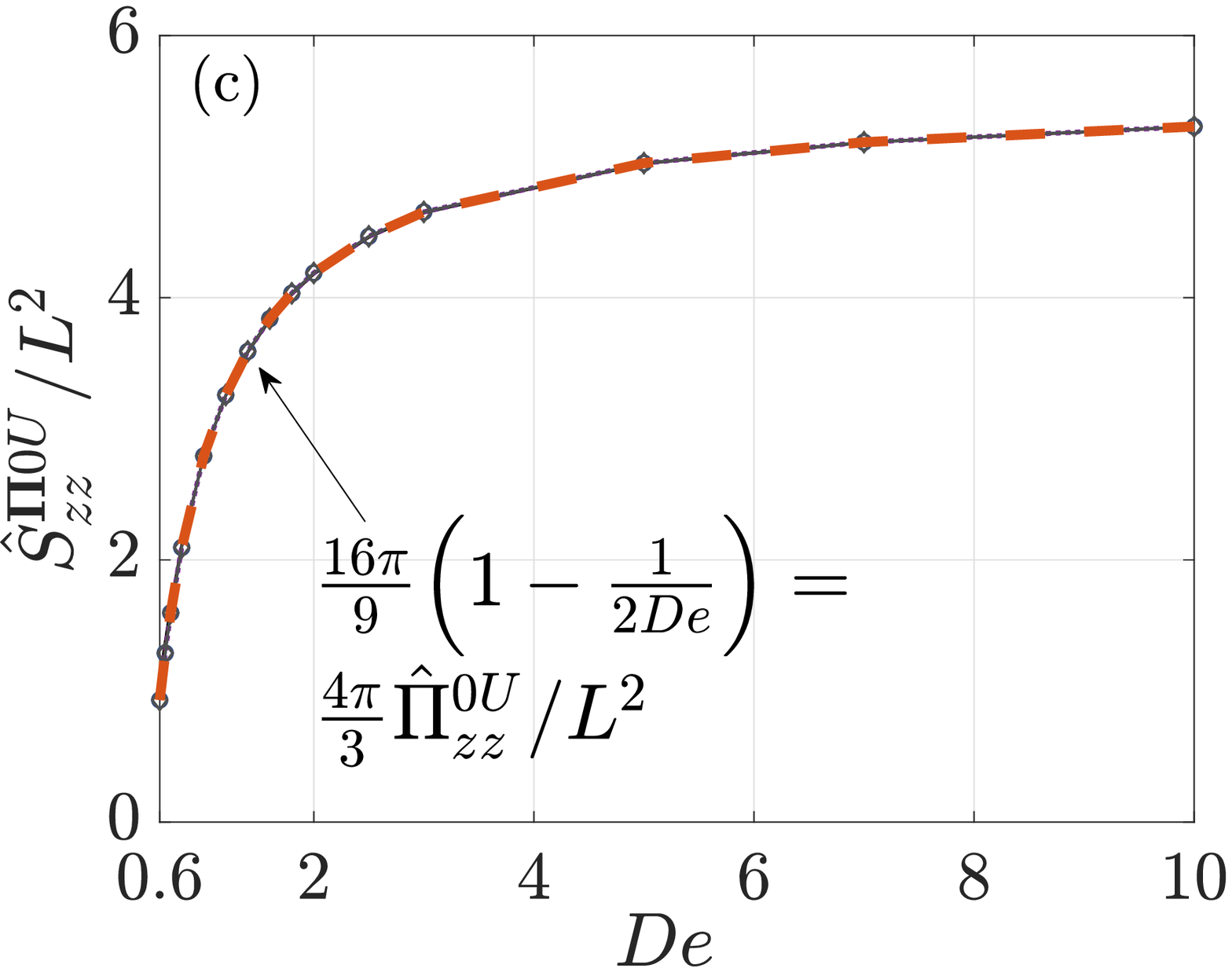}\label{fig:S1Pi0ULargeDe}}\hfill
	\subfloat{\includegraphics[width=0.33\textwidth]{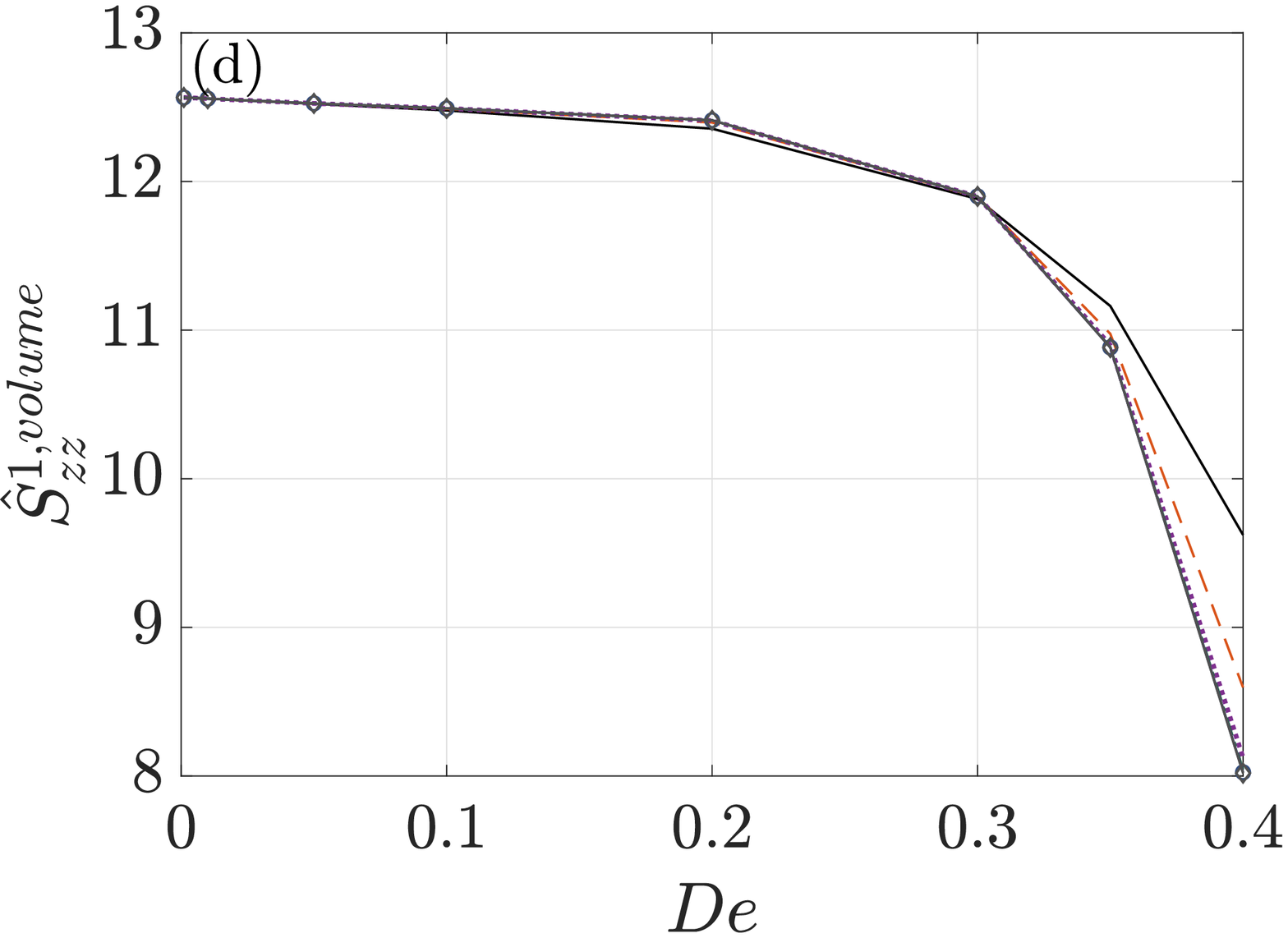}\label{fig:S1VolSmallDe}}\hfill
	\subfloat{\includegraphics[width=0.33\textwidth]{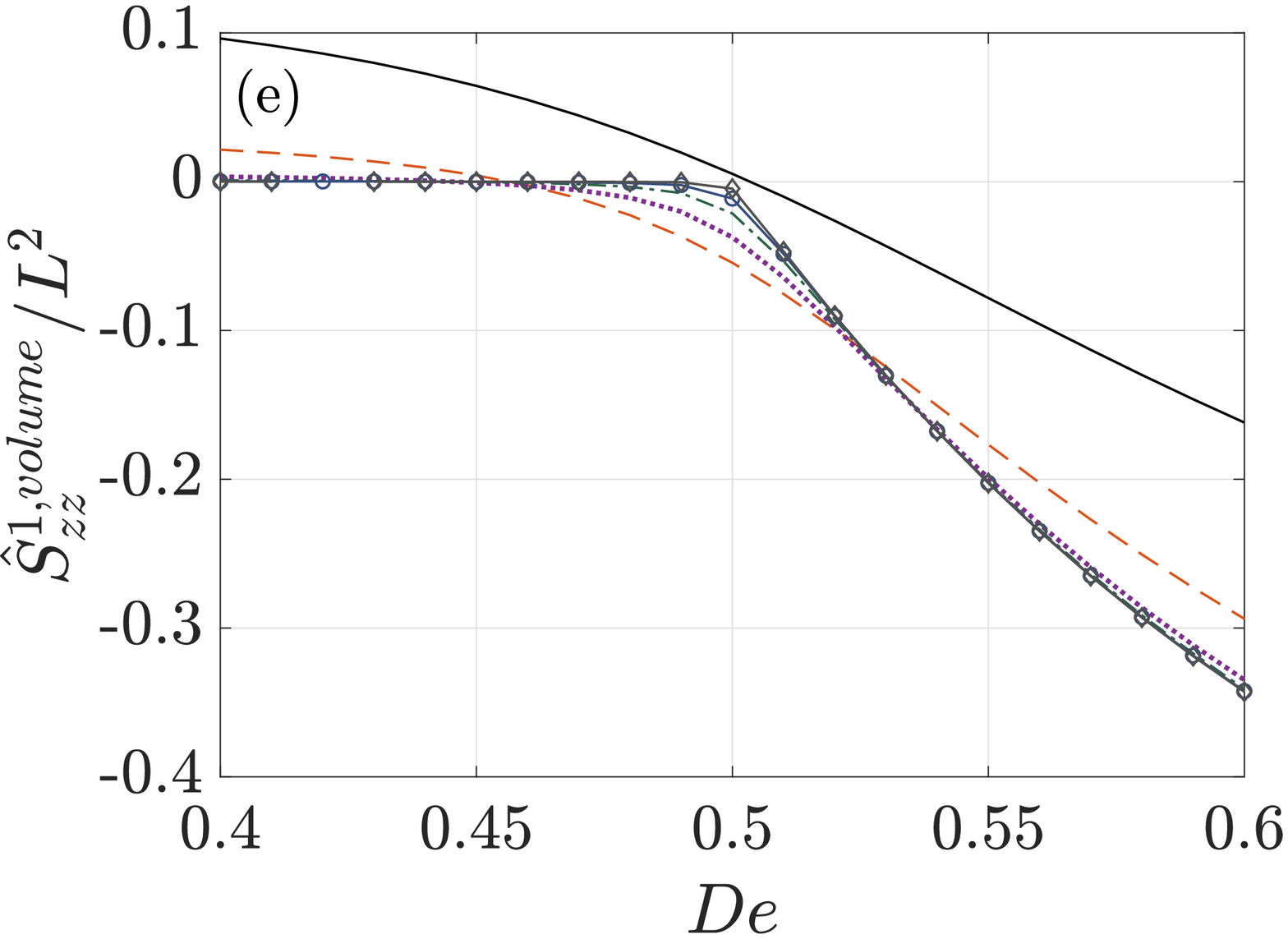}\label{fig:S1VolMediumDe}}\hfill
	\subfloat{\includegraphics[width=0.33\textwidth]{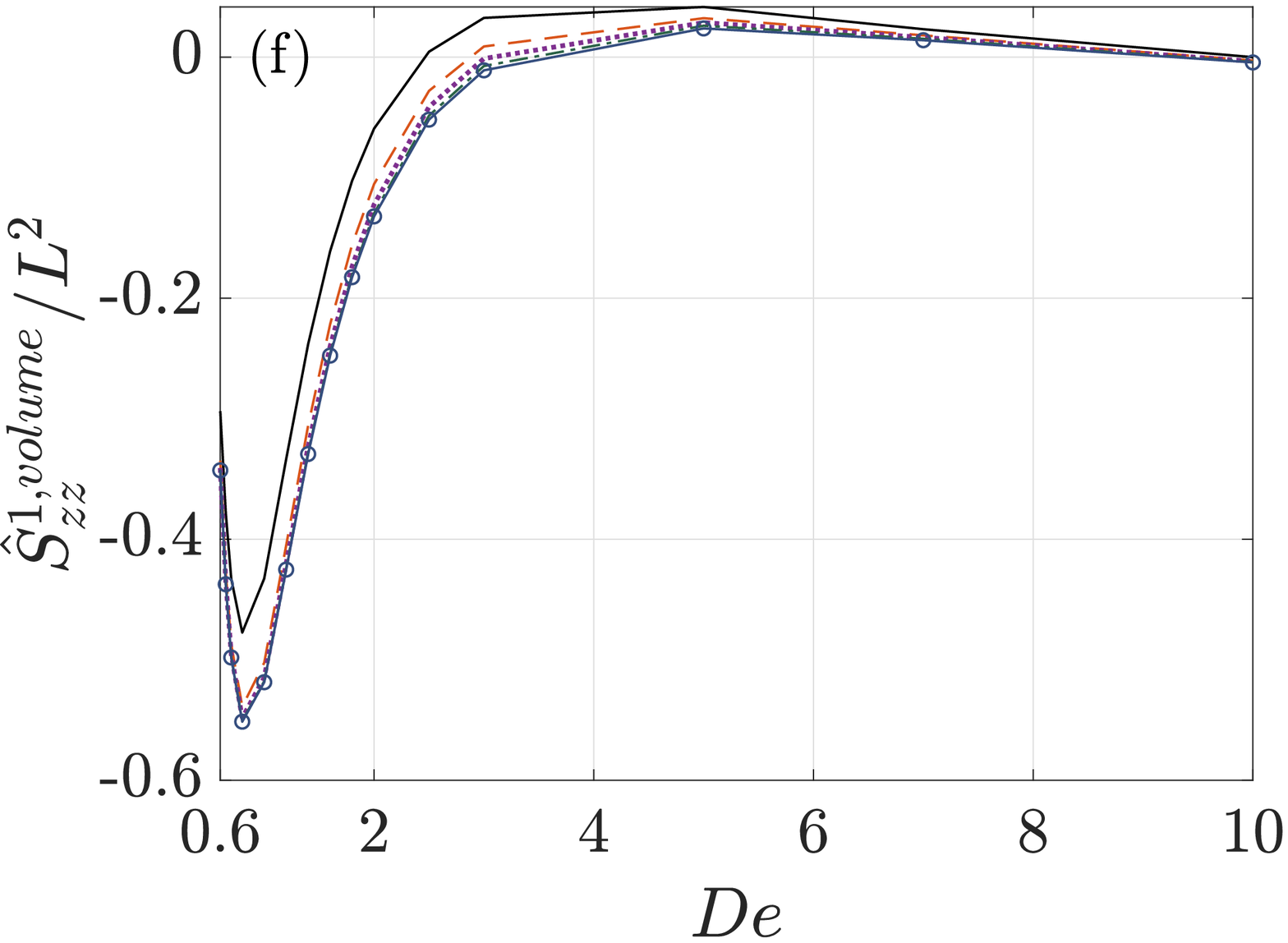}\label{fig:S1VolLargeDe}}
	\caption {Decomposition 2 of the interaction stresslet $\hat{\text{S}}^{(1)}_{zz}$ into the contribution due to the undisturbed stresslet, $\hat{\text{{S}}}^{\boldsymbol{\Pi} 0U}_{zz}$ ((a) to (c)) and the volumetric stresslet, $\hat{\text{S}}^{1,\text{volume}}_{zz}$ ((d) to (e)) for 6 different $L$ in $10\le L\le500$ for $De\le0.4$ ((a),(d)), $0.4< De<0.6$ ((b),(e)) and $De\ge0.6$ ((c),(f)). For the latter two $De$ regimes the stresses are scaled with $L^2$. All figures share the legend shown in (a). \label{fig:Stresses_Decomposition_Stresslet2}}
\end{figure}
As mentioned earlier, the interaction stresslet, $\hat{\text{S}}^{1}_{zz}$, follows the qualitative behavior of the undisturbed polymer stress. This can be observed by comparing subplots of figures \ref{fig:Stresses_Undisturbed} and plots (d)-(f) of figure \ref{fig:Stresses_Decomposition_Stresslet_PIPS} for each $De$ regime. We discussed two possible decompositions of the stresslet in section \ref{sec:RheologyTheory}. One is the usual decomposition into the stresslet (such as used by \cite{koch2006stress,koch2016stress}) arising from different sources of stress:  $\hat{\text{S}}^{1}_{zz}=\hat{\text{S}}^{1\boldsymbol{\Pi}}_{zz}+\hat{\text{S}}^{1\boldsymbol{\tau}}_{zz}$. $\hat{\text{S}}^{1\boldsymbol{\Pi}}_{zz}$ is the stresslet arising from the polymeric stress and $\hat{\text{S}}^{1\boldsymbol{\tau}}_{zz}$ is from the $\mathcal{O}(c)$ solvent stress on the particle's surface. This decomposition is shown in figure \ref{fig:Stresses_Decomposition_Stresslet1}. In section \ref{sec:SomeStressletRelations} (see equation \eqref{eq:approximate_polymer_stresslet} and figure \eqref{fig:polymerstresslet}) we showed that the polymeric stresslet remains approximately constant at 4$\pi$ at low to moderate. (This can also be observed from figures \ref{fig:S1PiSmallDeIPSSmallDe} and \ref{fig:S1PiMediumDe}). At larger $De$, the magnitude of the  polymeric stresslet reduces and $\hat{\text{S}}^{1\boldsymbol{\tau}}_{zz}$ (solvent stresslet) is the dominant component. Throughout the $De$ range shown, the solvent stresslet qualitatively explains the behavior of the total interaction stresslet (compare the subplots (a)-(c) of figure \ref{fig:Stresses_Decomposition_Stresslet1} with subplots (d)-(f) of figure \ref{fig:Stresses_Decomposition_Stresslet_PIPS} to see this qualitative similarity). In the $De>0.6$ regime $\hat{\text{S}}^{1\boldsymbol{\tau}}_{zz}$ fully captures $\hat{\text{S}}^{1}_{zz}$ as they are both equal to $16\pi/9(1-1/(2De))={4\pi}/{3}\hat{\Pi}_{zz}^{0U}/L^2$ (the reason for this scaling is indicated by equation \ref{eq:StressletLargeDe}). The magnitude of the $L^2$ scaled polymeric stresslet in the large $De$ regime is small because in this regime the polymers around the surface are almost collapsed relative to their undisturbed configuration. This is shown by the (blue) region of collapsed polymers around the particle surface in figures \ref{fig:polymerstretch_mediumDe} and \ref{fig:polymerstretch_largeDe}. Understanding the variation of $\hat{\text{S}}^{1\boldsymbol{\tau}}_{zz}$ with $De$ in plots (a)-(c) of figure \ref{fig:Stresses_Decomposition_Stresslet1} in terms of the behavior of polymers around or at the particle surface is not straightforward  This is because the polymer configuration and solvent stress $\boldsymbol{\tau}^{(1)}$ are indirectly coupled through the momentum equation. Since $\hat{\text{S}}^{1}_{zz}$ is dominated by $\hat{\text{S}}^{1\boldsymbol{\tau}}_{zz}$ at large $De$, the first decomposition is not able to provide physical insight into the variation of the interaction stresslet.

Therefore, we turn to the second stresslet decomposition into the undisturbed ($\hat{\text{{S}}}^{\boldsymbol{\Pi} 0U}_{zz}$) and volumetric ($\hat{\text{S}}^{1,\text{volume}}_{zz}$) stresslet shown in figure \ref{fig:Stresses_Decomposition_Stresslet2}.
The variation of the undisturbed stresslet, $\hat{\text{{S}}}^{\boldsymbol{\Pi} 0U}_{zz}$, with $De$ explains the qualitative variation of the total interaction stresslet. This can be checked by comparing the subplots (a)-(c) of figure \ref{fig:Stresses_Decomposition_Stresslet2} with subplots (d)-(f) of figure \ref{fig:Stresses_Decomposition_Stresslet_PIPS}. The undisturbed stresslet is the stresslet on a unit sphere in the far-field. It is directly proportional to the undisturbed polymer stress ($\hat{\Pi}_{zz}^{(0U)}$) as also shown in equation \eqref{eq:stresslet_new}. In the $De\ge0.6$ regime $\hat{\text{{S}}}^{\boldsymbol{\Pi} 0U}_{zz}={4\pi}/{3}\hat{{\Pi}}^{(0U)}_{zz}$ as also shown by a good match between numerical and analytical estimate in figure \ref{fig:S1Pi0ULargeDe}.

To understand why the undisturbed stresslet fully explains the interaction stresslet at large $De$ we describe why the volumetric stresslet ($\hat{\text{S}}^{1,\text{volume}}_{zz}$) is small. The integrand of the volumetric stresslet is proportional to the difference in polymer stress from its undisturbed value in the fluid region around the particle (equation \eqref{eq:StressletDecomp2}). The volumetric stresslet is positive in the $De\le0.4$ regime due to the wake of highly stretched polymers. However, its magnitude decreases with $De$ in figure \ref{fig:S1VolSmallDe} within the $De\le0.4$ regime due to the appearance of a (blue) region of collapsed polymers around the particle surface as $De$ is increased in figure \ref{fig:polymerstretch_smallDe}. The collapse of polymers is relative to the far-field or undisturbed polymers. Therefore, the collapse becomes more intense near the undisturbed coil-stretch transition, $0.4\le De\le 0.6$, as the highly stretched undisturbed polymers relax to a near equilibrium state in the low stretching (low velocity and its gradients) region around the particle surface. This is shown as the blue region in various plots of  figure \ref{fig:polymerstretch_mediumDe}. The volumetric stresslet therefore undergoes a stretch-to-coil transition at $De\approx0.5$ in figure \ref{fig:S1VolMediumDe}. In the $De\ge0.6$ regime the volumetric stresslet in figure \ref{fig:S1VolLargeDe} is negative and increases in magnitude up to $De\approx0.8$. This occurs because the region of collapsed polymers becomes more intense as $De$ is increased within these values (figure \ref{fig:polymerstretch_mediumDe}). Upon further increase in $De$ the region of collapsed polymers is concentrated closer to the particle surface (figures \ref{fig:polymerstretch_mediumDe} and \ref{fig:polymerstretch_largeDe}) and the volumetric stresslet starts becoming less negative in figure \ref{fig:S1VolLargeDe}. The volumetric stresslet becomes slightly positive for a small range of $De$ around 2 due to the small region of stretched polymers around the extensional axis (light red region in $De=1.5$ and 3 plots of figure \ref{fig:polymerstretch_largeDe}). At large $De$, the polymer stretch (figure \ref{fig:polymerstretch_largeDe}) and hence the polymer stress is similar to the undisturbed values everywhere, except in a thin layer near the surface of the sphere, where the polymers collapse to  a near equilibrium configuration (figure \ref{fig:polymerstretch_largeDe}). Thus, the volumetric stresslet also vanishes at large $De$ in figure \ref{fig:S1VolLargeDe} and the undisturbed stresslet (figure \ref{fig:S1Pi0ULargeDe}) fully captures the interaction stresslet (figure \ref{fig:TotalStressletLargeDe}).

We found an expression for the variation of the undisturbed stresslet in the large $De$ regime in equation \eqref{eq:far_field_fluid_Stresslet_largeDe}. Based on the above discussion of the components of the interaction stresslet, we can conclude that,
\begin{equation}
\hat{\text{{S}}}^{(1)}_{zz}\approx\hat{\text{{S}}}^{1\boldsymbol{\tau}}_{zz}\approx\hat{\text{{S}}}^{\boldsymbol{\Pi} 0U}_{zz}=\frac{4\pi}{3}\hat{{\Pi}}^{(0U)}_{zz}, \hspace{0.2in}De>0.5.\label{eq:StressletLargeDe}
\end{equation}
The analytical estimate, $\hat{\text{{S}}}^{\boldsymbol{\Pi} 0U}_{zz}={16\pi}/{9}(1-{1}/{(2De)})L^2$ from equation \eqref{eq:far_field_fluid_Stresslet_largeDe} fits well with the numerical solutions, as shown in figures \ref{fig:TotalStressletLargeDe}, \ref{fig:S1TauLargeDe} and \ref{fig:S1Pi0ULargeDe}. Additionally, we observe,
\begin{equation}
\hat{\Pi}^{\text{PP}}_{zz}\approx -10.35\Big(1-\frac{1}{2De}\Big)L^2=-7.76{\hat{{\Pi}}^{(0U)}_{zz}}, \hspace{0.2in}De\gtrsim 0.6,\label{eq:pi_largeDe}
\end{equation}
as shown in figures \ref{fig:Total_extra_polymer_large_de}. Combining the expressions of equations \eqref{eq:StressletLargeDe} and \eqref{eq:pi_largeDe} we find the particles' contribution to extensional viscosity, from equation \eqref{eq:ExtensionalViscosityParticles}, in the large $De$ regime to be
\begin{equation}
\mu^\text{part}=(2.5-0.85\mu^{poly})\phi, \hspace{0.2in}De \gtrsim 0.6
\end{equation}
where $\mu^{poly}$ is the extensional viscosity due to polymers and is given in equation \eqref{eq:ExtenPolymers}.

From figure \ref{fig:Stresses_LargeDe_undisturbed} or equation \eqref{eq:approximatelargeDe}, if $L\sim\mathcal{O}(100)$, $\hat{\Pi}^{(0U)}_{zz}\sim\mathcal{O}(10^4)$ in the large $De$ regime. Therefore, $2.5-0.85\mu^\text{poly}$ can become negative even at $c\sim\mathcal{O}(10^{-4})$. For a polymer with $L=100$, a polymer concentration as small as 0.0009 for $De=1$ and 0.0006 for $De=2$ allows the net extensional viscosity due to the particles, $\mu^\text{part}$, to be negative. Hence, adding particles to a low $c$ polymeric fluid reduces its extensional viscosity for $De\gtrsim0.6$, or in other words allows it to be stretched more easily if either the relaxation time of the polymers or the imposed extension rate is large. Within this regime, the reduction in the extensional viscosity increases with $c$, $De$ and $\phi$. Therefore, for a high throughput industrial process involving a large extension rate of a low $c$ polymeric fluid, such as fiber spinning or extrusion molding, adding a small concentration of spherical particles can be beneficial in reducing the operating cost as the stress required to be overcome can be reduced.

\section{Conclusion}\label{sec:Conclusion}
Our aim in this study has been to find the first effects of particle-polymer interactions on the extensional rheology of a dilute suspension of spheres in a dilute polymer solution. We find that when the polymer relaxation time and the imposed extension rate are small, the interaction leads to an increase in the extensional viscosity as compared to particle-free polymeric fluid. Interestingly, the particle-polymer interaction lowers the extensional viscosity of the suspension if the product of the polymer relaxation time and the imposed extension rate is large.

We characterize the $\mathcal{O}(c)$ (polymer concentration) behavior of the interaction between a spherical particle and a dilute polymer solution in a Newtonian solvent, in an imposed extensional flow. The FENE-P constitutive equation is used to model the polymer stress, as it captures the qualitative trends observed in  rheology experiments of particle-free viscoelastic fluids \cite{anna2008effect}. In the small $c$ limit, the leading order polymer configuration is driven by the Newtonian velocity field around the sphere, and the divergence of the polymer configuration induces an $\mathcal{O}(c)$ perturbation to the fluid velocity and pressure. The leading order polymer configuration around a sphere in an extensional flow is evaluated using the method of characteristics from the analytically known Newtonian velocity field around a sphere, thus making the method semi-analytical. This method was first demonstrated on the flow around a sphere in a simple shear flow of an Oldroyd-B fluid by Koch et al. \cite{koch2016stress}. In a dilute particle suspension particle-particle interactions are negligible and the $\mathcal{O}(\phi)$ (particle volume fraction) stress due to particle-polymer interactions within the suspension is that between an isolated particle and polymers in an infinite expanse of polymeric fluid. Therefore, using ensemble averaging \cite{koch2006stress,koch2016stress}, we use the polymeric field around an isolated particle in an infinite expanse of fluid to calculate these stresses.

The particle-polymer interaction stress consists of the particle-induced polymer stress (PIPS) and the stresslet. The stresslet is the stress due to surface traction on the particles and PIPS is the extra fluid stress due to the disturbance in polymer configuration by the presence of particles. Therefore, particle-polymer interaction is a two-way interaction. The stresslet is the symmetric part of the first moment of the force on the particle surface \cite{batchelor1970stress} and in a polymeric fluid has previously \cite{koch2016stress} been decomposed into the stresslet due to the polymer and the solvent stress at the particle surface. At $\mathcal{O}(c)$, the latter is also created by the polymer, albeit indirectly. Using a generalized reciprocal theorem and the divergence theorem, Koch et al. \cite{koch2016stress} provide a mathematical framework to obtain the $\mathcal{O}(c)$ solvent stresslet directly from the leading order polymer configuration. Further analysis of that derivation allows us to propose another decomposition, interpretable in terms of just the polymer stress around the particle. According to this, the total stresslet is a sum of the stresslet on a fluid volume equivalent to the particle in the far-field, and a volumetric contribution due to the extra polymer stress in the presence of the particle. The first part of this decomposition is analytically evaluated from the undisturbed (particle-free) polymer stress.

Previously in \cite{greco2007rheology,housiadas2009rheology} the stress equivalent to PIPS has been approximated as a volume average of the polymer stress instead of an ensemble average. Ensemble averaging is the suitable method because in an experiment, the stress in a homogeneous particle suspension is an average over the ensemble of all possible particle configurations. Mathematically, volume averaging leads to divergent integrals as shown by Koch et al. \cite{koch2016stress} for an Oldroyd-B fluid and in this paper for the FENE-P model. The polymer stress contribution that leads to divergent integrals scales as $1/r^3$ at large distances, $r$, from the particle. This contribution is part of the linear perturbation to the polymer stress due to fluid velocity perturbations caused by the particle. The governing equation for this linear polymer configuration is obtained by linearizing the relevant polymer constitutive equation about the undisturbed values for fluid velocity and polymer configuration (section \ref{sec:RheologyTheory}). It is shown in section \ref{sec:RheologyTheory} (also \cite{koch2016stress}) that although its slow decay leads to a divergent integral in volume averaging, the ensemble average of this linear polymer stress is zero. Therefore, to obtain convergent integrals and evaluate PIPS from the calculation of the polymer stress field around a single particle, the linear polymer stress is removed from the total polymer stress before the ensemble average is approximated in terms of a volume integral around an isolated particle by invoking the diluteness of particle concentration in the suspension.

In the absence of the particles, polymers undergo a coil-stretch transition at a Deborah number of $De=0.5$, above which the polymers stretch close to their maximum extensibility, $L$. The polymer stress for $De<0.5$ is independent of $L$, and for $De>0.5$ it scales as $L^2$ (figure \ref{fig:Stresses_Undisturbed}). We find the contribution from the particle-polymer interaction to respect the same scalings for $L\gtrsim 50$ (figures \ref{fig:Stresses_SmallDe_disturbed}, \ref{fig:Particle_extra_medium_de} and \ref{fig:Total_Fluid_extra_polymer_large_de}).

At small $De$ ($De\lesssim 0.52$), the particle-polymer interaction is positive and increases with $De$ (figures \ref{fig:Stresses_SmallDe_disturbed} and \ref{fig:Particle_extra_medium_de}). Contributions from both the stresslet (figure \ref{fig:TotalStressletSmallDe}) and the PIPS (figure \ref{fig:Total_extra_polymer_small_de}) are positive and significant. As $De$ is increased beyond this, the major stresslet contributions remain positive and keep increasing with $De$ (figures \ref{fig:TotalStressletMediumDe} and \ref{fig:TotalStressletLargeDe}).

At large $De$ (figure \ref{fig:S1TauLargeDe} and \ref{fig:S1PiLargeDe}), the total stresslet is almost entirely from the solvent stress at the particle surface in terms of the original decomposition \cite{koch2016stress}, or from the stresslet on a fluid element equivalent to the particle in the far-field in terms of the new decomposition as the volumetric contribution of the extra stress vanishes (figures \ref{fig:S1Pi0ULargeDe} and \ref{fig:S1VolLargeDe}).
The increase in the stresslet when viewed through the second decomposition is explained by the increase in the undisturbed polymer stress with $De$ (figure \ref{fig:Stresses_Undisturbed}).

However, the PIPS becomes increasingly negative with increasing $De$, after $De\gtrsim0.5$ (figures \ref{fig:Total_extra_polymer_medium_de} and \ref{fig:Total_extra_polymer_large_de}), and it has the dominant impact on the suspension rheology at large $De$.  For $L\gtrsim 50$, the overall particle-polymer interaction stress goes to zero at $De\approx0.55$ and becomes increasingly negative as $De$ is increased above this value (figure \ref{fig:Particle_extra_medium_de}).  At smaller $L$, the cross-over from positive to negative particle-polymer interaction stress occurs at a larger value of the Deborah number.  The extensional viscosity is half of the constant of proportionality between the deviatoric stress and the applied rate of strain tensor (equation \eqref{eq:ExtVisc2}). While at small $De$ adding particles leads to an increase in the extensional viscosity of the suspension due to their interaction with the polymers, at large $De$ particles and polymer interact to reduce the suspension's extensional viscosity. This is likely to have a large impact on the industrial processes mentioned in section \ref{sec:Introduction} and suggests the possibility of designing a fluid suspension with a desired range of extensional viscosities for a particular application.

The increasing value of the PIPS  with $De$ at small $De$ is due to the wake of extra stretched polymers along the extensional axis of the individual particles that becomes intensified as the $De$ increases (figure \ref{fig:polymerstretch_smallDe}). This occurs because the high extension rate regions due to the presence of particles stretch the undisturbed polymers near the particle surface. The polymers remain stretched in the wake downstream of the particle due to their finite relaxation time. For $De>0.5$, the undisturbed polymer stress is large, but the polymers contract in a region around the particle. This region is in the form of a wake of unstretched polymers and also covers the particle surface for moderate $De$, $0.5\lesssim De\lesssim 1.25$ (figure \ref{fig:polymerstretch_mediumDe}). Highly stretched undisturbed polymers beyond the coil-stretch transition collapse due to small velocity gradients in the low speed region around the incoming stagnation line on the particle. The polymers are then stretched as they convect downstream and ultimately recover to their large undisturbed stretch values. A deficit of polymer stress occurs in this recovery region. For even larger $De$, $De\gtrsim1.5$, the large negative contribution to the particle-induced polymer stress arises only from a very thin region of collapsed polyners near the particle surface (figure \ref{fig:polymerstretch_largeDe}). The thickness of this  region reduces with $De$ as the recovery is faster at larger extension rate ($De$). Using tools from non-linear dynamics, in section \ref{sec:FTLE}, to analyze kinematics of the velocity field we describe, in section \ref{sec:Fullpolymerconfiguration}, the qualitative analogy between the effect of a sphere in changing the steady-state polymer stretch and changing the transient stretch of a line of dye released in the flow, a finite time ago. Through this analogy, it is found that polymers with larger maximum extensibility have conformations resulting from longer Lagrangian stretching histories.

Anna and McKinley \cite{anna2008effect} showed that the FENE-P equations qualitatively capture the transient behavior of viscoelastic fluids with dilute polymer concentration (without particles) observed experimentally. The fluid in their experiments was pre-sheared orthogonal or parallel to the extensional axis of the subsequent uniaxial extensional flow. During the extensional flow of particle-filled polymeric fluids, a polymer molecule traveling around the particle experiences simple shear close the particle, followed by extension along the extensional axis. Thus, we expect our results to qualitatively match the extensional rheology experiments of dilute suspension of spherical particles in viscoelastic fluids with small polymer concentration.

Currently, there are no published experimental studies of the steady state extensional rheology of dilute particle suspension in dilute polymeric fluids with which we can systematically and quantitatively compare our theoretical and numerical predictions. There are, however, some preliminary results on transient rheology available from experiments described in ref. \cite{SHERE_Report,soulages2010extensional,hall2009preliminary,SHERE2_Report,jaishankar2012shear} conducted using a filament stretching extensional rheometer \cite{mckinley2002filament}. These experiments involved a polymeric fluid constituting of 0.025 wt\% of narrow polydispersity high molecular weight polystyrene (polymer) in oligomeric styrene oil (Newtonian solvent), yielding a Boger fluid \cite{rothstein1999extensional}. From the experimentally measured ratio of the polymer contribution to the zero-shear-rate viscosity to the solvent viscosity, the polymer concentration is 0.09. Thus, we can qualitatively apply our low $c$ theory. In the experiments performed at large $De$ from about 5 to 15, adding 3.5\% volume fraction of 6$\mu m$ diameter poly(methyl methacrylate) spherical particles leads to a reduction in extensional viscosity at large times. However, steady state is not achieved in any of the experiments and hence further experiments are required to fully confirm our findings. Nevertheless, this is in qualitative alignment with our large $De$ steady-state predictions.

As outlined in section \ref{sec:Introduction}, the FENE-P model over-predicts the polymer stretch during the transient phase of a uniaxial extensional flow when compared with Brownian simulations of the un-averaged FENE model. {Furthermore, by comparing Brownian simulations of a polymer molecule modelled as a freely jointed chain of beads and rods with Brownian simulations of a FENE dumbbell, Doyle et al. \cite{doyle1997dynamic} found that the two models predict similar transient behavior during the coil-stretch transition in a steady extensional flow. However, when the FENE dumbell or the bead and rod chain is fully extended after the coil-stretch transition, the FENE model predicts a higher extensional viscosity. Since the FENE model captures only the Brownian stress and the bead-rod chain experiences both viscous and Brownian stresses \cite{doyle1997dynamic} the discrepancy between the two models beyond the coil-stretch transition could be due to the limiting of the chain's stretch by the viscous stress. Neither the un-averaged FENE, the bead and rod, nor the FENE-P model yields precise quantitative predictions.} The qualitative features of our findings (made using the FENE-P model) hinge around the coil-stretch transition of polymers and hence we expect our findings to match those from future experiments at least qualitatively. The experimental evidence suggests that the FENE-P model strain hardens slower, due to modeling of the polymer as a single spring instead of multiple modes in the more realistic scenario \cite{anna2000interlaboratory}. {Slower strain hardening indicates slower transient stretching of the FENE-P polymer as compared to experiments. Ignoring the effect of viscous stresses on the limiting of the polymer stretch, the absence of viscous stresses in FENE-P polymers implies a smaller polymeric stress for a given polymer configuration in the FENE-P polymers than in experiments.}

Based on the preceding discussion of the relationship between FENE-P predictions and experimental or more complete physical modelling of polymers in transient extensional flow, we can anticipate a number of ways in which future experiments that probe the steady extensional rheology of particle-filled dilute polymer solutions may differ from the present predictions.
Compared to our findings, we expect polymers in an experiment to stretch along a streamline faster and have a larger stress for a given configuration/ polymer stretch. Therefore, we hypothesize that the wake of highly stretched polymers in the low $De\lessapprox0.52$ regime will be more intense. This would cause the positive particle-polymer interaction contribution at these $De$ to be larger in the future experiments. In the large $De\gtrapprox0.55$ regime, we expect the region of collapsed polymers around the particle to be thinner as the polymers in the experiments may stretch faster to recover the stretch deficit relative to the undisturbed state. However, due to the larger stress for a given polymer stretch we expect the undisturbed stress in the experiments at large $De$ to be larger. Thus making the stress deficit due to the presence of particles more intense in the collapsed region. Therefore, if the viscous stress of the polymer in an experiment is negligible, the negative particle-polymer interaction stress at large $De\gtrapprox0.55$ would be of a smaller magnitude, due to a thinner collapsed region, than we report here. However, if the rate of polymer stretch in the experiments is comparable to the FENE-P model but the viscous stresses in the polymer are significant we expect the experimental interaction stress at large $De$ to be even more negative.

{Although the present study considered a dilute suspension in which particle-particle interactions were neglected, it is of interest to speculate as to the possible importance of particle interactions.  It is useful to note that for both extensional flow and simple shear flow, which was studied by Koch et al. \cite{koch2016stress}, both the PIPS and the interaction stresslet contribute to the stress at small $De$ while the PIPS is the dominant form of the interaction stress at large $De$. It is reasonable to expect particle-particle interaction to influence the wake of highly stretched polymers in the low $De$ regime thereby modifying  the particle-polymer interaction stress. However, the particle's influence on the polymer stretch is confined to a small region near the particle in the large $De$ regime, so that particle interactions may not influence the PIPS and the polymer-particle interaction stress at moderate particle volume fractions.    A similar hypothesis was made and subsequently justified by Jain \& Shaqfeh \cite{jain2021transient} for shear rheology of a suspension of spheres in a viscoelastic fluid by comparing numerical results from immersed boundary simulations with multiple particles with the isolated particle simulations. They observed that particle-particle interaction only affects the per-particle interaction stresslet and not the per-particle PIPS, which, as observed by Jain \& Shaqfeh \cite{jain2021transient} and also Koch et al. \cite{koch2016stress} is the dominant component of the interaction stress in shear rheology at large $De$.}

A possible extension of this work is to consider a temporally evolving flow such as startup uniaxial extensional flow (similar to \cite{jain2019extensional}, but over a wider parameter regime to obtain analytical rheological expressions such as equation \eqref{eq:far_field_fluid_Stresslet_largeDe} and \eqref{eq:pi_largeDe}).  It would also be of interest to consider more complex time-varying flows such as a period of simple shear followed by uniaxial extension (similar to \cite{anna2008effect}, but with particles). This will allow a more faithful modeling of scenarios in hydraulic fracturing or extrusion molding, where the particle-filled viscoelastic fluid undergoes a series of linear flows for a finite time. Also, replacing the spherical particles with ellipsoids could lead to interesting results. In a uniaxial extensional flow, a prolate ellipsoid aligns its major axis with the extensional direction \cite{leal1980particle}. This could have implications for the thickness and extent of the region of stretched or unstretched polymers in the fluid surrounding the particles, and hence the rheology of the suspension. Taking the limit of highly eccentric prolate ellipsoids, suspensions of fibers can be studied and slender body theory \cite{batchelor1970slender} is likely to be a useful technique to obtain analytical insights. Recent experiments \cite{han2015extensional} provide a source of validation for such a study. This direction of work may allow extensional rheology of a suspension in a viscoelastic fluid to be tuned by changing the particle shape, which is an attractive proposition for many applications.

\section*{Acknowledgement}
This work was supported by NSF grants 1803156 and 2206851.

\bibliographystyle{vancouver}
\bibliography{SteadyState}

\end{document}